\documentclass[12pt,prd, aps, showpacs, superscriptaddress,floatfix]{revtex4}
\usepackage{graphicx}
\usepackage{bm}
\usepackage{multirow}
\usepackage{axodraw}
\usepackage{epsfig}
\usepackage{psfrag}
\usepackage{soul}
\usepackage{multirow}
\usepackage{mathptmx}
\usepackage{mathrsfs}
\usepackage{amsmath, amssymb}
\usepackage{dcolumn}
\usepackage{epstopdf}
\usepackage{color}
\usepackage{hhline}
\usepackage{changebar}
\usepackage{subfigure}

\newcommand\riken{RIKEN-BNL Research Center, Brookhaven National
  Laboratory, Upton, NY 11973, USA}
\newcommand\bnl{Brookhaven National Laboratory, Upton, NY 11973, USA}
\newcommand\edinb{SUPA, School of Physics, The University of
  Edinburgh, Edinburgh EH9 3JZ, UK}
\newcommand\epcc{EPCC, School of Physics, The University of
  Edinburgh, Edinburgh EH9 3JZ, UK}
\newcommand\cu{Physics Department, Columbia University, New York,
  NY 10027, USA}

\newcommand\uconn{Physics Department, University of Connecticut,
  Storrs, CT 06269-3046, USA}
\newcommand\soton{School of Physics and Astronomy, University of
  Southampton,  Southampton SO17 1BJ, UK}
\newcommand\kek{Institute of Particle and Nuclear Studies,
  KEK, Tsukuba, Ibaraki, 305-0801, Japan}
\newcommand\sokendai{Department of Particle and Nuclear Physics, Sokendai Graduate University of Advanced Studies, Hayama, Kanagawa 240-0193, Japan}
\newcommand\regensburg{Institute for Theoretical Physics, University of Regensburg, 93040 Regensburg, Germany}
\newcommand\yale{Department of Physics, Yale University, Sloane Physics Laboratory, New Haven, CT 06511, USA}
\newcommand\virginia{Dept. of Physics,
University of Virginia, 382 McCormick Rd. Charlottesville,
VA 22904-4714}
\newcommand{\julich}{J\"ulich Supercomputing Centre, Institute for Advanced Simulation, Forschungszentrum J\"ulich GmbH, 52425 J\"ulich, Germany}

\newcounter{Outline}
\newcounter{Introduction}
\newcounter{ChPT}
\newcounter{SimDetail}
\newcounter{LatticePS}
\newcounter{Baryons}
\newcounter{SuTwo}
\newcounter{Bk}
\newcounter{Vector}
\newcounter{CombinedChiralFits}
\newcounter{Conclusions}
\newcounter{Acknowledgments}
\newcounter{Appendix}
\newcounter{Tables}
\newcounter{Figures}
\newcommand{\ba}{\begin{eqnarray}}
\newcommand{\ea}{\end{eqnarray}}
\newcommand{\bas}{\begin{eqnarray*}}
\newcommand{\eas}{\end{eqnarray*}}
\newcommand{\be}{\begin{equation}}
\newcommand{\ee}{\end{equation}}
\newcommand{\bes}{\begin{equation*}}
\newcommand{\ees}{\end{equation*}}
\newcommand{\bi}{\begin{itemize}}
\newcommand{\ei}{\end{itemize}}

\newcommand{\bcentre}{\begin{center}}
\newcommand{\ecentre}{\end{center}}

\newcommand{\Tr}{\mathrm{Tr}}

\font\tenmsb=msbm10 scaled\magstep1
\font\sevenmsb=msbm7 scaled\magstep1
\font\fivemsb=msbm5 scaled\magstep1
\newfam\msbfam
\textfont\msbfam=\tenmsb
\scriptfont\msbfam=\sevenmsb
\scriptscriptfont\msbfam=\fivemsb

\newcommand{\MSb}{\overline{\rm MS}}

\newcommand{\mres}{m_{\rm res}}

\def\msbar{\overline{\mbox{\scriptsize MS}}}

\def\rmsub#1#2{#1_{\mbox{\tiny #2}}}	    

\def\ln{\mathop{\rm ln}}		    
\def\det{\mathop{\rm det}}		    

\def\su3{SU(3)}

\def\tD{\mbox{D}\kern-0.65em\raise0.15ex\hbox{/}\kern0.15em} 
\def\sD{\mbox{\scriptsize D}\kern-0.5em\raise0.15ex\hbox{\scriptsize/}}
\def\ssD{\mbox{\tiny D}\kern-0.42em\raise0.15ex\hbox{\tiny/}}

\def\dslash{\hbox{\(\partial\)}\kern-0.5em\raise0.15ex\hbox{/}} 

\def\su3{SU(3)}

\def\mom{p}

\def\mres{\rmsub{m}{res}}

\def\nicefrac#1#2{\leavevmode\kern.1em\raise.5ex\hbox{\the\scriptfont0 #1}\kern-
.1em/\kern-.15em\lower.25ex\hbox{\the\scriptfont0 #2}}

\newcommand{\ftwo}{f}
\newcommand{\ltwo}[1]{L_{#1}^{(2)}}

\newcommand{\SU}{\mathrm{SU}}

\begin{document}
\bibliographystyle{apsrev}

\title{Continuum Limit Physics from 2+1 Flavor Domain Wall QCD}

\author{Y.~Aoki}\affiliation{\riken}
\author{R.~Arthur}\affiliation{\edinb}
\author{T.~Blum}\affiliation{\uconn}\affiliation{\riken}
\author{P.A.~Boyle}\affiliation{\edinb}
\author{D.~Br\"ommel}\affiliation{\soton}\affiliation{\julich}
\author{N.H.~Christ}\affiliation{\cu}
\author{C.~Dawson}\affiliation{\virginia}
\author{J.M.~Flynn}\affiliation{\soton}
\author{T.~Izubuchi}\affiliation{\riken}\affiliation{\bnl}
\author{X-Y.~Jin}\affiliation{\cu}
\author{C.~Jung}\affiliation{\bnl}
\author{C.~Kelly}\affiliation{\edinb}
\author{M.~Li}\affiliation{\cu}
\author{A.~Lichtl}\affiliation{\riken}
\author{M.~Lightman}\affiliation{\cu}
\author{M.F.~Lin}\affiliation{\yale}
\author{R.D.~Mawhinney}\affiliation{\cu}
\author{C.M.~Maynard}\affiliation{\epcc}
\author{S.~Ohta}\affiliation{\kek}\affiliation{\sokendai}\affiliation{\riken}
\author{B.J.~Pendleton}\affiliation{\edinb}
\author{C.T.~Sachrajda}\affiliation{\soton}
\author{E.E.~Scholz}\affiliation{\regensburg}
\author{A.~Soni}\affiliation{\bnl}
\author{J.~Wennekers}\affiliation{\edinb}
\author{J.M.~Zanotti}\affiliation{\edinb}
\author{R.~Zhou}\affiliation{\uconn}
\collaboration{RBC and UKQCD Collaborations}
%
%
\mbox{}\hfill\noaffiliation{CU-TP-1195, Edinburgh 2010/13, KEK-TH-1380, RBRC-847, SHEP-1027}


\maketitle

\centerline{ABSTRACT}

We present physical results obtained from simulations using 2+1 flavors of domain wall quarks and
the Iwasaki gauge action at two values of the lattice spacing $a$, ($a^{-1}$=\,1.73\,(3)\,GeV and $a^{-1}$=\,2.28\,(3)\,GeV). On the coarser lattice, with $24^3\times 64\times 16$ points (where the 16 corresponds to $L_s$, the extent of the $5^{\textrm{th}}$ dimension inherent in the domain wall fermion (DWF) formulation of QCD), the analysis of ref.\,\cite{Allton:2008pn} is extended to approximately twice the number of configurations. The ensembles on the finer $32^3\times 64\times 16$ lattice are new. We explain in detail how we use lattice data obtained at several values of the lattice spacing and for a range of quark masses in combined continuum-chiral fits in order to obtain results in the continuum limit and at physical quark masses. We implement this procedure for our data at two lattice spacings and with unitary pion masses in the approximate range 290--420\,MeV (225--420\,MeV for partially quenched pions). We use the masses of the $\pi$ and $K$ mesons and the $\Omega$ baryon to determine the physical quark masses and the values of the lattice spacing. While our data in the mass ranges above are consistent with the predictions of next-to-leading order SU(2) chiral perturbation theory, they are also consistent with a simple analytic ansatz leading to an inherent uncertainty in how best to perform the chiral extrapolation that we are reluctant to reduce with model-dependent assumptions about higher order corrections. In some cases, particularly for $f_\pi$, the pion leptonic decay constant, the uncertainty in the chiral extrapolation dominates the systematic error. Our main results include $f_\pi=124(2)_{\rm stat}(5)_{\rm syst}$\,MeV, $f_K/f_\pi=1.204(7)(25)$ where $f_K$ is the kaon decay constant, $m_s^{\overline{\textrm{MS}}}(2\,\textrm{GeV})=
(96.2\pm 2.7)$\,MeV and $m_{ud}^{\overline{\textrm{MS}}}(2\,\textrm{GeV})=
(3.59\pm 0.21)$\,MeV\, ($m_s/m_{ud}=26.8\pm 1.4$) where $m_s$ and $m_{ud}$ are the mass of the strange-quark and the average of the up and down quark masses respectively, $[\Sigma^{\msbar}(2 {\rm GeV})]^{1/3} =
  256(6)\; {\rm MeV}$, where $\Sigma$ is the chiral condensate, the Sommer scale $r_0=0.487(9)$\,fm and $r_1=0.333(9)$\,fm.

\refstepcounter{section}
\setcounter{section}{0}


\newpage
\section{Introduction}
\label{sec:Introduction}

\ifnum\theIntroduction=1
%

For several years now, the RBC and UKQCD Collaborations have been undertaking a major programme of research in particle physics using lattice QCD with Domain Wall Fermions (DWF) and the Iwasaki gauge action. In the series of papers \cite{Antonio:2006px,Allton:2007hx, Allton:2008pn}, we studied general properties of ensembles with an inverse lattice spacing of $a^{-1}=1.73(3)\,$GeV (corresponding to $\beta=2.13$) and with unitary pion masses $m_\pi\ge 330$\,MeV (partially quenched $m_\pi\gtrsim 240$\,MeV). The number of points in these ensembles are $16^3\times 32\times 8$~\cite{Antonio:2006px}, $16^3\times 32\times 16$~\cite{Allton:2007hx} and $24^3\times 64\times 16$~\cite{Allton:2008pn}, where the fifth dimension is a feature of DWF and is not visible to low-energy physics which remains four-dimensional. We do not review the properties of DWF here, beyond underlining their physical chiral and flavor properties which we exploit in much of our wider scientific programme. We have used these ensembles to investigate a broad range of physics, including studies of the hadronic spectrum, mesonic decay constants and light-quark masses~\cite{Allton:2008pn}, the evaluation of the $B_K$ parameter of neutral-kaon mixing~\cite{Antonio:2007pb,Allton:2008pn}, the calculation of the form-factors of $K_{\ell 3}$ decays~\cite{Boyle:2007qe,Boyle:2010bh}, studies in nucleon structure~\cite{Yamazaki:2008py,Yamazaki:2009zq,Aoki:2010xg} and proton decay matrix elements~\cite{Aoki:2008ku} and very recently the first lattice study of the masses and mixing of the $\eta$ and $\eta^\prime$ mesons~\cite{Christ:2010dd} as well as a determination of the matrix elements relevant for neutral $B$-meson mixing in the static limit~\cite{Albertus:2010nm}. A key limiting factor in the precision of these results was that the simulations were performed at a single lattice spacing. In this paper we remove this limitation, by presenting results for the spectrum, decay constants and quark masses obtained with the same lattice action using ensembles generated on a $32^3\times 64\times 16$ lattice at a second value of the lattice spacing corresponding to $\beta=2.25$, for which we will see below that $a^{-1}=2.28(3)$\,GeV. Now that we have results for the same physical quantities with the same action at two values of the lattice spacing we are able to perform a continuum extrapolation and below we will present physical results in the continuum limit.

Since the most precise results at $\beta=2.13$ were obtained on the $24^3\times 64\times 16$~\cite{Allton:2008pn} lattices, as a shorthand throughout this paper we will refer to these lattices as the $24^3$ ensembles and label the new lattices at $\beta=2.25$ as the $32^3$ ensembles.

The new $32^3$ ensembles at $\beta=2.25$ will, of course, be widely used also in our studies of other physical quantities. In this first paper however, we discuss their properties in some detail (see Sec.\,\ref{sec:SimulationDetails}). In this section we also discuss \textit{reweighting} which allows us to eliminate one source of systematic uncertainty. While at present we cannot simulate with physical $u$ and $d$ quark masses, there is no reason, in principle, why we cannot simulate with the physical strange quark mass. The difficulty however, is that we don't know a priori what this mass is and so in practice the simulations are performed with a strange quark mass which is a little different from the physical one. As explained in Section\,\ref{subsec:reweighting}, the technique of reweighting allows us to correct a posteriori for the small difference in the simulated and physical strange quark masses. In Section\,\ref{sec:24cubed}, we present updated raw results for the pion and kaon masses and decay constants and the mass of the $\Omega$-baryon on the $24^3$ ensembles which have been extended beyond those discussed in ref.\cite{Allton:2008pn}. Section\,\ref{sec:32cubed} contains the corresponding results on the $32^3$ ensembles. In these two sections we also present the raw results for the masses of the nucleon and $\Delta$ baryons from the two ensembles, but in contrast to the mesonic quantities a description of their chiral behaviour and extrapolation to the continuum limit are postponed to a future paper.

The price we pay for using a formulation with good chiral and flavor properties is the presence of the fifth dimension and the corresponding increase in computational cost. The lightest unitary pion which we have been able to afford to simulate has a mass of 290\,MeV and so, in addition to the continuum extrapolation we need to perform the chiral extrapolation in the quark masses. In Sec.\,\ref{sec:CombinedChiralFits} we present a detailed explanation of how we combine the chiral and continuum extrapolations in an attempt to optimize the precision of the results, exploiting the Symanzik effective theory approach as well as chiral perturbation theory and other ansatze for the mass dependence of physical quantities. Having explained the procedure, we then proceed in Section\,\ref{subsec:results} to discuss the results, to determine the physical bare masses and lattice spacings as well as to make predictions for the pion and kaon decay constants. In particular we find that the ratio of kaon and pion decay constants~\endnote{In this Introduction we combine the statistical and systematic errors in the results. The separate errors are presented in the following sections.}
\begin{equation}\label{eq:fkfpiintro}
\frac{f_K}{f_\pi}=1.204\pm 0.026\,,
\end{equation}
where the error is largely due to the uncertainty in the chiral behaviour of $f_\pi$ as explained in Sec.\,\ref{subsubsec:chiralfpi}. From the chiral behaviour of the masses and decay constants we determine the corresponding Low Energy Constants (LECs) of SU(2) Chiral Perturbation Theory (ChPT).

Among the most important results of this paper are those for the average $u$ and $d$ quark mass and for the strange quark mass which are obtained in Sec.\ref{sec:quark_masses}:
\begin{equation}\label{eq:massesintro}
m_{ud}^{\msbar}(2 {\rm GeV})=(3.59\pm 0.21)\,\textrm{MeV}\quad\textrm{and}\quad
m_{s}^{\msbar}(2 {\rm GeV})=(96.2\pm 2.7)\,\textrm{MeV}\,.
\end{equation}
The masses are presented in the $\overline{\textrm{MS}}$ scheme at a renormalization scale of 2\,GeV, after the renormalization to symmetric momentum schemes has been performed non-perturbatively\,\cite{Aoki:2007xm,Sturm:2009kb} and the conversion to the $\overline{\textrm{MS}}$ scheme has been done using very recent two-loop results~\cite{Gorbahn:2010bf,Almeida:2010ns}.

Section\,\ref{sec:Topology} contains a discussion of the topological charge and susceptibility of both the $24^3$ and $32^3$ ensembles and in Sec.\ref{sec:Conclusions} we summarise our main results and present our conclusions. There are three appendices. Appendix~\ref{sec:appendix:separate_fits} contains the chiral extrapolations performed separately on the $24^3$ and $32^3$ ensembles. This is in contrast with the procedure described in Section\,\ref{subsec:results} in which the chiral and continuum extrapolations were performed simultaneously with common fit parameters at the two spacings. Appendix~\ref{sec:appendix:za} contains a detailed analysis of a subtle issue, the normalization of the partially conserved axial current. For domain wall fermions this is expected to deviate from the conventionally normalized continuum current by terms of order $am_{\textrm{res}}$, where $a$ is the lattice spacing and $m_{\textrm{res}}$ is the residual mass\,\cite{Sharpe:2007yd,Allton:2008pn}. Current simulations are now becoming sufficiently precise that these effects need to be understood and quantified and the method proposed in appendix~\ref{sec:appendix:za}, in which the $O(am_{\rm res})$ effects are absent, is implemented in the numerical analyses throughout the paper. Finally Appendix~\ref{sec:reweighting_appendix} contains a discussion of the expected statistical errors when reweighting is performed on Monte Carlo data to obtain results with a different action from that used to generate the data.

We end the Introduction with an explanation of our notation for quark masses~\cite{Allton:2008pn}. When discussing unitary computations, with the valence and sea quarks degenerate, we call the bare light ($u$ or $d$) quark mass $m_l$ and the bare heavy (strange) quark mass $m_h$. $m_{ud}$ and $m_s$ refer to the physical values of these masses (we work in the isospin limit so that the up and down quarks are degenerate). For the partially quenched computations we retain the notation $m_l$ and $m_h$ for the sea-quark masses, but use $m_x$ and $m_y$ for the valence quarks. A tilde over the mass indicates that the \textit{residual mass} has been added, $\widetilde{m}_q=m_q+m_{\rm res}$; it is $\widetilde{m}$ which is multiplicatively renormalizable.

\fi

\section{Simulation Details and Ensemble Properties}
\label{sec:SimulationDetails}

As described in Ref.~\cite{Allton:2007hx,Allton:2008pn,Antonio:2007tr}, we generate
ensembles using a combination of the DWF formulation of Shamir~\cite{Shamir:1993zy}
and the Iwasaki gauge action~\cite{Iwasaki:1983ck}.  For the fermionic action we
use a value of 1.8 for the ``domain wall height'' $M_5$ and an extension of the $5^{\textrm{th}}$ dimension of
$L_s=16$.  In addition to the new
ensembles generated on a $32^3\times 64$ lattice volume and a gauge coupling
$\beta=2.25$, we have also significantly extended the $24^3\times 64$, $\beta=2.13$
ensembles generated in our previous study~\cite{Allton:2008pn}.  As indicated in
Tab.~\ref{table:para} we have extended the $m_l=0.005$, $24^3\times 64$ ensemble
from 4460 to 8980 MD units while the $m_l=0.01$ ensemble has been extended from
5020 to 8540 MD units.  The three $32^3\times 64$ ensembles that are first reported
here are also shown in Tab.~\ref{table:para} and those with light quark masses of
0.004, 0.006 and 0.008 contain 6856, 7650 and 5930 MD units respectively.

\subsection{Ensemble Generation}

For the generation of both the $24^3\times 64$ and $32^3\times 64$ ensembles,
we employ the ``RHMC II'' algorithm described in Ref.~\cite{Allton:2008pn}. More
specifically, the simulation of two light quarks and one strange quark is carried out
using a product of three separate strange quark determinants each evaluated
using the rational approximation.  The 2 flavors of light quarks are
preconditioned by the strange quark determinant~\cite{Hasenbusch:2002ai}.
While the preconditioning mass does not have to be the same as the strange-quark
mass, we found that the strange-quark mass is close to being optimal in DWF
simulations in tests on smaller volumes.

Using the notation ${\cal D}(m_l) =  D^\dagger_{DWF}(M_5,m_l) D_{DWF}(M_5,m_l)$,
the fermion determinant including the contribution from the Pauli-Villars fields
and evaluated on a fixed gauge configuration can be written as
\begin{eqnarray}
\mbox{det}
 \left[ \frac{ {\cal D}(m_s)^{1/2}{\cal D}(m_l)} {{\cal D}(1)^{3/2}} \right] && \nonumber \\
&& \hskip -1.5in =  \mbox{det} \left[\frac{ {\cal D}(m_s)} {{\cal D}(1)} \right]^{3/2} \cdot
  \mbox{det} \left[\frac{ {\cal D}(m_l)} {{\cal D}(m_s)} \right] \\
&& \hskip -1.5in = \mbox{det} \left[ {\cal R}_{\frac12} \left(\frac{ {\cal D}(m_s)} {{\cal D}(1)}\right) \right] \cdot
\mbox{det}  \left[ {\cal R}_{\frac12}  \left( \frac{  {\cal D}(m_s)} {{\cal
D}(1)}\right)\right] \cdot
\mbox{det} \left[ {\cal R}_{\frac12} \left(\frac{ {\cal D}(m_s)} {{\cal
D}(1)}\right) \right] \cdot
 \mbox{det} \left[\frac{ {\cal D}(m_l)} {{\cal D}(m_s)} \right]\,.
\label{eq:rational_approx}
\end{eqnarray}
In the third line we explicitly show how this ratio of determinants is
implemented using the rational approximation.  Here ${\cal R}_{a}(x)$
denotes $x^a$ evaluated using the rational approximation and each determinant
is evaluated using a separate set of pseudofermion fields.  An
Omelyan integrator \cite{Takaishi:2005tz} with the Omelyan parameter
$\lambda=0.22$ was used in each part of evolution.

Given the disparate contributions to the molecular dynamics force coming from
the gauge action and the different factors in Eq.~(\ref{eq:rational_approx})
we follow the strategy of Ref.~\cite{Urbach:2005ji} and increase performance
by simulating these different contributions with different molecular dynamics
time step granularities.  In particular, the suppression of the force from the
light quark determinant that results from the Hasenbusch preconditioning
allows us to evaluate the computationally expensive force from the light
quark using the largest time step among the different terms, decreasing
the computational cost significantly.  As a result, we divide our
simulation in such a way that $\Delta t_{\rm light} : \Delta t_{\rm heavy} :
\Delta t_{\rm gauge} = 1: 1: 1/6$ which gave a good performance, measured in flops
per accepted trajectory in tuning runs performed separately.  (Note, the nature
of the Omelyan integrator makes $\Delta t_{heavy}$ effectively half of
$\Delta t_{light}$.)  This ratio of time steps was used for all the ensembles
studied here.  However $\Delta t_{light}$ was varied from ensemble to ensemble
to reach an approximate acceptance of 70\%.   The precise numbers that were
used are listed in Tab.~\ref{table:para}.

In addition, we chose to simulate with a trajectory length $\tau = 2$ for the $32^3$ ensembles, twice that used for the $24^3$ ensembles. While a longer trajectory length may be expected to reduce the autocorrelation between configurations, the time for a trajectory scales very nearly linearly in the trajectory length.  In comparisons between $\tau = 1$ and $\tau = 2$ trajectory lengths we were not able to recognize any statistically significant reduction in autocorrelations, especially in those for the topological charge, in terms of wall-clock time used to generate the configurations.

\begin{table}[hbt]
	\begin{tabular}{c|c|c|c|c|c|c|c|c}
		\hline
  		$m_sa$ & $m_la$ & $\tilde{m_s}/\tilde{m_l}$  & $\Delta
t_{light}$ & $\tau$(Ref.\cite{Allton:2008pn}) &$\tau$(MD) & Acceptance & $\langle P\rangle$& $\langle \bar{\psi} \psi (m_l)\rangle$ \\
		\hline
		\hline
		\multicolumn{9}{c}{$ V/a=24^3 \times 64$, $L_s=  16,\ \beta=2.13,
a^{-1} = 1.73(3)$ GeV, $m_{res} a = 0.003152(43), \tau / \mbox{traj} = 1$ } \\
		\hline
		\multirow{2}{*}{0.04} & 0.005 & 5.3 & 1/6 & 4460 & 8980 & 73\% & 0.588053(4) & 0.001224(2) \\
		& 0.01  & 3.3 & 1/5 &5020 & 8540 & 70\% & 0.588009(5) & 0.001738(2) \\
		\hline
		\multicolumn{9}{c}{$V/a = 32^3 \times 64$, $L_s= 16,\ \beta=2.25,
a^{-1} = 2.28(3)$ GeV, $m_{res} a = 0.0006664(76), \tau  / \mbox{traj} = 2$} \\
		\hline
		\multirow{3}{*}{0.03} & 0.004 & 6.6
& 1/8 & --- &6856 & 72\% & 0.615587(3) & 0.000673(1)\\
		 & 0.006 & 4.6
& 1/8 & --- &7650& 76\% & 0.615585(3)& 0.000872(1)\\
		 & 0.008 & 3.5
& 1/7 & --- &5930& 73\% & 0.615571(4) & 0.001066(1) \\
\hline
\end{tabular}
\caption{Simulation parameters as well as the average acceptance, plaquette 
($\langle P \rangle$) and value for the light-quark chiral condensate 
($\langle \bar{\psi} \psi (m_l)\rangle $) for the ensembles studied in this 
paper.  The fifth column shows the number of time units in the ensembles 
that were included from Ref.~\cite{Allton:2008pn}. 
The residual masses given explicitly and those appearing in
the ratio $\widetilde{m}_l/\widetilde{m}_s$ are taken from
Table \ref{tab-mreschiral} appearing in Section III below.
\label{table:para}}
\end{table}

A final optimization was used for the simulations run on the IBM BG/P machines
at the Argonne Leadership Computing Facility(ALCF).  Instead of using double
precision throughout, the BAGEL-generated assembly routines~\cite{Boyle:2009bagel}
keep the spin-projected spinors in single precision in the conjugate gradient(CG)
inverters during the molecular dynamics evolution to decrease the amount of
communication needed per CG iteration.  (Full precision is used in the
accept-reject step.)  While this kind of improvement is expected to make the
molecular dynamics integrator unstable for sufficiently large volumes, the effect on
the acceptance turned out to be minimal for all the ensembles presented in this
paper while improving the performance of the CG by up to 20\% compared to a
full double precision CG with the same local volume.

\subsection{Ensemble properties}

\begin{figure}
\hspace{-0.05\textwidth}
\begin{minipage}{0.5\textwidth}
\begin{center}
\includegraphics[angle=270,width=3.3in]{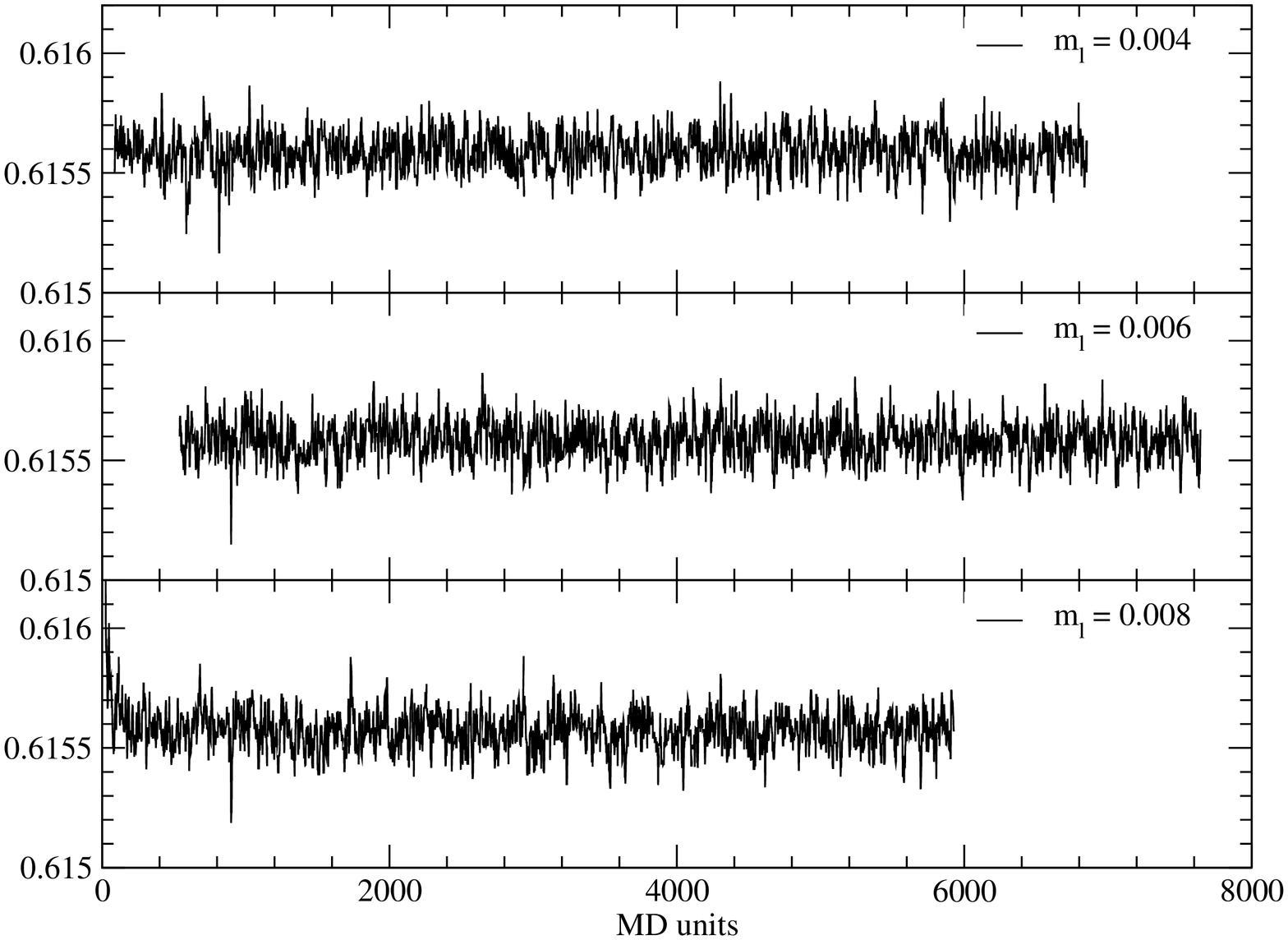}
\end{center}
\end{minipage}
\begin{minipage}{0.5\textwidth}
\begin{center}
\includegraphics[angle=270,width=3.3in]{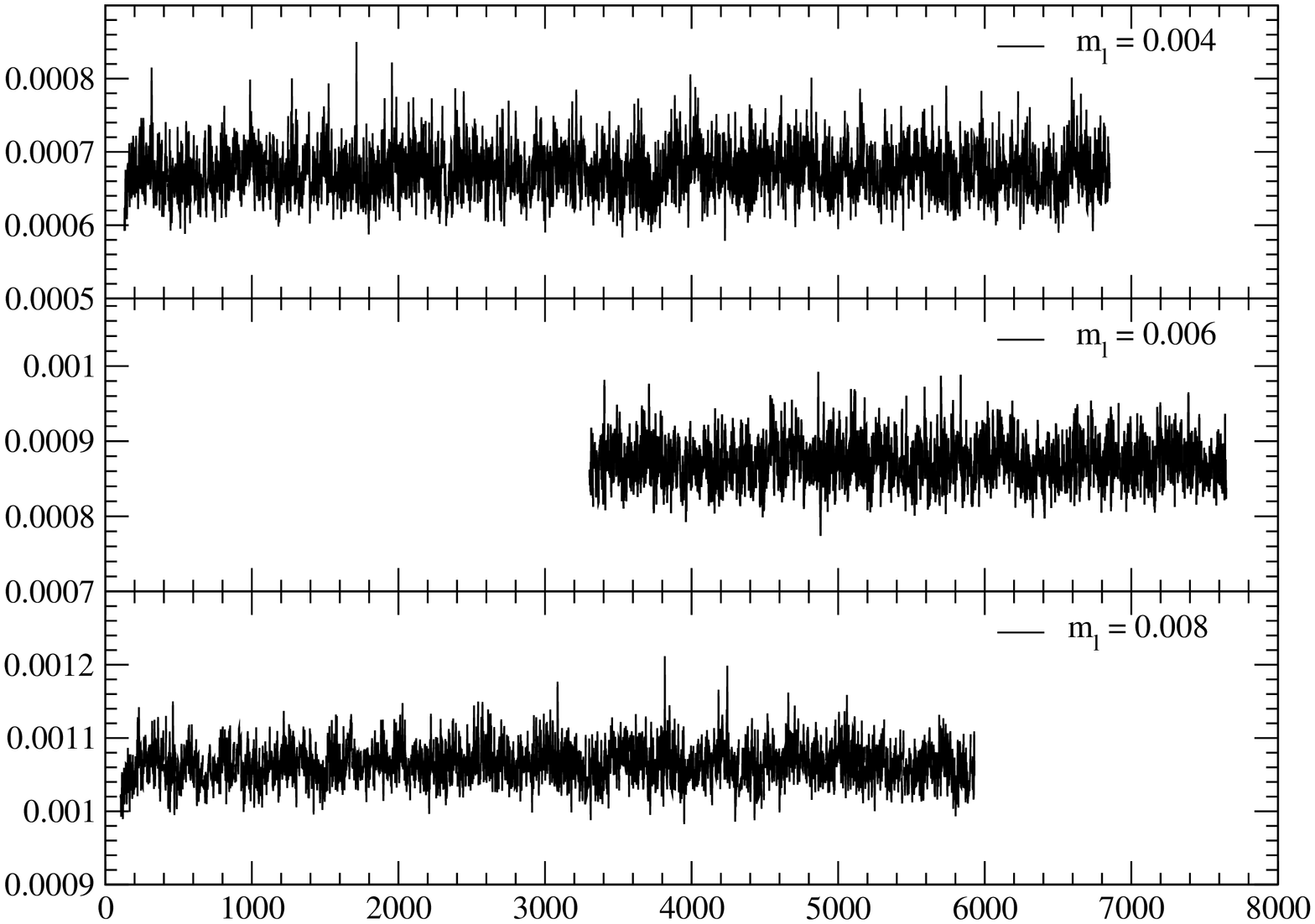}
\end{center}
\end{minipage}
\caption{Evolution of the average plaquette (left panel) and the chiral condensate (right panel) for the
$\beta=2.25,\ 32^3\times64$,\ $L_s=16$ ensembles. The chiral condensate is normalized
such that $\langle \bar\psi \psi \rangle \sim$ 1/m in the heavy quark limit.}
\label{fig:evol}
\end{figure}

In Fig.~\ref{fig:evol} we show the evolution of the plaquette and the chiral
condensate for the $32^3$ ensembles.  Both quantities suggest that 500 MD
units is enough for the thermalization of each of the $32^3$ ensembles.  We
have thus begun measurements at 1000 MD units for $m_l=0.006$ (except for the measurements of the
chiral condensate which started after 3304 MD units)
and  520 MD units for the other $32^3$ ensembles.
(The starting points for measurements on the three $24^3\times 64$ ensembles
are given in Tab.~I of Ref.~\cite{Allton:2008pn}.)

\begin{figure}
\includegraphics[angle=270,width=0.65\textwidth]{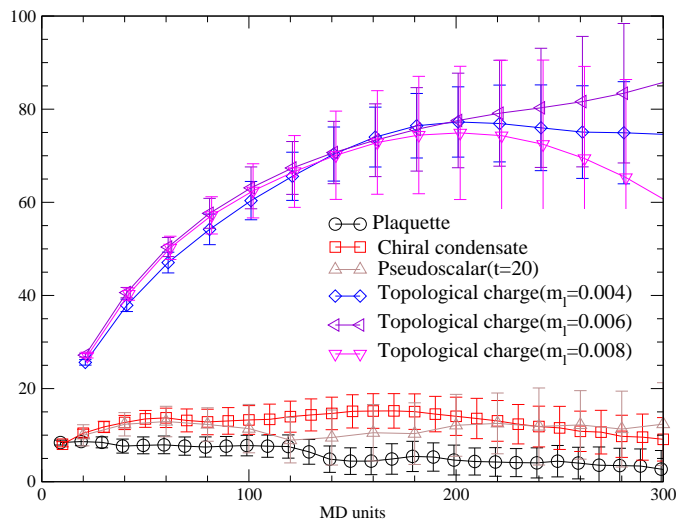}
\caption{The integrated autocorrelation time is shown for the average
plaquette, chiral condensate $\langle \bar\psi \psi\rangle$, pseudoscalar
propagator at time separation 20 from a Gaussian source and point sink, all
computed from the $32^3,\ m_l= 0.004$ ensemble and the global topological
charge for all three $32^3$ ensembles.  The chiral condensate and plaquette
are measured every two MD units and the averages within sequential blocks
of 10 MD units have been analyzed. The topological charge is measured every 4 MD units and
the averages within sequential blocks of 20 MD units have been analyzed. All other quantities
were measured every 20 MD units and no averaging has been performed.  Further
discussion of the topological charge is given in section \ref{sec:Topology}.}
\label{fig:ic_jk}
\end{figure}

Figure~\ref{fig:ic_jk} shows the integrated autocorrelation time for various
quantities measured on the $32^3$ ensembles.  As can be seen the plaquette,
chiral condensate and even the light pion propagator for a separation of
20 time units show a short autocorrelation time of 5-10 MD units.  However,
the measured autocorrelation times for the topological charge are much
larger, on the order of 80 MD units.  In fact, as is discussed in
Section~\ref{sec:Topology}, the evolutions shown in Fig.~\ref{fig:top_history}
suggest even longer autocorrelation times implying that the autocorrelation
times shown in Fig.~\ref{fig:ic_jk} may be underestimated because of
insufficient statistics.

In Section~\ref{sec:Topology} this issue of the autocorrelation time for the topological charge
is discussed in greater detail
and the $\beta=2.13$ and 2.25 evolutions are compared.  The $32^3$, $\beta=2.25$
ensembles (with finer lattice spacing) are shown to evolve topology more
slowly.  This suggests that the change from the DBW2 gauge action used in
earlier 2-flavor work~\cite{Aoki:2004ht} to the Iwasaki gauge action used
here may have been a wise one.  While the DBW2 gauge action gives smaller
residual DWF chiral symmetry breaking, it does this by suppressing the
tunneling which changes topological charge.  Thus, the use of the DBW2 gauge
action may have resulted in a topological charge evolution for our current
finest lattice spacings that would have been unacceptably slow.

\subsection{Fitting procedure}

In the analysis described in this paper it is important to take into
account the fact that the various quantities computed on a single gauge
configuration may be correlated. To do this we apply the
jackknife technique to simple uncorrelated fits. While there is no proof, or
even expectation, that this is an optimal procedure, the jackknife will
provide a good estimate of the error except in the unlikely event of large
deviations of our result from a normal distribution. While we could attempt to
perform a ``text-book'' correlated fit (again, using a jackknife procedure),
this would not be sensible: such fits assume that the data should exactly
follow the functional form used in the fit. In the case of a fit to chiral
perturbation theory or a simpler analytic ansatz for the quark-mass dependence of physical quantities we know that this is not the case.
While this complaint applies to both correlated and uncorrelated fits, for
the highly correlated lattice data with which we are dealing, small deviations
(which in this procedure are assumed to be statistical, but in our case are
likely to be systematic) are penalized by many orders of magnitude
more for the correlated than uncorrelated fits.  Nevertheless, we have
performed correlated fits, where the correlation matrix is obtained by taking
increasing numbers of the leading eigenvectors.  Within our limited ability
to estimate the correlation matrix, we find no significant difference in the
results and errors with those obtained using uncorrelated fits. Therefore,
in this paper (as was also the case in Ref.~\cite{Allton:2008pn}) we present
our main results from the uncorrelated fits, but with a full jackknife
procedure for estimating the errors. However, it must be borne in mind that for such uncorrelated fits the
resulting $\chi^2$ may not be a reliable indicator of goodness of fit.
Therefore, we present a sample set of our fits graphically.

\def\calO{{\cal O}}
\newcommand\vev[1]{\langle #1\rangle}
\newcommand\vevxi[1]{\langle\langle #1\rangle\rangle}
\def\mhsim{m_h^\text{(sim)}}
\def\mhtarget{m_h}
\newcommand{\Nconf}{N_{\textrm{conf}}}

\subsection{Reweighting in the mass of the sea strange-quark}
\label{subsec:reweighting}

The sea strange quark mass value used in our ensemble generation, $\mhsim$,
differs from the one in nature, which we determine only \emph{after}
performing our final analysis.  In this subsection, we describe the reweighting
method used to correct this strange quark mass from $\mhsim$ to
the target mass $\mhtarget$.  Various target heavy quark masses are determined in
Section~\ref{sec:CombinedChiralFits} through
interpolation/extrapolation to yield meson masses which match either
unphysical values present in a different ensemble or which reproduce
those from experiment.  Recently, several
large-scale QCD simulations have been reported using a
reweighting technique~\cite{Luscher:2008tw, Hasenfratz:2008fg, Jung:2010jt}.
The various uses of this method include obtaining sea quark mass derivatives
in Ref.~\cite{Ohki:2009mt}, tuning the light and strange quark masses in
Ref.~\cite{Aoki:2009ix}, tuning the strange and charm quark masses in
Ref.~\cite{Baron:2008xa} and going to larger $L_s$ for the DWF action in
Ref.~\cite{Ishikawa:2010tq}.

An observable, such as the meson propagator, at the target strange sea quark
mass $\mhtarget$ is obtained by measuring that observable on the ensemble
generated using $\mhsim$, multiplied by the reweighting factor $w$:
\begin{align}
\vev{\calO}_{\mhtarget} =
{ \vev{\calO w}_{\mhsim} \over \vev{w}_{\mhsim}}~~.
\label{eq:rew observable}
\end{align}
Here the reweighting factor $w[U_\mu]$ for a particular ensemble of gauge links
$U_\mu$ is the ratio of the square root of the two-flavor Dirac
determinant evaluated at the mass $\mhtarget$
divided by that same rooted determinant evaluated at $\mhsim$,
\begin{align}
w[U_\mu] =  { \det {\cal D}(\mhtarget)^{1/2} \over \det {\cal D} (\mhsim)^{1/2}}~~.
\label{eq:det ratio}
\end{align}
This factor must be calculated for each configuration
on which measurements will be performed in the ensemble generated using
the sea strange mass $\mhsim$.

Among the many possible ways of computing the
determinant ratio in Eq.~(\ref{eq:det ratio}), we have chosen to use
the Hermitian matrix
$\Omega(\mhtarget, \mhsim)$, whose determinant is $w[U_\mu]$,
\begin{align}
\Omega(\mhtarget, \mhsim) =
\left[ D(\mhsim)^\dagger\right]^{1/2}
\left[ D(\mhtarget)^\dagger\right]^{-1/2}
\left[D(\mhtarget)\right]^{-1/2}
\left[D(\mhsim)\right]^{1/2}~~.
\end{align}
The square root of these matrices is implemented
using the same rational polynomial approximation, ${{\cal R}_{1\over2}(x)}$,
and multi-shift conjugate gradient algorithm,
which are used in the ensemble generation.  The order of the matrix
products in $\Omega$ assures that in the limit of $\mhtarget\to \mhsim$, $\Omega$
goes to the unit matrix, so that the method described below for
evaluating $w$ has vanishing stochastic error in this
limit.

To obtain $w$ on each configuration, the determinant of $\Omega$ is
stochastically evaluated using a complex random Gaussian
vector $\xi$ of dimension $L_s\times 12$.
Each complex element is drawn from a random distribution
centered at zero with width $\sigma_\xi$ in both the real
and imaginary directions:
\begin{align}
w = \vevxi{e^{- \xi^\dagger [ \Omega -1/(2\sigma_\xi^2) ]\xi }}_\xi
\equiv
{\int {\cal D} \xi {\cal D} \xi^\dagger\,
e^{- \xi^\dagger [ \Omega -1/(2\sigma_\xi^2) ]\xi } e^{-\xi^\dagger\xi/(2\sigma_\xi^2)}
\over
\int {\cal D} \xi {\cal D} \xi^\dagger e^{-\xi^\dagger\xi/(2\sigma_\xi^2)}
}
 ~~.
\label{eq:eval det ratio}
\end{align}
We set $\sigma_\xi^2=1/2$ and sample using $N_\xi$
Gaussian vectors per configuration. For one sample, two multi-mass inversions,
one for $\mhtarget$ and another for $\mhsim$, are performed.

One needs to be careful in evaluating Eq. (\ref{eq:eval det ratio})
to avoid a large and difficult to estimate
statistical error.  When the eigenvalues of $\Omega$,
$\lambda_\Omega$, are far from $1/(2\sigma_\xi^2)$, the large shift
in the width of the Gaussian in the integrand will cause
poor sampling in this stochastic evaluation of $w$, as can be seen
if Eq.~\ref{eq:eval det ratio} is rewritten with $\Omega$ diagonal:
\begin{align}
w =
\prod_{\lambda_\Omega\in\text{spect}(\Omega)} \int d \xi_\lambda \xi_\lambda^\dagger
e^{-\xi_\lambda^\dagger [ \lambda_\Omega - 1/(2\sigma_\xi^2)]\xi_\lambda}
e^{-\xi_\lambda^\dagger \xi_\lambda/(2\sigma_\xi^2)}
/
\prod_{\lambda_\Omega\in\text{spect}(\Omega)} \int d \xi_\lambda \xi_\lambda^\dagger
e^{-\xi_\lambda^\dagger \xi_\lambda/(2\sigma_\xi^2)}
~~.
\label{eq:eval det ratio eigen}
\end{align}
The  first exponential function in the integrand
(\ref{eq:eval det ratio eigen}) will be a rapidly
decreasing function of $\xi^\dagger \xi$ when
$[\lambda_\Omega - 1/(2\sigma_\xi^2)]$ is large,
with most of the Gaussian samples
generated according to the second exponential function in Eq.~(\ref{eq:eval det ratio eigen})
falling in a region where the first factor is very small.
In this sense, Eq.~(\ref{eq:eval det ratio})
may provide a statistically noisy estimate of the ratio of the determinants
in Eq.~(\ref{eq:det ratio}).  The fluctuations in this estimate will
be rapidly reduced when $[\lambda_{\Omega}- 1/(2\sigma_\xi^2)]\to 0$ or,
for our choice of $\sigma_\xi$, when $\Omega$
becomes close to the unit matrix, $\Omega\to 1$.

To reduce the stochastic noise in our estimate, $\det \Omega$ is divided into
$N_{rw}$ factors \cite{Hasenfratz:2008fg}
\begin{align}
w = \det \Omega = \prod_{i=0}^{N_{rw}-1} \det \Omega_i=
\prod_{i=0}^{N_{rw}-1} \vevxi{e^{- \xi_i^\dagger [ \Omega_i -1/(2\sigma_\xi^2) ]\xi_i }}_{\xi_i}
~~.
\label{eq:eval det ratio step}
\end{align}
Each of $\Omega_i$ needs to be close to unit matrix while keeping
the determinant of the product the same as the original determinant.
Each factor $\det \Omega_i$ in the product,
is evaluated using Eq.~(\ref{eq:eval det ratio}) with
$N_\xi$ Gaussian vectors.  We note that all Gaussian vectors, $\xi_i$, must
be {\it statistically independent} otherwise there
will be unwanted correlation among contribution from the $N_{rw}$
steps.  A similar decomposition of the reweighting
factor is also possible by using the $n^{th}$ root of the
operators\cite{Ishikawa:2010tq}.

In this work, $\Omega_i$ is chosen by uniformly
dividing the interval $[\mhtarget,\mhsim]$ into smaller pieces:
\begin{align}
&\Omega_i = \Omega\left(\mhtarget^{(i+1)},\mhtarget^{(i)}\right)~~,\\
&\mhtarget^{(i)} = \mhsim + i {\mhtarget - \mhsim \over N_{rw}}~~,
(i=0,1,\cdots, N_{rw})~~.
\end{align}
In that way, reweighting factors for the
intermediate masses $m_h^{(i)}$ are also obtained, which will be
used in our analysis too.

\begin{figure}
\includegraphics[clip,width=0.6\textwidth]
{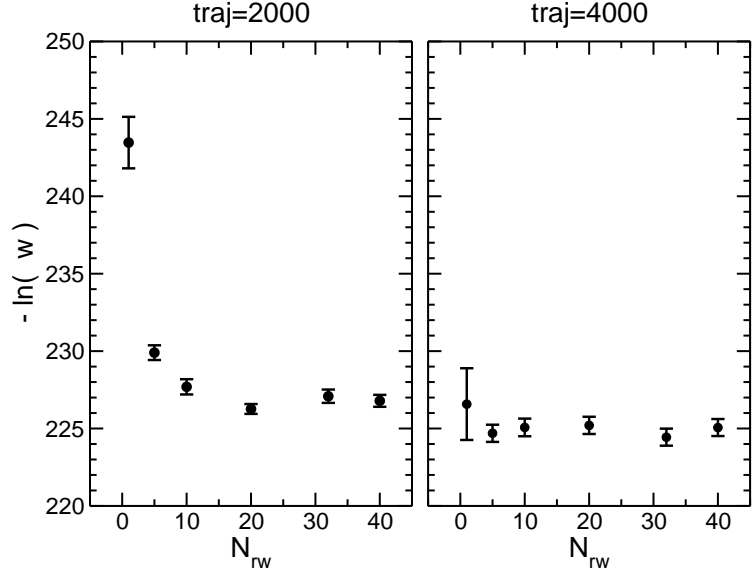}
\caption{
Logarithm of the reweighting factor, $-\ln(w)$, as a function of the
number of divisions in the strange quark mass, $N_{rw}$ on
the $\beta=2.13,\ 24^3\times 64,\ m_l=0.005$ lattices, the
2,000th trajectory on the left panel and the 4,000th trajectory
on the right panel.  The target and simulation quark masses are
$\mhtarget=0.035$ and $\mhsim=0.040$.  For $N_{rw}=1,\,5,\,10,\,20,\,32,\,40$, the
number of Gaussian samples per mass steps is set
to $N_\xi = 40,\,8,\,4,\,4,\,2,\,2$, respectively.  The error bars shown are
the standard deviations resulting from $N_{rw}\times N_\xi$ samples
for $\det \Omega_i$.  We interpret the inconsistency between the values
for $N_{rw} = 1$, 5 and 10 and those with larger $N_{rw}$ in the left-hand panel as
resulting from insufficient statistics leading to under-estimated errors for
these three cases where the stochastic sampling is very poor.}
\label{fig:Nrw check}
\end{figure}

For a given difference between the target
and the simulation masses, $\mhtarget-\mhsim$, $N_{rw}$ needs to be
sufficiently large that $\Omega_i$ is close to the
unit matrix, suppressing the
statistical noise in estimating
each of the determinants.  We have checked whether $N_{rw}$
is large enough in our calculation of the reweighting factor.
Figure~\ref{fig:Nrw check} shows the logarithm of the full
reweighting factor, $-\ln(w)$, as a function of the
number of divisions in strange quark mass, $N_{rw}$,
on the $\beta=2.13, 24^3\times 64, m_l=0.005$ lattices,
the 2,000th trajectory in the left panel and the
4,000th trajectory in the right panel.  The target and simulation
quark masses are $\mhtarget=0.035$ and $\mhsim=0.040$.

For $N_{rw}\leq 10$, the reweighting factor $w$ appears
inconsistent with the results obtained for larger $N_{rm}$ by
a large amount (note that $- \ln(w)$ is plotted) for the left
case (2,000th trajectory).  We believe this is caused by the
poor stochastic sampling in our method to compute $w$ when
$N_{rw}\leq 10$ and that for these cases the statistics are
insufficient to estimate the error accurately.

\begin{figure}
\includegraphics[clip,width=0.6\textwidth]
{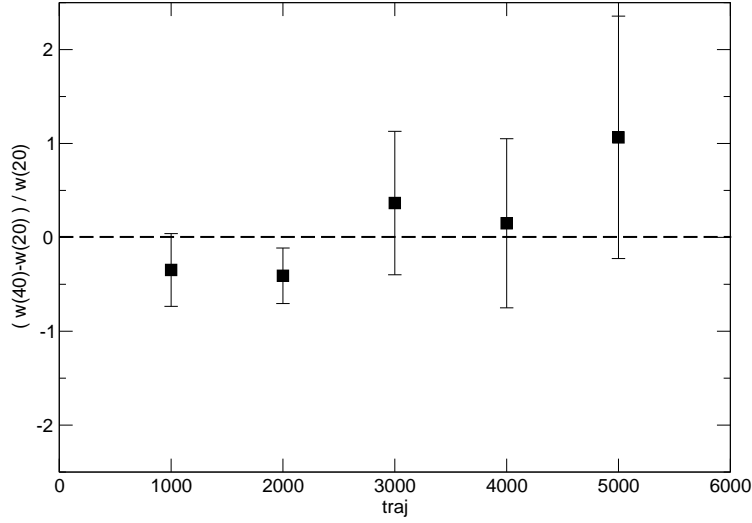}
\caption{
The relative differences between the reweighting
factors for $N_{rw}=20, N_\xi=4$ and $N_{rw}=40, N_\xi=2$
on five lattices.
The target and simulation quark masses are
$\mhtarget=0.035$ and $\mhsim=0.040$.
}
\label{fig:Nrw-20vs40-comparison}
\end{figure}

We also check the relative difference between the reweighting
factors for $N_{rw}=20$ and $N_{rw}=40$ in
Fig.~\ref{fig:Nrw-20vs40-comparison} for five lattices.  This plot
indicates that $N_{rw}=20$ is sufficient to
estimate the reweighting factor and its error
for changing from $\mhsim=0.040$ to $\mhtarget=0.035$ on this
ensemble.
We summarize the values of $N_{rw}$ and $N_{\xi}$ used in estimating
the reweighting factors for the sea strange quark mass in Tab.~\ref{tab:rewparam}.

\begin{table}
\begin{center}
\begin{tabular}{ccccc}
ensemble & $\mhsim$ & $\mhtarget$ & $N_{rw}$ & $N_{\xi}$\\
\hline
$32^3\times 64$ & 0.030 & 0.025 & 10 & 4\\
$24^3\times 64$ & 0.040 & 0.030 & 40 & 2\\
\hline
\end{tabular}
\end{center}
\caption{Parameters chosen for the sea strange quark mass
reweighting calculation.}
\label{tab:rewparam}
\end{table}

Is the $N_{rw}$ dependence, described above, all one needs to check
to assure the correctness of the reweighting procedure?
The answer is clearly no. So far, we have only established
that Eq.~(\ref{eq:eval det ratio step}) estimates $w$
to some degree of accuracy, on {\it each configuration} for large $N_{rw}$.
One needs further checks to see whether or not the reweighted observable in
Eq.~(\ref{eq:rew observable}) has
an accurately estimated statistical error.  A highly inaccurate estimate
of the statistical errors could easily result from a poor
overlap between the reweighted ensemble and the original ensemble generated
by the RHMC simulation. In addition, because the reweighted observable in
Eq.~(\ref{eq:rew observable}) is given by a ratio of averages it is a
biased estimator of the observable of interest.  In this circumstance, a
large statistical error, even if well determined, may lead to a systematic
error of order $1/\Nconf$ enhanced by this large statistical error.

\begin{figure}
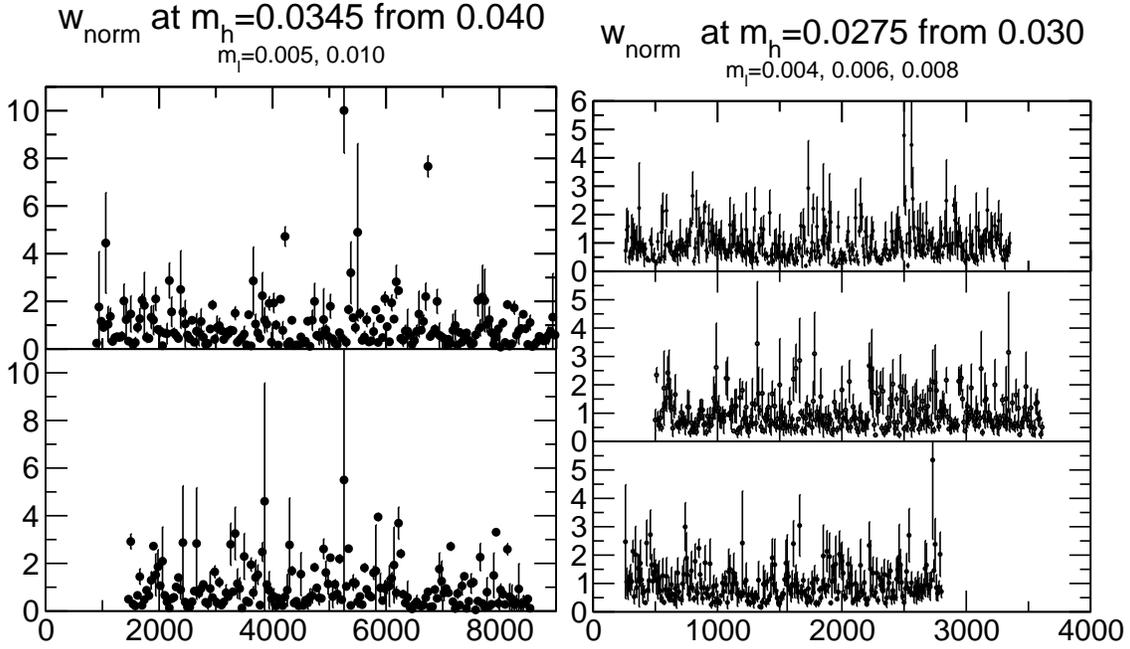

\begin{center}
\includegraphics[width=0.45\textwidth,clip]{fig/Fig5a_rw24-normalized.eps}
\includegraphics[width=0.45\textwidth,clip]{fig/Fig5b_rw32-normalized.eps}
\end{center}
\caption{The normalized reweighting factor
$\hat{w_i}$ as a function of trajectory number
$i$ for the $24^3\times 64,\ m_l=0.005,\, 0.010$ ensembles (left-hand plot)
and the $32^3\times 64,\ m_l=0.004,\, 0.006,\, 0.008$ ensembles (right-hand plot).
The sea quark masses $m_l$ are plotted in ascending order from
top to bottom. The target sea strange quark mass and
that of simulation are $\mhtarget=0.0345,\, \mhsim=0.040$
($\mhtarget=0.0275,\, \mhsim=0.030$) for the left-hand (right-hand) plot.
}
\label{fig:rew vs traj}
\end{figure}

We have attempted the following checks:
In Fig.~\ref{fig:rew vs traj}, $w$ is plotted as a function of trajectory.
If the fluctuation among different configurations is large,
Eq.~(\ref{eq:rew observable}) might be dominated
by a small number of measurements made on those configurations
with large $w$, and the measurement efficiency for the
reweighted observable would be very poor.
Using the reweighting factor,  $w_i$, obtained on the $i^{th}$ configuration,
the reweighted observable ${\cal O}$ can be written from
Eq.~(\ref{eq:rew observable}) as,
\begin{align}
&\vev{{\cal O}}_{m_s} = \sum_{i=1}^{\Nconf} O_i \hat{w_i}~~,
\label{eq:rew obs conf index}
\\
&\hat{w_i} = {w_i \over \sum_{i=1}^{\Nconf} w_i}~~.
\end{align}

Because the process of reweighing selectively samples
the original distribution, even with precisely determined
reweighting factors we should expect the effective number of
samples to be reduced and the statistical errors to increase.
In Appendix~\ref{sec:reweighting_appendix} this effect is
analyzed in the case that correlations between the data and
the reweighting factors can be neglected when estimating
these statistical errors, including the effects of
autocorrelations.  For the case of no autocorrelations,
we obtain the following expression for the effective number of
configurations after reweighting:
\begin{equation}
N_{\rm eff} = \frac{ \Bigl(\sum_{n=1}^{\Nconf} w_n\Bigr)^2}{\sum_{n=1}^{\Nconf} w_n^2}
\label{eq:Eff_def}.
\end{equation}
The quantity $N_{\rm eff}$ goes to $\Nconf$ if there is no
fluctuation in the $w_i$ while it goes to $1$ if the largest $w_i$
completely dominates the reweighted ensemble.  We summarize the
statistical features of the reweighting factors for each ensemble
in Tab.~\ref{tab:rewfactor}.  For completeness we also compare
the definition of $N_{\rm eff}$ given in Eq.~(\ref{eq:Eff_def}) with
the more pessimistic estimate used in Ref.~\cite{Ogawa:2005jn}:
\begin{align}
N_{\rm eff}^* = {\sum_{i=1}^{\Nconf} w_i \over \text{max}_j (w_j) }~~.
\label{eq:Eff_def_old}
\end{align}
As can be seen from Tab.~\ref{tab:rewfactor}, our choice gives a somewhat more
optimistic view of the effects of reweighting on the effective size
of our ensembles.

\begin{table}
\begin{center}
\begin{tabular}{cccccc}
\hline
ensemble                     & $\text{max}(w_i)$
                                    & $\text{min}(w_i)$
                                           & $N_{\rm Eff}$
                                                  & $N_{\rm Eff}^*$
                                                          & $\Nconf$\\
\hline
$24^3\times 64, m_l = 0.005$ & 10.0 & 0.078 & 90.3 & 20.3 & 203\\
$24^3\times 64, m_l = 0.010$ & 5.50 & 0.049 & 97.0 & 32.4 & 178\\
\hline
$32^3\times 64, m_l = 0.004$ & 4.77 & 0.17 & 228  & 63.9 & 305\\
$32^3\times 64, m_l = 0.006$ & 3.45 & 0.23 & 234  & 90.4 & 312\\
$32^3\times 64, m_l = 0.008$ & 5.36 & 0.16 & 183  & 47.0 & 252\\
\hline
\end{tabular}
\end{center}
\caption{The maximum and minimum reweighting factors, the effective number
of samples, $N_{\rm eff}$, according to the formula derived in this paper,
(Eq.~(\ref{eq:Eff_def})), the corresponding number, $N_{\rm eff}^*$ given
by the formula of Ref.~\cite{Ogawa:2005jn} (defined in Eq.~(\ref{eq:Eff_def_old}))
and the actual number of configurations $\Nconf$ in each ensemble.  The
target sea strange quark mass and that of the simulation are
$\mhtarget=0.0345,\ \mhsim=0.040$ ($\mhtarget=0.0275,\ \mhsim=0.030$)
for $24^3\times 64$ ($32^3\times 64$).}
\label{tab:rewfactor}
\end{table}

As the numbers in Tab.~\ref{tab:rewfactor} indicate, for our ensemble and
reweighting settings, the ensembles are not overwhelmed by a small number of
configurations.  

\begin{figure}
\begin{center}
\includegraphics[width=0.45\textwidth,clip]{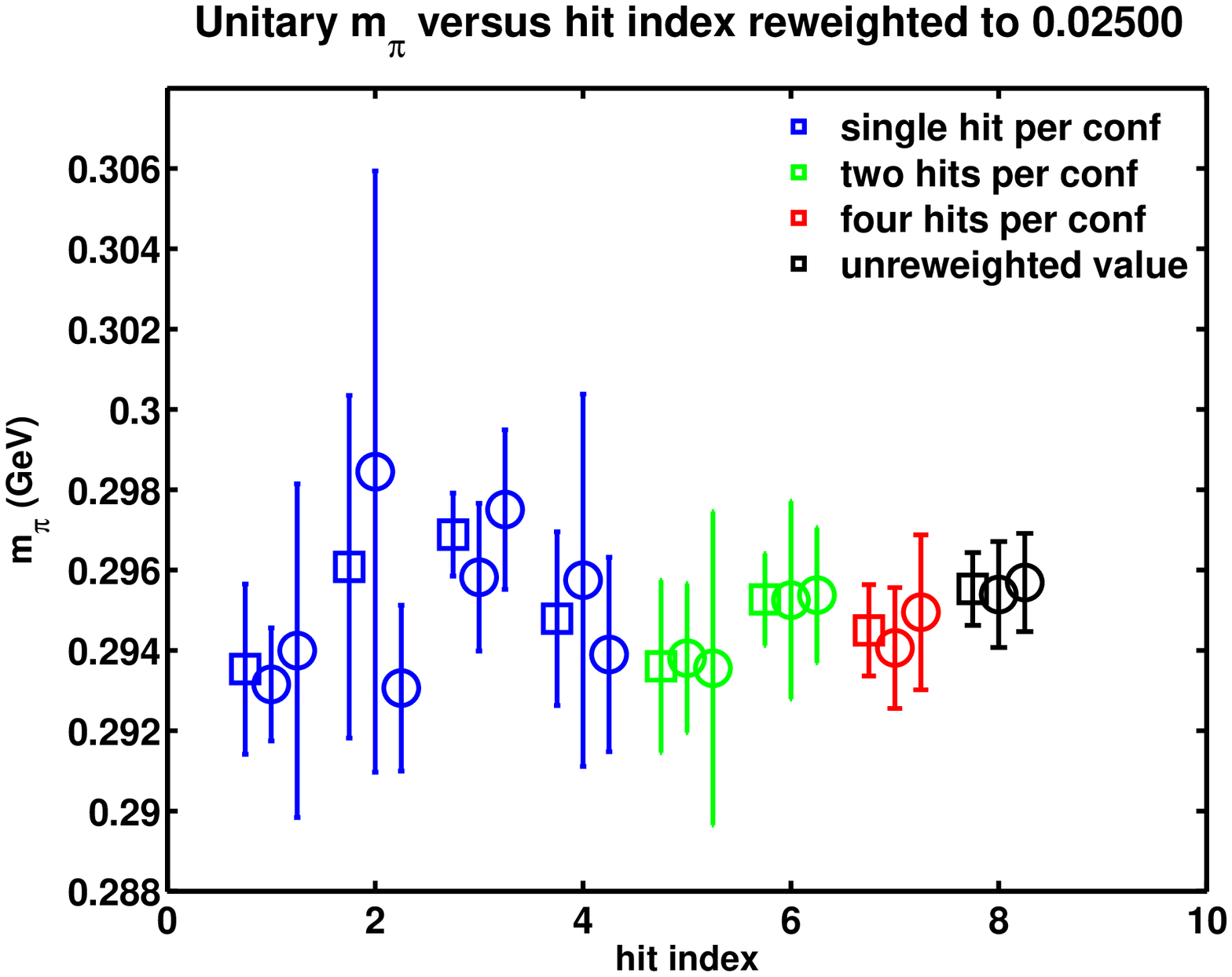}
\includegraphics[width=0.45\textwidth,clip]{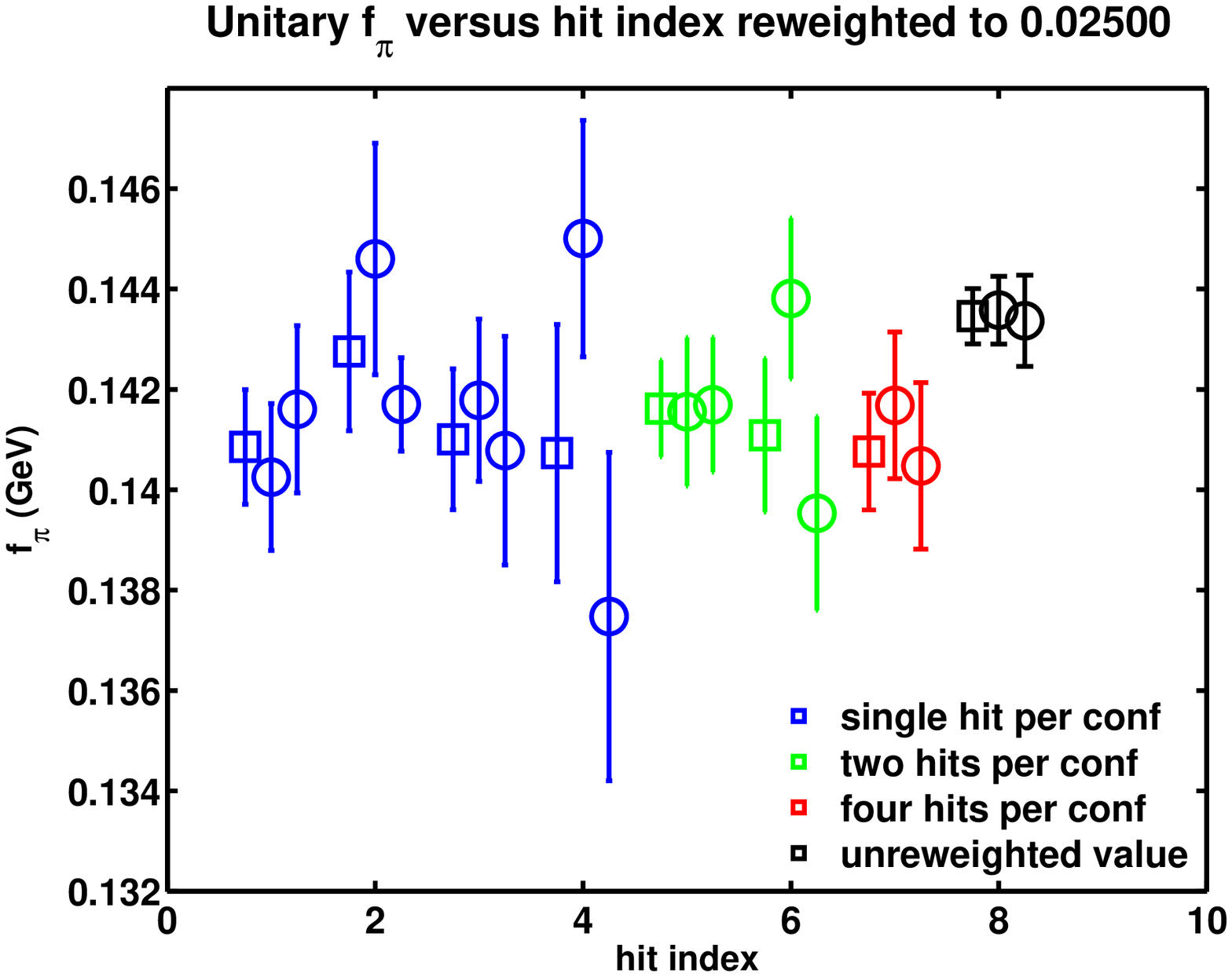}
\end{center}
\caption{
Reweighted values for $m_\pi$ (left) and $f_\pi$(right)
for various numbers of reweighting hits,
$N_\xi=1$ (blue), $N_\xi=2$ (green), $N_\xi=4$ (red) )
on each ensemble.
The squares are for the full data set (300 configurations)
and the circles are for the first and second half of the
data (150 configurations.)
The data is from the $32^3 \times 64\times16,\ (m_l,m_h)=(0.004,0.03)$
ensemble with a light valence quark of mass 0.004.
The black symbols are the unreweighted observables.
}
\label{tab:rew hit dependence}
\end{figure}

The efficiency of the reweighting procedure is also observable dependent.
It is influenced by the fluctuations of the reweighted observable within
the ensemble and the strength of the correlation between the reweighted
observable and the reweighting factor.
Sanity checks of the statistical properties of
the most important observables,
$m_\pi$ and $f_\pi$, have been performed and are summarized in
Fig.~\ref{tab:rew hit dependence}.
The observables reweighted to $\mhtarget=0.0250$
from $\mhsim=0.030$ are calculated using the first half and the second half
of the ensemble (circle symbols), which are compared to
that of the full statistics (square symbols). The number of
the Gaussian vectors, $N_\xi$, is also varied from $N_\xi=1$
(blue symbols) to $N_\xi=4$ (red symbols) in the same plot.
In the case of $m_\pi$, all the statistical samples are within $1\times\sigma$,
while for $f_\pi$ the deviations are less than $\sim 2\times\sigma$.

\begin{figure}
\begin{center}
\includegraphics[width=0.45\textwidth,clip]{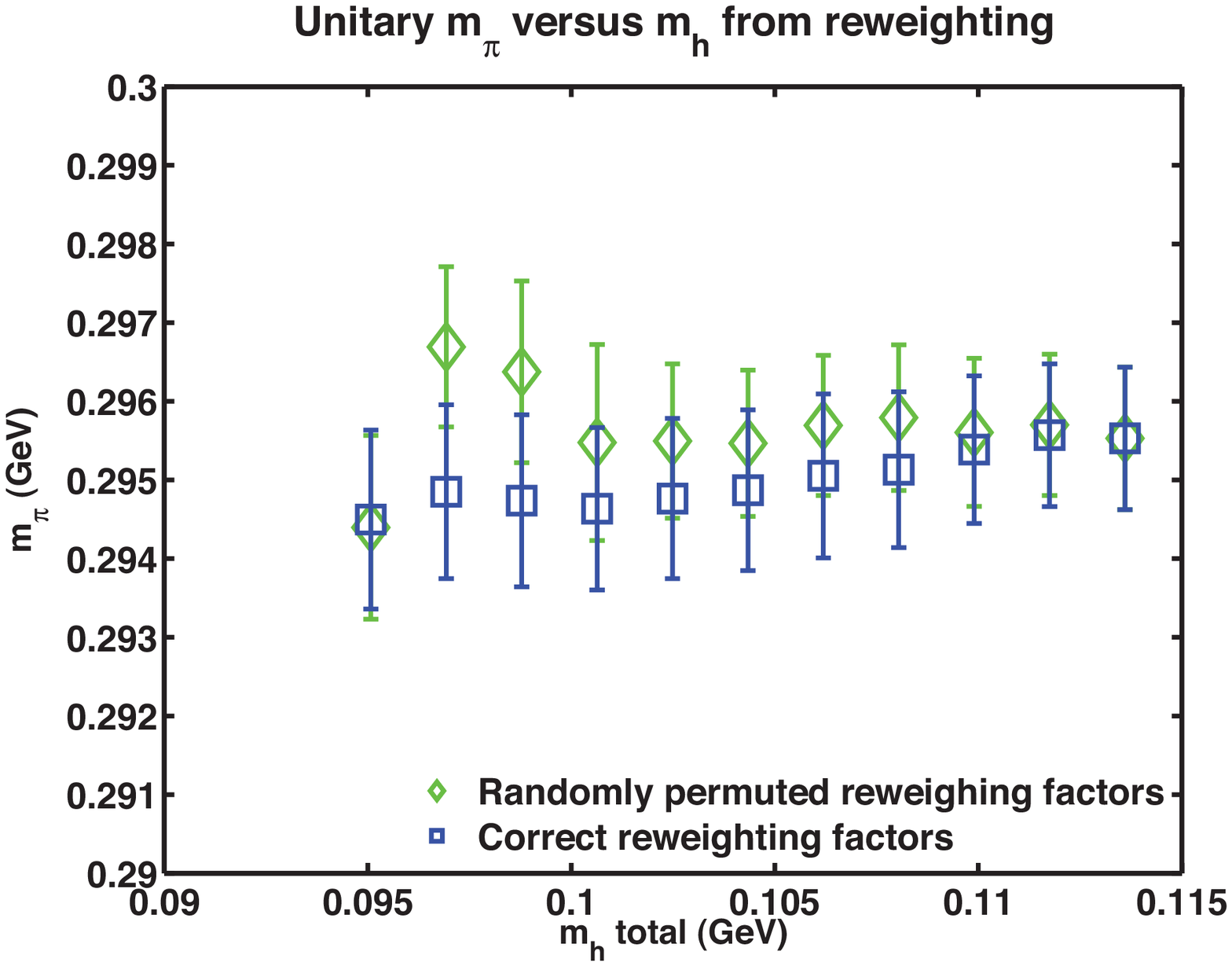}
\includegraphics[width=0.45\textwidth,clip]{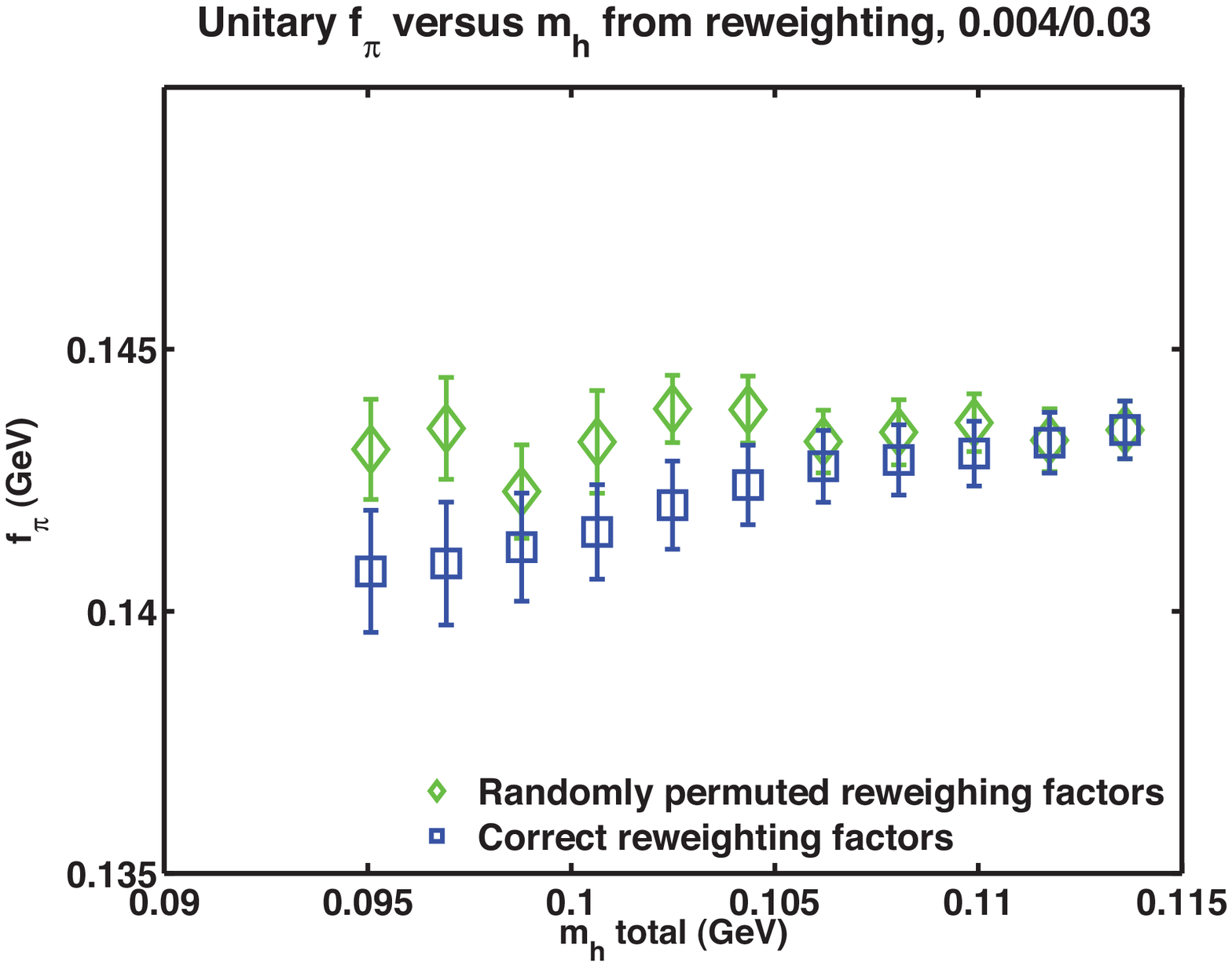}
\end{center}
\caption{
The left figure gives $m_\pi$ with correct reweighting
factors (blue squares) and with randomly permuted reweighting
factors (green diamonds).  The right figure is the same but for $f_\pi$.}
\label{fig:rew random permutation test}
\end{figure}

\begin{figure}
\begin{center}
\includegraphics[width=0.45\textwidth,clip]{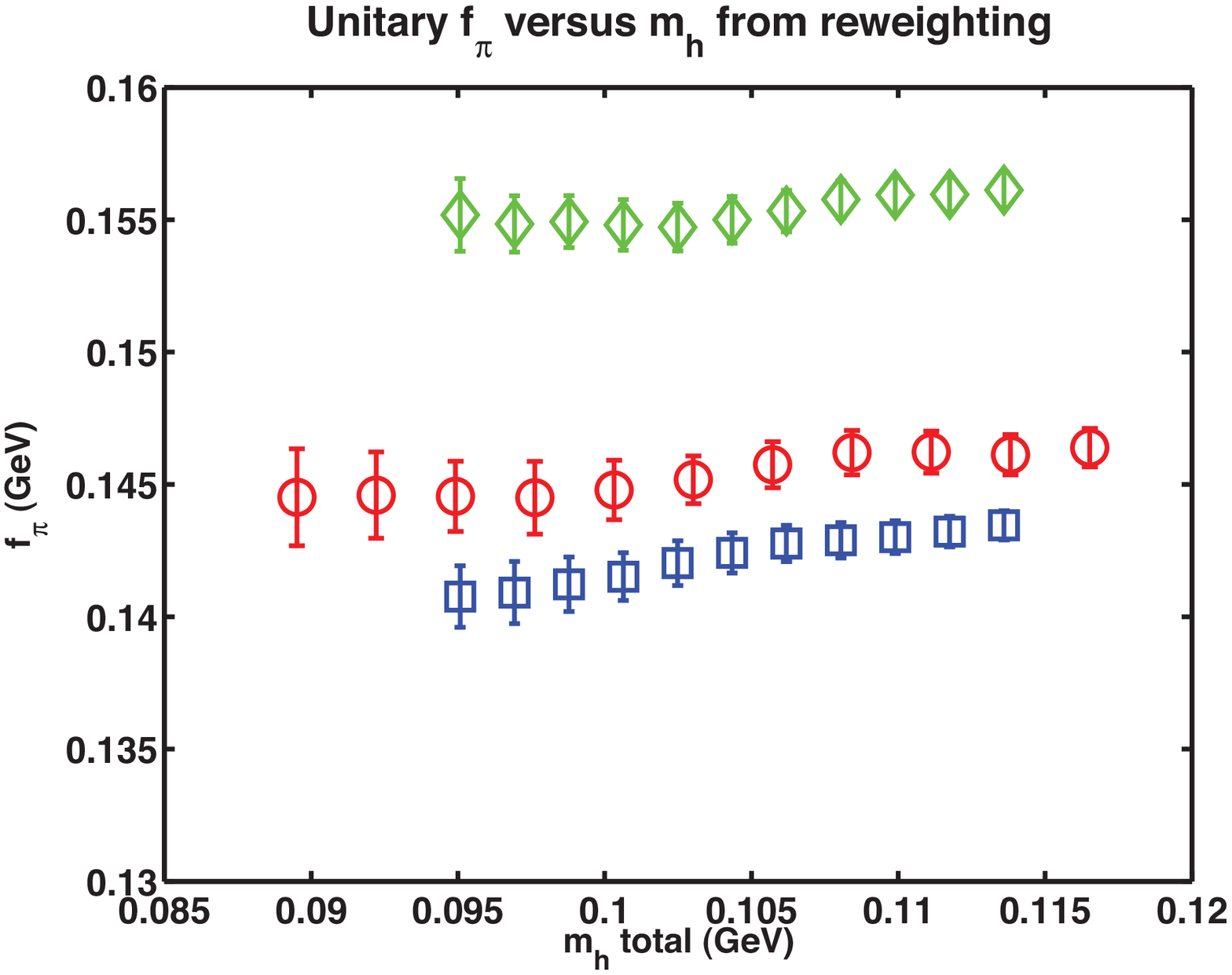}
\includegraphics[width=0.45\textwidth,clip]{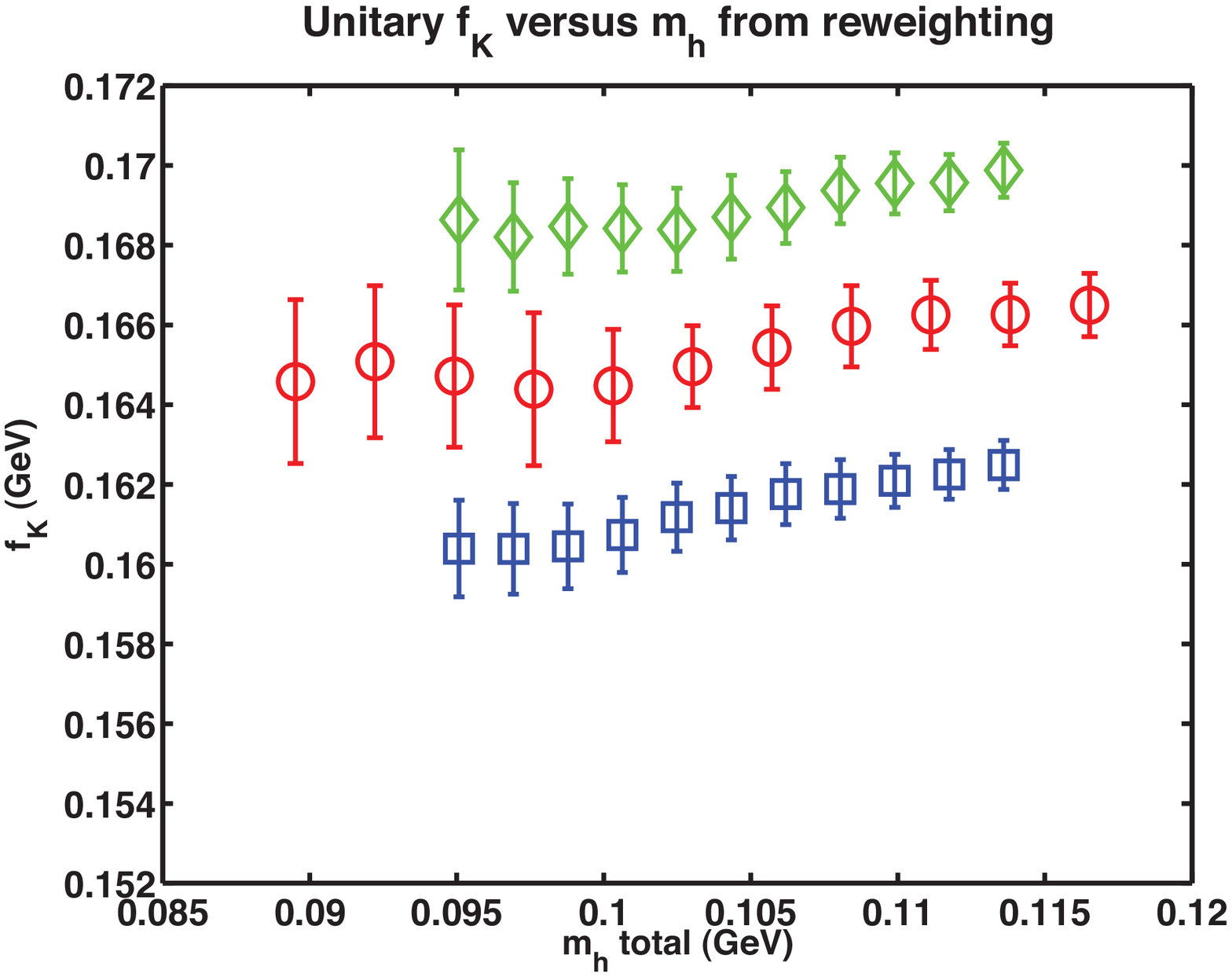}
\end{center}
\caption{Reweighted results for $f_\pi$ (left) and $f_K$ (right) as functions
of $m_h$ at three parameter sets $(\beta, m_l)$: green diamonds: (2.25, 0.008),
red circles: (2.13, 0.005),
blue squares: (2.25, 0.004).
}
\label{fig: rew sample fk fpi}
\end{figure}

To probe the $m_h$ dependence of the observables, we show in
Fig.~\ref{fig:rew random permutation test} the correctly reweighted
$m_\pi$ and $f_\pi$ as a function of $\mhtarget$ along
with the results obtained from randomly
permuting the $\{w_i\}$ in
Eq.~(\ref{eq:rew obs conf index}).  The random permutation is done for each
reweighted mass $\mhtarget$ to show the difference from the correctly
reweighted observables. While the randomly reweighted observables are almost flat
in $\mhtarget$, the correctly reweighted observables have a positive slope
in $\mhtarget$.  Finally in Fig.~\ref{fig: rew sample fk fpi} we plot
the reweighted observables $f_\pi$ and $f_K$ as a function of the
target reweighted mass $m_h$ for three example parameter points.
Note that in both Figs.~\ref{fig:rew random permutation test} and
\ref{fig: rew sample fk fpi} we observe an increase in statistical
errors which appears roughly consistent with what should be expected
from the decrease in $\sqrt{N_{\rm eff}}$.  We should emphasize that
further careful studies may be needed to establish a more accurate
estimate of possible errors
in the reweighting procedure.

\section{Updated Results from the $24^3$ Ensembles}
\label{sec:24cubed}

In this section we update the results presented on the $24^3$ ensembles
in \cite{Allton:2008pn} to the extended data set described in
Sec.\,\ref{sec:SimulationDetails}, and in
Table\,\ref{table:para} in particular. For this extended data set we make measurements of pseudoscalar
quantities on a total of 203  configurations for the $m_l=0.005$ ensemble
and 178 configurations for the $m_l=0.01$ ensemble. These configurations
were separated by 40 trajectories as documented
in the first two rows of Table\,\ref{tab:simulation_details}. In our previous work we used 92 of these measurements on each ensemble~\cite{Allton:2008pn,Antonio:2007pb}. Before performing the analyses we binned the data into blocks of either 80 or 400 trajectories and the measurements from each bin were then treated as being statistically independent. No statistically significant increase in the error was observed with the
analysis using bins of 400 trajectories compared to that with bins of 80 trajectories.

\begin{table}
\begin{center}
\begin{tabular}{ccccc}
Volume & $(m_l,m_h)$ & Total MD time & Measurement range &
  Measurement total \\ \hline
$24^3$ & (0.005,\,0.04) & 0-8980 &  900-8980 every 40 & 203 \\
$24^3$ & (0.01,\,0.04) & 1455-8540 & 1460-8540 every 40 & 178 \\
$32^3$ & (0.004,\,0.03) & 0-6756 &  520-6600 every 20 & 305 \\
$32^3$ & (0.006,\,0.03) & 0-7220 &  1000-7220 every 20 & 312\\
$32^3$ & (0.008,\,0.03) & 0-5930 & 520-5540 every 20 & 252\\
\end{tabular}
\caption{Summary of the five ensembles used in this work.
\label{tab:simulation_details}
}
\end{center}
\end{table}

In the following sections the results from the $24^3$ lattices,
combined with those obtained on the $32^3$ ensembles, will be input into
global chiral and continuum fits in order to determine physical
quantities; here we simply tabulate the fitted pseudoscalar masses and decay constants as obtained directly from the correlation functions at our simulated quark masses. In addition, since we use the mass of the $\Omega$ baryon in the definition of the scaling trajectory, we also present the results for $m_{hhh}$ here together with those for the Sommer scale $r_0$ and also the scale $r_1$. Finally, in Sec.\ref{subsec:baryons24} we give the results for the masses of the nucleons and $\Delta$ baryons from the $24^3$ ensembles, although the chiral and scaling behaviour of these masses will not be studied in this paper. We present these baryon masses partly for completeness and partly to share our experience in the use of different sources.

On the $24^3$ lattices discussed in this section, the measurements are presented for the two values of the sea light-quark mass, $m_l=0.005$ and 0.01, and for the full
range of valence quark masses $m_{x,y}=0.001,\,0.005,\,0.01,\,0.02,\,0.03$
and $0.04$. The ensembles with $m_l=0.02$ and 0.03, presented in \cite{Allton:2008pn},
are not included in this paper because such values of $m_l$
were found to be too large for SU(2) chiral perturbation theory to describe our data. The value of the sea strange-quark mass in these simulations is $m_h=0.04$. After completing the global chiral and continuum fits described in Section\,\ref{sec:CombinedChiralFits} below, we find that the physical value of the bare strange-quark mass, obtained using the chiral perturbation theory ansatz, is $m_s=0.0348(11)$. In this section we anticipate this result and use reweighting to obtain results also at this value of the strange-quark mass.

For the $24^3$ ensembles, we placed
Coulomb gauge-fixed wall sources at $t=5$ and at $t=57$.  From each source,
we calculated two quark propagators, one with periodic and the other with
anti-periodic boundary conditions.  From the periodic propagators
for the two sources, denoted by $D^{-1}_{P,5}$ and $D^{-1}_{P,57}$, and
the anti--periodic propagators, written as $D^{-1}_{A,5}$ and $D^{-1}_{A,57}$,
we form the combinations
\begin{equation}
  D^{-1}_{P+A,5} = \frac{1}{2} \left( D^{-1}_{P,5} + D^{-1}_{A,5}
    \right)\quad\textrm{and}\quad
    D^{-1}_{P+A,57} = \frac{1}{2} \left( D^{-1}_{P,57} + D^{-1}_{A,57}
  \right) \label{eq:24cubed:src}\,.
\end{equation}
The use of periodic plus anti-periodic boundary conditions in the time
direction doubles the length of the lattice in time, which markedly
reduces the contamination from around-the-world propagation in the time
direction.  For two point functions, such as the propagator of
a pseudoscalar meson given by
\begin{equation}
  \langle \pi(t) \pi(0) \rangle =
  \sum_{\vec{x}}
  \; {\rm Tr} \, \left\{
  \left[D^{-1}_{P+A,5}(t,\vec{x})\right]^\dagger D^{-1}_{P+A,5}(t,\vec{x})
  \right\}\,,
 \label{eq:24cubed:pipi:def}
\end{equation}
on a lattice of time extent $N_t$ the time dependence of the contribution of the ground state is given by
\begin{equation}
  \langle \pi(t) \pi(0) \rangle
  = A \left[ \exp(-m_\pi(t-5)) + \exp(-m_\pi (2N_t -(t-5) ) \right]\,.
  \label{eq:24cubed:pipi:form}
\end{equation}
Here $A$ is a $t$-independent constant. For our $24^3$ ensembles, we find that around-the-world propagation is not visible in two-point functions. This is not the case however, for three-point functions, as we now explain (although we do not analyze three-point functions in this paper, they are being evaluated in the computation of $B_K$, for example~\cite{bkpaper}).

For three-point functions of the form $\langle P(x) O(y) P(z) \rangle$,
where $P(x)$ and $P(z)$ are pseudoscalar interpolating fields and $O(y)$
is an operator whose matrix element we wish to measure, we use the
wall source at $t=5$ as the source for $P(z)$ and the wall source at
$t=57$ as the source for $P(x)$.  We only consider $y_0$ in the range
$5 \leq y_0 \leq 57$, so we do not perform any measurements in the doubled
lattice.  The doubling of the lattice is important to reject
around-the-world propagation in time for such measurements.  For kaons, we found that
a time separation of 52 between the sources gave us a broad plateau,
with sufficiently small errors. This measurement strategy was chosen to optimise
the measurement of the kaon bag parameter \cite{bkpaper,Antonio:2007pb}.

\begin{figure}[hbt]
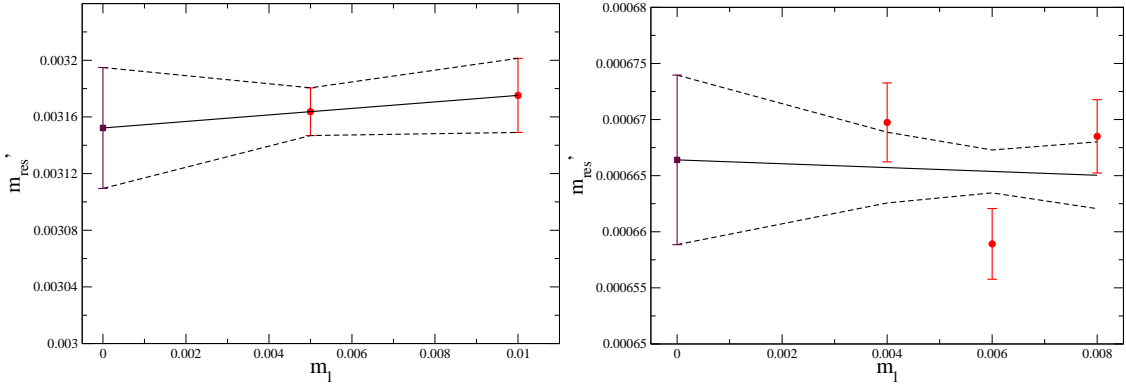

\centering
\includegraphics*[width=0.45\textwidth]{fig/Fig9a_mres_24_chiral_0.04.eps}
\includegraphics*[width=0.45\textwidth]{fig/Fig9b_mres_32_chiral_0.03.eps}
\caption{
Chiral extrapolation of the unitary values of $m_{\rm res}^\prime$ for
the $24^3$ (left) and $32^3$ (right) ensembles. While the
fit is only marginally acceptable for the $32^3$ lattices,
an additional uncertainty of O($5\times 10^{-6}$) is
negligible.
\label{fig:mres_chiral}
}
\end{figure}

Before presenting our results for masses, decay constants and $r_0$ and $r_1$, we discuss the values of the residual mass and the renormalization constant of the local axial current. The residual mass $m_\mathrm{res}^\prime(m_f)$  at each partially quenched valence mass used in this work is measured using the ratio~\footnote{We use the convention that the prime $^\prime$ in $m_\mathrm{res}^\prime(m_f)$ implies that the corresponding residual mass has been determined at a particular value of the light quark mass. $m_\mathrm{res}$ (without the $^\prime$) is defined by $m_\mathrm{res}\equiv m^\prime_\mathrm{res}(0)$.}
\begin{equation}
  \mres^\prime(m_f) = \frac{ \langle 0 | J^a_{5q} | \pi \rangle}
          {\langle 0 | J^a_5 | \pi \rangle}  \label{eq:mres_prime1}, \\
\end{equation}
where $J^a_{5q}$ is the usual DWF mid-point pseudoscalar density composed
of fields of each chirality straddling the mid-point in the fifth dimension,
and $J^a_{5}$ is the physical pseudoscalar density at the surfaces of
the fifth dimension composed of surface fields in the fifth dimension.
The results are given in Table~\ref{tab-mresprime24unrw}.
For completeness we also present the corresponding residual masses obtained after reweighting
to the physical strange mass in
Table~\ref{tab-mresprime24physmh}. The residual mass in the two-flavor chiral limit
$m_\mathrm{res} = m_\mathrm{res}^\prime(m_x=m_l=0)$  is given in
Table\,\ref{tab-mreschiral}
and in the left-hand plot of Figure~\ref{fig:mres_chiral}.

\begin{table}[tp]
\centering
\begin{tabular}{c|cc}
\hline
$m_x$ & \multicolumn{2}{c}{$m_l$}\\
\hline
      &     0.005     &     0.01\\
\hline
0.001 & 0.003194(16) & 0.003286(28) \\
0.005 & 0.003154(15) & 0.003259(26) \\
0.01 & 0.003079(14) & 0.003187(24) \\
0.02 & 0.002939(12) & 0.003042(21) \\
0.03 & 0.002822(12) & 0.002919(19) \\
0.04 & 0.002725(11) & 0.002818(17) \\
\end{tabular}
\caption{$m_\mathrm{res}^\prime(m_x)$ measured on the $24^3$ ensembles at the simulated strange quark mass $m_h=0.04$.
\label{tab-mresprime24unrw}
}
\end{table}

\begin{table}[tp]
\centering
\begin{tabular}{c|cc}
\hline
$m_x$ & \multicolumn{2}{c}{$m_l$}\\
\hline
      &     0.005     &     0.001\\
\hline
0.001 & 0.003146(27) & 0.003224(33) \\
0.005 & 0.003099(27) & 0.003191(32) \\
0.01 & 0.003025(26) & 0.003120(31) \\
0.02 & 0.002889(24) & 0.002981(26) \\
0.03 & 0.002774(23) & 0.002863(23) \\
0.04 & 0.002680(21) & 0.002765(21) \\
\end{tabular}
\caption{$m_\mathrm{res}^\prime(m_x)$ on the $24^3$ ensembles at the physical strange quark mass.
\label{tab-mresprime24physmh}
}
\end{table}

\begin{table}[pt]
\centering
\begin{tabular}{c|ccc}
\hline
$m_h$ & $m_\mathrm{res}^{24^3}$& $m_\mathrm{res}^{32^3}$\\
\hline
$m_h^\mathrm{sim}$ & 0.003152(43)& 0.0006664(76) \\ 
$m_h^\mathrm{phys}$ & 0.003076(58)& 0.0006643(82)\\ 
\end{tabular}
\caption{$m_\mathrm{res}$ in the two-flavor chiral limit on the $24^3$ and $32^3$ ensembles 
at the simulated and physical strange sea-quark masses.
\label{tab-mreschiral}
}
\end{table}

\begin{figure}[hbt]
\centering
\includegraphics*[width=0.45\textwidth]{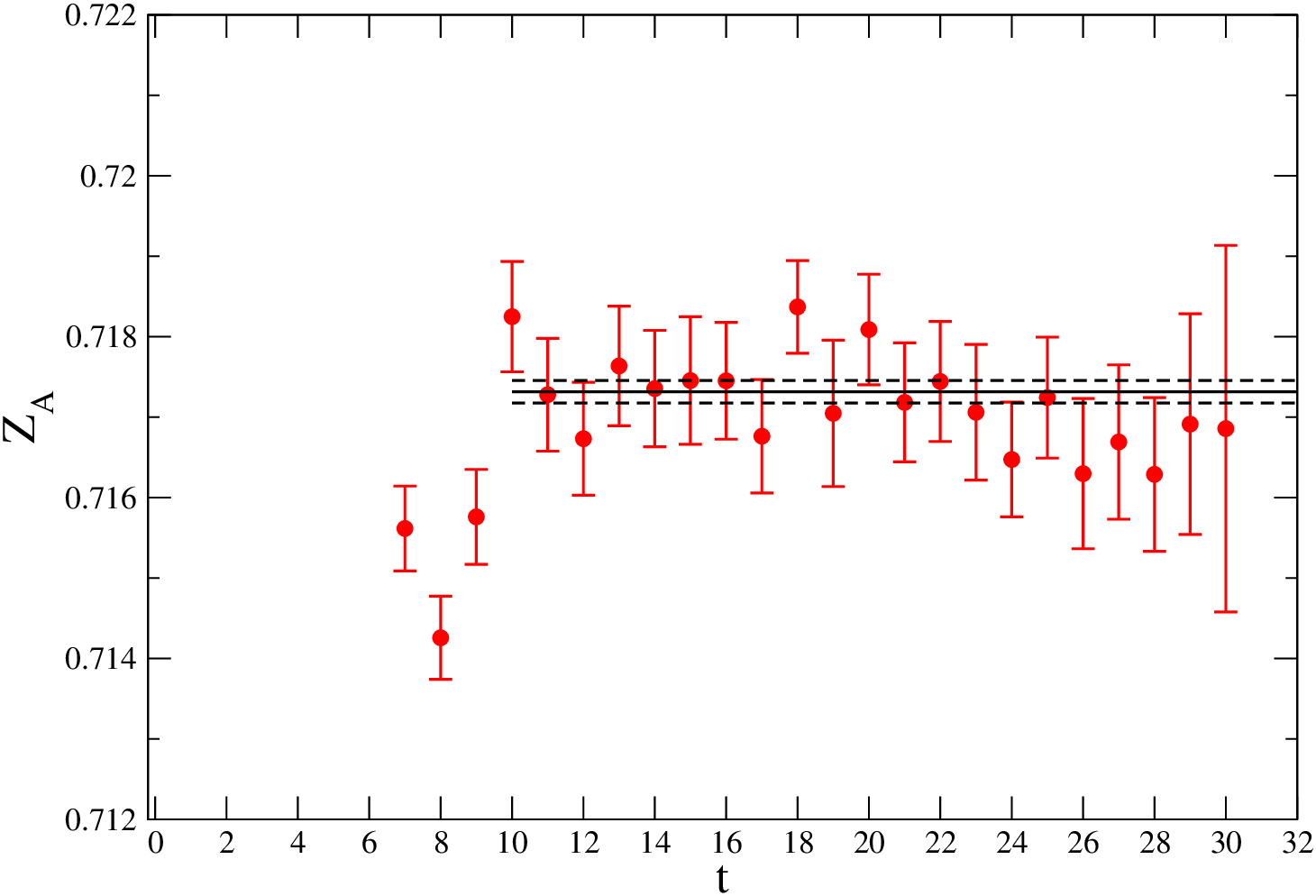}
\includegraphics*[width=0.45\textwidth]{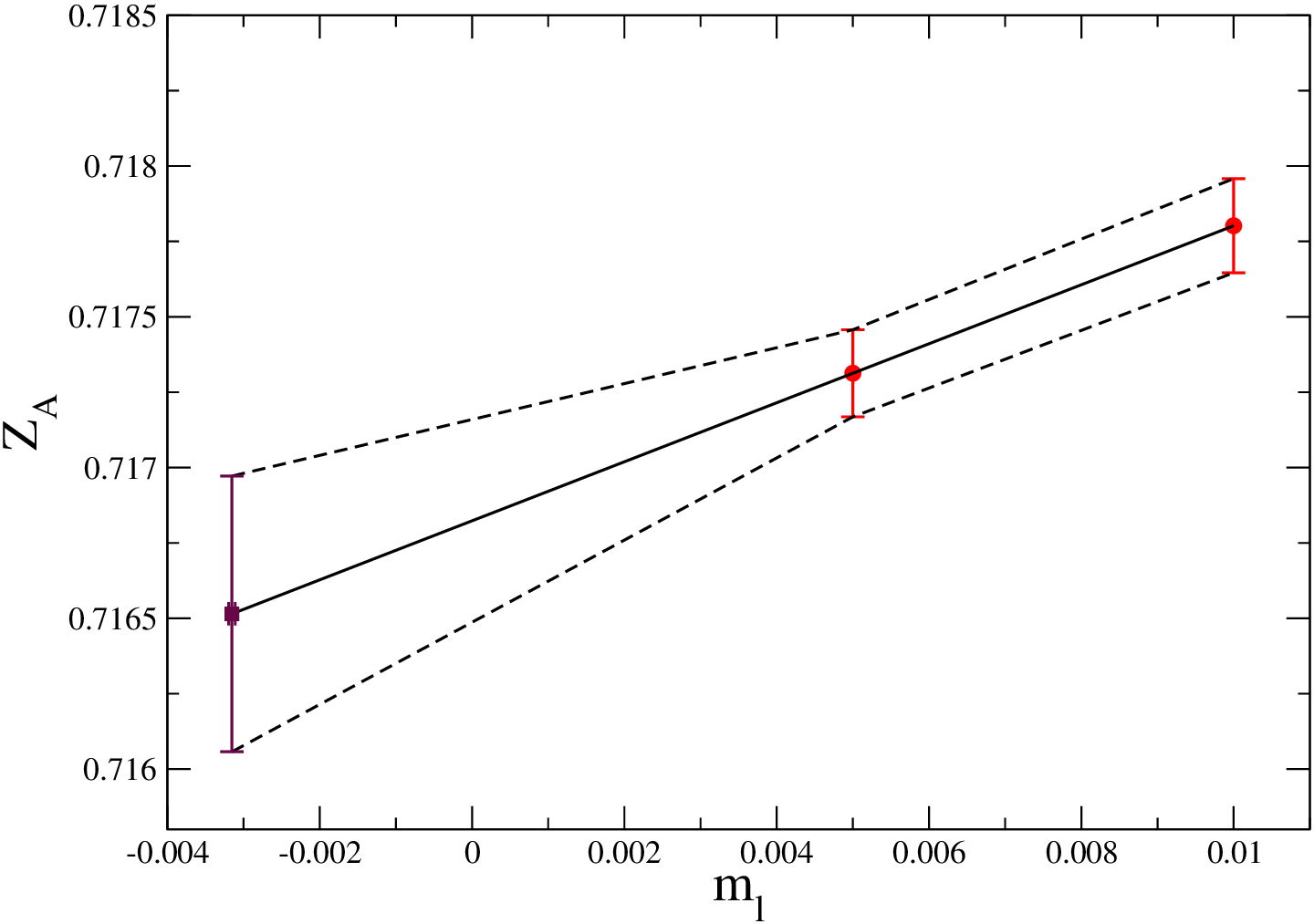}
\caption{
Measurement of $Z_A$ for $m_f=0.005$ on the $m_l=0.005$, $m_h=0.04$
ensemble (left panel) and the unitary chiral extrapolation of $Z_A$ for
the $24^3$ ensembles (right panel). The results do not change significantly
under reweighting to the physical strange mass.
\label{fig:za24}
}
\end{figure}

We define $Z_A$ to be the renormalization constant of the local axial current, $A_\mu$, composed of the physical surface fields. Here we have determined $Z_A$ through two methods.  In the first, $Z_A$ is determined for each valence mass using the improved ratio~\cite{Blum:2001xb} of the matrix element $\langle {\cal A}_4(t)
P(0) \rangle$ to $\langle A_4(t) P(0) \rangle$, where ${\cal A}_\mu$ is the conserved DWF axial current and the results are presented in Table~\ref{tab:za24}.  This method assumes $Z_{\cal A} = 1$, and we find $Z_A = 0.71651(46)$ in the two-flavor chiral limit with the simulated sea strange mass, and $Z_A = 0.71689(51)$ when reweighted to the nearby physical strange mass.  This determination of $Z_A$ is illustrated in the plots of Figure~\ref{fig:za24}.  As pointed out in \cite{Allton:2008pn}, we expect $Z_{\cal A} = 1 + O(a\,m_{\textrm{res}})$, and in \cite{Allton:2008pn} we added a $\sim 1$\% error to account for the size of this correction. As part of our current work, we have investigated the consequences of this correction, which is discussed in detail in appendix\,\ref{sec:appendix:za}.  From this analysis, we find $Z_A = 0.7041(34)$, a 1.8\% difference from the result with our previous method.  Although, as we will see, this error is smaller than our current combined errors on the decay constants and other physical quantities, we choose to use this value of $Z_A = 0.7019(26)$, coming from $Z_V/Z_{{\cal V}}$ as defined in Equation\,(\ref{eq:Z_V_calc}), as the normalization factor for the local axial current when quoting all our central values below. Here $V$ and ${\cal V}$ are the local and conserved vector currents.

\begin{table}[pt]
\begin{center}
\begin{tabular}{c|ccc}
\hline
$m_h$ & $Z_A$(chiral)& $Z_A(m_l=0.005)$ & $Z_A(m_l=0.01)$ \\
\hline
$m_h^\mathrm{sim} = 0.04$ & 0.71651(46) & 0.71732(14) & 0.71783(15) \\
$m_h^\mathrm{phys}$ &  0.71689(51) & 0.71746(17) & 0.71781(17) \\
\end{tabular}
\caption{\label{tab:za24}
$Z_A$ on the $24^3$ ensembles at the simulated and physical strange sea-quark masses.}\end{center}
\end{table}

\begin{figure}[hbt]
\centering
\includegraphics*[width=0.48\textwidth]{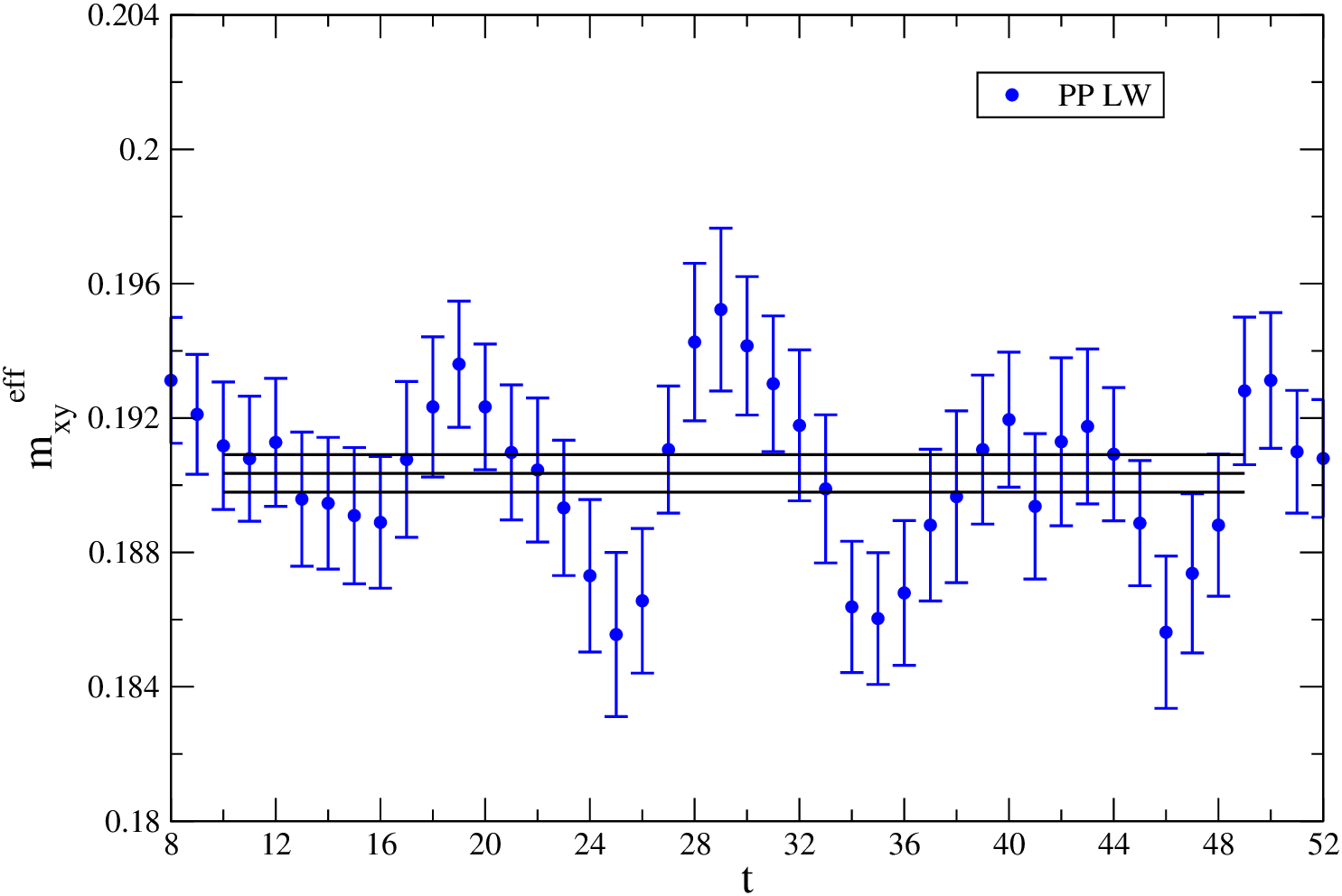}
\includegraphics*[width=0.48\textwidth]{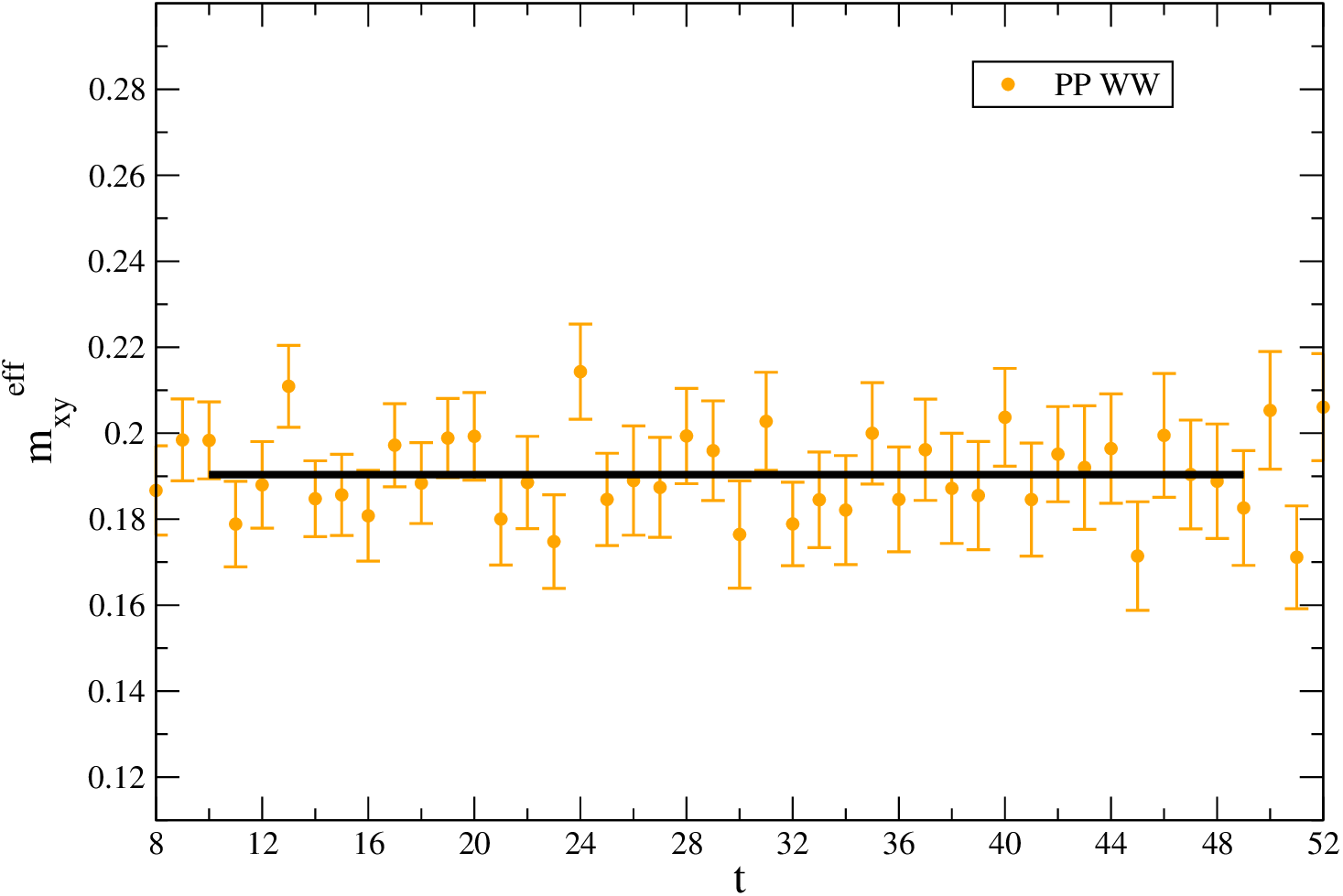}\\[0.1in]
\includegraphics*[width=0.48\textwidth]{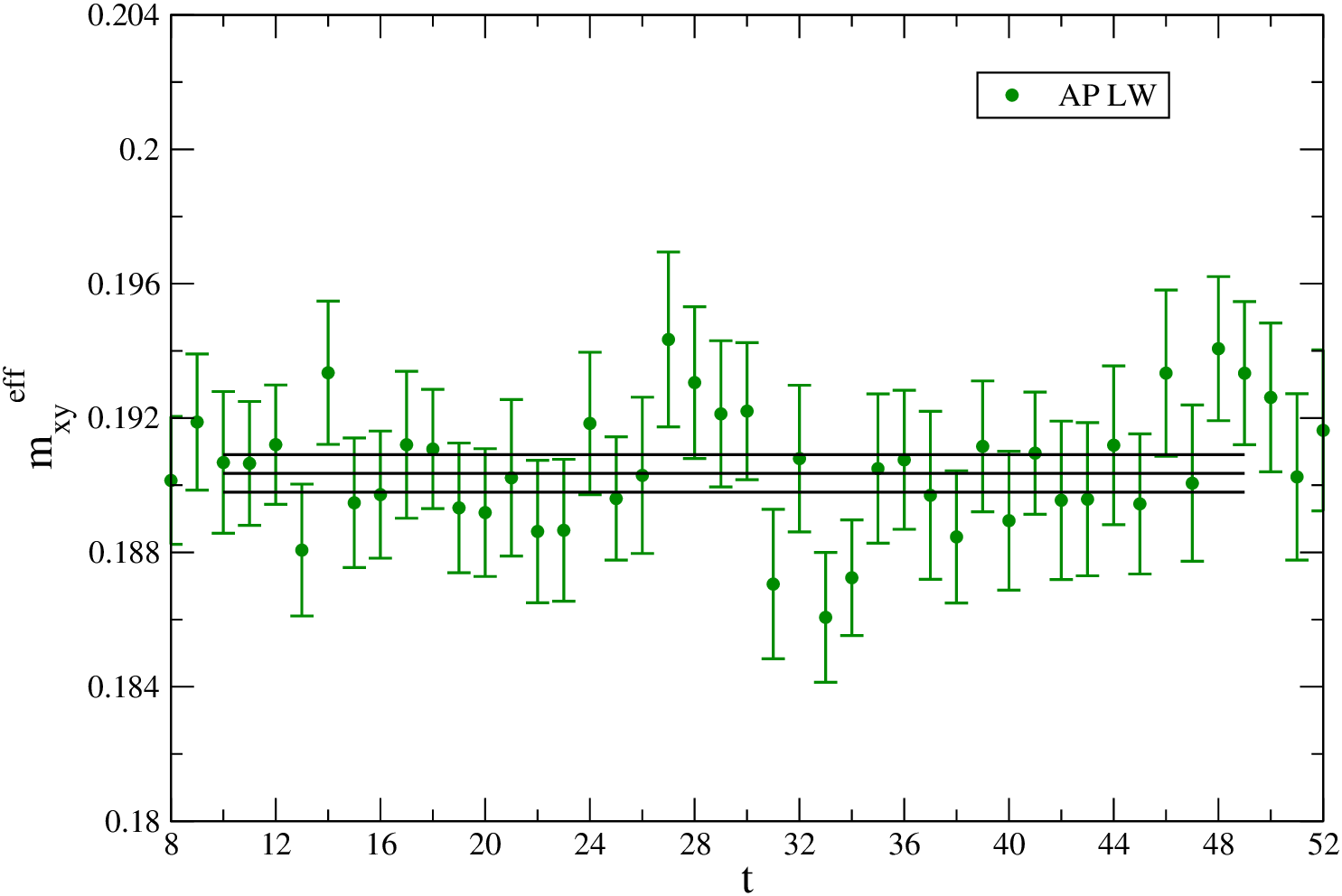}
\includegraphics*[width=0.48\textwidth]{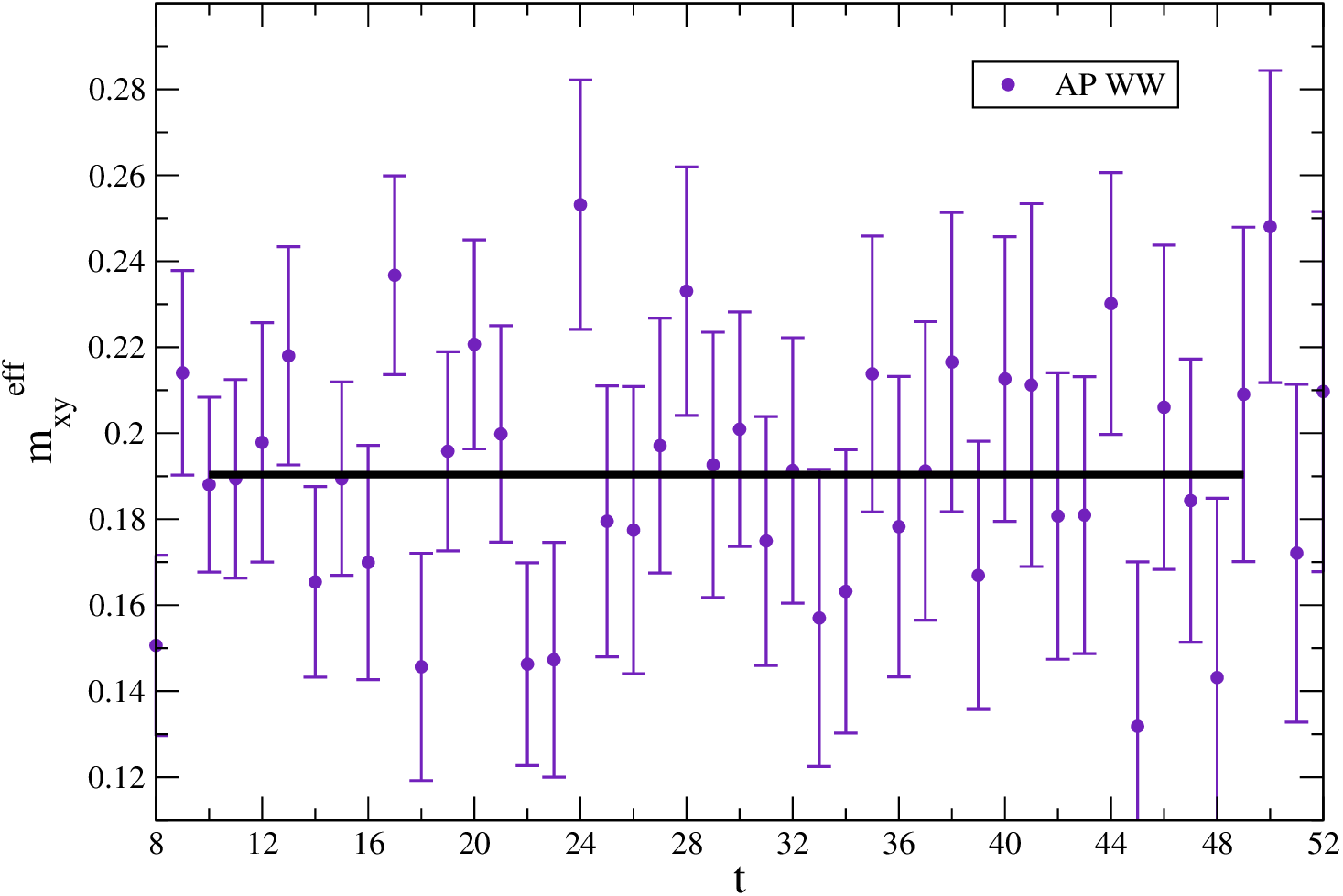}\\[0.1in]
\includegraphics*[width=0.48\textwidth]{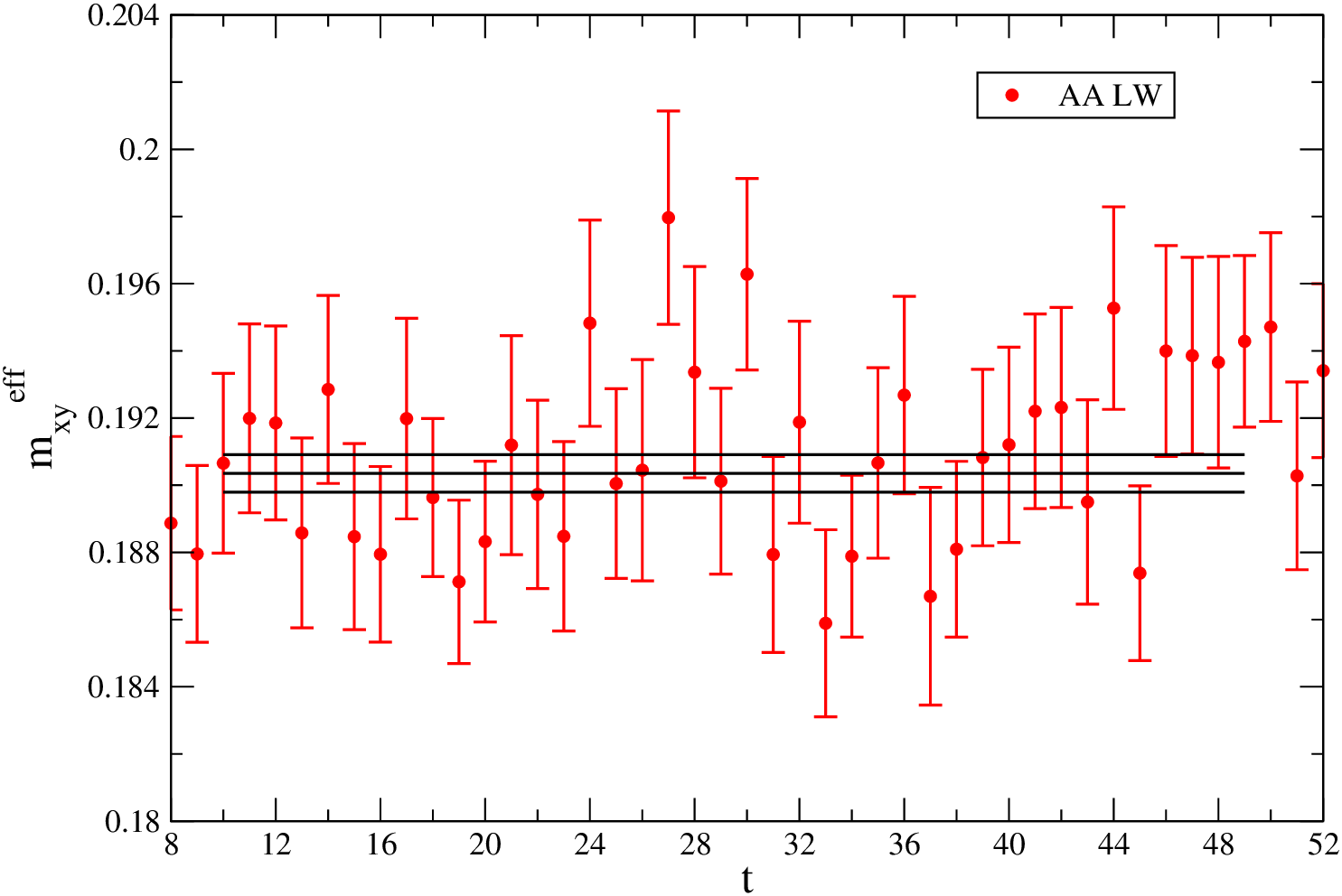}
\caption{
\label{fig:pionmeff}
Effective pion masses from the
PP LW correlator (top left), PP WW correlator (top right),
AP LW correlator (center left), AP WW (center right) and
AA LW correlator (bottom). Note the different vertical scale for the WW correlators. The horizontal bands represent the result for the mass from a simultaneous fit.
}
\end{figure}

\begin{figure}[hbt]
\centering
\includegraphics*[width=0.45\textwidth]{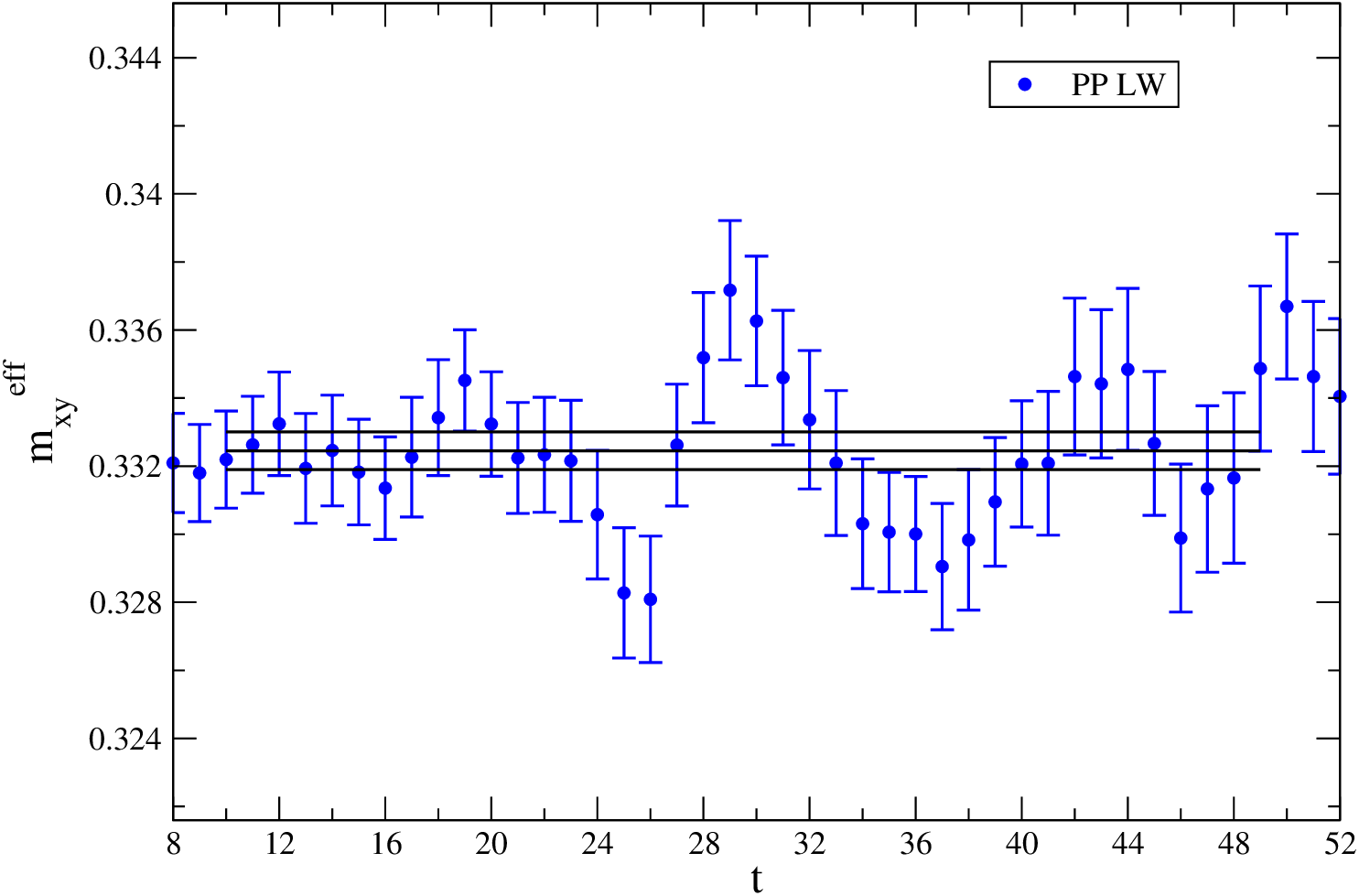}
\includegraphics*[width=0.45\textwidth]{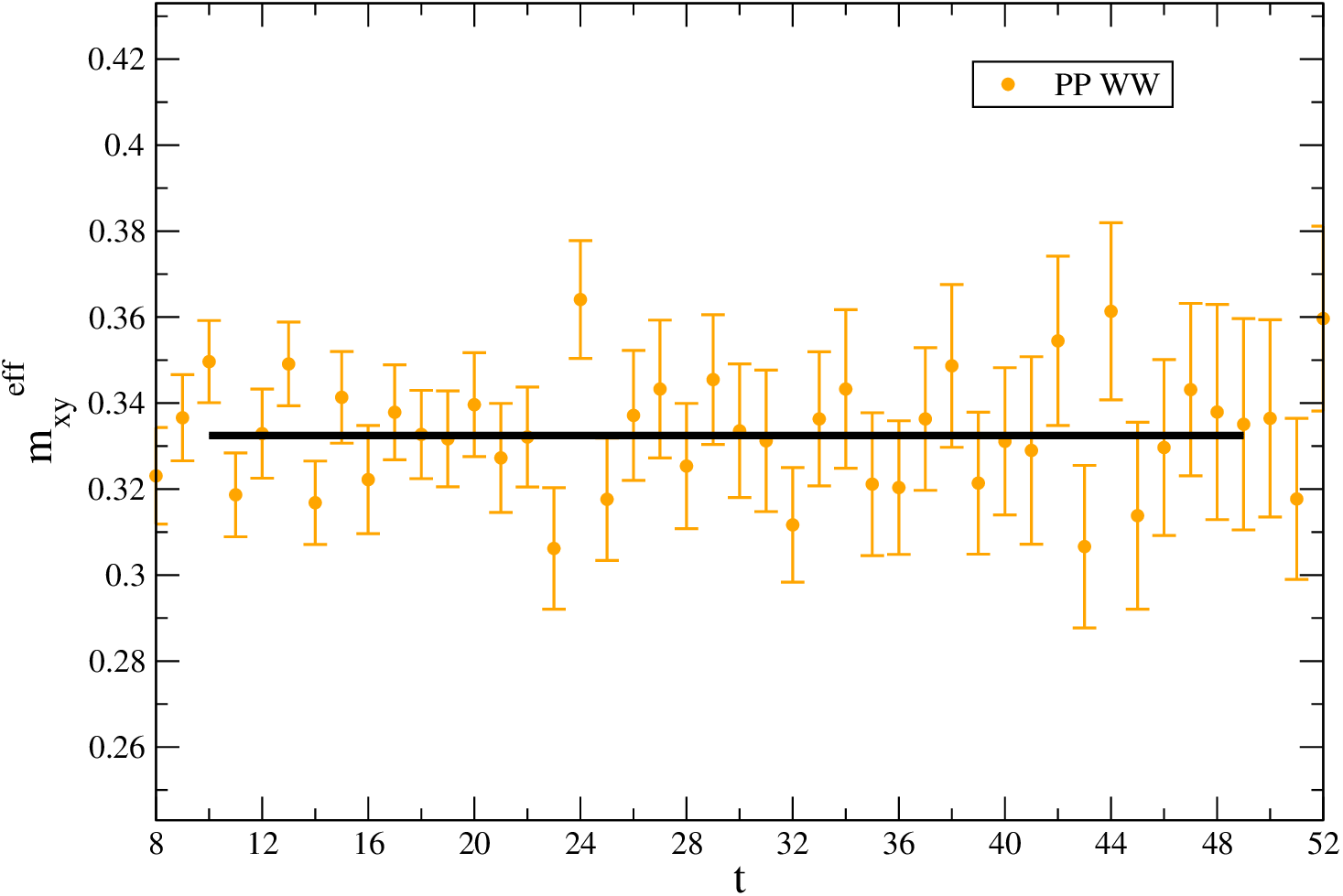}\\[0.1in]
\includegraphics*[width=0.45\textwidth]{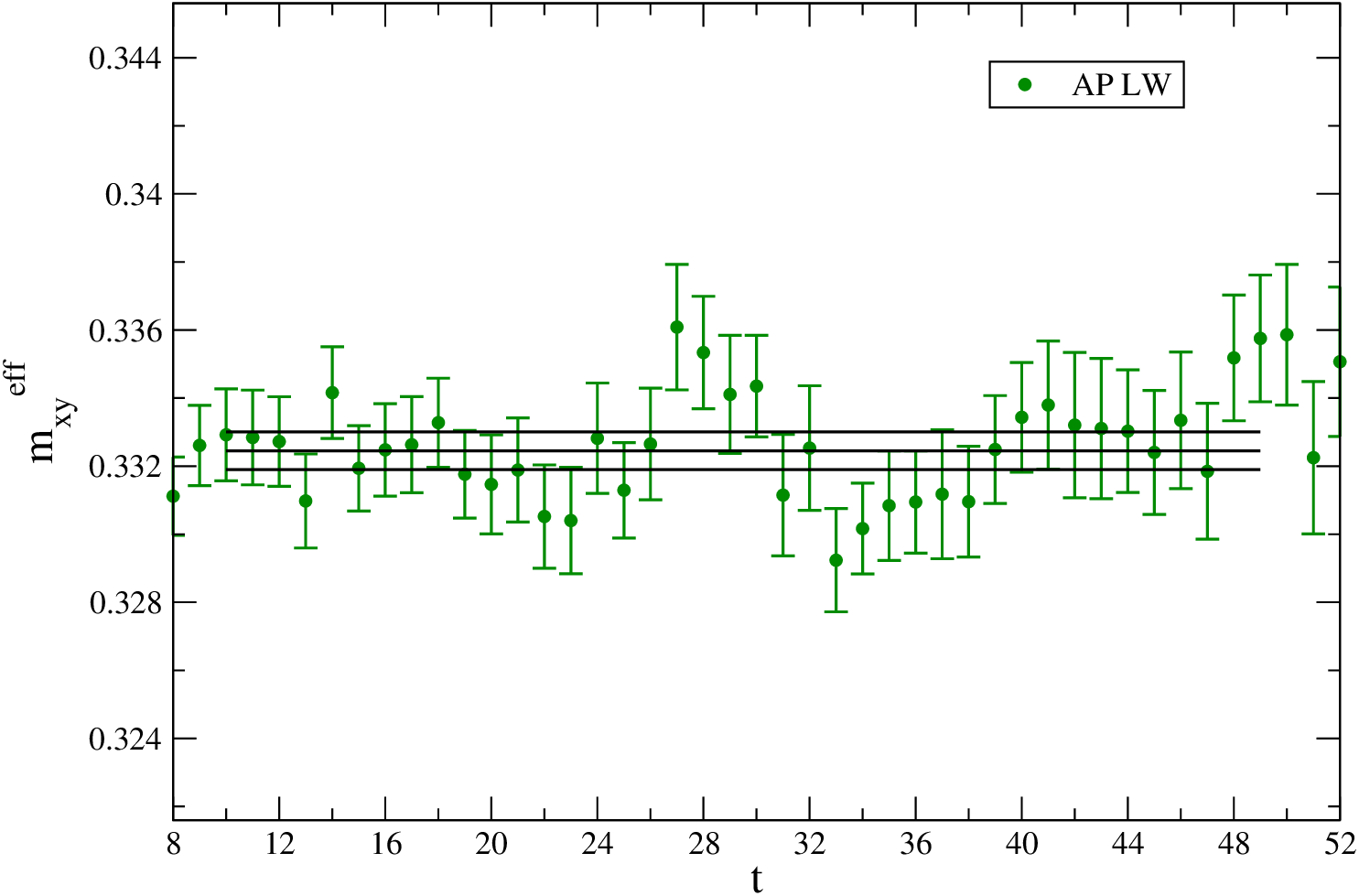}
\includegraphics*[width=0.45\textwidth]{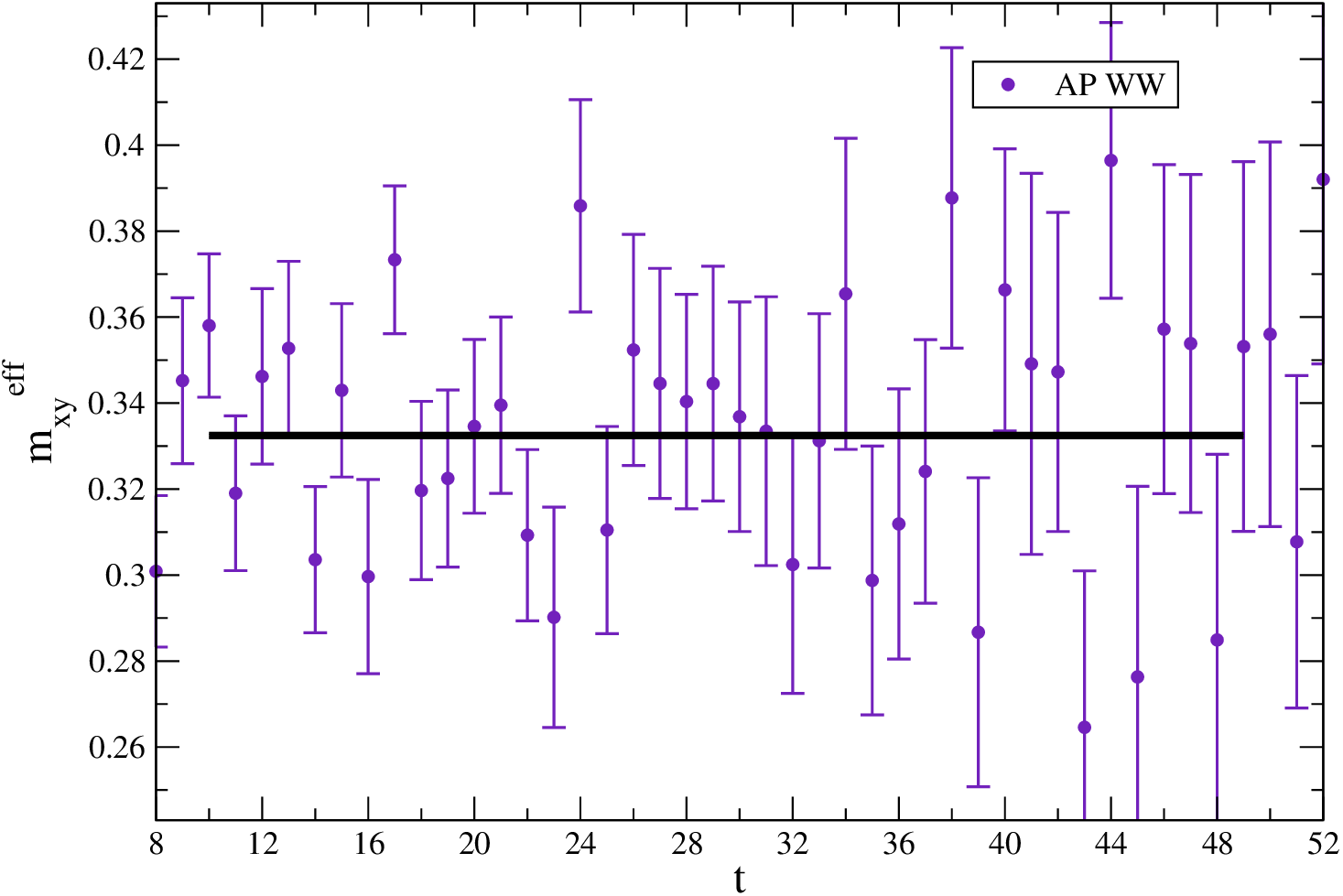}
\includegraphics*[width=0.45\textwidth]{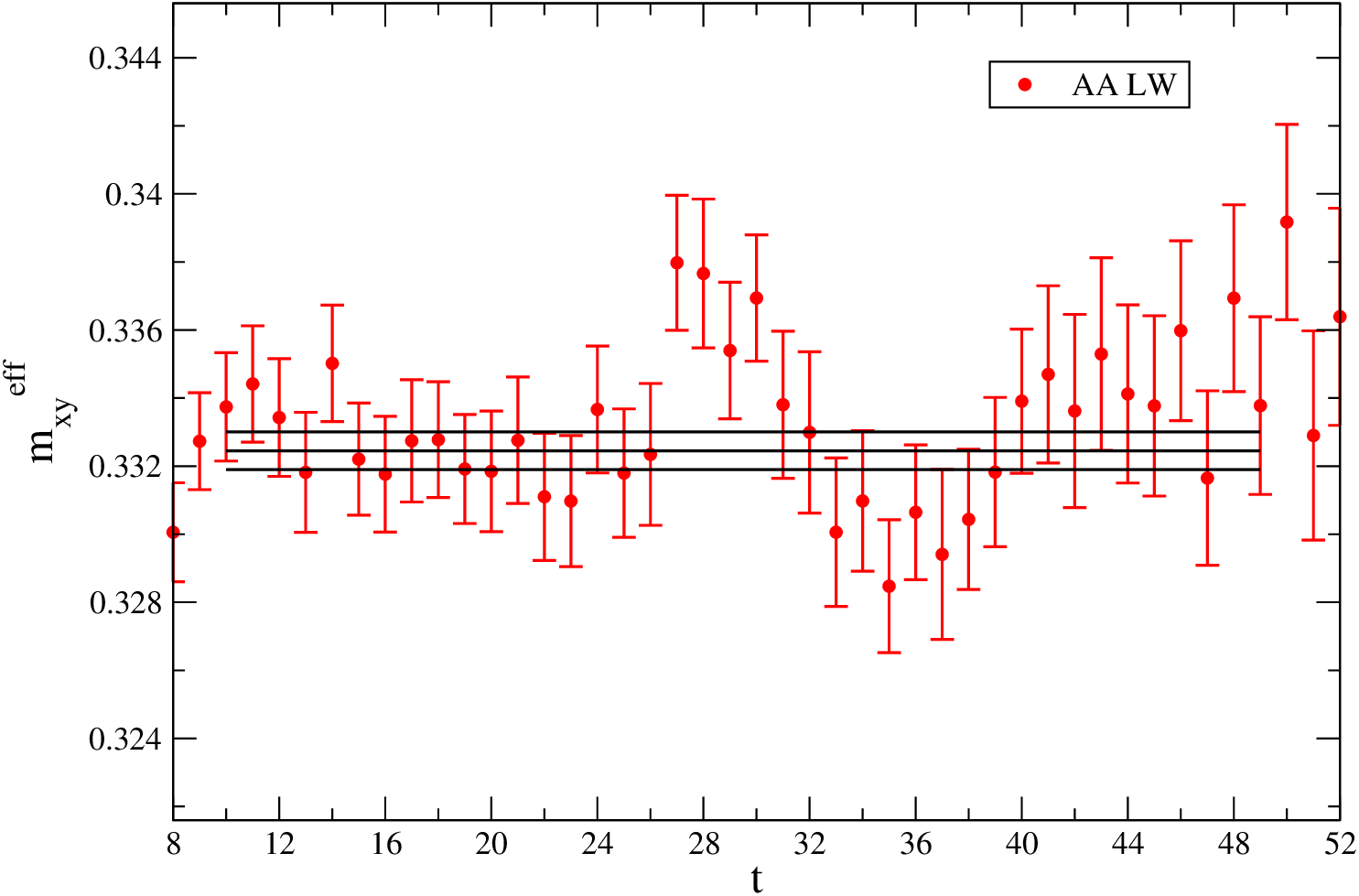}
\caption{
\label{fig:kaonmeff}
Effective kaon masses from the
PP LW correlator (top left), PP WW correlator (top right),
AP LW correlator (center left), AP WW (center right) and
AA LW correlator (bottom).
Note the different vertical scale for the WW correlators. The horizontal bands represent the result for the mass from a simultaneous fit.
}
\end{figure}

\begin{figure}[hbt]
\centering
\includegraphics*[width=0.45\textwidth]{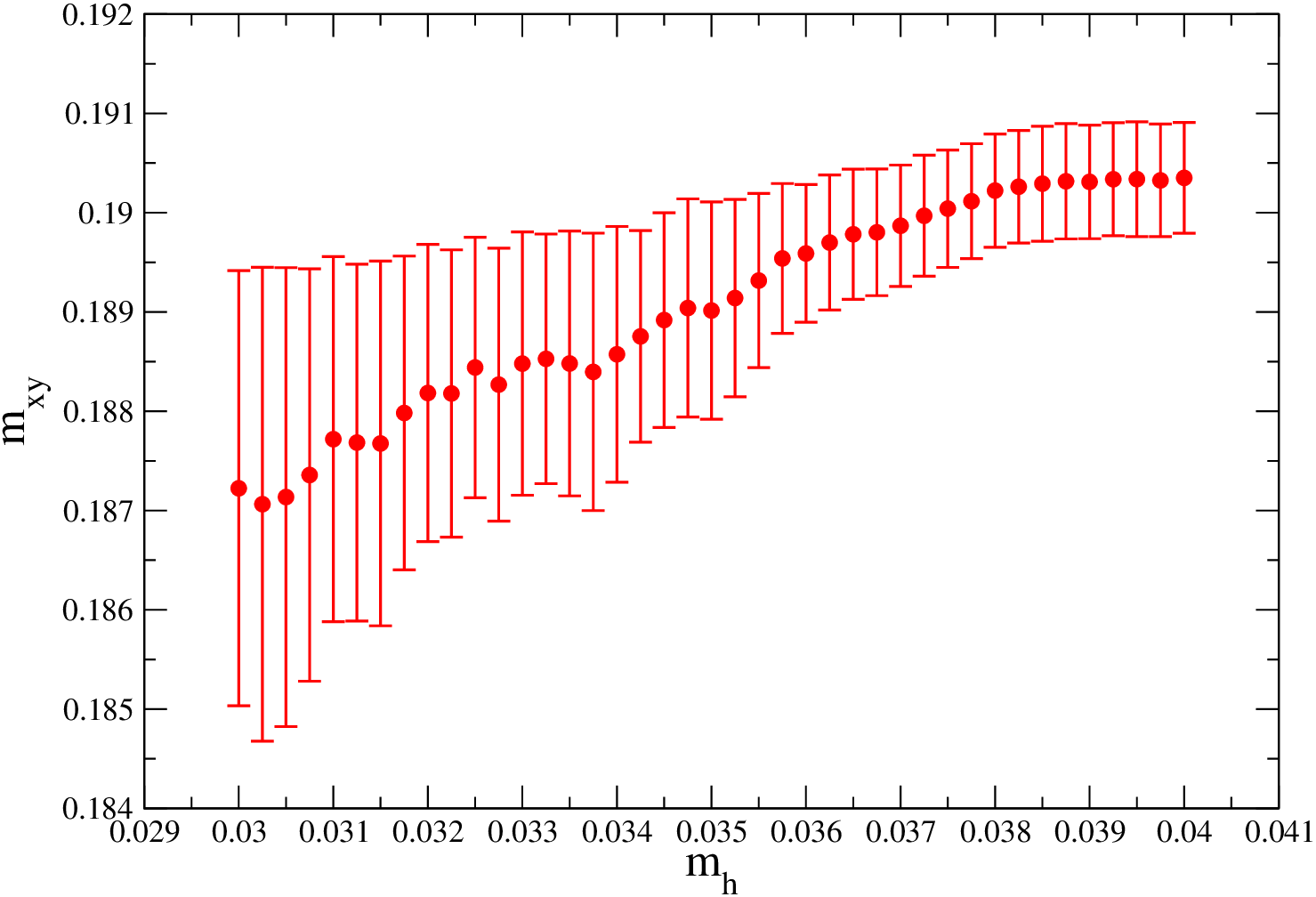}
\includegraphics*[width=0.45\textwidth]{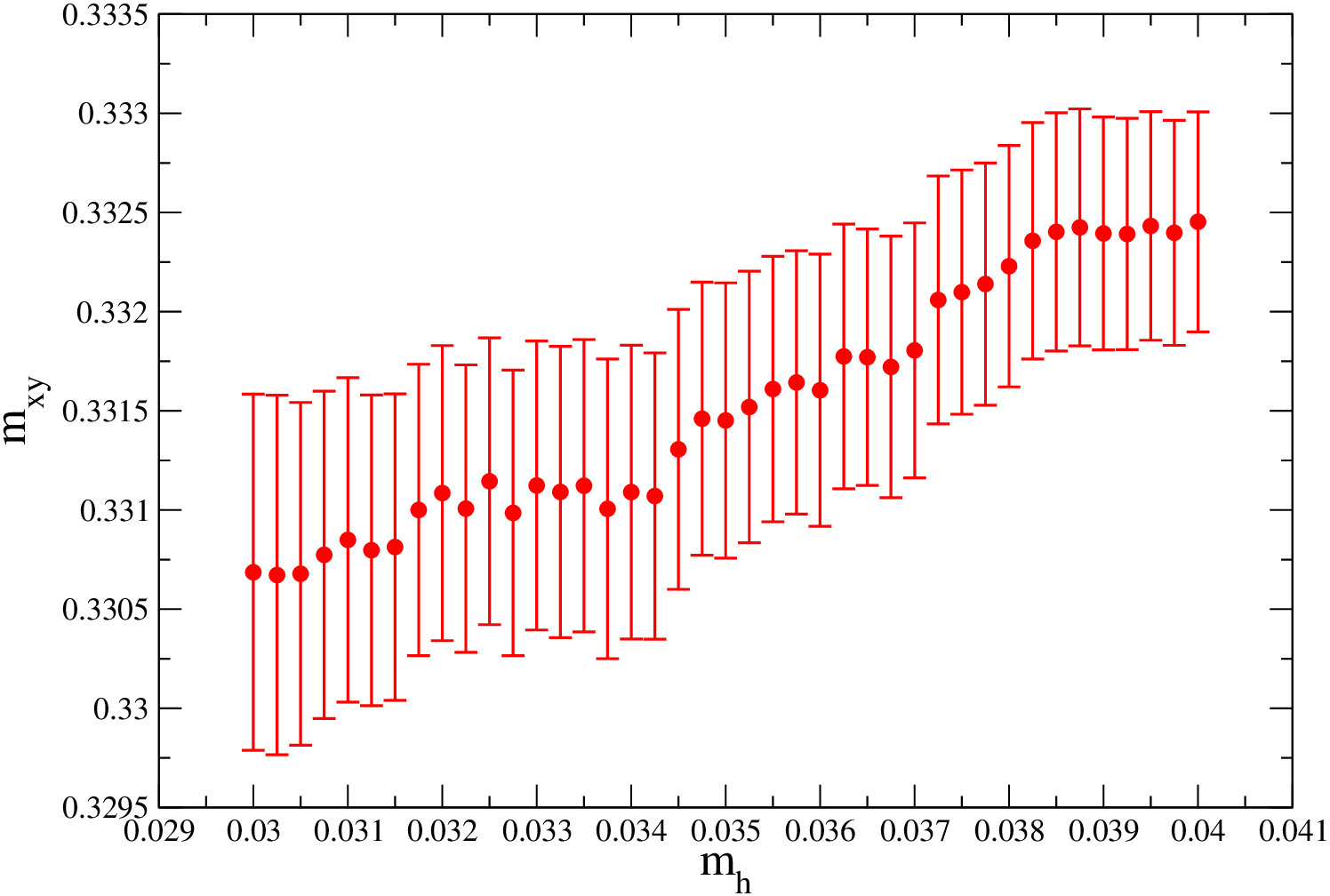}
\caption{
\label{fig:pionreweight}
We illustrate the $m_h$ dependence of the unitary pion (left panel) and kaon (right panel) masses
on the $m_l=0.005$,  $24^3$ ensemble. The values are obtained by reweighting around the simulated value
($m_h=0.04$).
}
\end{figure}

We now turn to the measurements of the meson masses and decay constants. In order to illustrate the quality of the fits, we start by presenting some sample plots for the unitary pion and kaon on the $m_l=0.005$, $m_h=0.04$ ensemble.
The pion effective masses obtained using different sources and sinks are shown in Figure\,\ref{fig:pionmeff}.
The mass and decay constant is obtained from a simultaneous
fit with a single, constrained mass to five correlation functions.
These are the $\langle P | P \rangle$,
$\langle A | A \rangle$
and $\langle A | P \rangle$
correlation functions (denoted in the figure by PP, AA and AP respectively) with gauge-fixed wall sources and local (LW) or wall (WW) sinks (we do not use the AA-WW combination because it is noisier).
The long time extent $N_t=64$ on our lattices together with the noise properties
of pseudoscalar states allow for long plateaux and the results are insensitive
to the choice of $t_{{\rm min}}$, the starting point of the fits.
Figure~\ref{fig:kaonmeff} displays the effective masses for the unitary kaon, together with the results obtained from a simultaneous constrained fit. We give an example of the $m_h$ dependence of the unitary
pion and kaon masses in figure~\ref{fig:pionreweight}. This dependence is obtained by reweighting.

We normalize the states so that, for periodic boundary conditions, the time dependence of the
correlators for large times is given by
\begin{equation}
        {\cal C}_{O_{1}O_{2}}^{s_{1}s_{2}} (t)
        = \frac{\langle 0 | O_{1}^{s_1} | \pi \rangle \langle \pi | O_{2}^{s_2}
| 0 \rangle }{2 m_{xy} V} \left [ e^{-m_{xy}t} \pm e^{-m_{xy}(2N_t-t)} \right ],
\label{eq:corr-t}
\end{equation}
where the superscripts specify the type of smearing and the
subscripts denote the interpolating
operators. The sign in the square brackets in Eq.\,(\ref{eq:corr-t}) is + for $PP$ and $AA$ correlators and $-$ for $AP$ ones.
We therefore define the amplitude of the correlator to be
\begin{equation}
{\cal N}_{O_1 O_2}^{s_{1}s_{2}} \equiv \frac{\langle 0 |
O_1^{s_1} | \pi \rangle \langle \pi | O_2^{s_2} | 0 \rangle }{2 m_{xy} V}.
\end{equation}
For each correlator included in the simultaneous fit
\[
{\cal N}_{AA}^{LW}, {\cal
  N}_{PP}^{LW}, {\cal N}_{AP}^{LW}, {\cal N}_{PP}^{WW} \ \ {\rm and} \ \ {\cal
  N}_{AP}^{WW},
\label{eq:amps}
\]
we determine the amplitude and obtain the decay constant $f_{xy}$ using
\begin{equation}
        f_{xy} = Z_A \sqrt{  \frac{2}{m_{xy}} \frac{{{\cal
        N}_{AP}^{LW}}^2}{{\cal N}_{PP}^{WW} } }. \label{eq:fpi-calc}
\end{equation}

Table~\ref{tab-24nt64-mpi-unreweighted} contains the measured
pseudoscalar masses and decay constants at the simulated strange-quark mass
$m_h=0.04$. After reweighting to the estimated physical strange-quark mass
$m_s=0.0348(11)$ the masses and decay constants of the pions are presented
in Table\,\ref{tab-24nt64-mpi-physmh} and those for the kaons in
Table\,\ref{tab-24nt64-mk-physmh}.


\begin{table}[t]
\centering
\begin{tabular}{cc|cc|cc}
\hline
$m_x$ & $m_y$ & \,$m_{xy}(0.005)$  & \,$m_{xy}(0.01)$& \,$f_{xy}(0.005)$  & $f_{xy}(0.01)$\\
\hline
0.04 & 0.04 & 0.4317(4) & 0.4344(4)& 0.1063(6) & 0.1087(6)\\
0.03 & 0.04 & 0.4051(4) & 0.4080(4)& 0.1034(6) & 0.1059(6)\\
0.02 & 0.04 & 0.3772(5) & 0.3802(4)& 0.1002(5) & 0.1028(5)\\
0.01 & 0.04 & 0.3478(5) & 0.3509(5)& 0.0967(5) & 0.0996(6)\\
0.005 & 0.04 & 0.3325(6) & 0.3358(5)& 0.0949(5) & 0.0982(6)\\
0.001 & 0.04 & 0.3199(7) & 0.3233(7)& 0.0937(6) & 0.0975(7)\\
0.03 & 0.03 & 0.3771(4) & 0.3800(4)& 0.1006(5) & 0.1031(5)\\
0.02 & 0.03 & 0.3472(5) & 0.3502(4)& 0.0974(5) & 0.1001(5)\\
0.01 & 0.03 & 0.3152(5) & 0.3184(4)& 0.0939(5) & 0.0969(5)\\
0.005 & 0.03 & 0.2983(5) & 0.3016(5)& 0.0920(5) & 0.0954(6)\\
0.001 & 0.03 & 0.2843(6) & 0.2877(6)& 0.0908(6) & 0.0946(6)\\
0.02 & 0.02 & 0.3149(5) & 0.3179(4)& 0.0943(5) & 0.0971(5)\\
0.01 & 0.02 & 0.2794(5) & 0.2826(5)& 0.0908(5) & 0.0938(5)\\
0.005 & 0.02 & 0.2603(5) & 0.2636(5)& 0.0889(5) & 0.0923(5)\\
0.001 & 0.02 & 0.2440(6) & 0.2475(6)& 0.0876(5) & 0.0915(6)\\
0.01 & 0.01 & 0.2389(5) & 0.2422(5)& 0.0872(5) & 0.0905(5)\\
0.005 & 0.01 & 0.2161(5) & 0.2195(5)& 0.0853(5) & 0.0889(5)\\
0.001 & 0.01 & 0.1960(6) & 0.1997(6)& 0.0840(5) & 0.0879(5)\\
0.005 & 0.005 & 0.1904(6) & 0.1940(6)& 0.0834(5) & 0.0871(5)\\
0.001 & 0.005 & 0.1669(6) & 0.1709(6)& 0.0819(5) & 0.0858(5)\\
0.001 & 0.001 & 0.1391(6) & 0.1434(7)& 0.0802(5) & 0.0840(5)
\end{tabular}
\caption{Pseudoscalar masses $m_{xy}(m_l)$ and decay constants $f_{xy}(m_l)$ on the $24^3$ ensembles at the 
simulated strange-quark mass ($m_h=0.04$).}
\label{tab-24nt64-mpi-unreweighted}
\end{table}


\begin{table}[t]
\centering
\begin{tabular}{cc|cc|cc}
\hline
$m_x$ & $m_y$ & $m_{xy}(0.005)$  & $m_{xy}(0.01)$& $f_{xy}(0.005)$  & $f_{xy}(0.01)$\\
\hline
0.01 & 0.01 & 0.2378(8) & 0.2420(7)& 0.0867(5) & 0.0900(6)\\
0.005 & 0.01 & 0.2149(9) & 0.2192(7)& 0.0848(6) & 0.0882(6)\\
0.001 & 0.01 & 0.1948(10) & 0.1994(8)& 0.0833(6) & 0.0871(6)\\
0.005 & 0.005 & 0.1891(10) & 0.1936(8)& 0.0828(5) & 0.0863(6)\\
0.001 & 0.005 & 0.1656(11) & 0.1704(8)& 0.0813(6) & 0.0850(6)\\
0.001 & 0.001 & 0.1377(12) & 0.1427(9)& 0.0796(6) & 0.0832(7)
\end{tabular}
\caption{Pion masses $m_{xy}(m_l)$ and decay constants $f_{xy}(m_l)$ on the $24^3$ ensembles at the physical 
strange-quark mass $m_s = 0.0348(11)$.}
\label{tab-24nt64-mpi-physmh}
\end{table}

\begin{table}
\centering
\begin{tabular}{c|cc|cc}
\hline
$m_x$ & $m_{xh}(0.005)$  & $m_{xh}(0.01)$& $f_{xh}(0.005)$  & $f_{xh}(0.01)$\\
\hline
0.01 & 0.330(4) & 0.334(4)& 0.0947(7) & 0.0978(8)\\
0.005 & 0.314(4) & 0.318(4)& 0.0928(7) & 0.0963(9)\\
0.001 & 0.301(4) & 0.305(4)& 0.0915(8) & 0.0955(10)
\end{tabular}
\caption{Kaon masses $m_{xh}(m_l)$ and decay constants $f_{xh}(m_l)$ on the $24^3$ ensembles at 
the physical strange-quark mass $m_s = 0.0348(11)$.}
\label{tab-24nt64-mk-physmh}
\end{table}

\begin{figure}[t]
\centering
\includegraphics*[width=0.45\textwidth]{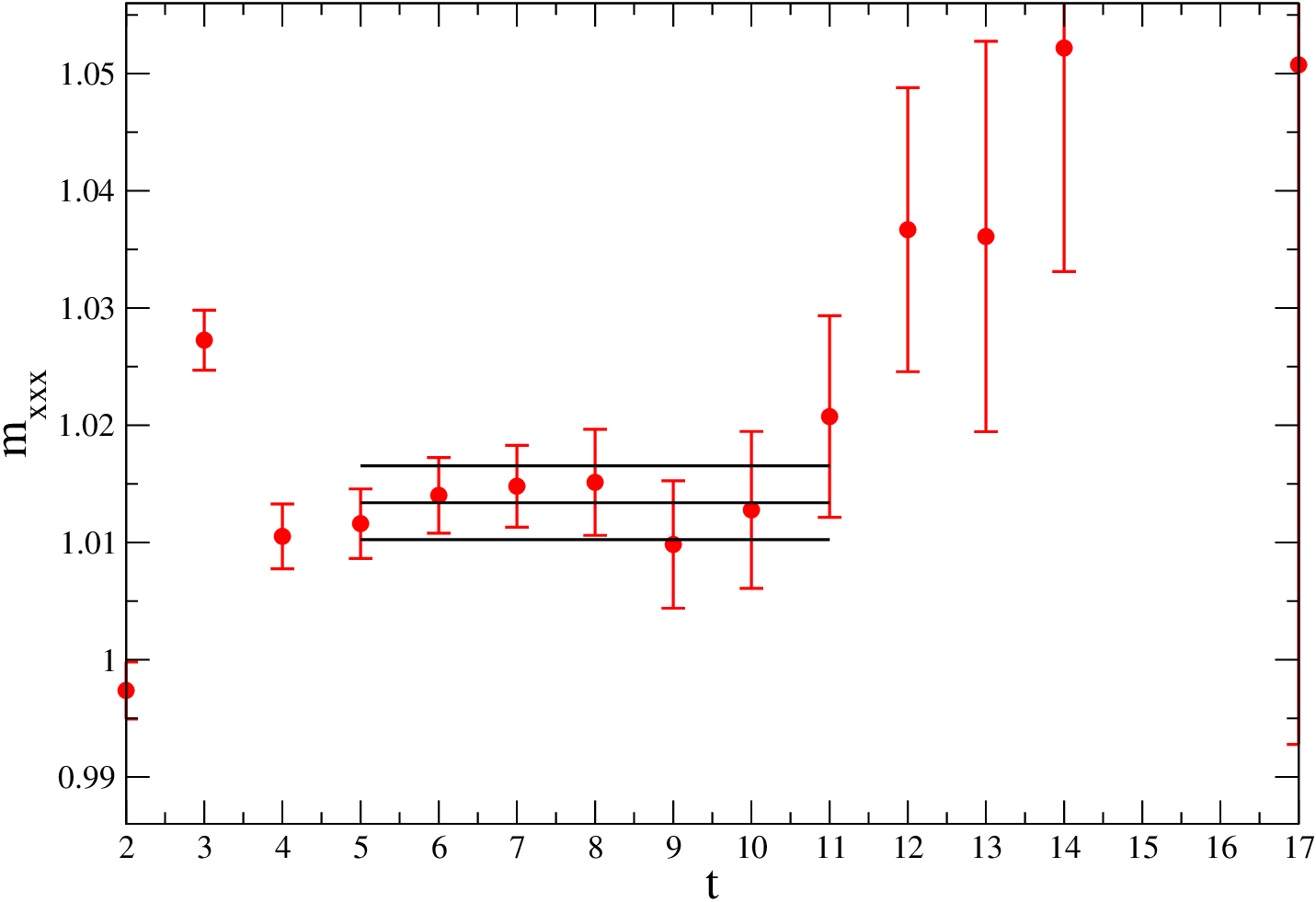}
\includegraphics*[width=0.45\textwidth]{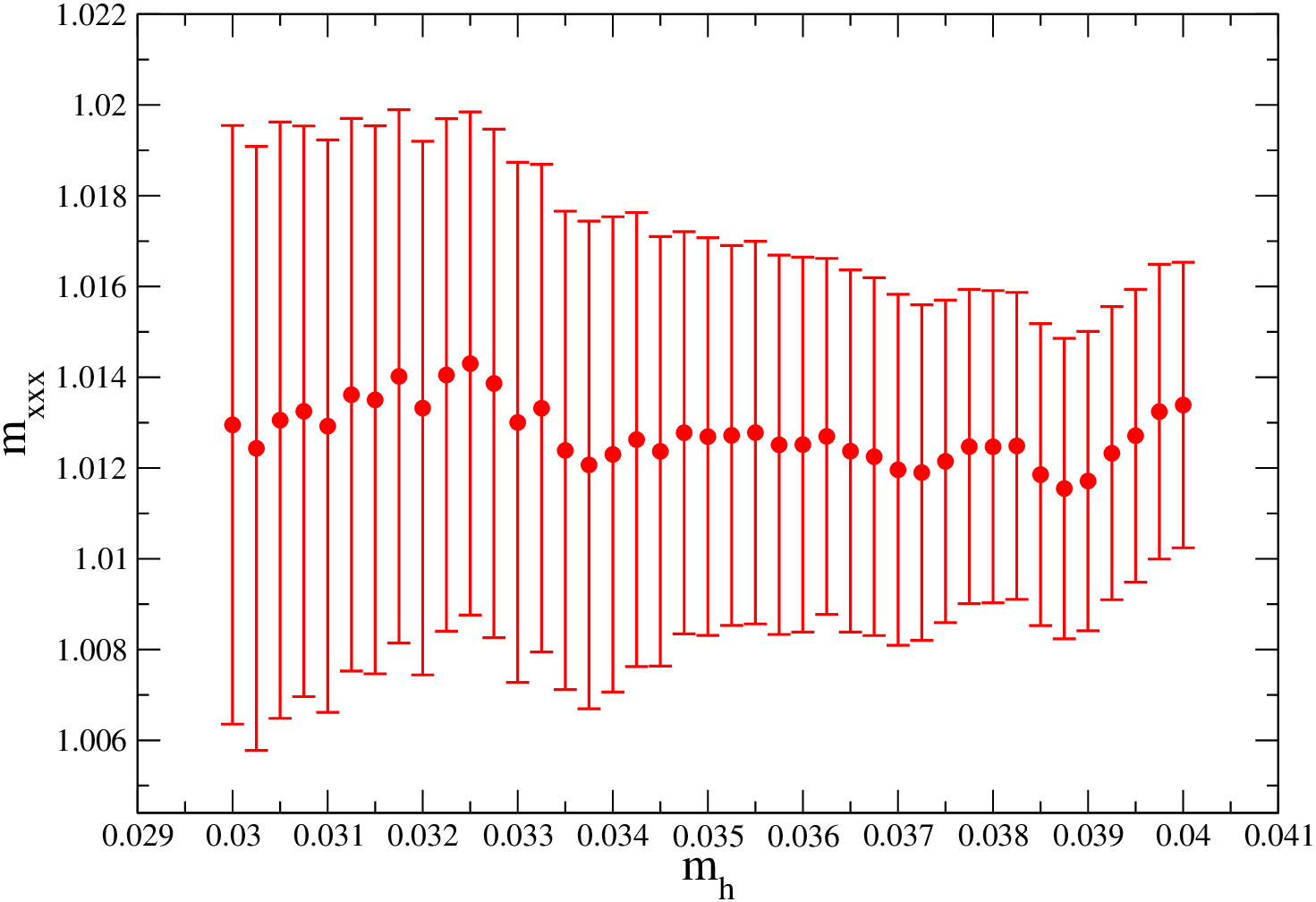}
\caption{
\label{fig:omegaeff}
Fit to the $\Omega$ baryon mass with valence
strange mass $m_x=0.04$ on the $m_l=0.005$, $m_h=0.04$, $24^3$ ensemble
showing the quality of the fit with our box source (left panel).
We also show the weak dependence of the $\Omega$ baryon mass with
fixed valence mass $m_x=0.04$ on our simulated $m_h$ inferred
by the reweighting procedure on the $m_l=0.005$, $24^3$ ensemble
(right panel).
}
\end{figure}

The $\Omega$ baryon, being one of
the quantities included in the definition of our scaling trajectory (see Section\,\ref{sec:CombinedChiralFits}),
plays an important r\^ole in our analysis.
We have performed  measurements on the same configurations
using a gauge-fixed box source
of size 16 lattice units that gives a good plateau for the $\Omega$-state for valence quark masses $m_x=0.04 $ and $m_x=0.03$ to enable
interpolation to the physical strange-quark mass.
We display the fit to the $m_x=0.04$ $\Omega$ baryon mass
on the $m_l=0.005$, $m_h=0.04$ ensemble in figure~\ref{fig:omegaeff},
along with the dependence of this mass on the dynamical strange mass
using reweighting.

\begin{table}[h] \centering \begin{tabular}{cc|cc} \hline 
$m_y$ & $m_h$ & $m_{\Omega}(0.005)$ & $m_{\Omega}(0.01)$
\\  \hline 
0.04 & 0.04&   1.013(3) & 1.028(4)\\ 
0.03 & 0.04& 0.963(4) & 0.978(4)\\ 
0.0348 & 0.0348& 
0.988(9) & 1.001(7)\end{tabular} 
\caption{Omega baryon masses on the $24^3$ ensembles 
at the simulated strange quark mass $m_h=0.04$  (first two 
rows) and at the physical strange-quark mass (third 
row).\label{tab-24nt64-momega-r0}} \end{table} 

\begin{table}[tp]
\centering
\begin{tabular}{c|cc|cc}
\hline
\multirow{2}{*}{Quantity}&\multicolumn{2}{c|}{$m_h=0.04$}&\multicolumn{2}{c}{$m_h=0.0348$}\\ 
& $Q(0.005)$ & $Q(0.01)$ &$Q(0.005)$ & $Q(0.01)$\\
\hline
$r_0$& 4.16(2) & 4.10(2) & 4.15(2) & 4.12(3) \\
$r_1$& 2.82(3) & 2.70(2)& 2.83(3) & 2.72(3) \\
$r_1/r_0$& 0.678(8) & 0.657(6) & 0.682(9) & 0.661(10)
\end{tabular}
\caption{The quantities $r_0,\, r_1$ and $r_1/r_0$ at the simulated ($m_h=0.04$) 
and physical ($m_h=0.0348$) strange quark masses on the $24^3$ ensembles. $Q(m_l)$ 
denotes the quantity measured with light-quark mass $m_l$.}
\label{tab:r024}
\end{table}

\begin{figure}[tp]
\centering
\includegraphics*[width=0.45\textwidth]{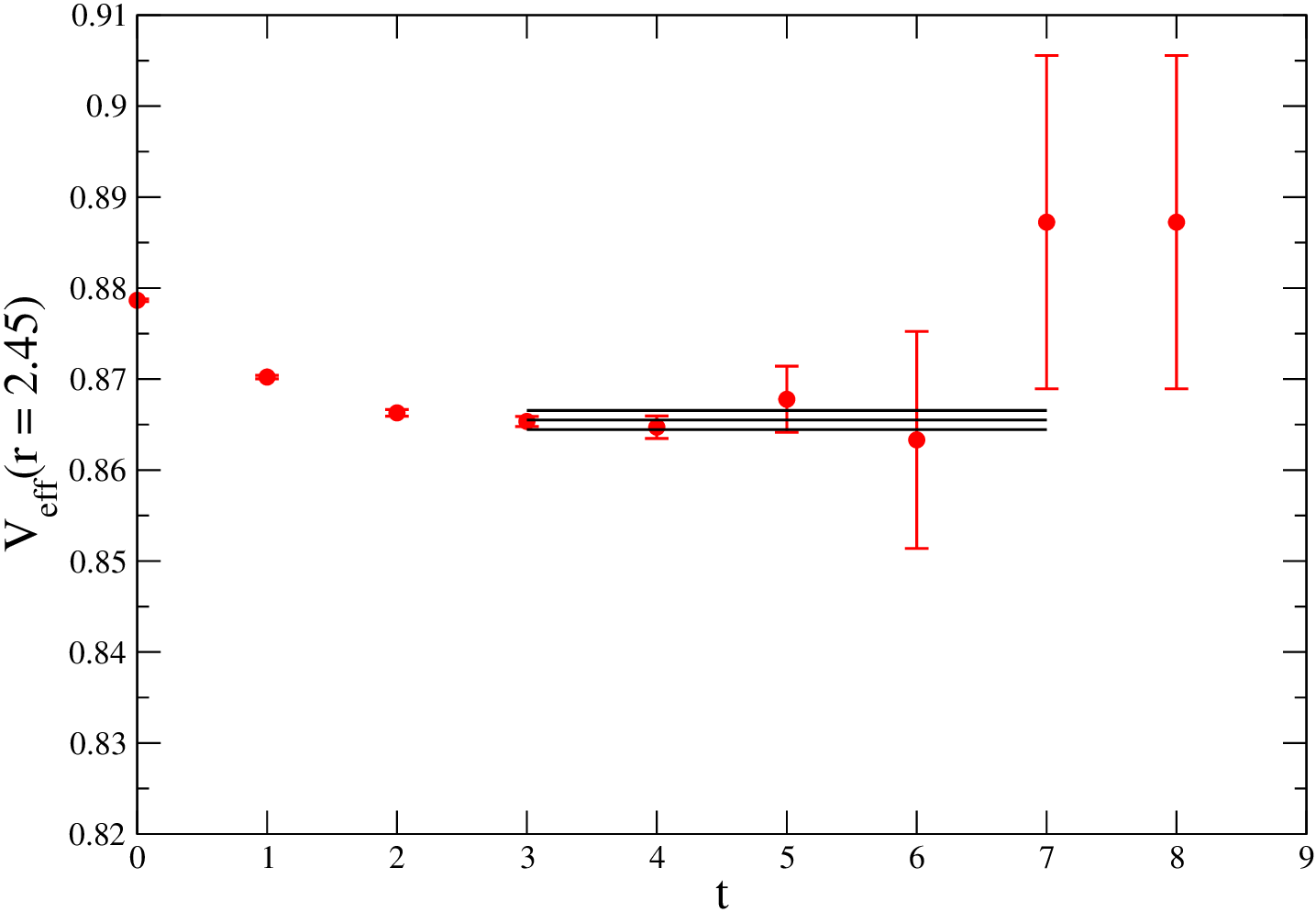}
\includegraphics*[width=0.45\textwidth]{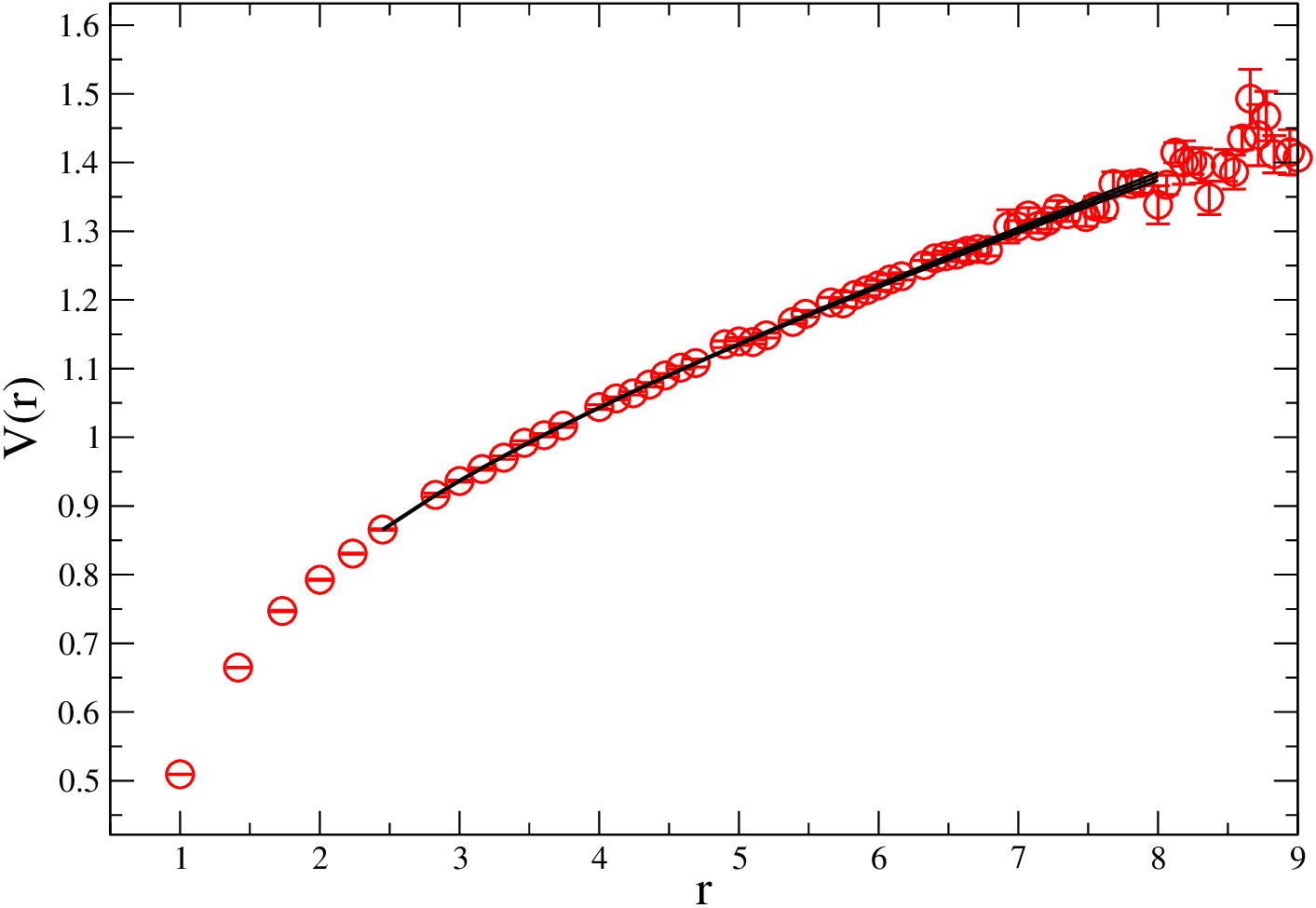}
\caption{
\label{fig-staticpot24}
The effective potential of the Wilson loops with a spatial extent of $r=2.45$ on the $24^3$, $m_l=0.005$ ensemble at the physical strange-quark mass, overlaid by the fit to the range $t=3-7$ (left panel). The right panel shows the static inter-quark potential $V(r)$ on this ensemble, again at the physical strange-quark mass, as a function of the spatial extent of the Wilson loops, overlaid by the fit to the Cornell form over the range $r=2.45-8$.}
\end{figure}

The results for the $\Omega$ mass, $m_{hhh}$, obtained
directly at the simulated
strange-quark mass ($m_h=0.04$) with valence strange-quark masses
$m_y=0.04$ and 0.03 are presented in Table\,\ref{tab-24nt64-momega-r0}.
In this table we also present the results for $m_{hhh}$ obtained after
reweighting to the physical strange-quark mass. In Table\,\ref{tab:r024} we display the values of the Sommer scale $r_0$, $r_1$ and their ratio at both the simulated and physical strange-quark masses. These quantities were determined using Wilson loops formed from products of temporal gauge links with Coulomb gauge-fixed closures in spatial directions, with an exponential fit to the time-dependence of the Wilson loop $W(r,t)$ from $t = 3$ to $t = 7$ for each value of the separation $r$. The resulting potential $V(r)$ was then fit over the range $r = 2.45-8$ to the Cornell form\,\cite{Eichten:1978tg}
\begin{equation}\label{eq:cornell}
V(r)=V_0-\frac{\alpha}{r}+\sigma\,r\,,
\end{equation}
where $V_0$, $\alpha$ and $\sigma$ are constants. These fits are illustrated in Figure~\ref{fig-staticpot24}, which shows the fit to the time dependence of the Wilson loop $W(r=2.45,t)$ at the physical strange-quark mass, and also the subsequent fit over the potential. The strange-quark mass dependence of the scales $r_0$ and $r_1$ is small and cannot be resolved within our statistics.

\subsection{Nucleon and $\Delta$ Masses}
\label{subsec:baryons24}

A detailed study of the baryon mass spectrum, including the continuum and chiral extrapolations, is postponed to a separate paper. The one exception is the $\Omega$ baryon, whose mass is used in the definition of the scaling trajectory and which is therefore studied in detail together with the properties of pseudoscalar mesons. In this subsection we briefly discuss our experiences in extracting the masses of the nucleons and $\Delta$-baryons using different sources and present the results for these masses on each ensemble, starting here with those from the $24^3$ ensembles. The baryon spectrum from the $32^3$ ensembles will be discussed in Sec.\,\ref{subsec:baryons32}.
We start however, with some general comments about our procedures which are relevant to both sets of ensembles.

We use the standard operator, \(N = \epsilon_{abc} (u_a^T C \gamma_5d_b)u_c\), to create and annihilate nucleon states and \(\Delta = \epsilon_{abc} (u^T_a C\gamma_{\mu} u_b) u_c\) for the flavor decuplet \(\Delta\) states.
On an anti-periodic lattice of size \(N_t\) in the time direction, the zero-momentum two-point correlation function, \(C(t)\), calculated with one of these baryonic operators at its source and sink, takes the following asymptotic form for sufficiently large time, \(t\),
\begin{equation}\label{eq:baryonc}
C(t) = Z[(1+\gamma_4) e^{-Mt} - (1-\gamma_4) e^{-M(N_t-t)}] ,
\end{equation}
corresponding to particle and anti-particle propagation, respectively.
Conventionally one chooses an appropriate range in time where the excited-state contributions can be neglected so that this form is valid, and extracts the ground-state mass, \(M\), by fitting the numerical data to the function in Eq.\,(\ref{eq:baryonc}). This is indeed what we do to extract baryon masses from the \(24^3\) ensembles.
Alternatively we can try to fit the correlation function to a sum of two exponentials, representing the ground- and excited-state contributions. As will be reported in Sec.\,\ref{subsec:baryons32}, this is the method we use for the \(32^3\) ensembles.

The determination of baryon masses can be made more effective by an appropriate choice of smearing at the source and/or sink. We use several different choices of the smearing of these operators, wall, box, and gauge-invariant Gaussian~\cite{Alexandrou:1992ti, Berruto:2005hg},
in an attempt to obtain a better overlap with the ground state; our choices are summarized in Table \ref{tab:baryon_ensembles}.
\begin{table*}[tb]
\begin{tabular}{clllll}
\hline
 size  &  $m_l$ & source type & correlators & source time slices & configurations \\
\hline
\hline
$24^3$ & 0.005 & Gaussian & \(N\) & 0,8,16,19,32,40,48,51 &  647\\
	 & 	0.005 & Box & \(\Delta\), \(\Omega\) & 0,32 &  90 \\
	 & 0.01& Gaussian & \(N\) & 0,8,16,19,32,40,48,51 &  357\\
	 & 0.01& Box & \(\Delta\), \(\Omega\) & 0,32 &  90 \\
	 & 0.02 & Gaussian & \(N\) & 0,8,16,19,32,40,48,51 &  99\\
	 & 0.02 & Box & \(\Delta\), \(\Omega\) & 0,32 &  43 \\
	 & 0.03 & Gaussian & \(N\) & 0,8,16,19,32,40,48,51 &  106\\
	 & 0.03 & Box & \(\Delta\), \(\Omega\) & 0,32 &  44 \\
\hline
$32^3$ & 0.004 & Gaussian & \(N\), \(\Delta\) & 10, 26, 42, 58 &  264 \\
	& 0.004 & Wall & \(N\), \(\Delta\) & 0, 16, 32, 48 & 305 \\
	& 0.006 & Wall &  \(N\), \(\Delta\) & 0,16,32, 48 &  224 \\
	& 0.008 &  Gaussian & \(N\), \(\Delta\) & 10, 26, 42, 58 &  169 \\
	& 0.008 &  Wall & \(N\), \(\Delta\) & 0, 16, 32, 48 &  254 \\
\hline
\end{tabular}
\caption{Summary of the configurations used in the calculation of the baryon spectrum.}
\label{tab:baryon_ensembles}
\end{table*}
The wall source, used for the \(32^3\) ensembles, is Coulomb-gauge fixed.
A box source of size 16, also Coulomb-gauge-fixed, is used for the \(24^3\) ensembles.
The Gaussian-source radius is set to 7 lattice units and 100 smearing steps are used for the \(24^3\) ensembles, while the radius is 6 in the \(32^3\) ensembles: these choices are optimized for our nucleon-structure calculations \cite{Yamazaki:2008py,Yamazaki:2009zq,Aoki:2010xg}.

As can also be seen from the table, several steps are taken to reduce the statistical error.
For each configuration, as many as four different time slices are used for the sources, usually separated by 16 lattice units, but occasionally fewer.
Measurements are made as frequently as every tenth trajectory and are averaged into bins of 40 hybrid Monte Carlo time units.

\begin{figure}[tb]
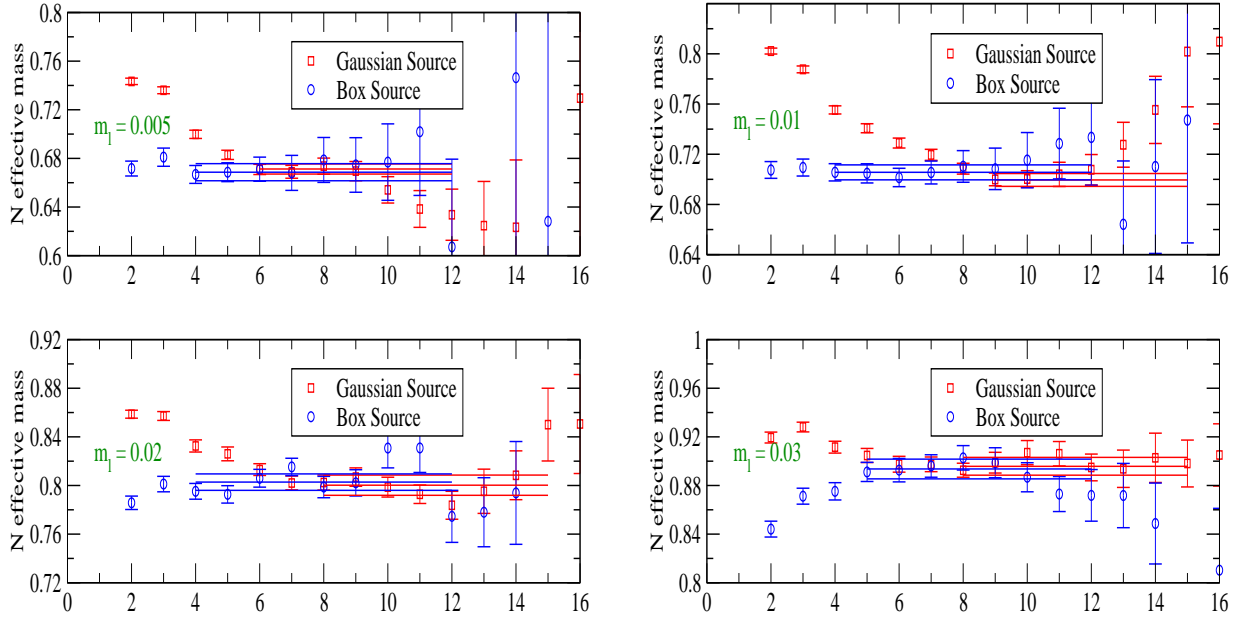

\begin{center}
\includegraphics[width=0.47\hsize, height=1.5in, clip]{fig/Fig16a_N-effm-005.eps}\qquad
\includegraphics[width=0.47\hsize, height=1.5in, clip]{fig/Fig16b_N-effm-01.eps}\\[0.2in]
\includegraphics[width=0.47\hsize, height=1.5in, clip]{fig/Fig16c_N-effm-02.eps}\qquad
\includegraphics[width=0.47\hsize, height=1.5in, clip]{fig/Fig16d_N-effm-03.eps}
\end{center}
\caption{Nucleon effective mass plots from the \(24^3\) ensembles. Results obtained using the Gaussian source are marked by red squares and those from the box source by blue circles. The four plots correspond to unitary light-quark masses 0.005 (top-left), 0.01 (top-right), 0.02 (bottom-left) and 0.03 (bottom-right).}
\label{fig:N_effective_mass24}
\end{figure}

\begin{figure}[tb]
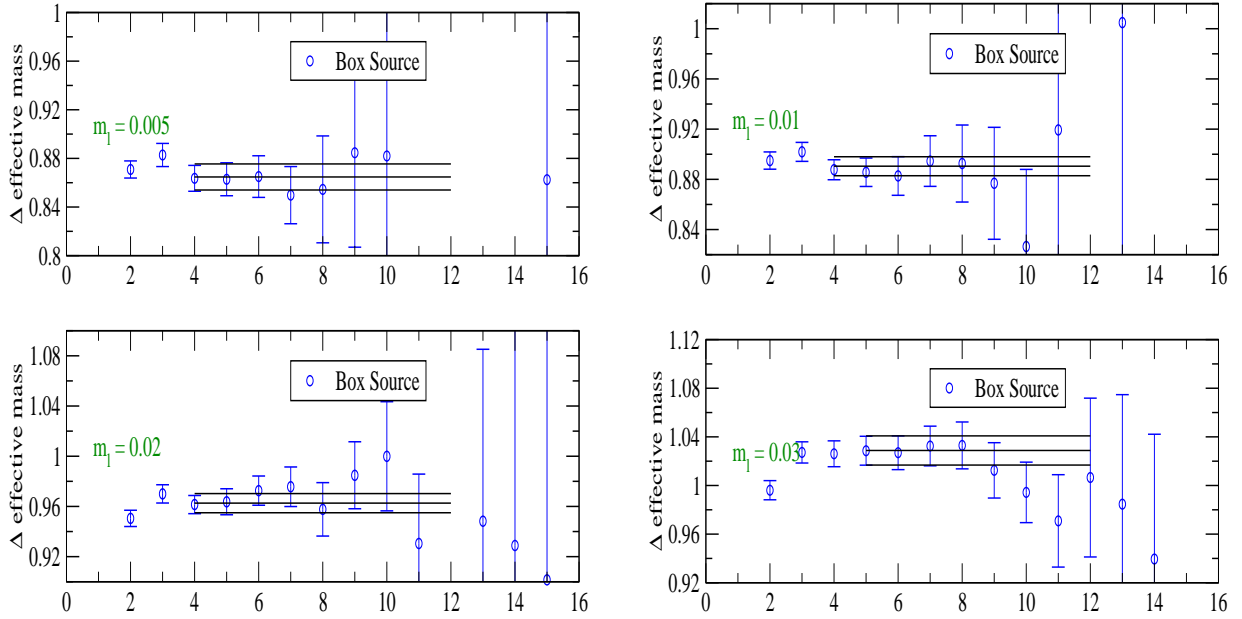

\begin{center}
\includegraphics[width=0.47\hsize, height=1.5in, clip]{fig/Fig17a_D-effm-005.eps}\qquad
\includegraphics[width=0.47\hsize, height=1.5in, clip]{fig/Fig17b_D-effm-01.eps}\\[0.2in]
\includegraphics[width=0.47\hsize, height=1.5in, clip]{fig/Fig17c_D-effm-02.eps}\qquad
\includegraphics[width=0.47\hsize, height=1.5in, clip]{fig/Fig17d_D-effm-03.eps}
\end{center}
\caption{Effective mass plots for the $\Delta$ baryon from the \(24^3\) ensembles. The results were obtained using the box source. The four plots correspond to unitary light-quark masses 0.005 (top-left), 0.01 (top-right), 0.02 (bottom-left) and 0.03 (bottom-right).}
\label{fig:Delta_effective_mass24}
\end{figure}

We now turn to the results obtained specifically on the $24^3$ ensembles.
The unitary nucleon and \(\Delta\) effective masses are plotted in Figs.~\ref{fig:N_effective_mass24}
and~\ref{fig:Delta_effective_mass24} for each choice of quark mass. For the nucleon, both Gaussian and box sources are shown. Plateaus for the effective masses obtained with the box source appear quickly, suggesting a strong overlap with the ground state. The corresponding plateaus obtained with the Gaussian source appear more slowly, from above. Both sets of results agree reasonably well for sufficiently large $t$. For the $\Delta$ the correlators were only computed using the box source and the plateaus for the effective masses again appear quickly.
The results for the masses, obtained using fully correlated fits, are summarized in Table \ref{tab:baryon_spectrum_24}.
Note such fully correlated fits work well for extracting baryon masses as the procedure involves much shorter ranges in time than for the meson observables discussed in the rest of this paper.
As expected from the effective mass plots, nucleon masses obtained using different sources agree fairly well when the fits are performed over appropriate ranges.
All values of $\chi^2$/d.o.f. are close to 1 or smaller, except for the box-source nucleon fit at $m_f=0.02$ which is about 2.5.
\begin{table}[tb]
\begin{center}
\begin{tabular}{lllllll}
\hline
\(m_l\)  &
\multicolumn{2}{c}{\(N\) (Gaussian)}  &
\multicolumn{2}{c}{\(N\) (Box)} &
\multicolumn{2}{c}{\(\Delta\) (Box)} \\
 \hline
  0.005&  0.671(4) & \{6-12\} &0.669(7)& \{4-12\} & 0.865(11) & \{4-12\}  \\
  0.01  &  0.699(5) & \{9-15\} &0.706(6)& \{4-12\} & 0.891(8)  &\{4-12\}\\
  0.02  &  0.800(8) &\{8-15\} & 0.803(7) & \{4-12\} &  0.963(8) & \{4-12\} \\
  0.03  &  0.896(7)  &\{8-15\}& 0.894(8) & \{5-12\} & 1.029(12)  &\{5-12\}  \\
  \hline
\end{tabular}
\end{center}
\caption{Baryon mass in lattice units from the $\beta=2.13$, \(24^3\) ensembles.
\{\} denotes fit range.
}
\label{tab:baryon_spectrum_24}
\end{table}

Some of these results have been reported earlier at Lattice 2008 \cite{Blum:2008zzb}, and also partially in related papers on nucleon structure \cite{Yamazaki:2009zq,Aoki:2010xg}.
A preliminary report on a bootstrap correlated analysis with frozen correlation matrix was presented at Lattice 2009 \cite{Maynard:2010wg} and the results agree with the updated ones given here. 

\section{Results from the $32^3$ Ensembles}
\label{sec:32cubed}

The results for masses, decay constants, $r_0$ and $r_1$ obtained directly on
the $32^3$ lattice are presented in the same format as those from the
$24^3$ ensembles in Section\,\ref{sec:24cubed} and the available
measurements are also summarised in table~\ref{tab:simulation_details}. The results are presented
for three values of the sea light-quark mass $m_l=0.004,\,0.006$ and 0.008
which correspond to unitary pion masses in the range 290\,MeV\,--\,400\,MeV
which we had found to be consistent with SU(2) chiral
perturbation theory on the $24^3$ lattice~\cite{Allton:2008pn}. The valence-quark masses used
in the analysis are $m_{x,y}=0.002,\,0.004,\,0.006,\,0.008,\,0.025$ and
$0.03$. For pseudoscalar quantities we use 305, 312 and 252 measurements separated by
20 trajectories on the 0.004, 0.006 and 0.008 ensembles respectively (see Table\,\ref{tab:simulation_details}).
For the $32^3$ lattices, we have used a single-source technique for our
measurements of pseudoscalar quantities, which differs from the two-source
method for the $24^3$ ensembles. Recall that for the $24^3$ ensembles,
as discussed in Section\,\ref{sec:24cubed}, we placed
Coulomb gauge-fixed wall sources at $t=5$ and at $t=57$.  For the
$32^3$ ensembles we have used a single source and calculated
both periodic and anti-periodic propagators from this one source.
The source is placed at $t = 0$ on the first configuration used for measurements,
and the position of the source is then increased by 16 for every subsequent measurement so that
$t_{\rm src} = 16n \; {\rm mod} \; 64$ where $n$ is the measurement
index, which starts from zero.  Moving the source in this way helps
to decorrelate measurements.  We always place the anti-periodic boundary
condition on the links in the time direction going from the hyperplane with
$t = t_{\rm src} - 1$ to $t = t_{\rm src}$.  Clearly the number of
propagators to calculate for the single source method is half that for
the two-source method.

For meson two-point functions, as given in Eq.\,(\ref{eq:24cubed:pipi:def}),
the single-source method is identical to the two-source method,
except for having half the number of measurements per configuration.
For the light-quark masses on our $32^3$ ensembles we
do see around-the-world effects at the fraction of a percent level,
so fits of the form in Eq.\,(\ref{eq:24cubed:pipi:form}) must be used.
We also perform measurements using three-point functions of the type $\langle
P(x) O(y) P(z) \rangle$, where $P(x)$ and $P(z)$ are pseudoscalar
interpolating fields and $O(y)$ is an operator whose matrix element
we wish to measure.  Here $P(x)$ is made out of propagators of the
form $D^{-1}_{P+A,0}  =  1/2 \left( D^{-1}_{P,0} + D^{-1}_{A,0}
\right)$ in the notation of Eq.\,(\ref{eq:24cubed:src}) and $P(z)$
is composed of $D^{-1}_{P-A,0}  =  1/2 \left( D^{-1}_{P,0} -
D^{-1}_{A,0} \right)$ propagators.  This means that the time
separation between $P(x)$ and $P(z)$ is $N_t$, the time extent of
our lattice.  We performed tests on our $24^3$ ensembles, comparing the
single-source and two-source methods and found that, for the same number of inversions,
the single-source methods gave at least
as small an error as the two-source methods.  The single-source method
allows us to measure on more configurations for the same
computer time and so we chose this method.  Although we do not discuss
three-point measurements in this paper, sharing propagators between them
and the two-point measurements discussed here has helped to define
our measurement strategy.

The measured values of the residual mass $m_\mathrm{res}^\prime$ at each pair of valence and sea light-quark masses ($m_x,m_l$) used in this work are given in table~\ref{tab-mresprime32unrw}; in this table the strange-quark mass is the one used in the simulation $m_h=0.03$. Table~\ref{tab-mresprime32physmh} contains the corresponding results obtained after reweighting to the
physical strange mass ($m_s=0.0273(7)$) determined later in the analysis and presented in Section\,\ref{sec:CombinedChiralFits}. The residual mass in the unitary two-flavor chiral
limit is given in table~\ref{tab-mreschiral} and figure~\ref{fig:mres_chiral}.

\begin{table}[tp]
\centering
\begin{tabular}{c|ccc}
\hline
$m_x$ & \multicolumn{3}{c}{$m_l$}\\
\hline
      &     0.004     &     0.006     &    0.008\\
\hline
0.002 & 0.0006761(35) & 0.0006688(34) & 0.0006822(37) \\
0.004 & 0.0006697(34) & 0.0006651(31) & 0.0006791(36) \\
0.006 & 0.0006622(33) & 0.0006589(30) & 0.0006736(35) \\
0.008 & 0.0006550(32) & 0.0006524(29) & 0.0006676(34) \\
0.025 & 0.0006090(24) & 0.0006089(21) & 0.0006218(25) \\
0.03 & 0.0005993(23) & 0.0005997(20) & 0.0006115(24) \\
\end{tabular}
\caption{$m_\mathrm{res}^\prime$ on the $32^3$ ensemble set at the simulated strange quark mass $m_h=0.03$.}
\label{tab-mresprime32unrw}
\end{table}

\begin{table}[tp]
\centering
\begin{tabular}{c|ccc}
\hline
$m_x$ & \multicolumn{3}{c}{$m_l$}\\
\hline
      &     0.004     &     0.006     &    0.008\\
\hline
0.002 & 0.0006718(39) & 0.0006671(36) & 0.0006781(44) \\
0.004 & 0.0006658(39) & 0.0006633(33) & 0.0006751(42) \\
0.006 & 0.0006586(37) & 0.0006569(31) & 0.0006696(40) \\
0.008 & 0.0006515(36) & 0.0006503(30) & 0.0006636(39) \\
0.025 & 0.0006063(26) & 0.0006058(24) & 0.0006180(31) \\
0.03 & 0.0005967(24) & 0.0005966(22) & 0.0006080(29) \\
\end{tabular}
\caption{$m_\mathrm{res}^\prime$ on the $32^3$ ensemble set at the physical strange quark mass.}
\label{tab-mresprime32physmh}
\end{table}

\begin{figure}[t]
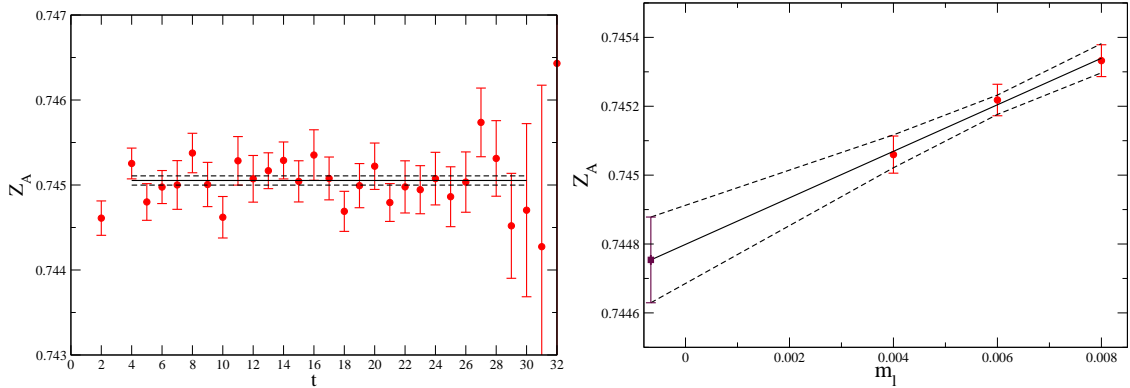

\centering
\includegraphics*[width=0.45\textwidth]{fig/Fig18a_za_32_0.004_0.03.eps}
\includegraphics*[width=0.45\textwidth]{fig/Fig18b_za_32_chiral_0.03.eps}
\caption{
Measurement of $Z_A$ for $m_f=0.004$ on the $m_l=0.004$, $m_h=0.03$
ensemble (left panel) and the unitary chiral extrapolation of $Z_A$ for
the $32^3$ ensemble set (right panel). The results do not change significantly
under reweighting to the physical strange mass.
\label{fig:za32}
}
\end{figure}

The results for $Z_A$ for the $32^3$ ensembles obtained from the ratios of matrix elements of ${\cal A}_4$ and $A_4$ are given in table~\ref{tab-za32unrw}. We obtain $Z_A = 0.74475(12)$ in the chiral limit with the simulated
sea strange mass and $Z_A = 0.74468(13)$ when reweighted to the nearby
physical strange mass. This is illustrated in figure~\ref{fig:za32}. As explained in Section\,\ref{sec:24cubed} and appendix\,\ref{sec:appendix:za} however, in this paper we use $Z_V/Z_{{\cal V}}=0.7396(17)$ as the normalization factor for the local axial current when calculating the central values of physical quantities.

\begin{table}[pt]
\centering
\begin{tabular}{c|cccc}
\hline
$m_h$ &$Z_A(chiral)$& $Z_A(m_l=0.004)$ & $Z_A(m_l=0.006)$ & $Z_A(m_l=0.008)$ \\
\hline
$m_h^\mathrm{sim} = 0.03$ & 0.74475(12) & 0.745053(54) & 0.745222(45) & 0.745328(48) \\
$m_h^\mathrm{phys}$ & 0.74469(13) & 0.745059(52) & 0.745239(47) & 0.745384(56)\\
\end{tabular}
\caption{\label{tab-za32unrw}
$Z_A$ on the $32^3$ ensembles at the simulated and physical strange sea-quark masses.}
\end{table}

\begin{figure}[hbt]
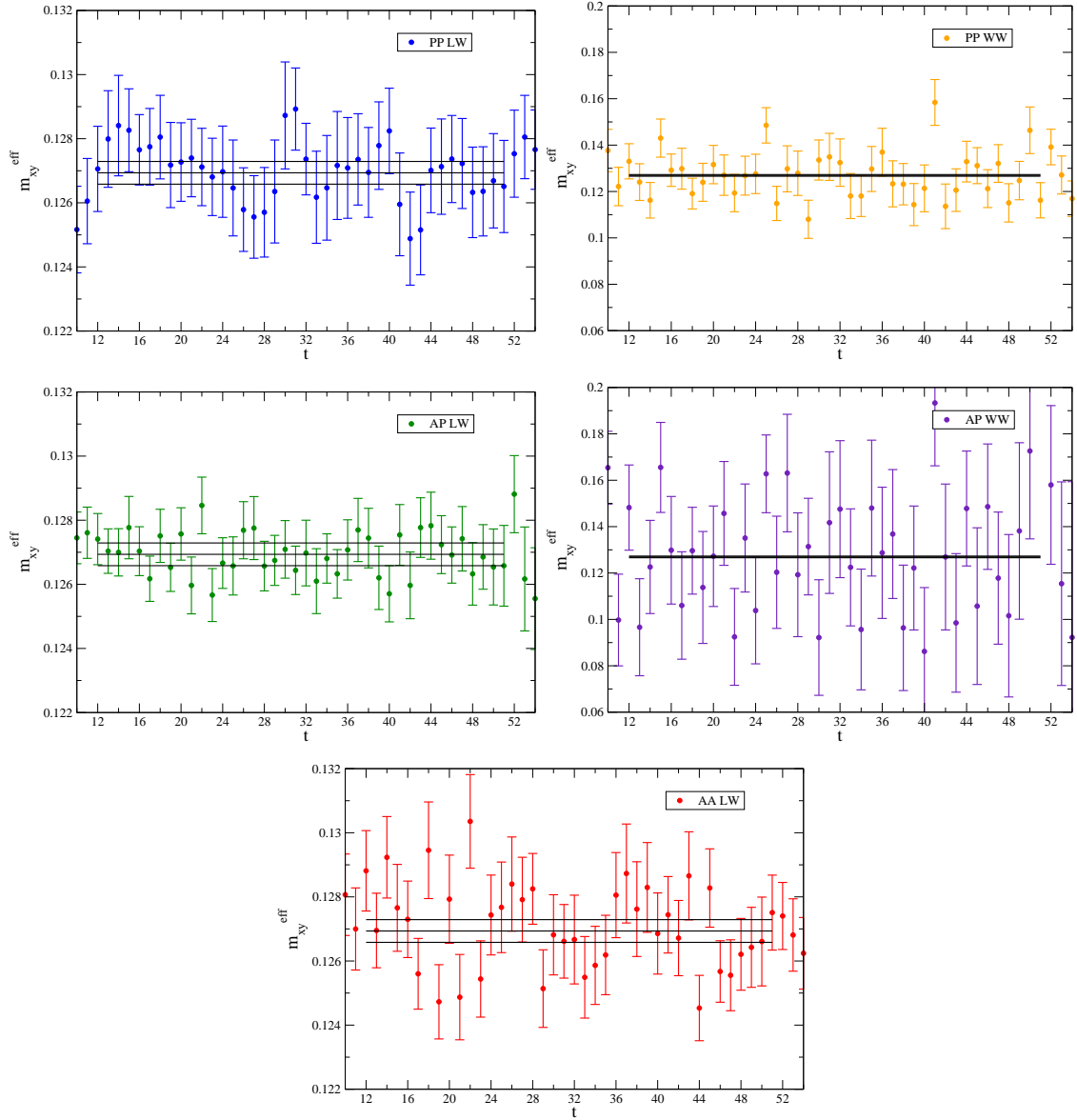

\centering
\includegraphics*[width=0.45\textwidth]{fig/Fig19a_mpi_32_0.004_0.03_ppwl.eps}
\includegraphics*[width=0.45\textwidth]{fig/Fig19b_mpi_32_0.004_0.03_ppww.eps}\\[0.1in]
\includegraphics*[width=0.45\textwidth]{fig/Fig19c_mpi_32_0.004_0.03_apwl.eps}
\includegraphics*[width=0.45\textwidth]{fig/Fig19d_mpi_32_0.004_0.03_apww.eps}\\[0.1in]
\includegraphics*[width=0.45\textwidth]{fig/Fig19e_mpi_32_0.004_0.03_aawl.eps}
\caption{
\label{fig:pionmeff32}
Effective pion masses from the
PP LW correlator (top left), PP WW correlator (top right),
AP LW correlator (center left), AP WW (center right) and
AA LW correlator (bottom). Note the different vertical scale for the WW correlators. The horizontal bands represent the result for the mass from a simultaneous fit.
}
\end{figure}

\begin{figure}[hbt]
\centering
\includegraphics*[width=0.45\textwidth]{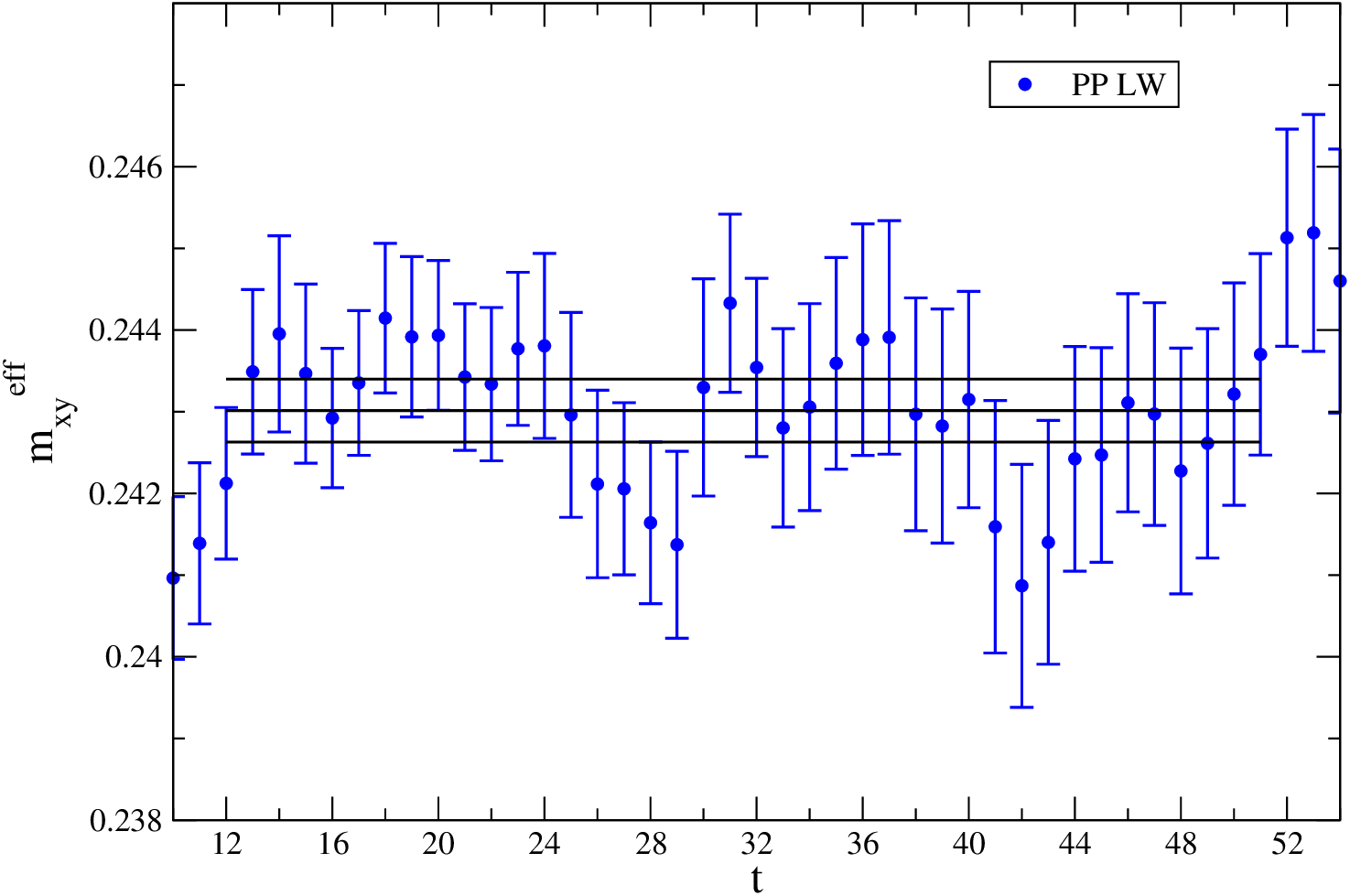}
\includegraphics*[width=0.45\textwidth]{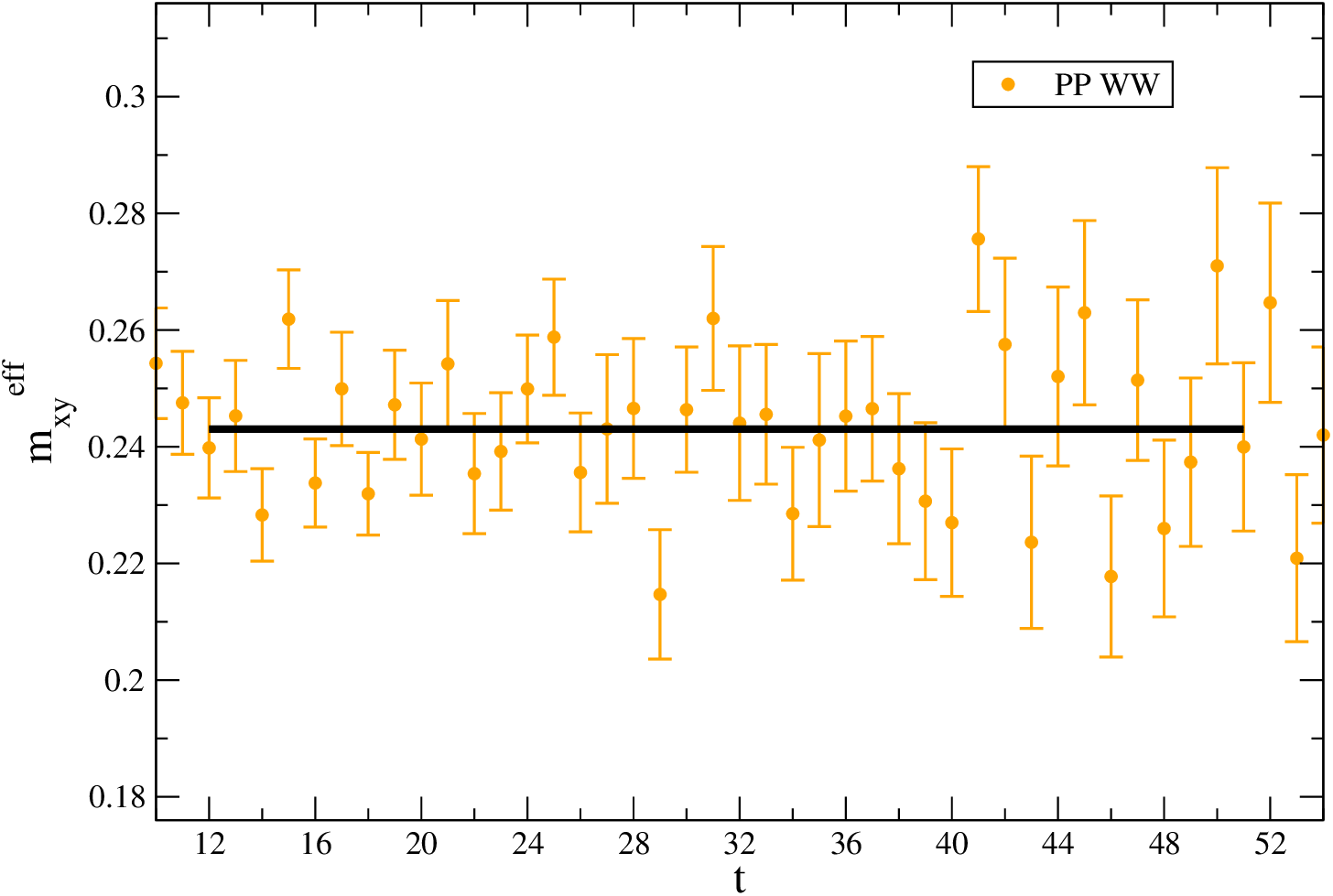}\\[0.1in]
\includegraphics*[width=0.45\textwidth]{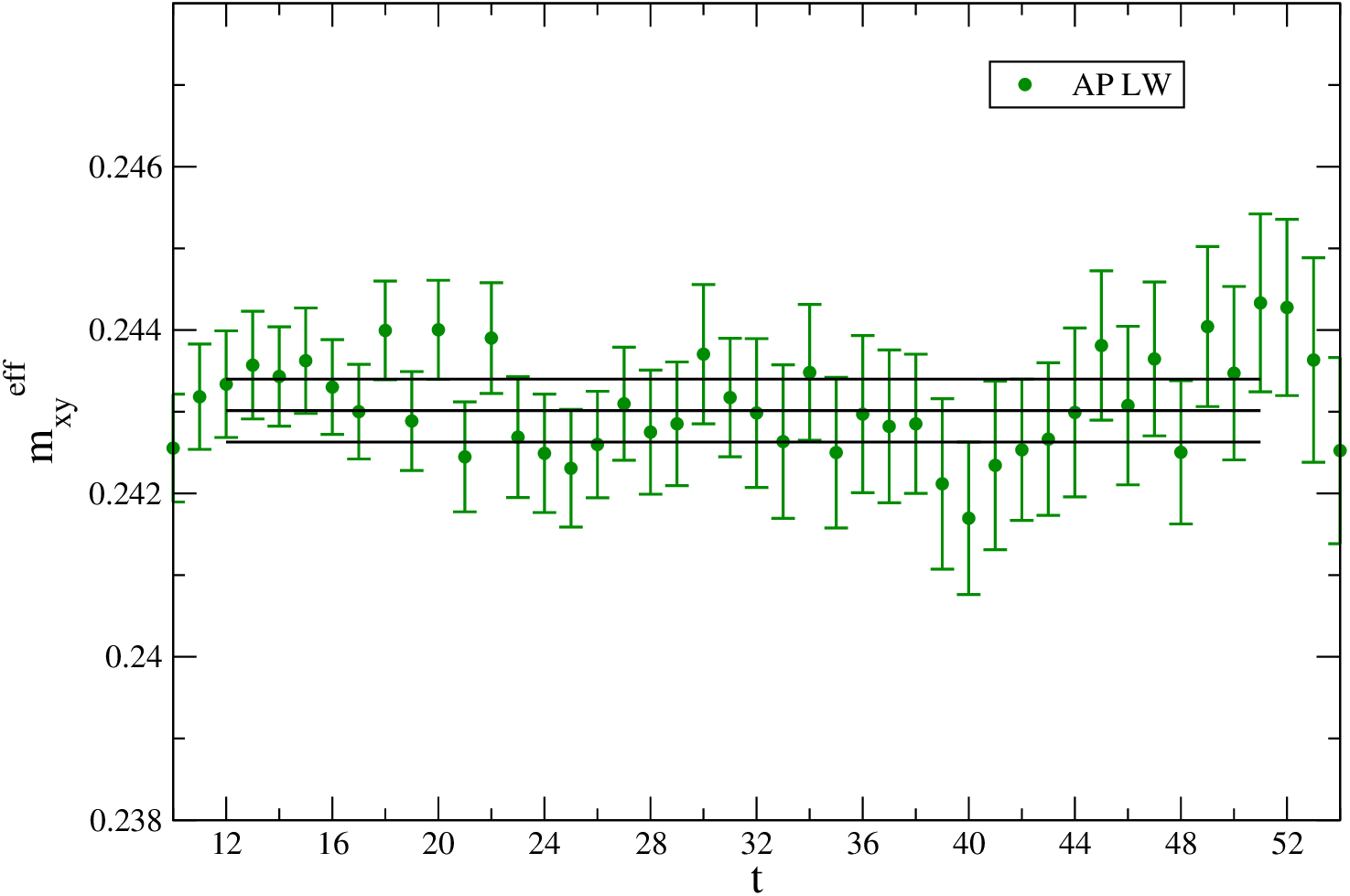}
\includegraphics*[width=0.45\textwidth]{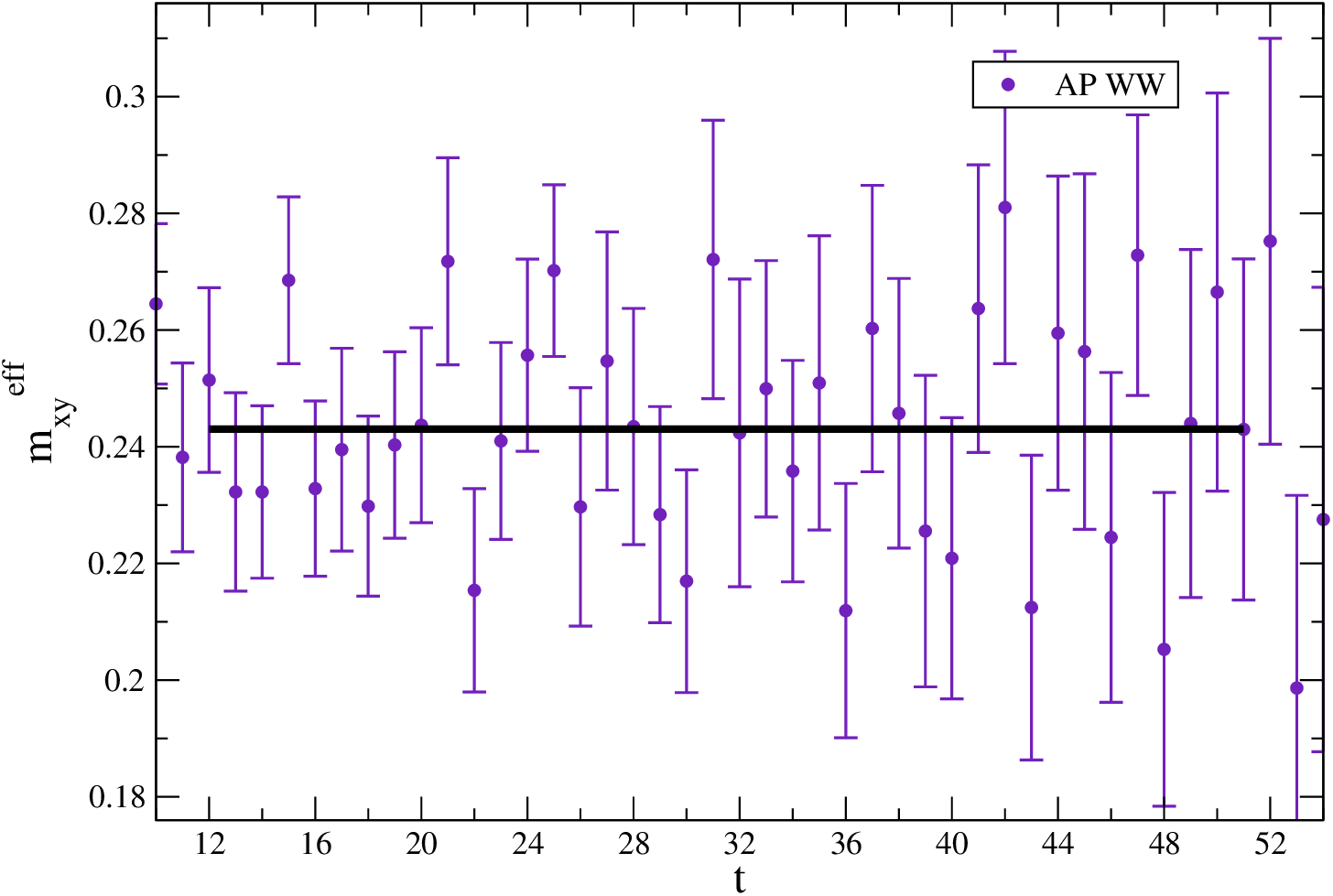}\\[0.1in]
\includegraphics*[width=0.45\textwidth]{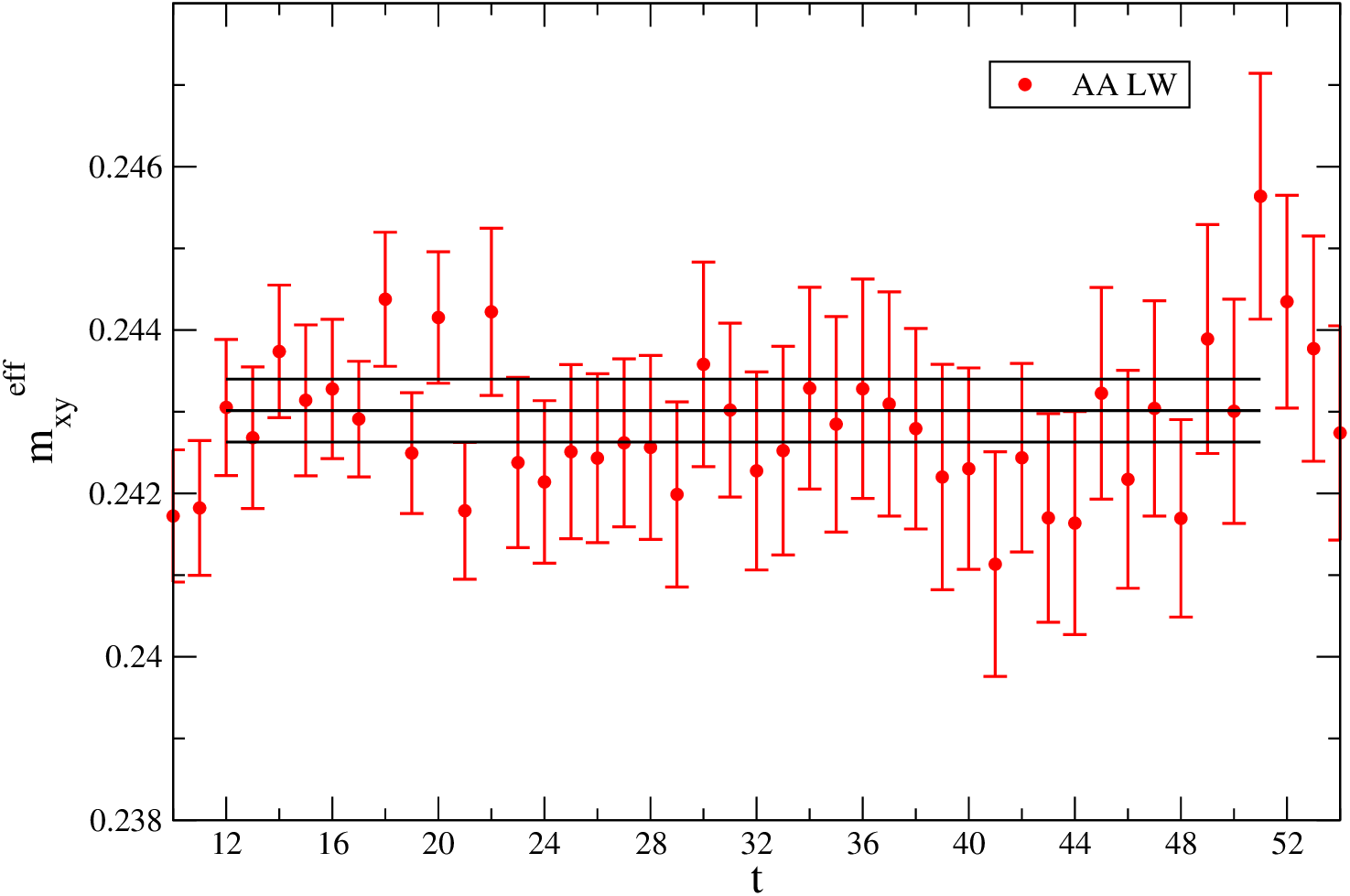}
\caption{
\label{fig:kaonmeff32}
Effective kaon masses from the
PP LW correlator (top left), PP WW correlator (top right),
AP LW correlator (center left), AP WW (center right) and
AA LW correlator (bottom). Note the different vertical scale for the WW correlators. The horizontal bands represent the result for the mass from a simultaneous fit.
}
\end{figure}

\begin{figure}[hbt]
\centering
\includegraphics*[width=0.45\textwidth]{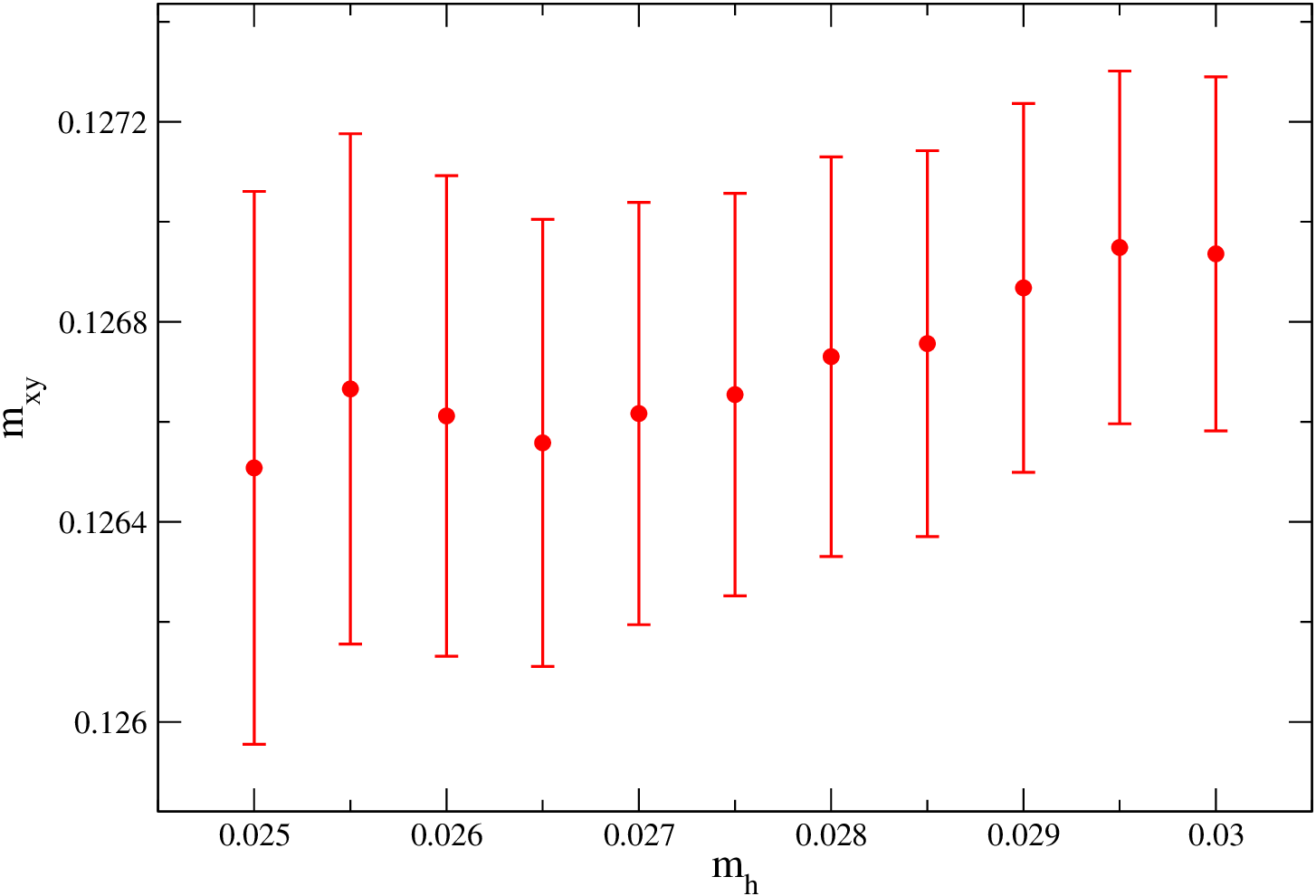}
\includegraphics*[width=0.45\textwidth]{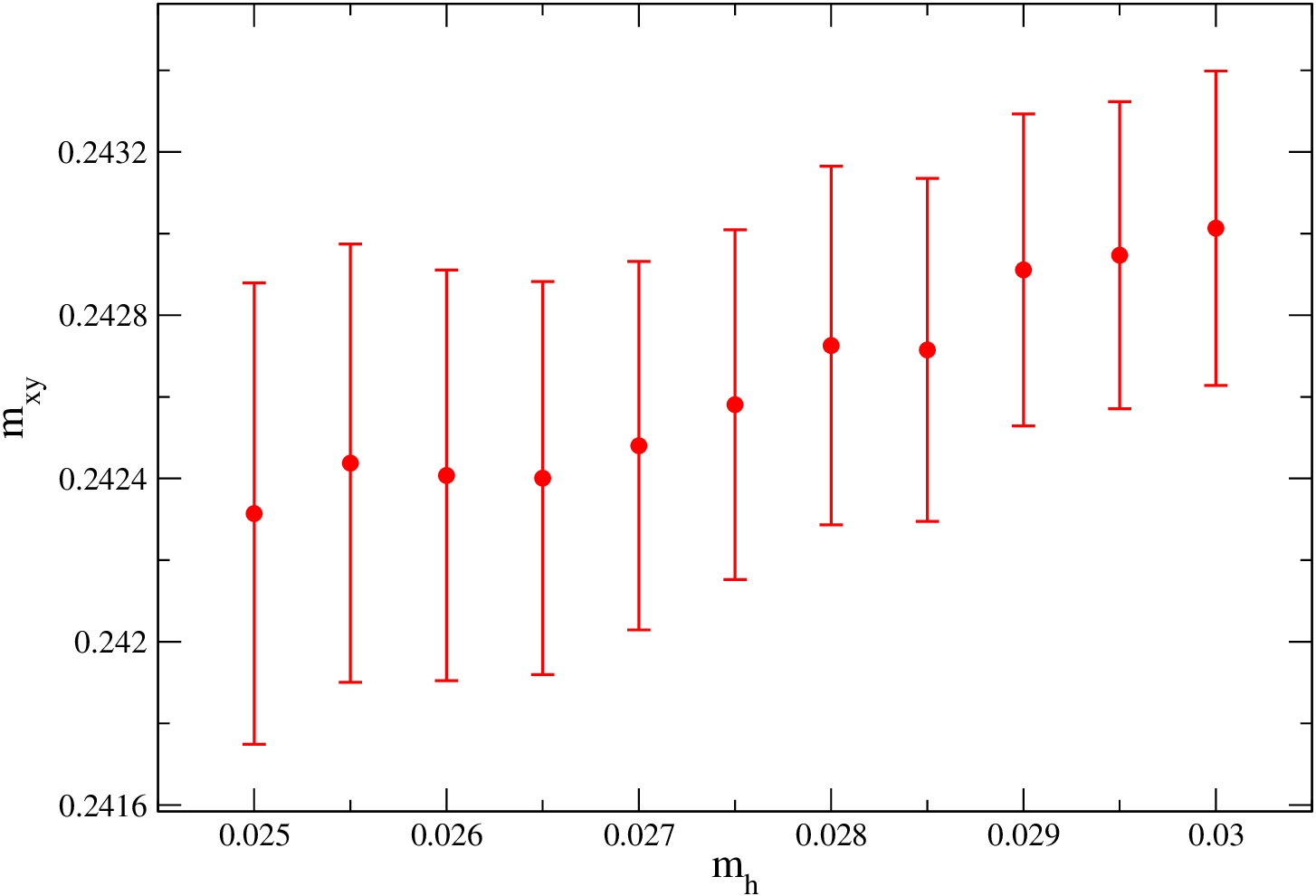}
\caption{
\label{fig:pionreweight32}
We illustrate the $m_h$ dependence of the unitary pion (left panel) and kaon (right panel) masses
on the $m_l=0.004$,  $32^3$ ensemble. The values are obtained by reweighting around the simulated value
($m_h=0.03$).}
\end{figure}

In order to illustrate the quality of the fits, we present sample effective mass plots for the unitary simulated
pion on the $m_l=0.004$, $m_h=0.03$ ensemble in
figure~\ref{fig:pionmeff32} and for the kaon in
Figure~\ref{fig:kaonmeff32}. The analysis is
performed as a simultaneous constrained fit to the five pseudoscalar
channels as for the $24^3$ ensembles (see Section\,\ref{sec:24cubed}).
The fits are performed between $t_{{\rm min}}=12$ and $t_{{\rm max}}=51$.
We give an example of the reweighted $m_h$ dependence of the unitary
pion and kaon masses in figure~\ref{fig:pionreweight32}.

Table~\ref{tab-32nt64-mpi-unreweighted} contains the measured
pseudoscalar masses and decay constants at the simulated strange-quark mass
$m_h=0.03$. Reweighting to the estimated physical strange-quark mass
$m_h=0.0273(7)$, we obtain the masses and decay constants of the pions
and kaons in Tables\,\ref{tab-32nt64-mpi-physmh} and
\ref{tab-32nt64-mk-physmh} respectively.

\begin{figure}[hbt]
\centering
\includegraphics*[width=0.45\textwidth]{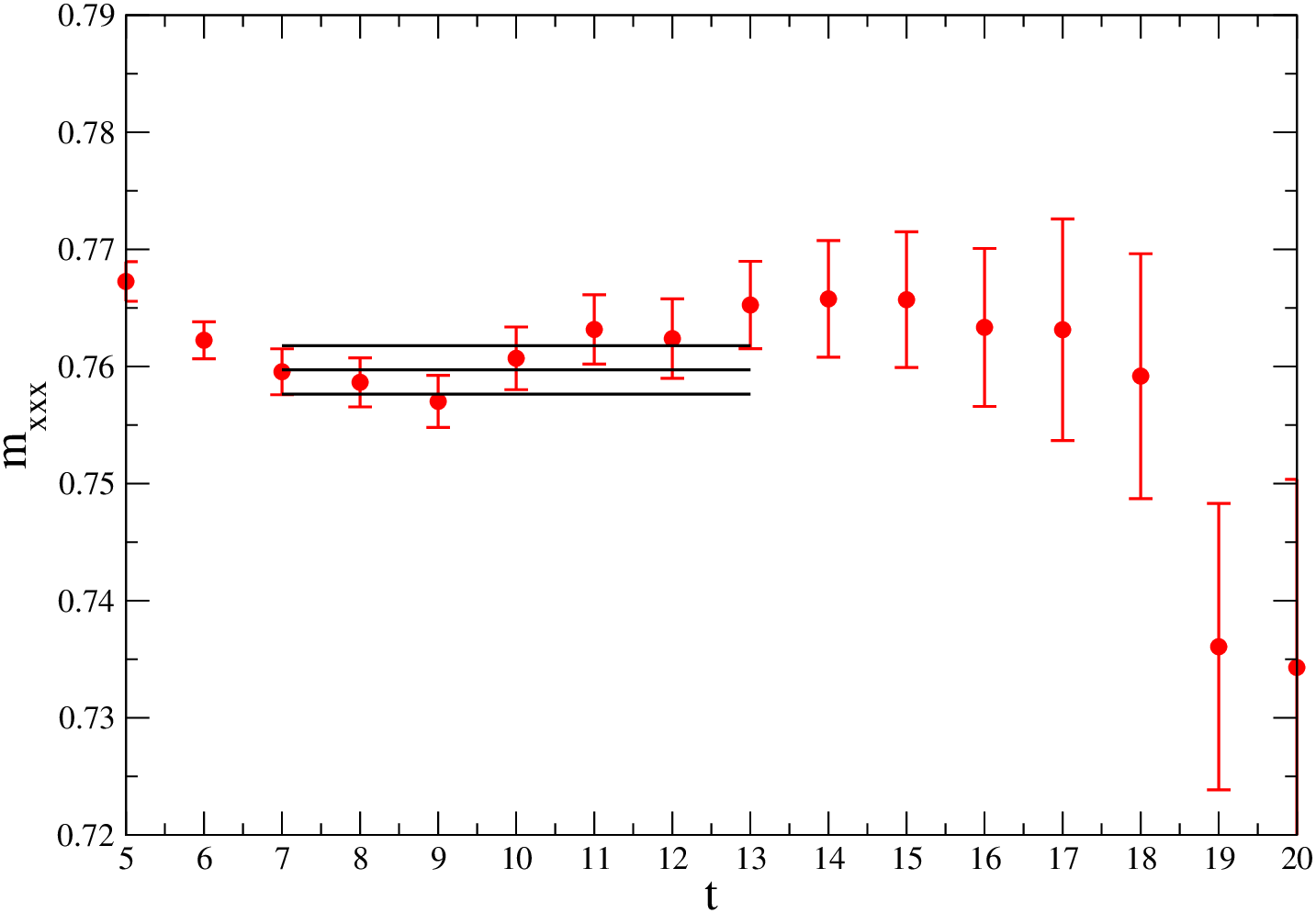}
\includegraphics*[width=0.45\textwidth]{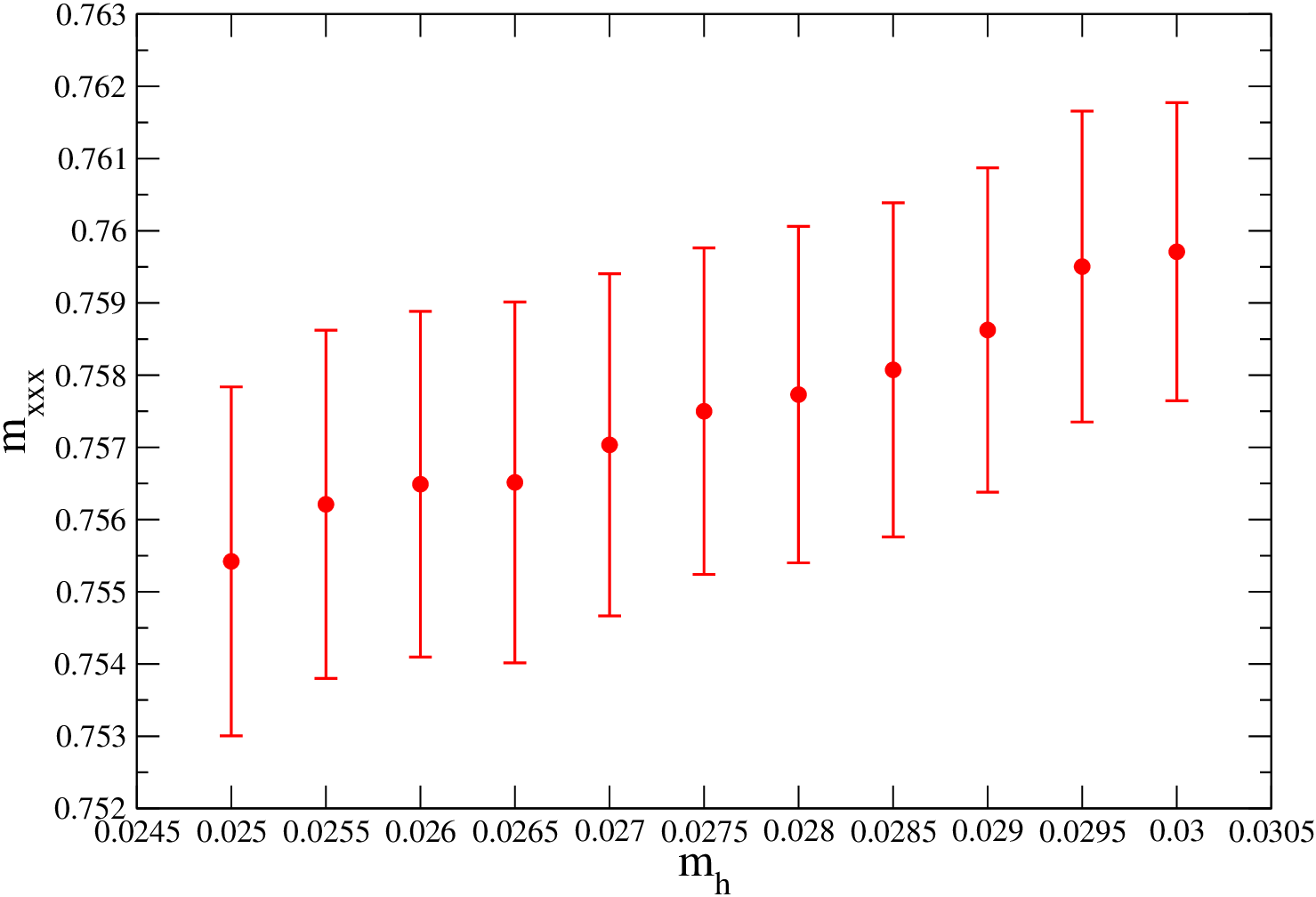}
\caption{
\label{fig:omegaeff32}
We display the fit to the $\Omega$ baryon mass with valence
strange mass $m_x=0.03$ on the $m_l=0.004$, $m_h=0.03$, $32^3$ ensemble
showing the quality of the fit with our box source (left panel).
We also show the weak dependence of the $\Omega$ baryon mass with
fixed valence mass $m_x=0.03$ on our simulated $m_h$ inferred
by the reweighting procedure on the $m_l=0.004$, $32^3$ ensemble
(right panel).
}
\end{figure}

We use a gauge fixed box source of size 24 for the $\Omega$ baryon
using the same configurations as for our pion measurements with valence strange-quark masses $m_x=0.03$ and $m_x=0.025$ to enable an interpolation to
the physical strange-quark mass.
We display the fit to the $m_x=0.03$ $\Omega$ baryon mass
on the $m_l=0.004$, $m_h=0.03$ ensemble in figure~\ref{fig:omegaeff32},
along with the dependence of this mass on the dynamical strange mass
under reweighting. We take our fitting range
between $t_{{\rm min}}=7$ and $t_{{\rm max}}=13$.

\begin{table}[h]
\centering
\begin{tabular}{cc|ccc|ccc}
\hline
$m_x$ & $m_y$ & $m_{xy}(0.004)$  & $m_{xy}(0.006)$ & $m_{xy}(0.008)$&
$f_{xy}(0.004)$  & $f_{xy}(0.006)$ & $f_{xy}(0.008)$\\
\hline
0.03 & 0.03 & 0.3212(3) & 0.3216(2) & 0.3224(3)& 0.0801(3) & 0.0804(3) & 0.0809(3)\\
0.025 & 0.03 & 0.3073(3) & 0.3078(2) & 0.3086(3)& 0.0786(3) & 0.0789(3) & 0.0794(3)\\
0.008 & 0.03 & 0.2561(3) & 0.2565(2) & 0.2579(4)& 0.0723(3) & 0.0729(3) & 0.0738(3)\\
0.006 & 0.03 & 0.2496(3) & 0.2500(3) & 0.2516(4)& 0.0715(3) & 0.0721(3) & 0.0731(3)\\
0.004 & 0.03 & 0.2430(4) & 0.2434(3) & 0.2452(5)& 0.0707(3) & 0.0714(3) & 0.0725(3)\\
0.002 & 0.03 & 0.2363(5) & 0.2367(3) & 0.2388(6)& 0.0701(3) & 0.0709(4) & 0.0723(4)\\
0.025 & 0.025 & 0.2930(3) & 0.2934(2) & 0.2943(3)& 0.0770(3) & 0.0775(3) & 0.0780(3)\\
0.008 & 0.025 & 0.2392(3) & 0.2396(2) & 0.2410(4)& 0.0709(3) & 0.0715(3) & 0.0724(3)\\
0.006 & 0.025 & 0.2323(3) & 0.2327(3) & 0.2342(4)& 0.0701(3) & 0.0707(3) & 0.0717(3)\\
0.004 & 0.025 & 0.2252(4) & 0.2256(3) & 0.2273(5)& 0.0693(3) & 0.0700(3) & 0.0711(3)\\
0.002 & 0.025 & 0.2180(4) & 0.2184(3) & 0.2203(5)& 0.0686(3) & 0.0695(3) & 0.0708(4)\\
0.008 & 0.008 & 0.1708(3) & 0.1714(2) & 0.1727(4)& 0.0649(3) & 0.0657(3) & 0.0666(3)\\
0.006 & 0.008 & 0.1610(3) & 0.1616(3) & 0.1629(4)& 0.0641(3) & 0.0648(3) & 0.0659(3)\\
0.004 & 0.008 & 0.1506(3) & 0.1513(3) & 0.1526(4)& 0.0633(3) & 0.0640(3) & 0.0651(3)\\
0.002 & 0.008 & 0.1395(4) & 0.1403(3) & 0.1417(4)& 0.0625(3) & 0.0634(3) & 0.0646(4)\\
0.006 & 0.006 & 0.1505(3) & 0.1512(3) & 0.1525(4)& 0.0633(3) & 0.0640(3) & 0.0651(3)\\
0.004 & 0.006 & 0.1393(3) & 0.1400(3) & 0.1413(4)& 0.0624(3) & 0.0632(3) & 0.0643(3)\\
0.002 & 0.006 & 0.1271(4) & 0.1280(3) & 0.1293(4)& 0.0615(3) & 0.0624(3) & 0.0637(4)\\
0.004 & 0.004 & 0.1269(4) & 0.1278(3) & 0.1291(4)& 0.0614(3) & 0.0623(3) & 0.0634(3)\\
0.002 & 0.004 & 0.1133(4) & 0.1144(3) & 0.1156(4)& 0.0605(3) & 0.0614(3) & 0.0627(4)\\
0.002 & 0.002 & 0.0976(4) & 0.0989(4) & 0.1001(5)& 0.0595(3) & 0.0603(3) & 0.0617(4)\\
\end{tabular}
\caption{Pseudoscalar masses $m_{xy}(m_l)$ and the decay constants $f_{xy}(m_l)$ on the $32^3$ ensembles  
at the simulated strange-quark mass ($m_h=0.03$).}
\label{tab-32nt64-mpi-unreweighted}
\end{table}


\begin{table}[t]
\centering
\begin{tabular}{cc|ccc|ccc}
\hline 
$m_x$ & $m_y$ &\, $m_{xy}(0.004)$  & $m_{xy}(0.006)$ & $m_{xy}(0.008)$
&\, $f_{xy}(0.004)$  & $f_{xy}(0.006)$ & $f_{xy}(0.008)$\\ 
\hline
0.008 & 0.008 & 0.1706(3) & 0.1711(3) & 0.1725(5) & 0.0645(3) & 0.0653(3) & 0.0662(4)\\
0.006 & 0.008 & 0.1608(4) & 0.1613(3) & 0.1628(5) & 0.0636(3) & 0.0645(4) & 0.0654(4)\\
0.004 & 0.008 & 0.1503(4) & 0.1510(3) & 0.1526(5) & 0.0628(4) & 0.0636(4) & 0.0647(4)\\
0.002 & 0.008 & 0.1392(4) & 0.1401(3) & 0.1417(5) & 0.0620(4) & 0.0630(4) & 0.0641(4)\\
0.006 & 0.006 & 0.1503(4) & 0.1509(3) & 0.1524(5) & 0.0628(4) & 0.0636(4) & 0.0646(4)\\
0.004 & 0.006 & 0.1390(4) & 0.1398(3) & 0.1414(5) & 0.0619(4) & 0.0628(4) & 0.0638(4)\\
0.002 & 0.006 & 0.1268(4) & 0.1278(3) & 0.1295(5) & 0.0611(4) & 0.0620(4) & 0.0632(4)\\
0.004 & 0.004 & 0.1267(4) & 0.1276(3) & 0.1292(5) & 0.0609(4) & 0.0618(4) & 0.0630(4)\\
0.002 & 0.004 & 0.1131(4) & 0.1142(4) & 0.1158(5) & 0.0601(4) & 0.0610(4) & 0.0622(4)\\
0.002 & 0.002 & 0.0974(4) & 0.0988(4) & 0.1003(5) & 0.0590(4) & 0.0598(4) & 0.0612(5)\\
\end{tabular}
\caption{Pion masses $m_{xy}(m_l)$ and decay constants $f_{xy}(m_l)$ computed on the $32^3$ ensembles  
at the physical strange-quark mass $m_h = 0.0273(7)$.}
\label{tab-32nt64-mpi-physmh}
\end{table}

\begin{table}[hbt]
\centering
\begin{tabular}{c|ccc|ccc}
\hline
$m_x$ &\,$m_{xh}(0.004)$  & $m_{xh}(0.006)$ & $m_{xh}(0.008)$ & 
\,$f_{xh}(0.004)$  & $f_{xh}(0.006)$ & $f_{xh}(0.008)$\\ 
\hline
0.008 & 0.247(2) & 0.247(3) & 0.249(3) & 0.0712(4) & 0.0718(5) & 0.0727(5)\\
0.006 & 0.240(2) & 0.240(3) & 0.242(3) & 0.0703(4) & 0.0710(5) & 0.0720(5)\\
0.004 & 0.233(3) & 0.234(3) & 0.235(3) & 0.0695(4) & 0.0703(5) & 0.0713(5)\\
0.002 & 0.226(3) & 0.227(3) & 0.229(3) & 0.0687(5) & 0.0698(5) & 0.0710(6)
\end{tabular}
\caption{Kaon masses $m_{xh}(m_l)$ and decay constants $f_{xy}(m_l)$ on the $32^3$ ensembles  
at the physical strange-quark mass $m_h = 0.0273(7)$.}
\label{tab-32nt64-mk-physmh}
\end{table}

\begin{table}[h]
\centering
\begin{tabular}{cc|ccc}
\hline
$m_y$ & $m_h$&\,$m_{\Omega}(0.004)$  & $m_{\Omega}(0.006)$ & $m_{\Omega}(0.008)$
\\
\hline
0.03 & 0.03 & 0.760(2) & 0.765(2) & 0.766(3)\\
0.025 & 0.03& 0.733(2) & 0.739(2) & 0.740(3)\\ 
0.0273& 0.0273&0.743(6) & 0.749(5) & 0.753(4)
\end{tabular}
\caption{Omega baryon masses on the $32^3$ ensembles at the simulated strange quark mass $m_h=0.03$ (first two rows) and at the physical strange-quark mass (third row).}
\label{tab-32nt64-momega-r0}
\end{table}

\begin{table}[tp]
\centering
\begin{tabular}{c|ccc|ccc}
\hline
\multirow{2}{*}{Quantity}&\multicolumn{3}{c|}{$m_h=0.03$}&\multicolumn{3}{c}{$m_h=0.0273$}\\ 
& $Q(0.004)$ & $Q(0.006)$ &$Q(0.008)$ & $Q(0.004)$ & $Q(0.006)$ &$Q(0.008)$\\
\hline

$r_0$& 5.52(2) & 5.50(2) & 5.53(2)&5.52(2) & 5.52(2) & 5.55(2)\\
$r_1$& 3.738(9) & 3.718(8) & 3.707(9)&3.754(12) & 3.728(9) & 3.723(10) \\
$r_1/r_0$& 0.678(2) & 0.676(2) & 0.670(2)& 0.680(2) & 0.675(2) & 0.670(2)
\end{tabular}
\caption{The quantities $r_0,\, r_1$ and $r_1/r_0$ at the simulated ($m_h=0.03$) 
and physical ($m_h=0.0273$) strange quark masses on the $32^3$ ensembles. $Q(m_l)$ 
denotes the quantity measured with light-quark mass $m_l$.}
\label{tab:r032}
\end{table}

\begin{figure}[tp]
\centering
\includegraphics*[width=0.45\textwidth]{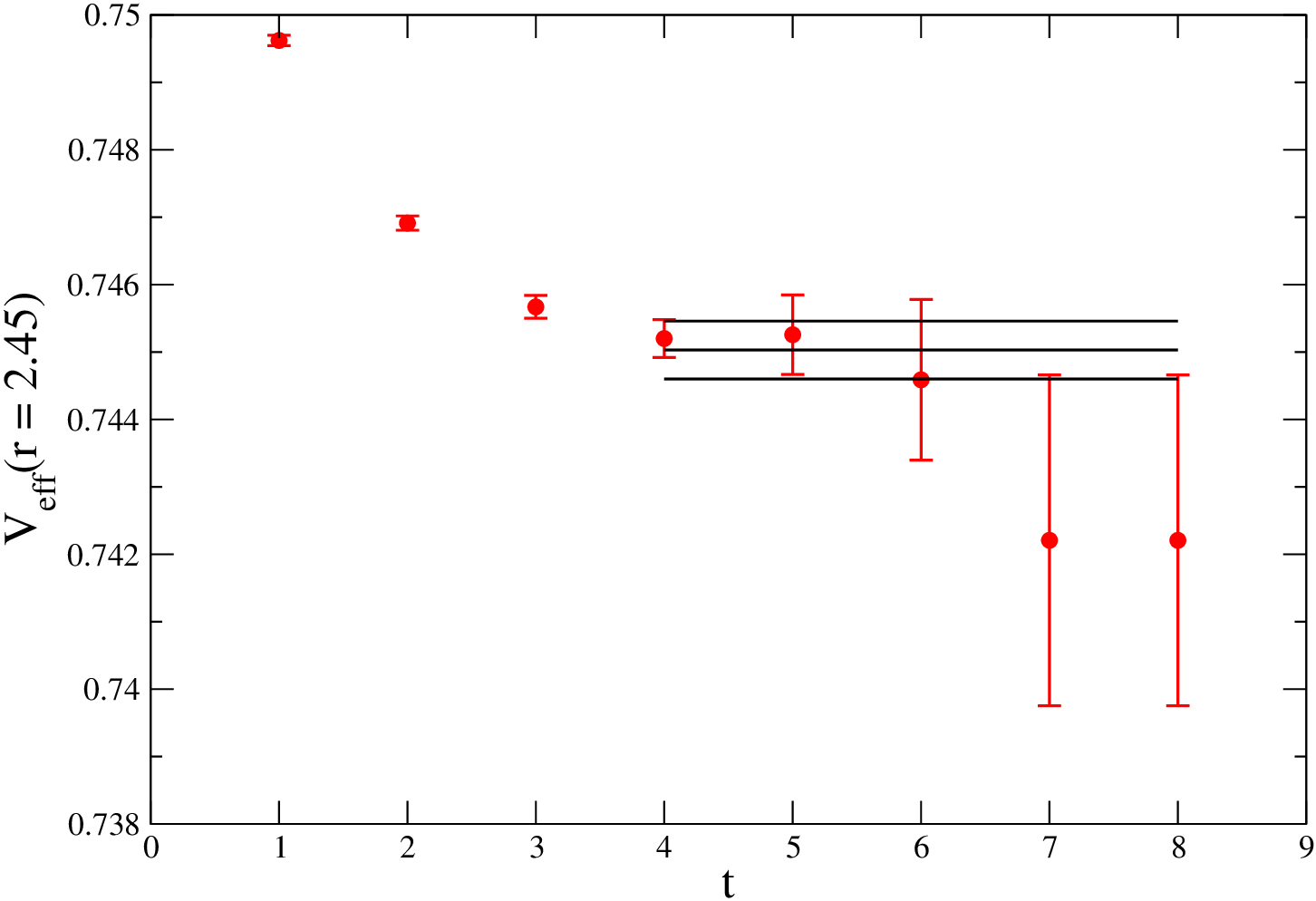}
\includegraphics*[width=0.45\textwidth]{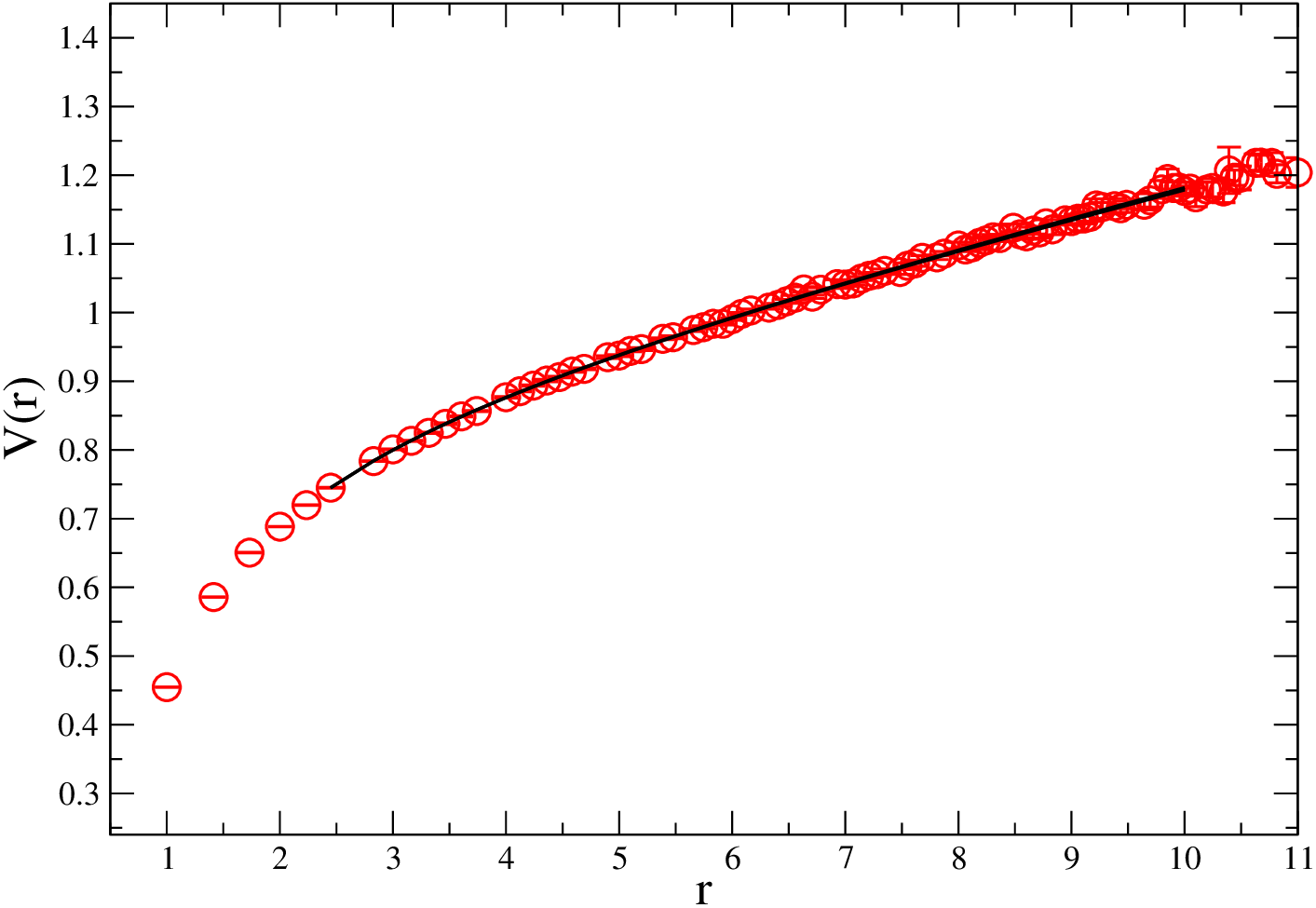}

\caption{
\label{fig-staticpot32}
The effective potential of the Wilson loops with a spatial extent of $r=2.45$ on the $m_l=0.004$ ensemble at the physical strange-quark mass, overlaid by the fit to the range $t=4-8$ (left panel). The right panel shows the static inter-quark potential $V(r)$ on this ensemble, again at the physical strange-quark mass, as a function of the spatial extent of the Wilson loops, overlaid by the fit to the Cornell form over the range $r=2.45-10$.}
\end{figure}

The results for the masses
of the $\Omega$ baryon and the scales $r_0$, $r_1$ and $r_1/r_0$ are given in
Table\,\ref{tab-32nt64-momega-r0} and \ref{tab:r032} respectively. $r_0$ and $r_1$ were determined again using
Wilson loops formed from products of temporal gauge links with
Coulomb gauge-fixed closures in spatial directions, with an exponential
fit from $t=4$ to $t=8$ and the resulting potential fit to the Cornell form
in the range $r=2.45-10$. An example of the fit to the time dependence of the Wilson loops at the physical strange-quark mass is given in Figure~\ref{fig-staticpot32}. This figure also shows the fit to the potential. On these ensembles, the strange-quark mass dependence of $r_0$ and $r_1$ can be resolved within the statistics, but remains small.

\subsection{Nucleon and $\Delta$ Masses}
\label{subsec:baryons32}

\begin{figure}[t]
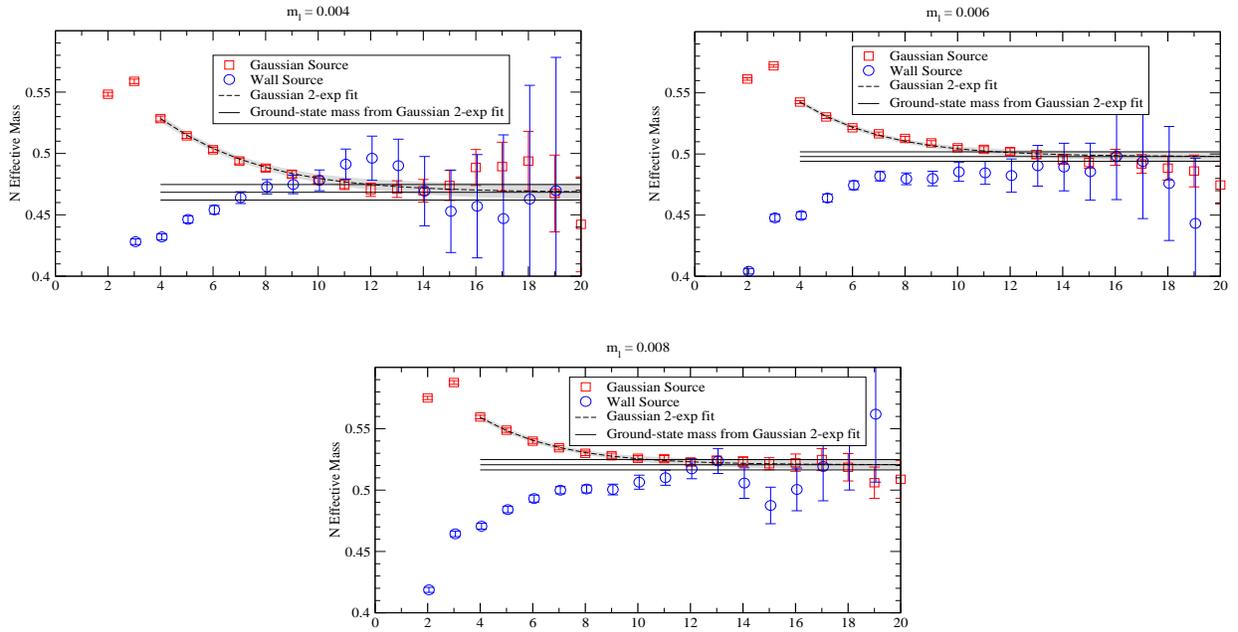

\begin{center}
\includegraphics*[width=0.47\hsize,height=1.5in]{fig/Fig24a_nucleon_paper_32c_004.eps}\qquad
\includegraphics*[width=0.47\hsize,height=1.5in]{fig/Fig24b_nucleon_paper_32c_006.eps}\\[0.25in]
\includegraphics*[width=0.47\hsize,height=1.5in]{fig/Fig24c_nucleon_paper_32c_008.eps}
\end{center}
\caption{Nucleon effective mass plots from the \(32^3\) ensembles.}
\label{fig:N_effective_mass32}
\end{figure}

\begin{figure}[t]
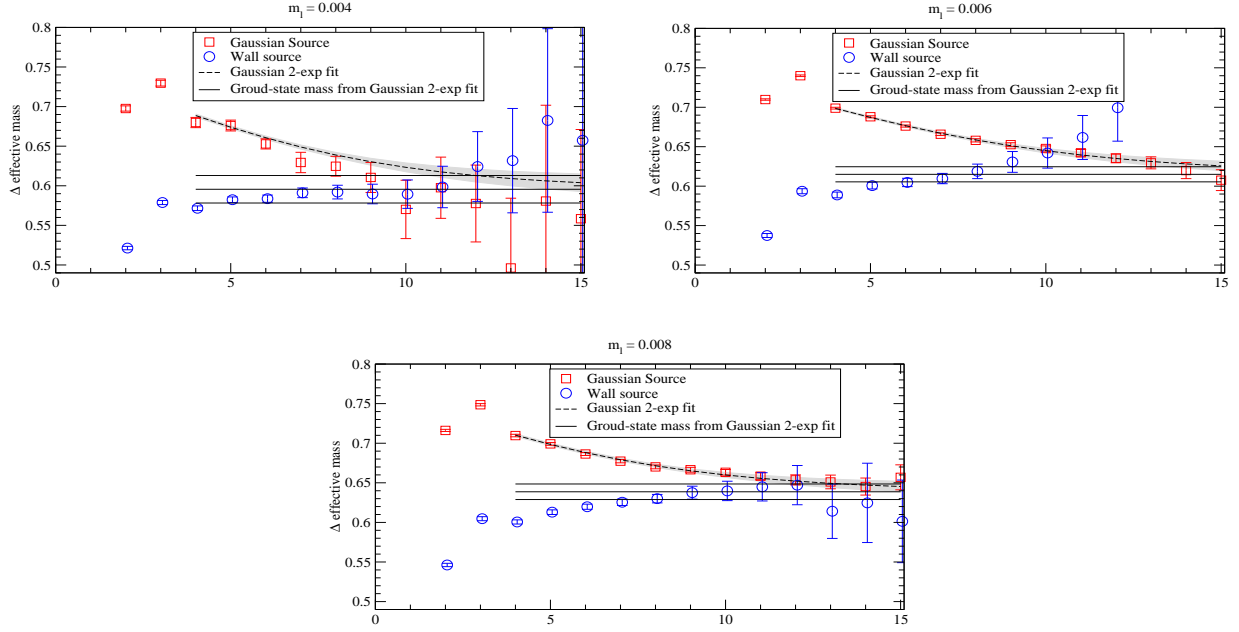

\begin{center}
\includegraphics[width=0.47\hsize,height=1.5in]{fig/Fig25a_delta_paper_32c_004.eps}\qquad
\includegraphics[width=0.47\hsize,height=1.5in]{fig/Fig25b_delta_paper_32c_006.eps}\\[0.25in]
\includegraphics[width=0.47\hsize,height=1.5in]{fig/Fig25c_delta_paper_32c_008.eps}
\end{center}
\caption{\(\Delta\) effective mass plots from the \(32^3\) ensembles.}
\label{fig:Delta_effective_mass32}
\end{figure}

Baryon effective masses from  the \(32^3\) ensembles are plotted in Fig.~\ref{fig:N_effective_mass32}
and \ref{fig:Delta_effective_mass32}.
The Gaussian-source correlators give good effective-mass signals, while the wall-source correlators are much noisier; indeed it is hard to identify a plateau in effective mass signals from the latter.
While for nucleons effective mass signals from the wall-source seem to eventually settle at the same values as from Gaussian source correlators, for the \(\Delta\) baryons a plateau cannot be identified from the wall source except for the lightest up/down mass.
Nevertheless fully correlated fits using two exponentials to represent the contributions of the ground and first-excited states can be performed for both the nucleon and \(\Delta\), yielding the results summarized in Table \ref{tab:baryon_spectrum_32}.
\begin{table}[t]
\begin{tabular}{lllll}
\hline
\multicolumn{1}{c}{\(m_l\)}  &
\multicolumn{2}{c}{\(N\)} &
\multicolumn{2}{c}{\(\Delta\)}
\\
 \hline
0.004 &  0.468(6) &\{4-20\} & 0.596(15) &\{4-15\} \\
0.006 &  0.498(4) &\{4-20\} & 0.615(9) &\{4-15\} \\
0.008 &  0.521(4) &\{4-20\} & 0.639(10)  &\{4-15\}\\
  \hline
\end{tabular}
\caption{Nucleon and $\Delta$ masses in lattice units from the \(32^3\) ensembles obtained by two-exponential correlated fits to Gaussian-source correlators. \{\} denotes fit range.
}
\label{tab:baryon_spectrum_32}
\end{table}
In addition to this fully-correlated two-exponential fit,  we have tried two other fit methods: uncorrelated and bootstrap correlated with frozen correlation matrix \cite{Maynard:2010wg}.
While those earlier analysis were conducted on smaller statistics, they agree with the two-state fully correlated fits
within two standard deviations (see Table \ref{tab:compare_LHP}.)
We use the results from the two-state fully correlated fits as our best values of the baryon masses.
\begin{table}[t]
\begin{center}
\begin{tabular}{lllll}
\hline\hline
\multicolumn{1}{c}{\(m_l\)} &
\multicolumn{1}{c}{full corr.} &
\multicolumn{1}{c}{uncorr.} &
\multicolumn{1}{c}{bootstrap\,$^a$} &
\multicolumn{1}{c}{LHP\,$^b$} \\
\hline
0.004 & 0.477(4) & 0.465(5) & 0.469(4) & 0.474(4) \\
0.006 & 0.498(2) & 0.486(10)& 0.489(7) & 0.501(2) \\
0.008 & 0.517(3) & 0.524(4) & 0.5254(16) & 0.522(2) \\
\hline\hline
\end{tabular}
\end{center}
\caption{Comparison of nucleon mass results from different analyses on the same \(32^3\) ensembles. Superscript $a$
denotes Ref.\,\cite{Maynard:2010wg}, where a frozen correlation matrix was used and superscript $b$ denotes
Ref.\,\cite{Syritsyn:2009mx}.
\label{tab:compare_LHP}}
\end{table}
They also broadly agree with an independent analysis of baryon masses from our ensembles by the
LHP collaboration \cite{Syritsyn:2009mx} within two standard deviations.

\section{Combined Continuum and Chiral Fits}
\label{sec:CombinedChiralFits}

\ifnum\theCombinedChiralFits=1
%
%

We now turn to the main objective of this paper which is to use the results obtained on the $24^3$ and $32^3$ ensembles, as discussed in the previous two sections, to determine physical hadron and quark masses and mesonic decay
constants in the continuum limit, for physical values
of the light and strange quark masses.  Since we are reporting our
first results obtained at a second lattice spacing, we present a
careful discussion of our approach to taking the continuum limit
and the relation between evaluating the continuum limit and determining
the physical quark masses. We start in Section~\ref{subsec:scaling} with a discussion of what we mean by a \textit{scaling trajectory} and explain in some detail the choice of scaling trajectory which we use in the following. In Section~\ref{subsec:chiral_scaling} we describe our power counting scheme, in which we treat the $O(a^2)$ terms in our two ensembles and the NLO terms in SU(2) chiral perturbation theory as being
of comparable size. In order to gain insights into the uncertainties associated with the chiral extrapolation, in addition to SU(2) chiral perturbation theory, we introduce an analytic ansatz which is a simple first-order Taylor expansion in the light-quark mass. This is explained in Section~\ref{subsec:analytic}.
We then discuss the specific fitting procedure which implements this
power counting strategy in Section~\ref{subsec:procedure} and in
Section~\ref{subsec:results} we present and discuss the results.

\subsection{Defining the scaling trajectory}
\label{subsec:scaling}

Although ultimately we will combine the continuum and chiral
extrapolations by performing \textit{global fits} as described
in subsection \ref{subsec:fitting_strategies} and in the following subsections, we start by focussing
on the approach to the continuum limit and discussing the
definition and choice of \textit{scaling trajectory.} For the
purposes of this subsection we imagine that we can perform lattice
computations for any choice of quark masses and envision
performing a series of lattice simulations for a range of values
of $\beta$, the inverse square of the bare lattice coupling.
As $\beta \rightarrow\infty$ the lattice spacing, measured in physical
units, will vanish along with all discretization errors.  We refer
to such a one-dimensional path through the space of possible
lattice theories as a scaling trajectory.  For 2+1 flavor QCD we
must vary the bare lattice mass $m_{ud}(\beta)$ of the up and down
quarks and $m_s(\beta)$ of the strange quark so that this trajectory
describes physically equivalent theories up to order $a^2$ errors.
The functions $m_{ud}(\beta)$ and $m_s(\beta)$ can be determined by
requiring two mass ratios (or two other dimensionless quantities)
to remain fixed as $\beta$ varies.  Because of the presence of
$O(a^2)$ discretization errors, using a different pair of mass
ratios will yield a different trajectory of lattice theories,
whose low-momentum Green's functions will be equivalent to those of the
first up to $O(a^2)$ corrections.

In ref.\,\cite{Allton:2008pn}, where we obtained results from simulations at a single value of $\beta$, we found that using the masses of the $\pi$ and $K$ mesons and the $\Omega$ baryon to determine the lattice spacing $a$ and the bare values of $m_{ud}$ and $m_s$ was an effective procedure.
A natural choice of scaling trajectory would therefore be to keep the ratios $m_\pi/m_\Omega$
and $m_K/m_\Omega$ fixed as $\beta$ varies. Thus these ratios would be
chosen to take their continuum values for all $\beta$ with no $a^2$
corrections.  This choice of scaling trajectory then fixes the functions
$m_{ud}(\beta)$ and $m_s(\beta)$.  In addition, we will identify an
inverse lattice spacing, expressed in GeV, with each point on this scaling
trajectory.  To do this we use the mass of the $\Omega^-$ baryon and
define $1/a =1.672/m_\Omega$ GeV where 1.672\,GeV is the physical mass of
this baryon and $m_\Omega$ is the mass of the $\Omega^-$ as measured
along our trajectory in lattice units.

Having defined the scaling trajectory and determined the lattice
spacing at each $\beta$ by fixing the ratios $m_\pi/m_\Omega$,
$m_K/m_\Omega$ and the mass of the $\Omega$ baryon to their
physical values, we are in a position to make predictions for
other physical quantities. The results obtained at a particular
value of $\beta$ will differ from the physical ones by terms of $O(a^2)$.
We imagine eliminating these artefacts by extrapolating results
obtained at several values of $\beta$ to the continuum limit.  In
order to discuss this continuum extrapolation it is convenient to
introduce some notation. Let us assume that we have performed lattice
calculations at a series of $N$ values of $\beta$,
$\{\beta^{\bf e}\}_{1 \le {\bf e} \le N}$ corresponding to points
along the scaling trajectory defined above (in the present study $N=2$).
This will determine a series of bare quark masses
$m_f^{\bf e} = m_f(\beta^{\bf e})$ where $f = ud$ or $s$. On each
of the lattices we compute a number of physical quantities, e.g.
the kaon leptonic decay constant $f_K^{\bf e}$, and our prediction for the physical
value of $f_K$ is the value obtained by extrapolating to the
continuum limit.

Of course, as already mentioned above, the scaling trajectory and the assigned value of the
lattice spacing at a particular $\beta$ are not unique. Had we
used three different physical quantities to calibrate the lattice
at each $\beta$ and then used the resulting bare quark masses and
lattice spacing to compute $m_\pi/m_\Omega$, $m_K/m_\Omega$ and
the mass of the $\Omega$ baryon, we would find results which
differed from the physical ones by terms of $O(a^2)$. Although
there is a choice of the quantities used to define and determine
the scaling trajectory and the value of the lattice spacing at
each $\beta$, for a 2+1 flavor theory the number of conditions is always $3N$, where $N$
is the number of different $\beta$ values used in the simulations
and the factor 3 corresponds to the fact that at each $\beta$
there are three parameters, the bare masses $m_{ud}$ and $m_s$
and the lattice spacing $a$.

In the above presentation we have tried to provide a pedagogical
introduction to the determination of scaling trajectories and
chose to decouple issues related to the extrapolations in the
mass of the light quark (chiral extrapolations) from the
discussion.  Of course, in practice at present we are unable to
perform simulations at physical quark masses, i.e. with masses
which give the physical values of $m_\pi/m_\Omega$ and
$m_K/m_\Omega$, and so chiral extrapolations are necessary. It
will therefore be useful in the following to discuss the scaling behavior of
a general 2+1 flavor theory in which the masses of the pion and
kaon differ from those in Nature.  Following the conventions defined
elsewhere in this paper, we will use $m_l$ and $m_h$ for the
quark masses in the DWF lattice action which correspond to the usual
$ud$ and $s$ quarks, and $\widetilde{m}_l$ and $\widetilde{m}_h$
for the corresponding multiplicatively renormalizable bare quark
masses $\widetilde{m}_l=m_l+m_{\textrm{res}}$ and
$\widetilde{m}_h=m_h+m_{\textrm{res}}$  specific to the DWF action.
In the next subsection we review the origin of the $a^2$ errors as
described by the Symanzik effective theory for DWF and in the
following subsection present our treatment of scaling for this
more general theory.

\subsubsection{Symanzik effective theory and $a^2 \rightarrow 0$
extrapolation}

Symanzik's effective theory provides a powerful framework in which to discuss
the approach to the continuum limit.  For any finite value of $\beta$
we expect the low-momentum Green's functions in our lattice theory
to agree with those in a corresponding effective continuum theory.
The effective action for this theory contains not only the usual dimension-3
and 4 terms standard in QCD but also higher-dimension operators.  If
the quark masses and the coefficients of these higher-dimension operators
are properly chosen then the low-energy Green's functions of the
lattice and effective theories will agree through $O(a^{d-4})$ provided
the effective theory includes all necessary terms of dimension up to
and including $d$.  This implies that the low-energy Green's functions
of the lattice theory and the usual continuum theory will differ by
the matrix elements of these dimension-5 and higher operators which of
course are not present in the standard continuum theory.

For the domain wall fermion calculation presented here the leading
corrections come from operators of dimension 6.  While the dimension-5
Pauli term $\overline{q}\sigma^{\mu\nu} F^{\mu\nu} q$ is present, its
chiral properties imply that it is generated by chirality violation due to
propagation between the left and right domain walls.  This same
residual breaking of chiral symmetry gives rise to the residual mass
$m_{\rm res}$, the coefficient of the dimension-3 mass term which
remains when the input quark mass is set equal to zero.  The largest value for
$m_{\rm res}$ found in our current calculation, $m_{\rm res} = 0.003152(43)$,
is suppressed from unity by more than two orders of magnitude.  Since a
similar suppression for this dimension 5 operator is expected, the
combination of chiral symmetry and the small value of
$a\Lambda_{\rm QCD}\sim 0.2$ suggest this term can be ignored and
that the largest finite lattice spacing errors that we should
expect are $O(a^2)$.

We require that for our choice of scaling trajectory the matrix elements
of these $O(a^2)$ Symanzik terms behave as $a^2$, allowing a linear
extrapolation in $a^2$ to give the continuum limit.  This implies
that the coefficients of these operators remain reasonably constant
along our trajectory.  This is typically achieved by varying only $\beta$
and quark masses along the trajectory so the only variation in the
coefficients of these $O(a^2)$ terms comes from the variations in
$\beta$ which are quite small in present scaling studies~\footnote{Of
course the varying quark masses $m_{ud}(\beta)$ and $m_s(\beta)$
will also appear in the coefficients of these $O(a^2)$ terms but when
expressed in physical units such mass dependence will be of order
$a^4$.}.

\subsubsection{Scaling and the quark masses}
\label{subsec:mass_scaling}

In the present calculation we obtain results using a number of light-quark
masses, all of which are significantly larger than the physical quark masses
that were used in the introductory remarks above to describe a
physical scaling trajectory in which $m_\pi/m_\Omega$, $m_K/m_\Omega$
and $m_\Omega$ were fixed at their physical values.  However,
we can easily generalize our notion of a scaling trajectory to include
families of choices for the parameters $(\beta,\widetilde{m}_l,\widetilde{m}_h)$
for which, in an obvious notation, the ratios $m_{ll}/m_{hhh}$ and
$m_{lh}/m_{hhh}$ are held fixed.  In the language used earlier, we
require that the $N$ triplets of parameters
$(\beta^{\bf e},\widetilde{m}_l^{\bf e}, \widetilde{m}_h^{\bf e})$,
$1 \le {\bf e} \le N$, lie on the same scaling trajectory if
\begin{eqnarray}
\frac{m_{ll}(\beta^{\bf e},\widetilde{m}_l^{\bf e},\widetilde{m}_h^{\bf e})}
     {m_{hhh}(\beta^{\bf e},\widetilde{m}_l^{\bf e},\widetilde{m}_h^{\bf e})}
&=& \frac{m_{ll}(\beta^{\bf e^\prime},\widetilde{m}_l^{\bf e^\prime},\widetilde{m}_h^{\bf e^\prime})}
         {m_{hhh}(\beta^{\bf e^\prime},\widetilde{m}_l^{\bf e^\prime},\widetilde{m}_h^{\bf e^\prime})} \\
\frac{m_{lh}(\beta^{\bf e},\widetilde{m}_l^{\bf e},\widetilde{m}_h^{\bf e})}
     {m_{hhh}(\beta^{\bf e},\widetilde{m}_l^{\bf e},\widetilde{m}_h^{\bf e})}
&=& \frac{m_{lh}(\beta^{\bf e^\prime},\widetilde{m}_l^{\bf e^\prime},\widetilde{m}_h^{\bf e^\prime})}
         {m_{hhh}(\beta^{\bf e^\prime},\widetilde{m}_l^{\bf e^\prime},\widetilde{m}_h^{\bf e^\prime})}
\end{eqnarray}
for each pair ${\bf e}$ and ${\bf e^\prime}$.
The ratio of lattice spacings for such a pair would be defined as
\begin{equation}
\frac{a^{\bf e}}{a^{\bf e^\prime}}
  =  \frac{m_{hhh}(\beta^{\bf e},\widetilde{m}_l^{\bf e},\widetilde{m}_h^{\bf e})}
          {m_{hhh}(\beta^{\bf e^\prime},\widetilde{m}_l^{\bf e^\prime},\widetilde{m}_h^{\bf e^\prime})}.
\end{equation}

The scaling trajectory determines two functions
$\widetilde{m}_l(\beta)$ and $\widetilde{m}_h(\beta)$, where these
bare masses are non-trivial functions of $\beta$.  While a portion
of their $\beta$ dependence should reflect their naive mass
dimension, these quantities also carry a logarithmic dependence on
$a$ characteristic of the anomalous dimension of the mass operator
$\overline{q}q$ in QCD.  Thus, even when expressed as dimensionless
ratios, {\it e.g.} $\widetilde{m}_l(\beta)/m_\Omega$ and
$\widetilde{m}_h(\beta)/m_\Omega$, these parameters will have singular
continuum limits (in fact, the sign of the anomalous dimension of $\overline{q}q$ is such that these ratios vanish in the continuum limit).

The mass parameters $\widetilde{m}_l$ and
$\widetilde{m}_h$ are short-distance quantities whose definition is
free of infrared singularities.  For example, they could be specified
by examining high-momentum, infra-red safe Green's functions with no
need to compute low-energy masses which are dependent upon the low-energy, non-perturbative behavior of QCD.  While the individual masses $\widetilde{m}_l(\beta)$ and $\widetilde{m}_h(\beta)$ do not
have a continuum limit, both the naive and anomalous scale dependence
cancels in their ratio $\widetilde{m}_l(\beta)/\widetilde{m}_h(\beta)$,
which is well-defined in the continuum limit and agrees with the corresponding
ratio in conventional renormalization schemes, such as RI/MOM or $\overline{\rm MS}$.

Let us now assume that we have performed lattice calculations
at a series of $N$ values of $\beta$,
$\{\beta^{\bf e}\}_{1 \le {\bf e} \le N}$, corresponding to points
along the scaling trajectory defined above.  This will determine a
series of quark masses
$\widetilde{m}_f^{\bf e} = \widetilde{m}_f(\beta^{\bf e})$
where $f = l$ or $h$.  It is natural to introduce a series of
factors which relate the lattice spacings and quark masses between
these $N$ ensembles.  For convenience, we identify a primary ensemble
$\bf 1$, and introduce $3(N-1)$ factors relating
each ensemble $\bf e$ to the ensemble $\bf 1$ as follows:
\begin{eqnarray}
R_a^{\bf e} &=& \frac{ a^{\bf 1} }{ a^{\bf e} }
                    = \frac{ m_{hhh}^{\bf 1} }{ m_{hhh}^{\bf e} }
\label{eq:R_a_def}\\
Z_f^{\bf e} &=& \frac{1}{ R_a^{\bf e} }\frac{ \widetilde{m}_f^{\bf 1} }
                         { \widetilde{m}_f^{\bf e} }
                  \quad \mbox{for} \quad f = l\; \mbox{or}\; h.
\label{eq:Z_f_def}
\end{eqnarray}

Since the ratio $\widetilde{m}_l/ \widetilde{m}_h$ is well-defined
in the continuum limit, the corresponding ratio for each of
these ensembles $\widetilde{m}_l^{\bf e}/ \widetilde{m}_h^{\bf e}$
differs from that limit by a term proportional to $(a^{\bf e})^2$.
This $O(a^2)$ correction represents the discrepancy between
our choice of scaling trajectory with $m_{ll}/m_{lh}$ fixed as we vary $\beta$ and an
alternative choice where instead $\widetilde{m}_l^{\bf e}/ \widetilde{m}_h^{\bf e}$ is held fixed.
Since these trajectories differ at $O(a^2)$, we expect that
\begin{equation}
\frac{\widetilde{m}_l^{\bf e}}{\widetilde{m}_h^{\bf e}}
    = \lim_{\beta \rightarrow \infty}
          \left( \frac{\widetilde{m}_l(\beta)}{\widetilde{m}_h(\beta)}\right)
          \left(1 + c_m (\Lambda_{\rm QCD} a^{\bf e})^2\right).
\label{eq:m_h_m_l_scaling}
\end{equation}
The term proportional to $c_m$ arises from the shifts in $m_{ll}^2$ and $m_{lh}^2$
caused by the first-order effects of dimension-6 terms in the Symanzik effective
action.  While $c_m$ must vanish as $\widetilde{m}_l^{\bf e}
\rightarrow\widetilde{m}_h^{\bf e}$, we prefer not to write $c_m$ as
proportional to the difference $\widetilde{m}_l^{\bf e} -
\widetilde{m}_h^{\bf e}$ because of possible non-analytic terms in the quark masses (e.g. possible
logarithms of $m^{\bf e}_l$)
that may appear in the low-energy matrix
elements of these dimension-6 operators.  If we divide
Eq.~(\ref{eq:m_h_m_l_scaling}) evaluated for our primary ensemble {\bf 1}
by the same equation applied to the ensemble {\bf e} and Taylor expand
in the lattice spacing, we obtain the following useful relation between $Z_h^{\bf e}$
and $Z_l^{\bf e}$:
\begin{equation}
Z_h^{\bf e} = Z_l^{\bf e}
    \left(1 + c_m \Lambda_{\rm QCD}^2
                        \left[(a^{\bf e})^2 - (a^{\bf 1})^2\right]\right)
\label{eq:Z_h_Z_l_scaling}
\end{equation}
implying the $2(N-1)$ $Z$ factors associated with the quark masses
actually depend on $N$ quantities through order $a^2$ (e.g. we can take the $(N-1)$ $Z_l^{\bf e}$ and $c_m$ as the independent quantities). The constraints
implied by Eq.~(\ref{eq:Z_h_Z_l_scaling}) do not simplify the $N=2$ case
addressed in the present paper where we would simply be trading the two
parameters $Z_h^{\bf 2}$ and $Z_l^{\bf 2}$ for the alternative pair of parameters $Z_l^{\bf 2}$
and $c_m$.

Equation~(\ref{eq:Z_h_Z_l_scaling}) provides an explicit estimate of how
scaling violations revise the standard expectation that all quark masses
will scale with a common $Z$ factor as the cut-off is varied.  As we
will see from our simulation results presented below, the terms proportional to $c_m$ are
small and difficult to resolve from zero given our statistical errors.

Since we are now using formulae in which the lattice spacing
$a^{\bf e}$ appears alone rather than in a ratio, e.g.
as $a^{\bf e}/a^{\bf e^\prime}$, it may be useful to explain
how we intend this is to be determined. It is natural to start by considering the physical scaling trajectory
discussed in Section\,\ref{subsec:scaling} on which $m_{ll}/m_{hhh} = m_\pi/m_\Omega$ and
$m_{lh}/m_{hhh} = m_K/m_\Omega$.  For this physical trajectory,
the actual value of the Omega mass measured in GeV can be used
to define the lattice spacing for any point $\beta^{\bf e}$ on that
trajectory using $a^{\bf e} = m^{\bf e}_{hhh}/(1.67245(0.29)\,{\rm GeV})$. In our present study, in order to reach the physical trajectory a chiral extrapolation must be performed from the quark masses used in our simulation.
Ultimately of course, when we present results for dimensionful quantities in physical units, it will be necessary to perform the chiral extrapolation and this is the subject of the following subsections. For the present discussion of scaling it is sufficient simply to imagine that the lattice spacing has been determined in this way and this is the most straightforward way of interpreting the $O((a^{\bf e})^2)$ terms appearing in equations in this subsection. We stress however, that even this is not strictly necessary.  We can consider a scaling trajectory defined by fixed, but unphysical, values of $m_{ll}/m_{hhh}$ and $m_{lh}/m_{hhh}$ and define the lattice spacing by assigning an arbitrary value to $M_{hhh}$, the mass of the $hhh$ baryon on the trajectory in ``physical" units, $a^{\bf e}\equiv m^{\bf e}_{hhh}/M_{hhh}$. While the value of $a^{\bf e}$ defined in this way depends, of course, on the choice of $M_{hhh}$, this arbitrariness is simply absorbed by a change in constants such as $c_m$ in (\ref{eq:m_h_m_l_scaling}). For the discussion in this subsection it is sufficient to note that such a definition of the lattice spacing is possible in principle, the numerical determination of $a^{\bf e}$ does not actually have to be performed.

In the analysis to follow we will examine a family of nearby scaling
trajectories in which $\widetilde{m}_l$ and $\widetilde{m}_h$ vary over limited ranges (specifically, $\widetilde{m}_l$ varies up to about 0.013 on our coarser lattice
and $\widetilde{m}_h$ varies by up to 20\% around $\widetilde{m}_s$). Consider two such trajectories, defined by keeping the ratios $m_{ll}/m_{hhh}$ and $m_{lh}/m_{hhh}$ fixed along each trajectory, but taking different values on the two trajectories. Let $m_{ll}/m_{hhh}=r_{ll}$ and $m_{lh}/m_{hhh}=r_{lh}$ on the first trajectory and $m_{ll}/m_{hhh}=r^\prime_{ll}$ and $m_{lh}/m_{hhh}=r^\prime_{lh}$ on the second. As
$\beta \rightarrow \infty$,
the ratio of bare quark masses on the two trajectories will approach
a limit up to $O(a^2)$ corrections:
\begin{equation}
\frac{\widetilde{m}_f^{\bf e}(r_{ll},r_{lh})}
     {\widetilde{m}_f^{\bf e}(r_{ll}^\prime,
                                    r_{lh}^\prime)}
    = \lim_{\beta \rightarrow \infty}
          \left( \frac{\widetilde{m}_f(\beta)}{\widetilde{m}^\prime_f(\beta)}\right)
          \left(1 + d_{m,f} (\Lambda_{\rm QCD} a^{\bf e})^2\right),
\label{eq:m_f_scaling}
\end{equation}
where $f$=$l$ or $h$, and $\widetilde{m}^{\bf e}_l(r_{ll},r_{lh})$ and $\widetilde{m}^{\bf e}_h(r_{ll},r_{lh})$ ($\widetilde{m}^{\bf e}_l(r^\prime_{ll},r^\prime_{lh})$ and $\widetilde{m}_h^{\bf e}(r^\prime_{ll},r^\prime_{lh})$) are the values of the bare quark masses on ensemble ${\bf e}$ such that $m_{ll}/m_{hhh}=r_{ll}$ and $m_{lh}/m_{hhh}=r_{lh}$ ($m_{ll}/m_{hhh}=r^\prime_{ll}$ and $m_{lh}/m_{hhh}=r^\prime_{lh}$).
The ratios $R_a=m_{hhh}^{\bf 1}(\widetilde{m}_{l}^{\bf 1}(r_{ll},r_{lh}),\widetilde{m}_{h}^{\bf 1}(r_{ll},r_{lh}))
/m_{hhh}^{\bf e}(\widetilde{m}_{l}^{\bf e}(r_{ll},r_{lh}),\widetilde{m}_{h}^{\bf e}(r_{ll},r_{lh}))$  and $R_a^\prime
=m_{hhh}^{\bf 1}(\widetilde{m}_{l}^{\bf 1}(r^\prime_{ll},r^\prime_{lh}),\widetilde{m}_{h}^{\bf 1}(r^\prime_{ll},r^\prime_{lh}))
/m_{hhh}^{\bf e}(\widetilde{m}_{l}^{\bf e}(r^\prime_{ll},r^\prime_{lh}),\widetilde{m}_{h}^{\bf e}(r^\prime_{ll},r^\prime_{lh}))$
each describe the change in lattice
scale as the bare coupling changes from $\beta^{\bf 1}$ to $\beta^{\bf e}$.
In the limit of small bare coupling, this change of scale can be determined
entirely from the short-distance part of the theory and must be the same
for our two trajectories up to order $a^2$ corrections since these two
trajectories differ only in the choice of quark masses.  Thus we can write
\begin{equation}
\frac{R_a}{R_a^\prime}
      = 1 + d_a \Lambda_{\rm QCD}^2\left((a^{\bf e})^2-(a^{\bf 1})^2\right)
\label{eq:a_scaling}
\end{equation}
where we have explicitly represented the fact that each ratio and hence
the ratio of ratios must approach unity as $a^{\bf e} \rightarrow a^{\bf 1}$.
Both the coefficients $d_{m,f}$ and $d_a$ will vanish when the primed and
unprimed trajectories that are being compared become identical.

Taking the ratio of two versions of Eq.~(\ref{eq:m_f_scaling}), one for
$\beta^{\bf e}$ and the other for our primary ensemble $\beta^{\bf 1}$
and using Eq.~(\ref{eq:a_scaling}), we obtain
an expression for the
change in the factors $Z_f$ between these two trajectories:
\begin{equation}
\frac{Z_f^{\bf e}}{Z_f^{{\bf e}\,\prime}} =
    \left(1 + (d_{m,f}+d_a) \Lambda_{\rm QCD}^2
                        \left[(a^{\bf 1})^2 - (a^{\bf e})^2\right]\right).
\label{eq:Z_f_scaling}
\end{equation}
Since the changes in $\widetilde{m}_l$ and $\widetilde{m}_h$ between
these two trajectories which we wish to compare are small,
the resulting coefficients $d_{m,f}$ and $d_a$ will also be small and we will neglect the
$O(a^2)$ correction on the right-hand side of Eq.~(\ref{eq:Z_f_scaling}).
Thus, we will use the same values for $Z_l$ and $Z_h$ for this family
of nearby trajectories, i.e. we drop lattice artefacts proportional to $\tilde{m}_l$ and $(\tilde{m}_h-\tilde{m}_s)$ and so neglect the mass dependence of $Z_l$ and $Z_h$ in this limited range of masses.  In the following we will refer to this range for $\widetilde{m}_l $ and $\widetilde{m}_h$ as their ``allowed range''.

\subsubsection{Fitting strategies}
\label{subsec:fitting_strategies}

We exploit the above relations between numerical results obtained at the
two values of $\beta$ for which we have performed simulations in two ways. The first
we label the ``fixed-trajectory'' method.  In this approach we determine
$R_a$, $Z_l$ and $Z_h$ by matching results obtained at a single pair
of equivalent quark masses~\footnote{Since $N$, the number of different $\beta$s for which we currently have results, is 2, there is only a single set of ratios $R_a^{\bf 2}$, $Z_l^{\bf 2}$ and $Z_h^{\bf 2}$. When specifically discussing our data, we therefore drop the superfix $^{\bf 2}$ and simply write $R_a$, $Z_l$ and $Z_h$.}.
For example, the masses used at one value of $\beta$ may correspond to values at which a simulation was actually
performed.  The corresponding set of masses for the other $\beta$
might be determined by linear interpolation to make the two ratios
$m_{ll}/m_{hhh}$ and $m_{lh}/m_{hhh}$ agree with those on the first
ensemble.  The ratio of lattice spacings and the two $Z_f$ factors are then
determined from Eqs.~(\ref{eq:R_a_def}) and (\ref{eq:Z_f_def}).  It will be important to recall that $Z_l$ and $Z_h$ are constant in the allowed range of quark masses. Finally,
knowing the three factors $R_a$, $Z_l$ and $Z_h$ we make a common fit to the mass dependence of physical quantities computed for both values of $\beta$.

In the final step, we adopt an ansatz for the mass dependence that is
expected to be accurate both for the points in our calculation and for
the physical values to which we wish to extrapolate, specifically a
NLO chiral expansion about the chiral limit or a simple Taylor
expansion about the physical point.  Each ansatz for the continuum
theory, when combined with the three scaling factors $R_a$, $Z_l$ and
$Z_h$ and with any required $a^2$ corrections, will then provide a
set of formulae which should describe all of our data for both
$\beta$ values.  For example, in the chiral fits described in the
next section we can use a common set of Low Energy Constants (LECs) to fit both sets of
data provided we scale the values used on one set by the required
factors of $R_a$, $Z_l$ and $Z_h$ before we use them on the other.
Where explicit $O(a^2)$ terms are required, these can be added with
unknown coefficients which are also scaled appropriately between our
two values of $\beta$.  In such a combined chiral and $a^2$ expansion
we adopt a power counting scheme, described below, so that only
effects of a similar minimum size are consistently included.

During the initial process of determining $R_a$, $Z_l$ and $Z_h$ we
cannot assign a physical value to the lattice spacing.  The original
trajectory being used does not correspond to physical masses so no
notion of ``GeV'' exists for that case.  Of course, the further fitting
to the quark mass dependence of the two ensembles is introduced to
allow extrapolation to physical values for the ratios $m_{ll}/m_{hhh}$
and $m_{lh}/m_{hhh}$.  When $m_\Omega$ is evaluated at this same physical
point, its value can be compared with 1.672 GeV to determine the lattice
scale.

This fixed trajectory method is intended to cover a wider range of
possible scaling trajectories than the example discussed above where
the trajectory passes precisely through one of the simulation points.
If we wish, we can adopt an ansatz for the quark mass dependence of
$m_\pi$, $m_K$ and $m_\Omega$ and perform this fixed trajectory scaling
with the parameters $R_a$, $Z_l$ and $Z_h$ allowed to vary and fix their
values from Eqs.~(\ref{eq:R_a_def}) and (\ref{eq:Z_f_def}) at values of $m_l$
and $m_h$ for which the ratios $m_{ll}/m_{hhh}$ and $m_{lh}/m_{hhh}$
take their physical values.

The second approach, termed ``generic scaling'', introduces the factors
$R_a$, $Z_l$ and $Z_h$ as parameters into the ansatz being used to fit the quark mass
dependence.  In this approach we perform a fit to all our data for $m_\pi$, $m_K$
and $m_\Omega$ over a range of quark masses for which the fitting ansatz
is accurate and for which the use of fixed values for $R_a$, $Z_l$ and
$Z_h$ is legitimate.  In this generic scaling approach, our choice of
scaling trajectory with fixed hadron mass ratios $m_{ll}/m_{hhh}$ and
$m_{lh}/m_{hhh}$ and with $m_{hhh}$ determining the lattice scale is
realized somewhat indirectly.  The
three conditions associated with this choice of scaling trajectory are
realized by omitting possible $a^2$ corrections from the expressions
used to fit $m_{ll}$, $m_{lh}$ and $m_{hhh}$.
The resulting trajectory can therefore be interpreted as being the one along which the masses of the pion, kaon and $\Omega$-baryon take their physical values, as was the case in the discussion of Section~\ref{subsec:scaling}. The difference of course, is that whereas in Section~\ref{subsec:scaling} we envisaged (unrealistically at present) being able to simulate directly at the physical value of $m_l$, we now reach the physical point after an extrapolation in quark masses. The detailed discussion
of the ChPT functions used in describing the quark mass dependence
of the pion and kaon masses is given in Subsection~\ref{subsec:chiral_scaling}
and those for the analytic ansatz in Subsection~\ref{subsec:analytic} below.  However, both our ChPT and Taylor expansion ans\"atze stipulate
that to the order being studied $m_{hhh}$ is a linear function of
$\widetilde{m}_l$ and $\widetilde{m}_h$.  It is instructive to explore this
case here.

Included among the equations used to determine the low energy constants
and the scaling factors $R_a$, $Z_l$ and $Z_h$ are two equations for
$m_{hhh}$ on our two ensembles:
\begin{eqnarray}
m_{hhh}^{\bf 1}(\widetilde{m}_l,\widetilde{m}_h)
                            &=& m_{hhh}^{\bf 1}(0,\widetilde{m}_{h0})
                             + c_{m_\Omega m_l}^{\bf 1} \widetilde{m}_l
                             + c_{m_\Omega m_h}^{\bf 1}(\widetilde{m}_h - \widetilde{m}_{h0})
\label{eq:generic_m_Omega_1} \\
m_{hhh}^{\bf 2}(\widetilde{m}_l, \widetilde{m}_h) &=& \frac{1}{R_a} m_{hhh}^{\bf 1}(R_a Z_l \widetilde{m}_l, R_a Z_h \widetilde{m}_h) \nonumber\\
&=& \frac{1}{R_a} \Big[m_{hhh}^{\bf 1}(0,\widetilde{m}_{h0}) + c_{m_\Omega m_l}^{\bf 1} R_a Z_l\widetilde{m}_l + c_{m_\Omega m_h}^{\bf 1}\Big(R_a Z_h \widetilde{m}_h-\widetilde{m}_{h0}\Big)\Big]\,.
\label{eq:generic_m_Omega_2}
\end{eqnarray}
Here ${\bf 1}$ is our primary ensemble, for us that is the one with $\beta =2.25$ and
the $32^3\times 64$ volume, while the second ensemble is the one with the coarser lattice spacing
and is labeled ${\bf 2}$. $m_{hhh}^{\bf e}(\widetilde{m}_l,\widetilde{m}_h)$ are the $hhh$-baryon masses corresponding to bare-quark masses $\widetilde{m}_l$ and $\widetilde{m}_h$ on ensemble ${\bf e}$.  Although we have written $\widetilde{m}_{h0}$ as a general constant, we have in mind to use the equations with $\widetilde{m}_{h0}$ in the allowed range of the physical bare strange quark mass in the primary ensemble. Equations~(\ref{eq:generic_m_Omega_1}) and (\ref{eq:generic_m_Omega_2}) define the three constants $m_{hhh}^{\bf 1}(0,\widetilde{m}_{h0})$, $c_{m_{\Omega}m_l}^{\bf 1}$ and $c_{m_{\Omega}m_h}^{\bf 1}$
which are related to the physical $\Omega^-$  mass and its ``physical''
dependence on the quark masses.
The absence of $O(a^2)$
corrections to Eqs.~(\ref{eq:generic_m_Omega_1}) and (\ref{eq:generic_m_Omega_2})
implements our choice that $m_\Omega$ is being used to set the scale and hence by construction
contains no finite lattice-spacing errors.  While part of a larger set of
equations which are being used to determine the low energy constants as
well as $R_a$, $Z_l$ and $Z_h$, the leading order effect of these two
equations is to determine $R_a$.  Note that this is identical to imposing
Eq.~(\ref{eq:R_a_def}) in the fixed trajectory method at the point $\widetilde{m}_l=0$,
$\widetilde{m}_h = \widetilde{m}_{h0}$. Since the variation of $R_a$ as $\widetilde{m}_l$ and $\widetilde{m}_h$ change
over their allowed range is of the same size as the variation of $Z_l$
and $Z_h$ over this same range it can also be neglected, so any particular
choice of $\widetilde{m}_h$ is equivalent to any other within this allowed range.

The fixed trajectory and generic scaling methods are similar in nature.  Both require that an ansatz be
adopted to allow the quark mass dependence of lattice quantities to be described in order
to define the scaling parameters $R_a$, $Z_l$ and $Z_h$ and to
extrapolate to the physical point.  Both assume that the scaling relations
between the two ensembles defined by $R_a$, $Z_l$ and $Z_h$ hold over
the allowed range of masses.  The fixed trajectory method corresponds
most closely to our original definition of a scaling trajectory and decouples the matching of the two lattices from the chiral extrapolation.
It requires however, the introduction of a convenient but arbitrary point at which the matching
between the two ensembles is performed.  The generic method avoids
this arbitrary choice and applies these assumptions uniformly over
the entire range of allowed masses.  The fixed trajectory method determines $R_a$, $Z_l$ and $Z_h$
in an iterative fashion as explained in Section\,\ref{subsec:procedure}.  The generic approach determines the coefficients in the adopted ansatz from a single $\chi^2$ minimization.  The physical
quark masses are then determined by inverting the resulting equations
which give $m_\pi$, $m_K$ and $m_\Omega$ in terms of $\widetilde{m}_l$
and $\widetilde{m}_h$.

The detailed discussion and results presented in this paper correspond to the fixed trajectory method; fits using the generic scaling approach were performed to monitor the consistency of the results and estimated errors.

\subsection{Scaling and chiral perturbation theory}
\label{subsec:chiral_scaling}

At the start of section\,\ref{subsec:scaling} we discussed the continuum extrapolation in an
idealized situation in which we can perform simulations at any value
of the quark mass $m_{l}$. In reality this is not the case; for example,
the lightest unitary pion appearing in the current study has mass
290~MeV. In order to compare our results with Nature we therefore need
to extrapolate to lighter quark masses and this was already acknowledged when discussing the fitting strategies in section\,\ref{subsec:fitting_strategies} above. We now explain how we combine
the continuum and chiral extrapolations in \textit{global fits}. We start in this section by using
SU(2) chiral perturbation theory for the mass dependence, with the expectation that the extrapolation will be
made more precise if constrained by the theoretically known behavior of
QCD in the chiral limit~\cite{Allton:2008pn}.  However, in order to
estimate possible systematic errors associated with this extrapolation
and to obtain a more complete understanding of the implications of our
calculation, we also examine a simpler analytic extrapolation
to physical quark masses~\cite{Lellouch:2009fg} and this is explained in the following subsection.
Although later we will perform extrapolations using partially quenched
ensembles, for the purposes of this introduction we restrict the
discussion to the unitary theory in which the valence and sea quark
masses are equal.

We now explain the power counting scheme
we employ to identify NLO corrections to the chiral and continuum
limits.  Since the pion mass and decay constant are central to
SU(2) ChPT, we begin by considering the predictions of continuum
NLO ChPT for these two quantities:
\begin{eqnarray}
m_{ll}^2 &=& \chi_l\, +\,
\chi_l\,\cdot\,\Bigg\{\,
         \frac{16}{\ftwo^2}\Big((2\ltwo{8}-\ltwo{5})
         +2(2\ltwo{6}-\ltwo{4})\Big)\chi_l\,
         +\,\frac{1}{16\pi^2\ftwo^2}\chi_l
         \log\frac{\chi_l}{\Lambda_\chi^2}\,\Bigg\}
\label{eq:chPTsu2:mPi} \\
f_{ll} &=& \ftwo \, + \, \ftwo\,\cdot\,\Bigg\{\,
   \frac{8}{\ftwo^2}(2\ltwo{4}+\ltwo{5})\chi_l\, -\,\frac{\chi_l}{8\pi^2\ftwo^2}
     \log\frac{\chi_l}{\Lambda_\chi^2}\,\Bigg\}.
\label{eq:chPTsu2:fPi}
\end{eqnarray}
Here $m_{ll}$ and $f_{ll}$ are the mass and decay constant of
the pseudoscalar meson composed of two light quarks, $f$,
$L_4$, $L_5$, $L_6$ and $L_8$ are the conventional low energy
constants and $\Lambda_\chi$ is the usual chiral scale.
The quantity $\chi_l$ comes directly from
the lowest order chiral symmetry breaking term in the effective
chiral theory and is proportional to the QCD light quark mass. It
is conventionally written $\chi_l = 2B \widetilde{m}_l$, where $B$ is another low-energy constant.

We now discuss how we apply these formulae to describe the
low energy behavior of lattice theories which lie on a scaling
trajectory.  For a sequence of ensembles $\{{\bf e}\}_{1 \le {\bf e} \le N}$
lying on such a scaling trajectory not only will the quark masses
and lattice units, $(\widetilde{m}_l^{\bf e}, \widetilde{m}_h^{\bf e},
a^{\bf e})$ be related, but also, when expressed in physical units,
the quantities $f$, $L_4$, $L_5$, $L_6$ and $L_8$ should take the
same values up to $O(a^2)$ corrections. The same is true for the renormalization independent combination
 $\chi_l=2B\widetilde{m}_l$ (see the discussion below). As detailed in
Ref.~\cite{Allton:2008pn}, chiral
perturbation theory at finite lattice spacing for domain wall
fermions involves a simultaneous expansion in the explicit bare
quark mass, $m_l$, the squared lattice spacing, $a^2$, and the
residual chiral symmetry breaking arising from the finite
separation, $L_s$, between the two four-dimensional walls in
the fifth dimension.  We will denote this last quantity by
$e^{-\lambda L_s}$, suggesting the exponential decrease in such
residual chiral symmetry breaking found in perturbation theory
for DWF.  (The actual behavior is a sum of exponential and
inverse power dependence on $L_s$.)  No new terms need to be
added to the resulting effective low energy theory to
describe the resulting Green's functions to NLO in the parameters
$\widetilde{m}_l$, $a^2$ and $e^{-\lambda L_s}$. Thus, we can
use equations with the form of Eqs.~(\ref{eq:chPTsu2:mPi}) and
(\ref{eq:chPTsu2:fPi}) to describe the lattice results for $m_{ll}$
and $f_{ll}$ along a scaling trajectory, provided we work to NLO
in a power counting scheme which treats the quantities
$\chi_l/(4\pi f)^2$, $a^2\Lambda^2_{\textrm{QCD}}$ and $e^{-\lambda L_s}$
as equivalent and keep a single power of any of these quantities as a
correction.  We must now determine how the parameters appearing in
these equations must be adjusted to describe lattice results at
finite $a^2$.

Since the scale $\Lambda_\chi$ can be freely varied if the other
analytic terms are appropriately changed, we will choose this quantity
to be constant if measured in physical units.  Thus, for each point
on our physical scaling trajectory we will choose
$\Lambda_\chi = m_\Omega \cdot 1/1.672$, giving it the value of 1 GeV.
Because of their proportionality to the NLO factor $\chi_l$ all of
the parameters which appear in the large curly brackets on the right
hand side of Eqs.~(\ref{eq:chPTsu2:mPi}) and (\ref{eq:chPTsu2:fPi}) can
be given their continuum values, dropping possible $O(a^2)$ terms as
being of NNLO in our power counting scheme.  Thus, within those brackets
the quantities $f$, $L_4$, $L_5$, $L_6$ and $L_8$, when expressed in
physical units, can be given identical values for the ensembles on the
scaling trajectory.

In contrast, when Eq.~(\ref{eq:chPTsu2:fPi}) is used to describe our
finite lattice spacing results, the LO quantity $f^{\bf e}$
determined on ensemble ${\bf e}$, expressed in physical units,
depends on $\beta^{\bf e}$.  However, it approaches its continuum
limit with $O(a^2)$ corrections and so we write
$f^{\bf e} = f + c_f (a^{\bf e})^2$.

Given the definition of a scaling trajectory, the variation of
the quantity $\chi_l^{\bf e}$ needed to apply Eq.~(\ref{eq:chPTsu2:mPi})
to the ensemble ${\bf e}$ is actually trivial.  Because our choice of
quark mass $\widetilde{m}_l^{\bf e}$ gives the same value for $m_{ll}$
for each ensemble $\bf e$ on our scaling trajectory, all of the
quantities in Eq.~(\ref{eq:chPTsu2:mPi}) with the possible exception
of the $\chi_l^{\bf e}$ which we are now considering, are the same when
expressed in physical units for all points on the scaling trajectory.
Thus, $\chi_l^{\bf e} = 2B^{\bf e} \widetilde{m}_l^{\bf e}/(a^{\bf e})^2$
must be a constant as well, where $B^{\bf e}$ and $\widetilde{m}_l^{\bf e}$
are explicitly left in lattice units.  Since we know how the quantities
$\widetilde{m}_l$ and $a^2$ are related between an ensemble ${\bf e}$ and
our primary ensemble ${\bf 1}$, we can determine the $N-1$ constants
$B^{\bf e}$ in terms of the single constant $B^{\bf 1}$:
\begin{equation}
B^{\bf e} = \frac{Z_l^{\bf e}}{R_a^{\bf e}} B^{\bf 1}
\end{equation}
without any $a^2$ corrections.  Because of the complex scaling behavior
of the mass, we will treat $B^{\bf 1}$ as one of the LEC's to be determined
in our fitting and not relate it to a ``physical'' continuum quantity whose
definition would require introducing a continuum mass renormalization scheme.

We conclude that our lattice results for light pseudoscalar masses
and decay constants obtained from a series of ensembles $\{{\bf e}\}$
can be described through NLO by the formulae:
\begin{eqnarray}
(m_{ll}^{\bf e})^2 &=& \chi_l^{\bf e}\, +\,
\chi_l^{\bf e}\,\cdot\,\Bigg\{\,\frac{16}{\ftwo^2}
     \Big((2\ltwo{8}-\ltwo{5})+2(2\ltwo{6}-\ltwo{4})\Big)
     \chi_l^{\bf e}\,+\,\frac1{16\pi^2\ftwo^2}\chi_l^{\bf e}
     \log\frac{\chi_l^{\bf e}}{\Lambda_\chi^2}\,\Bigg\}
\label{eq:chPTsu2:mPi:a2} \\
f_{ll}^{\,\bf e} &=& \ftwo\left[1 + c_f (a^{\bf e})^2 \right] \, + \,
\ftwo\,\cdot\,\Bigg\{\,\frac{8}{\ftwo^2}(2\ltwo{4}+\ltwo{5})\chi_l^{\bf e}\,
         -\,\frac{\chi_l^{\bf e}}{8\pi^2\ftwo^2}
              \log\frac{\chi_l^{\bf e}}{\Lambda_\chi^2}\,\Bigg\}
\label{eq:chPTsu2:fPi:a2} \\
\mbox{with} \nonumber \\
\chi_l^{\bf e} &=& \frac{Z_l^{\bf e}}{R_a^{\bf e}} \frac{B^{\bf 1}\widetilde{m}_l^{\bf e}}{(a^{\bf e})^2}
\label{eq:chPTsu2:chi:a2}
\end{eqnarray}
where all quantities in Eqs.~(\ref{eq:chPTsu2:mPi:a2}) and
(\ref{eq:chPTsu2:fPi:a2}) are expressed in physical units (except for $B^{\bf 1}$ and $\widetilde{m}_l^{\bf e}$ in Eq.\,(\ref{eq:chPTsu2:chi:a2}) which are given in lattice units).

Two important refinements should be mentioned.  First, for the case of a physical scaling
trajectory, i.e. one which terminates in the physical masses $m_\pi$, $m_K$ and
$m_\Omega$, these physical units are naturally GeV.
However, for other scaling trajectories appropriate ``physical'' units to use can be those
in which the Omega mass is unity.  Second, for simplicity in
Eqs.~(\ref{eq:chPTsu2:mPi}), (\ref{eq:chPTsu2:fPi}), (\ref{eq:chPTsu2:mPi:a2})
and (\ref{eq:chPTsu2:fPi:a2}) we have treated the
heavy quark mass as fixed and not displayed the dependence of the
quantities $f$, $B$, $L_4$, $L_5$, $L_6$ and $L_8$ on $m_h$.
In practice we can easily generalize these equations to describe the
dependence of $m_{ll}$ and $f_{ll}$ on $m_h$ as well.  Provided we limit
the variation of $m_h$ to a small range about an expansion point $\widetilde{m}_{h0}$,
this variation can be described by including a linear term in $m_h-\widetilde{m}_{h0}$
and treating this term as NLO in our power counting scheme.  Thus, such
extra linear terms will only be introduced into the leading order terms
in Eqs.~(\ref{eq:chPTsu2:mPi:a2}) and (\ref{eq:chPTsu2:fPi:a2}).

Next we present the corresponding formulae for the quantities $m_K$ and
$m_\Omega$ which are used in the determination of the scaling trajectory
and in the assignment of a lattice spacing at each value of $\beta$:
\begin{eqnarray}\label{eq:chPTsu2K:mPS:a2}
(m^{\bf e}_{lh})^2 &=&
\left(m^{(K)}\right)^2\,+\,
\left(m^{(K)}\right)^2\,\bigg\{\,\frac{\lambda_1+\lambda_2}{\ftwo^2}
             \,\chi_l^{\bf e}\bigg\}\\
m^{\bf e}_{hhh}&=&m^{(\Omega)}+m^{(\Omega)}\,c_{m_\Omega,m_l}\,\chi_l^{\bf{e}}\,.
\label{eq:chPTsu2:Omega:a2}
\end{eqnarray}
Here $m^{(K)}$ and $m^{(\Omega)}$ are the mass of the $lh$ meson and the $hhh$
baryon respectively in the SU(2) chiral limit, i.e. with $\widetilde{m}_l=0$,
for the value of $\widetilde{m}_h$ used in the simulation. Similarly the LECs $\lambda_{1,2}$
and $c_{m_\Omega,m_l}$ depend on $\widetilde{m}_h$ and we are using the notation for
the LECs $\lambda_{1,2}$ which we introduced in \cite{Allton:2008pn}. (Note that
$c_{m_\Omega,m_l}$, whose value is given in Table~\ref{tab-NLOsu2fitparameters} below, should be distinguished from the related parameter $c_{m_\Omega m_l}^{\bf 1}$ which appears in Equations~(\ref{eq:generic_m_Omega_1}) and (\ref{eq:generic_m_Omega_2}) above.) The
absence of any corrections of $O(a^2)$ on the right-hand sides of
Eqs.\,(\ref{eq:chPTsu2K:mPS:a2}) and (\ref{eq:chPTsu2:Omega:a2}) follows
from the same argument which justified omitting an $O(a^2)$ correction from
the right hand side of Eq.~(\ref{eq:chPTsu2:mPi:a2}).  For masses
$\widetilde{m}^{\bf e}_l$ and $\widetilde{m}^{\bf e}_h$ lying on a scaling
trajectory the left hand sides of these equations must all be the same
because of our definition of scaling trajectory.  Because of our power
counting scheme, no $a^2$ corrections need to be included in the NLO
terms proportional to $\chi^{\bf e}_l$ on the right hand side of these two
equations. Therefore the leading order terms $m^{(K)}$ and $m^{(\Omega)}$
must also be the same for all ensembles when expressed in physical units and
no $O(a^2)$ correction can appear.  As discussed above, these equations can
be generalized to describe the NLO dependence on $\widetilde{m}_h$ varying about an
expansion point $\widetilde{m}_{h0}$.  In fact, for the $\Omega$ baryon this  more general case for
Eq.~(\ref{eq:chPTsu2:Omega:a2}) was described in the previous subsection
in the equivalent Eqs.~(\ref{eq:generic_m_Omega_1}) and
(\ref{eq:generic_m_Omega_2}).

Note that the coefficient of the chiral logarithm in
Eq.\,(\ref{eq:chPTsu2:mPi:a2}) includes a factor which depends on $f$,
the pion decay constant in the SU(2) chiral limit (all other factors
of $f$ in Eqs.(\ref{eq:chPTsu2:mPi:a2}) and (\ref{eq:chPTsu2K:mPS:a2})
can be absorbed into a redefinition
of LECs which in any case are determined by fitting). This low energy
constant $f$ can be determined from the measured values of $f_{ll}$
using Eq.\,(\ref{eq:chPTsu2:fPi:a2}), but to NLO it can also be replaced
by the measured values of $f_{ll}$.

As described in Subsection~\ref{subsec:fitting_strategies}, these ChPT
formulae can now be used to determine physical results in the continuum
limit from those obtained on our two lattice spacings.  We can employ the fixed trajectory method, finding the ratios $Z_l$ and $Z_h$ which relate a specific
choice of quark masses on one ensemble to those on the other which lie on the
same scaling trajectory.  The corresponding ratio of values of $m_{hhh}$
determines $R_a$.  These three quantities then allow a single set of
LECs to be used to extrapolate the results of both ensembles to the
continuum limit and to the physical value of the light quark mass using
Eqs.~(\ref{eq:chPTsu2:mPi:a2}), (\ref{eq:chPTsu2:fPi:a2}), (\ref{eq:chPTsu2K:mPS:a2})
and (\ref{eq:chPTsu2:Omega:a2}).  As a result we learn the physical
values of $\widetilde{m}_{ud}(\beta^{\bf e})$, $\widetilde{m}_s(\beta^{\bf e})$
and $a^{\bf e}$ on our two ensembles.  In other words, we determine the quark
masses and lattice spacings for our two ensembles which lie on the
physical scaling trajectory.

Alternatively, we can use the generic fitting approach and introduce
the three parameters $(Z_l, Z_h, R_a)$ into the four equations
Eqs.~(\ref{eq:chPTsu2:mPi:a2}), (\ref{eq:chPTsu2:fPi:a2}), (\ref{eq:chPTsu2K:mPS:a2})
and (\ref{eq:chPTsu2:Omega:a2}) and obtain a fit to the lattice data from
both ensembles for which the quark masses lie in the allowed range.
The resulting values of the LECs and $(Z_l, Z_h, R_a)$ then determine
the functions $m_{ll}^{\bf e}(\widetilde{m}_l,\widetilde{m}_h)$,
$m_{lh}^{\bf e}(\widetilde{m}_l,\widetilde{m}_h)$ and
$m_{hhh}^{\bf e}(\widetilde{m}_l,\widetilde{m}_h)$.  The physical quark masses on each ensemble,
$m^{\bf e}_{ud}=m_{ud}(\beta^{\bf e})$ and $m^{\bf e}_{s}=m_{s}(\beta^{\bf e})$,
are then obtained by solving the equations:
\begin{equation}
\frac{m_{ll}^{\bf e}(\widetilde{m}^{\bf e}_{ud},\widetilde{m}^{\bf e}_s)}
{m_{hhh}^{\bf e}(\widetilde{m}^{\bf e}_{ud},\widetilde{m}^{\bf e}_s)}
  = \frac{m_\pi}{m_\Omega}\qquad\textrm{and}\qquad
\frac{m_{lh}^{\bf e}(\widetilde{m}^{\bf e}_{ud},\widetilde{m}^{\bf e}_s)}
{m_{hhh}^{\bf e}(\widetilde{m}^{\bf e}_{ud},\widetilde{m}^{\bf e}_s)}
  = \frac{m_K}{m_\Omega}\,,
\end{equation}
where on the right-hand sides the ratios take their physical values.

Having determined $m_{ud}(\beta^{\bf e})$, $m_{s}(\beta^{\bf e})$ and
$a^{\bf e}$ as described above, we are in a position to compute other
physical quantities. For example, at NLO in our power counting the
behaviour of the kaon decay constant $f_K$ is
\begin{eqnarray}\label{eq:chPTsu2K:fPS:a2}
f\,^{\bf e}_{lh} &=&
f^{(K)}\left[1+c_{f^{(K)}} (a^{\bf e})^2 \right]\,+\,
f^{(K)}\Bigg\{\,\frac{\lambda_3+\lambda_4}{\ftwo^2}\chi_l^{\bf e}\,
               \;-\frac1{(4\pi\ftwo)^2}\,\frac34
              \,\chi_l^{\bf e}
              \log\frac{\chi_l^{\bf e}}{\Lambda_\chi^2}
   \,\Bigg\}\,,
  \end{eqnarray}
where $f^{(K)}$ is the result in the SU(2) chiral limit ($\widetilde{m}_l=0$), $\lambda_{3,4}$ are $m_h$-dependent low-energy constants and $c_{f^{(K)}}$ is a constant. For each $\beta^{\bf e}$, having
determined $\widetilde{m}_s(\beta^{\bf e})$ we measure $f^{\bf e}_{lh}$
for $\widetilde{m}_h^{\bf e}=\widetilde{m}_s(\beta^{\bf e})$ as a function
of $\widetilde{m}_l$; fit the measured values at all $\beta^{\bf{e}}$ to
determine the LECs and $c_{f^{(K)}}$ in Eq.\,(\ref{eq:chPTsu2K:fPS:a2})
and finally obtain the physical value of $f_K$ by setting $a=0$ and
$\widetilde{m}_l=\widetilde{m}_{ud}$. Such a procedure is then generalized
to the other physical quantities we wish to compute.

\subsection{Scaling combined with an analytic ansatz for the chiral dependence} \label{subsec:analytic} \newcommand{\wm}{\widetilde{m}}

While we know that the ansatz based on chiral perturbation theory described in the previous subsection is valid in the limit of small $u$ and $d$ quark masses, we do not know the precision with which it holds over the range of masses which we analyze in this paper (corresponding to data in the range $240\,\textrm{MeV}\le m_\pi\lesssim 420$\,MeV). Indeed it is precise lattice simulations which will answer such questions. In order to obtain some understanding of the corresponding systematic uncertainties, in addition to the procedures based on chiral perturbation theory described in section\,\ref{subsec:chiral_scaling}, we
consider an ansatz based on a first-order Taylor expansion about a non-zero quark mass, in the style of
ref.\,\cite{Lellouch:2009fg,Durr:2008zz}. Within this approach, since we do not include chiral logarithms, we are not able to take the chiral limit and only assume the validity of the analytic ansatz between the physical point (to which we extrapolate) and the region where we have data.
In this work we only consider linear, first-order fits
and are therefore insensitive to the choice of expansion point which we
take to be the same as that at which we match the ensembles when using the fixed trajectory method. This simplifies the
discussion below of the simultaneous expansion in $a^2$ and mass differences.
Beyond first order, convergence may be improved by considering an expansion point
between the region in which we have data and the physical point, but this is beyond the scope of our current analysis.

Using the analytic ansatz for $m_\pi^2$ as a function of the quark mass $m_q$, we find numerically that the constant (mass independent) term is consistent with zero,
indicating that the tangent of $m_\pi^2(m_q)$ in the unitary case does
pass through the origin. Thus, at our statistical precision, no significant
chiral curvature is needed to satisfy Goldstone's
theorem, however we retain the view that we are indeed using a
model which is valid only in a restricted region of
non-zero quark masses.

Goldstone's theorem also applies in the partially quenched theory and the pion mass vanishes as the valence-quark masses are taken to zero while keeping the sea-quark masses fixed. In this case however,
our linear fit extrapolates to a non-zero pion mass for massless
valence quarks, and this naturally implies that some form of curvature is
required at smaller masses. This is consistent with enhanced chiral logarithms in the partially quenched theory.
However, the fits do not necessarily
imply that chiral logarithms at NLO correctly represent the
quark-mass dependence between the simulated range of masses and the physical
point. Instead, in this approach the sum over multiple orders of
chiral perturbation theory is assumed to be approximated by a
linear dependence in the relevant range of masses.
It is also possible of course that the simulated range of masses is outside the
useful domain of chiral perturbation theory and that, for example, phenomenological models based on
combining NLO chiral perturbation theory with arbitrary analytic subsets of terms which appear at NNLO and NNNLO are less well motivated than our linear ansatz.

For $m_\pi^2$ and $f_\pi$ it is convenient to define the average valence quark mass
$\wm_v = \frac{\wm_x+\wm_y}{2}$.
As in section\,\ref{subsec:chiral_scaling}, we apply a power counting rule in a double expansion
in $m_x-m^m$, $m_y-m^m$, $m_l-m^m$ and $a^2$, where $m^m$ is the
mass at which we match the ensembles which we also choose to be the
point around which we perform the Taylor expansion and we recall that $m_{x,y}$
and $m_l$ are the valence and sea light-quark masses respectively (here we allow for partial quenching). For the pion mass we use the ansatz
\begin{equation}
\label{eq:CemANmllEX}
m_{xy}^2 = C_0^{m_\pi} + C_1^{m_\pi}\left(\wm_v-\wm^m\right) + C_2^{m_\pi}(\wm_l-\wm^m)\,,
\end{equation}
where we use our standard notation in which the subscripts $xy$ imply that the two valence quarks have mass $m_x$ and $m_y$ respectively. By the definition of our scaling trajectory, there is no $O(a^2)$ term at the match point and so there is no correction
to $C_0^{m_\pi}$. Within our power counting we could equivalently use
\begin{equation}
\label{eq:CemANmll}
m_{xy}^2 = C_0^{m_\pi} + C_1^{m_\pi} \wm_v + C_2^{m_\pi} \wm_l\,,
\end{equation}
where for convenience we redefine $C_0^{m_\pi}$ between equations (\ref{eq:CemANmllEX}) and (\ref{eq:CemANmll}).

In searching for evidence of chiral logarithms it is conventional to
plot the ratio $m_{xy}^2/\wm_v$
as a function of the quark masses. With the ansatz proposed in Eq.\,(\ref{eq:CemANmll})
\begin{equation}
\frac{ m_{xy}^2}{\wm_v} = \frac{ C_0^{m_\pi}}{\wm_v} + C_1^{m_\pi} + \frac{C_2^{m_\pi} \wm_l}{\wm_v},
\end{equation}
and we note that an observed deviation of the mass dependence of
 $\frac{m_{xy}^2}{\wm_v}$
from a constant in the finite range of quark masses which can be
simulated, is not in itself unambiguous evidence of a non-analytic structure.

For decay constants, which do not vanish in the chiral limit, the
$O(a^2)$ term are not sensitive to the choice of expansion point:
\begin{eqnarray}
f_{xy} &=& C_0^{f_\pi} [1+C_{f_\pi} a^2] + C_1^{f_\pi}\left(\wm_v-\wm^m\right) + C_2^{f_\pi}(\wm_l-\wm^m)\\
&\equiv& C_0^{f_\pi} [1+C_f a^2] + C_1^{f_\pi} \wm_v + C_2^{f_\pi} \wm_l,\label{eq:CemANfll}
\end{eqnarray}
where again we have redefined $C_0^{f_\pi}$ between the first and second lines.

Following a similar argument, at a fixed strange-quark mass, we take the light-quark mass dependence
of the kaon mass and decay constant and the mass of the $\Omega$-baryon to be given by
\begin{eqnarray}
\label{eq:CemANmxh}
m_{xh}^2(a, m_l) &=& C_0^{m_K} + C_1^{m_K} \wm_x + C_2^{m_K} \wm_l\,,\\
\label{eq:CemANfxh}
f_{xh}(a,m_l) &=&
C_0^{f_K} [1+C_{f_K} a^2] + C_1^{f_K} \wm_x + C_2^{f_K}\wm_l\,.\\
m_{hhh}(a,m_l) &=& C_0^{m_\Omega} + C_2^{m_\Omega} \wm_l\,.
\label{eq:CemANmhhh}
\end{eqnarray}
We stress that the constants $C_n^{m_\pi}$, $C_n^{f_\pi}$, $C_f$, $C_n^{m_K}$, $C_n^{f_K}$, $C_{f_K}$ and $C_n^{m_\Omega}$ implicitly depend on the strange quark mass. 

\subsection{Procedure for combined scaling and chiral fitting} \label{subsec:procedure} Having introduced the theoretical framework behind our combined scaling and chiral fits in Sections~\ref{subsec:chiral_scaling} and \ref{subsec:analytic}
we now explain its practical implementation. The formulae given above which describe the combined behaviour
are valid only for a fixed strange-quark mass and we are faced with the problem that the
physical strange mass is not known a priori but is an output of the calculation. The procedure for performing the combined chiral-continuum fits is therefore necessarily iterative. As explained in more detail below, we start with some initial values for the lattice spacings and quark masses, perform the fits and then use linear
interpolations in $m_h$ to obtain updated estimates. The
process terminates when the updated estimates converge. During this iterative procedure we use reweighting
(see section\,\ref{subsec:reweighting}) to adjust all pionic observables
to the new strange-quark mass on each ensemble. For kaon and $\Omega$
observables a linear interpolation between the unreweighted
unitary measurement, and measurements with a second valence strange quark
(reweighted-to-be-unitary) suffice to obtain that
observable for $m_y = m_h = m_s^{\rm guess}$.

For the remainder of this subsection we explain further the procedure which we use to match lattices with different $\beta$ and present results for the ratios $R_a^{\bf e}$ and $Z_f^{\bf e}$ defined in Eqs.\,(\ref{eq:R_a_def}) and (\ref{eq:Z_f_def}) for our ensembles using the fixed trajectory method explained in Section~\ref{subsec:fitting_strategies}. We start by taking a specific value of $(m_l,m_h)^{\bf M}$ on the ensemble $\bf M$ to which the other ensembles are matched. We refer to this as the matching point. The ensemble set $\bf M$ may be the same as the primary ensemble {\bf 1}, but does not need to be. As discussed in section \ref{subsec:scaling}, the matching to other ensembles ${\bf e}\neq {\bf M}$ is performed by requiring that the ratios of hadronic masses $\frac{m_{ll}}{m_{hhh}}$ and $\frac{m_{lh}}{m_{hhh}}$ are the same on all lattices at the matching point.   Although the final physical predictions do not depend upon the choice of matching point, certain choices are favoured due to the quality of the data at the matching point and the range over which the data must be interpolated/extrapolated on the other ensembles to perform the matching. The ideal point has as small a statistical error as possible and lies within the range of simulated data on all of the matched ensembles such that only a small interpolation is required. In practice, the errors on the mass ratios at the matching point can be reduced by fitting to all partially quenched simulated data on the ensemble set $\bf M$ and interpolating to the matching point along the unitary curve. We use linear fitting functions for the light-quark mass dependence of the pseudoscalar mesons and the $\Omega$ baryon in these short interpolations:
\begin{eqnarray}
m_{xy}^2 &=& c_0 + c_l\,m_l + c_v (m_x+m_y)\,,\label{eq:mxyinterp}\\
m_{xh}^2 &=& d_0 + d_l\,m_l + d_v\,m_x\,,\label{eq:mxhinterp}\\
m_{hhh} &=& e_0 + e_l\,m_l\,,\label{eq:momegainterp}
\end{eqnarray}
where as elsewhere $x,y$ ($l$) represent the light valence (sea) quarks and $h$ represents the heavy quark. Equations (\ref{eq:mxyinterp})\,-\,(\ref{eq:momegainterp}) are written in lattice units. Although the linear behaviour in Eqs.\,(\ref{eq:mxyinterp})\,-\,(\ref{eq:momegainterp}) is similar to that used in the analytic ansatz, Eqs.\,(\ref{eq:CemANmll}), (\ref{eq:CemANmxh}) and (\ref{eq:CemANmhhh}), we stress that the meaning is different. When using the analytic ansatz we assume its validity in the full range of masses between the physical ones and those we simulate. Eqs.\,(\ref{eq:mxyinterp})\,-\,(\ref{eq:momegainterp}) on the other hand, are only assumed to represent the mass behaviour in the short intervals between the matching and simulated points on ensembles ${\bf e}\neq{\bf M}$, independently of whether we subsequently use chiral perturbation theory or the analytic ansatz to perform the chiral extrapolation.

Once a matching point has been chosen, the matching proceeds as follows:
\begin{enumerate}
\item For each set of ensembles $\bf e\neq \bf M$, we perform an independent partially-quenched linear fit to the
simulated pion, kaon and Omega masses using the forms given in Eqs.\,(\ref{eq:mxyinterp})\,-\,(\ref{eq:momegainterp}).
\item We make a first estimate of the pair of quark masses $(m_l,m_h)^{\bf e}$ on each ensemble set $\bf e\neq \bf M$ that corresponds to the matching point.
\item\label{enum-matchiterstart} We then interpolate the three hadronic masses to the estimated $m_l^{\bf e}$ for each value of the simulated unitary heavy quark mass.
\item We linearly interpolate each quantity to the estimated value of $m_h^{\bf e}$.
\item Next we calculate the ratios $R_l^{\bf e} = \displaystyle \frac{m_{ll}^{\bf e} }{m_{hhh}^{\bf e}}$ and $R_h^{\bf e} = \displaystyle \frac{m_{lh}^{\bf e}}{m_{hhh}^{\bf e}}$.
\item Using the measured slopes of $m_{ll}^{\bf e}$ and $m_{hhh}^{\bf e}$ with respect to $m_l^{\bf e}$, by comparing $R_l^{\bf e}$ to the corresponding value $R_l^{\bf M}$ at the matching point we obtain an updated estimate of $m_l^{\bf e}$.
\item Similarly, by comparing the ratio $R_h^{\bf e}$ to $R_h^{\bf M}$ we obtain an updated estimate of $m_h^{\bf e}$.
\item With these updated estimates of the quark masses $(m_l,m_h)^{\bf e}$, we return to step \ref{enum-matchiterstart} and iterate the steps until the process converges.
\end{enumerate}
Once this procedure has converged, we have a set of bare quark masses $(m_l,m_h)^{\bf e}$ which, in physical units, are equivalent to the masses $(m_l,m_h)^{\bf M}$. Following the discussion in Sec.\,\ref{subsec:mass_scaling}, we choose a primary ensemble $\bf 1$ and determine the ratios of quark masses $Z_f^{\bf e}$ in ensembles $\bf 1$ and $\bf e$ as in Eq.\,(\ref{eq:Z_f_def}) with the corresponding ratios of lattice spacing $R_a$ given in Eq.\,(\ref{eq:R_a_def}).

In the above we assumed that for each ensemble {\bf e} we had performed simulations at several values of $m_h^{\bf e}$. In our present study the simulations were performed at a single value of $m_h^{\bf e}$ and the dependence on the heavy-quark mass is obtained by reweighting as explained in Section\,\ref{subsec:reweighting}.

The above discussion was deliberately presented in a general case where there are an arbitrary number of ensembles. In our case we only have two sets, i.e. the $24^3$ and $32^3$ lattices. For the primary ensemble we choose the finer $32^3$ lattice. As we have only one other ensemble set ($24^3$), from now on we drop the superscript on the ratios of lattice spacings ($R_a$) and quark masses ($Z_l$ and $Z_h$).

In Table~\ref{tab:zlzhzacomparison} we give results for $Z_l$, $Z_h$ and $R_a$ obtained by matching at several matching points on both ensemble sets $\bf M \in \{24^3, 32^3\}$. Since we prefer to have a matching point within the range of simulated data on both ensembles, we can discard the first and last entries in the table. From the remaining 3 possibilities, we choose as our final values $Z_l=0.981(9)$, $Z_h=0.974(7)$ and $R_a=0.7583(46)$ from the second entry with $\bf M=32^3$ and $(m_l,m_h)^{32^3}=(0.006,0.03)$.

\begin{table}[t]
\centering
\begin{tabular}{ccc|ccccccc}
\hline\hline
$\bf M$ & $(am_l)^{\bf M}$& $(am_h)^{\bf M}$& $(am_l)^{\bf e}$& $(am_h)^{\bf e}$& $Z_l$    & $Z_h$   &$R_a$     \\
\hline
$32^3$ & 0.004           & 0.03            & 0.00313(13)     & 0.03812(80)     & 0.980(15)& 0.976(11)&0.7617(72)\\
$32^3$ & 0.006           & 0.03            & 0.00583(12)     & 0.03839(51)     & 0.981(9) & 0.974(7) &0.7583(46)\\
$32^3$ & 0.008           & 0.03            & 0.00860(19)     & 0.03869(64)     & 0.979(10)& 0.972(8) &0.7545(58)\\
$24^3$ & 0.005           & 0.04            & 0.00545(11)     & 0.03148(51)     & 0.985(12)& 0.978(9) &0.7620(57)\\
$24^3$ & 0.01            & 0.04            & 0.00897(18)     & 0.03074(57)     & 0.974(11)& 0.968(9) &0.7517(70)\\
\hline
\end{tabular}
\caption{Values of the quark mass ratios $Z_l$ and $Z_h$ and the lattice spacing ratio $R_a$ determined by matching at five points over both ensemble sets. The quark masses here are quoted without the additive $m_\mathrm{res}$ correction. The ensemble $\bf e\neq M$.}
\label{tab:zlzhzacomparison}
\end{table} 

\begin{figure}[t]
\includegraphics*[width=0.95\textwidth]{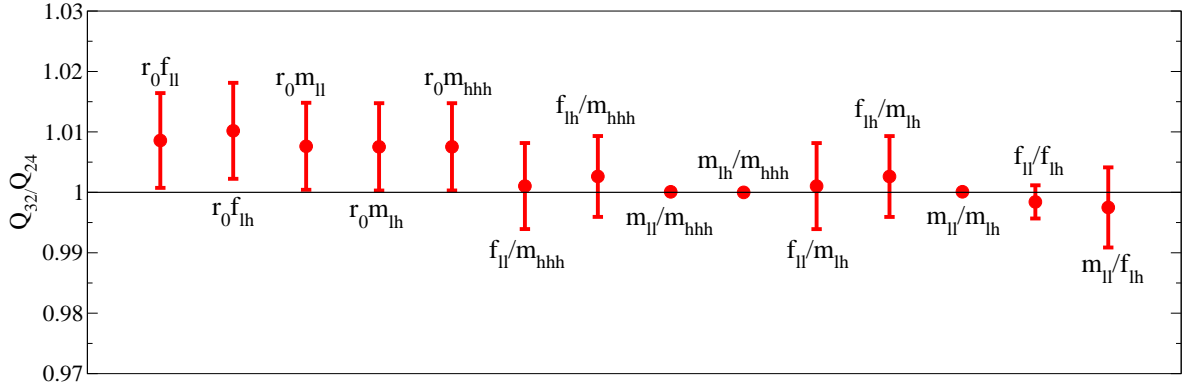}
\caption{Ratios of dimensionless combinations of lattice quantities $Q$ (listed in the figure) between the $32^3$ and $24^3$ lattices at the matching point corresponding to $m_l=0.006$, $m_h=0.03$ on the $32^3$ lattice. A value of unity indicates perfect scaling. The ratios $m_{ll}/m_{hhh}$ and $m_{lh}/m_{hhh}$ (and consequently $m_{ll}/m_{lh}$) are defined to scale perfectly at these quark masses as a consequence of our choice of scaling trajectory.}
\label{fig-beijingplot}
\end{figure}

Having chosen to perform the matching of the lattices at the two lattice spacings by requiring that $m_{ll}/m_{hhh}$ and $m_{lh}/m_{hhh}$ take the same values at the matching point, we expect to see lattice artefacts in ratios of other physical quantities. This is illustrated in Figure~\ref{fig-beijingplot} in which we show the ratios of several other dimensionless combinations of lattice quantities between the two lattices at the quark masses used in the matching procedure above. The figure shows that we can expect only small scaling violations on the order of $1$--$2\%$ for the other quantities used in our global fits, and also confirms that other dimensionless combinations of lattice quantities would be equally suitable choices for the definition of the scaling trajectory.

\subsection{Results of combined scaling and chiral fits} \label{subsec:results}

Using the matching factors $Z_l$, $Z_h$ and $R_a$ determined as described in the previous section we are ready to
perform a simultaneous fit of all our pion, kaon and $\Omega$ mass and decay constant data to
either the NLO forms in chiral perturbation theory, Eq.\,(\ref{eq:chPTsu2:mPi:a2}) to Eq.\,(\ref{eq:chPTsu2:Omega:a2}),
or the analytic forms Eq.\,(\ref{eq:CemANmll}) to Eq.\,(\ref{eq:CemANmhhh}). We also correct for
finite volume effects in NLO PQChPT
by substituting the chiral logarithms with the corresponding finite-volume
sum of Bessel functions \cite{Sharpe:1992ft}. The iterative procedure is the same for each of these three fit ans\"atze.
For each iteration $i$, we:
\begin{itemize}
\item[1] estimate the physical strange-quark masses, $m_s^i$, from the $(i-1)$th iteration;
\item[2] interpolate and reweight the data to $m_s^i$;
\item[3] fit the $m_x, m_y, m_l$ dependence of the
         light pseudoscalar mass and decay constant;
\item[4] fit the $m_x, m_l$ dependence of
	 kaon quantities at $m_h = m_s^i$;
\item[5] fit the $m_l$ dependence of the Omega mass for $m_h = m_s^i$;
\item[6] by comparing to the physical values of $m_\pi/m_\Omega$ and $m_K/m_\Omega$, determine the
iterated predictions for the physical strange quark masses $m_s^{i+1}$\,.
\end{itemize}
This process is repeated until it converges and a self consistent set of
quark masses, lattice spacings and results in the continuum limit are obtained.

For the fits based on NLO chiral perturbation theory we use
Eqs.\,(\ref{eq:chPTsu2:mPi:a2}) and (\ref{eq:chPTsu2:fPi:a2})
for the pion mass and decay constant respectively, and Eqs.\,(\ref{eq:chPTsu2K:mPS:a2}) and (\ref{eq:chPTsu2K:fPS:a2})
for the kaon mass and decay constant. In our earlier work~\cite{Allton:2008pn}
we found that we had to apply cuts to keep the pion mass below around 420 MeV in order for NLO SU(2) ChPT to
give an acceptable description of our data.
All the additional data introduced in this work satisfies this cut and
we include all the data for pions with valence masses
$m_x,m_y \le 0.01$ on the two $24^3$ ensembles and all data for pions
with valence masses $m_x,m_y \le 0.008$ for the three $32^3$ ensembles.
For kaons we include all the valence light-quark masses in the above range for each fixed strange-quark mass.
For this infinite-volume SU(2) NLO global fit the fitted parameters are presented in the second column of
table\,\ref{tab-NLOsu2fitparameters}. The $\chi^2/$dof for all the fits discussed here are
given in table\,\ref{tab:chisq}. We also perform the corresponding fits using the finite-volume chiral logarithm
composed of a sum of Bessel functions~\cite{Sharpe:1992ft}; resummed expressions
are not available for our partially quenched fits. The parameters of the fit are presented in the third column of
table~\ref{tab-NLOsu2fitparameters}. In terms of the conventional LECs $\bar{l}_3$ and $\bar{l}_4$ the results are
\begin{eqnarray}\label{eq:l3l4}
\bar{l}_3 = 2.82(16),\quad&\bar{l}_4  = 3.76(9)\qquad (\textrm{Infinite Volume ChPT})\\
\bar{l}_3 = 2.57(18),\quad&\bar{l}_4  = 3.83(9)\qquad (\textrm{Finite Volume ChPT})\,.\label{eq:l3l4fv}
\end{eqnarray}
\begin{table}[tp]
\centering
\begin{tabular}{c|c|c}
\hline
Parameter\ & No FV Corrections & With FV Corrections\\
\hline
$B$ & 4.12(7)\,GeV & 4.03(7)\,GeV\\
$f$ & 0.110(2)\,GeV & 0.112(2)\,GeV\\
$c_f$ & 0.05(7)\,GeV$^2$ & 0.04(7)\,GeV$^2$\\
$L_4^{(2)}$ & -0.00000(7) & -0.00005(7)\\
$L_5^{(2)}$ & 0.00050(5) & 0.00047(5)\\
$L_6^{(2)}$ & -0.00003(4) & -0.00005(4)\\
$L_8^{(2)}$ & 0.00055(2) & 0.00059(2)\\
$m^{(K)}$ & 0.4856(4)\,GeV & 0.4854(4)\,GeV\\
$f^{(K)}$ & 0.141(3)\,GeV & 0.143(3)\,GeV\\
$c_{f^{(K)}}$ & 0.01(6)\,GeV$^2$ & 0.01(6)\,GeV$^2$\\
$\lambda_1$ & 0.0043(9) & 0.0046(10)\\
$\lambda_2$ & 0.023(1) & 0.024(1)\\
$\lambda_3$ & -0.0018(9) & -0.0016(10)\\
$\lambda_4$ & 0.0058(2) & 0.0057(2)\\
$m^{(\Omega)}$ & 1.666(2)\,GeV &  1.666(2)\,GeV\\
$c_{m_\Omega,m_l}$ & 0.20(6)\,GeV$^{-2}$ & 0.20(6)\,GeV$^{-2}$
\end{tabular}
\caption{Parameters of the global fit to our ensembles using NLO ChPT without finite-volume corrections (second column) and 
with finite-volume corrections (third column). For the unitary theory the parameters are defined in 
Sect.\,\ref{subsec:chiral_scaling} and for the partially quenched theory in appendix B of 
Ref.\,\cite{Allton:2008pn}.}
\label{tab-NLOsu2fitparameters}
\end{table}

\begin{table}[hbt]
\centering
\begin{tabular}{ccc}
Ansatz &\quad & $\chi^2/\mathrm{dof}$ \\
\hline
NLO & &0.72(46) \\
NLO-fv &&  1.07(47)\\
Analytic& &  0.60(44)\\
\hline
\end{tabular}
\caption{Fit ansatze, mass ranges and uncorrelated $\chi^2/\mathrm{dof}$ obtained in our analyses. The fits were 
performed for pion masses less than 420\,MeV.\label{tab:chisq}}
\end{table}

In table~\ref{tab-analyticfitparameters} we present the parameters of the fit with the
analytic ansatz over the same mass range as for the fits using SU(2) chiral perturbation theory, as explained in the previous paragraph. We find that analytic fits including a larger range of pseudoscalar masses
give an acceptable uncorrelated $\chi^2/$dof but then the lightest data points were consistently missed by the fit
by about one standard deviation. The utility of such extended fits for extrapolating to the physical point was therefore compromised and we therefore decided to restrict the range of masses used in the analytic fits.
\begin{table}[tp]
\centering
\begin{tabular}{cc||cc}
\hline
Parameter &\ Value\ &\ Parameter\  &\ Value\ \\
\hline
$C_0^{m_\pi}$ & -0.001(1)\,GeV$^2$&$C_1^{m_K}$ & 3.67(4)\,GeV\\
$C_1^{m_\pi}$ & 7.45(9)\,GeV&$C_2^{m_K}$ & 0.7(1)\,GeV\\
$C_2^{m_\pi}$ & 0.43(8)\,GeV&$C_0^{f_K}$ & 0.149(2)\,GeV\\
$C_0^{f_\pi}$ & 0.123(2)\,GeV&$C_{f_K}$ & 0.02(6)\,GeV$^2$\\
$C_{f_\pi}$ & 0.04(7)\,GeV$^2$&$C_1^{f_K}$ & 0.34(1)\\
$C_1^{f_\pi}$ & 0.85(2)&$C_2^{f_K}$ & 0.52(10)\\
$C_2^{f_\pi}$ & 0.56(9)&$C_0^{m_\Omega}$ & 1.666(2)\,GeV\\
$C_0^{m_K}$ & 0.2353(8)\,GeV$^2$&$C_2^{m_\Omega}$ & 2.7(9)
\end{tabular}
\caption{Parameters of the global fit to our ensembles using the analytic ansatz. The parameters are defined 
in Eqs.\,(\ref{eq:CemANmll})\,--\,(\ref{eq:CemANmhhh}).
\label{tab-analyticfitparameters}}
\end{table}

The global fit to many ensembles of partially quenched data
is naturally a high dimensional space and so the exposition
of the fits is best performed by looking at portions of the data in turn.
In order to illustrate the quality of the fits, in the following subsections we display the fit and
data for each physical quantity in turn. In total we have analysed five ensembles at two lattice
spacings, and each ensemble has measurements at many partially
quenched valence-quark masses. As it is only feasible
to present a subset of possible plots, in the following we display the dependence of each quantity on the valence quark masses at the lightest sea-quark mass ($m_l=0.005$ for the $24^3$ ensembles and $m_l=0.004$ on the $32^3$ ensembles). The exception of course, is the mass of the Omega baryon $m_{hhh}$ which does not depend on the light
valence-quark masses. We also display the unitary subset of data on both lattice spacings
along with the mass dependence we infer from our fits in the unitary
continuum limit.

Before discussing the chiral and continuum behaviour of hadronic masses and decay constants in detail, we present in table\,\ref{tab:quarkmasseslatt} our results for the unrenormalised physical quark masses and the lattice spacings obtained from the three fits. In this table the quark masses are given in lattice units. The non-perturbative renormalization of the masses will be discussed in Sec.\,\ref{sec:quark_masses} where the values of the renormalized quark masses in the $\overline{{\rm MS}}$ scheme will be presented.
\begin{table}[hbt]
\centering
\begin{tabular}{c|c|c|c}
 & NLO & NLO fv & Analytic \\
\hline
$\tilde{m}_l(32^3)$ & 0.00100(3) & 0.00102(3) & 0.00105(6) \\
$\tilde{m}_s(32^3)$ & 0.0280(7)  & 0.0280(7) & 0.0279(7) \\
$a^{-1}(32^3)$ & $2.280(28)$\,GeV & $2.281(28)$\,GeV & $2.282(28)$\,GeV\\ 
\hline
$\tilde{m}_l(24^3)$ & 0.00134(4)  & 0.00136(4) &  0.00141(9)\\
$\tilde{m}_s(24^3)$ & 0.0379(11)  & 0.0379(11)  & 0.0378(11) \\
$a^{-1}(24^3)$ & $1.729(25)$\,GeV & $1.729(25)$\,GeV & $1.730(25)$\,GeV\\ 
\hline
\end{tabular}
\caption{
\label{tab:quarkmasseslatt}
Unrenormalised physical quark masses in lattice units and the values of the 
inverse lattice spacing $a^{-1}$ for the $32^3$ and $24^3$ ensembles.
}
\end{table}

\subsubsection{Chiral and continuum behaviour of the $\Omega$-baryon}

The $\Omega$ mass is fitted using Eq.\,(\ref{eq:chPTsu2:Omega:a2}) (or equivalently (\ref{eq:CemANmhhh})\,).
The fit form for the $\Omega$ baryon does not change
between the different ans\"atze and only very small differences arise
from the different estimates of physical quark masses and hence of the lattice spacings. For illustration,
Figure~\ref{fig:omegaextrap} shows the extrapolation of the $\Omega$ mass using the analytic ansatz.

\begin{figure}[t]
\centering
\includegraphics*[width=0.48\textwidth]{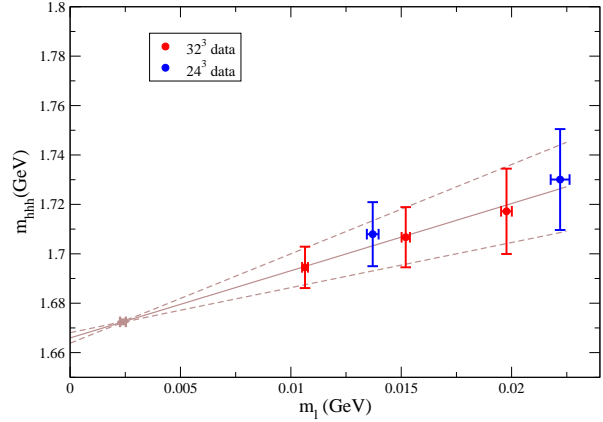}
\caption{
\label{fig:omegaextrap}
The fit to the light-quark mass behaviour of the
$\Omega$-baryon in the continuum limit obtained using the analytic ansatz. The corresponding plots using the infinite and finite-volume SU(2) ChPT ansatz are almost indistinguishable, differing only slightly in the estimates of the physical quark masses and the lattice spacings.}
\end{figure}

\subsubsection{Chiral and continuum behaviour of the pion mass}

We display the fits of the partially quenched pion
masses using infinite volume NLO SU(2)
partially quenched ChPT (i.e. to the partially quenched generalization of Eq.\,(\ref{eq:chPTsu2:mPi}) given in Eq.\,(B.32) of ref.\,\cite{Allton:2008pn})
in figure \ref{fig:mpi:on:m:NLO} for the lightest $24^3$ and $32^3$ ensembles.
As discussed in section~\ref{subsec:analytic},
we divide by the average valence-quark mass with the intention of
enhancing the visibility of chiral logarithms.
Figure \ref{fig:mpi:on:m:NLOfv} displays
the corresponding fit of the same data but including finite-volume corrections.

\begin{figure}[t]
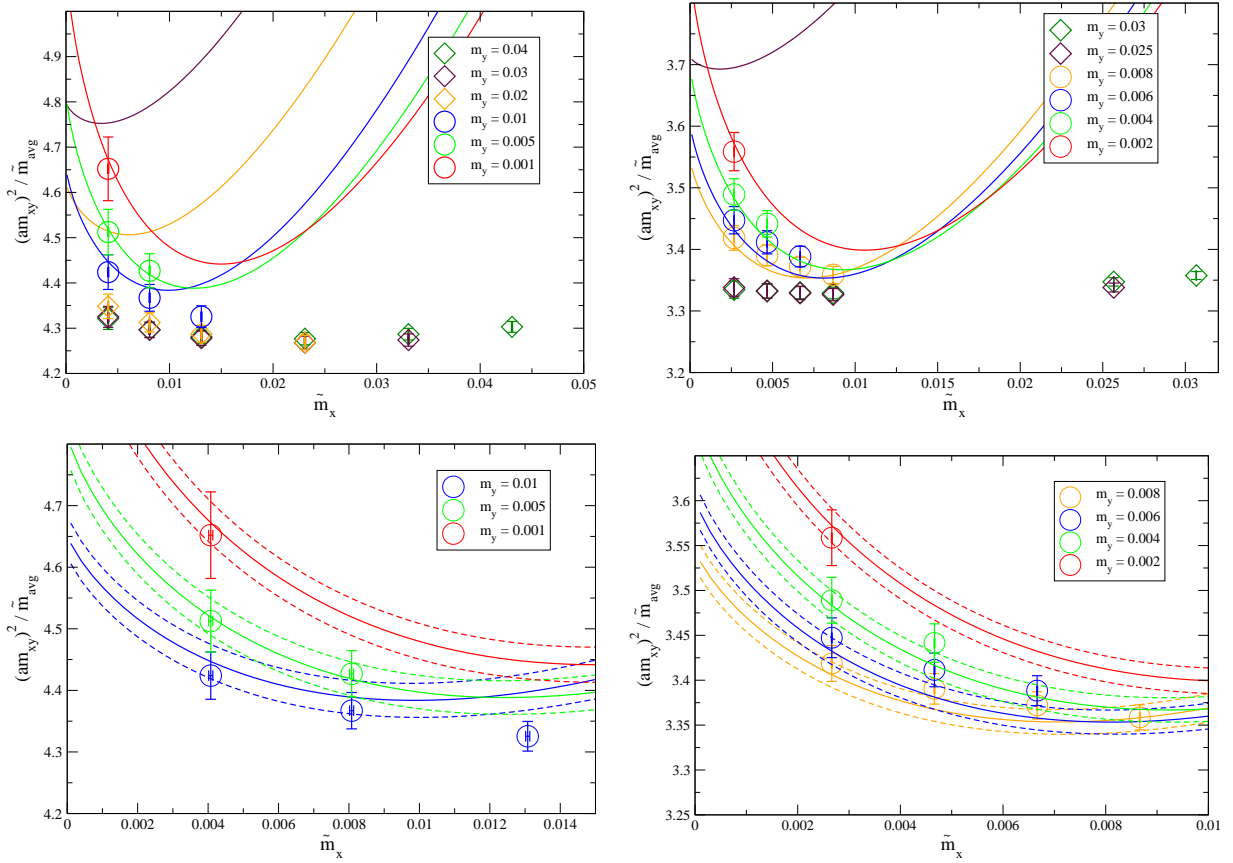

\centering
\includegraphics*[width=0.48\textwidth]{fig/Fig28a_mps2dml_PQ24_0.005_chpt.eps}\quad
\includegraphics*[width=0.48\textwidth]{fig/Fig28b_mps2dml_PQ32_0.004_chpt.eps}\\[0.1in]
\includegraphics*[width=0.48\textwidth]{fig/Fig28c_mps2dml_PQ24_0.005_chpt_closeup.eps}\quad
\includegraphics*[width=0.48\textwidth]{fig/Fig28d_mps2dml_PQ32_0.004_chpt_closeup.eps}
\caption{
\label{fig:mpi:on:m:NLO}
Global fits obtained using
infinite volume NLO SU(2) chiral perturbation theory for the pion mass. The top-left panel includes the partially quenched data from the $m_l=0.005$ ensemble on the $24^3$ lattice and the data points in the top-right panel are from the $m_l=0.004$ ensemble from the $32^3$ lattice. In each case the curves correspond to the appropriate value of the lattice spacing. The points marked by the circles were included in the fit, whereas those marked by the diamonds were not. In the bottom two panels we zoom into the low-mass region, illustrating the fits to the points which were included ($24^3$ points on the left and $32^3$ points on the right). (For fixed $\tilde{m}_x$, $m_y$ decreases as $(am_{xy})^2/\tilde{m}_{\textrm{avg}}$ increases.)}
\end{figure}
\begin{figure}[hbt]
\centering
\includegraphics*[width=0.48\textwidth]{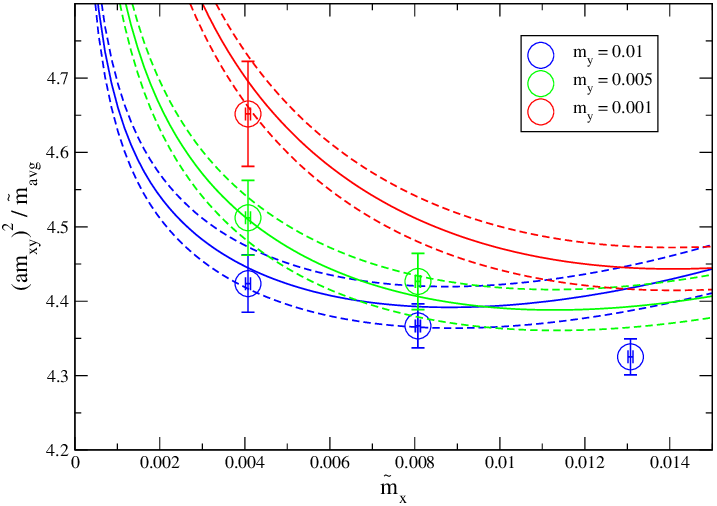}
\quad\includegraphics*[width=0.48\textwidth]{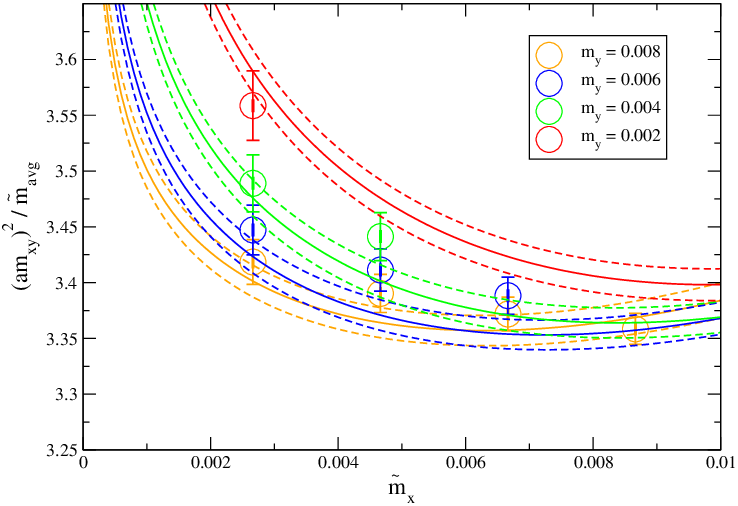}
\caption{
\label{fig:mpi:on:m:NLOfv}
Global fits for the pion mass obtained using
NLO SU(2) chiral perturbation theory with finite-volume corrections. In this case we only include the points which were included in the fit ($m_l=0.005$, $24^3$ points on the left and $m_l=0.004$, $32^3$ points on the right) since the finite-volume corrections at larger masses are small. (For fixed $\tilde{m}_x$, $m_y$ decreases as $(am_{xy})^2/\tilde{m}_{\textrm{avg}}$ increases.)}
\end{figure}

It is apparent that the infinite volume and finite
volume NLO fits diverge rapidly from
our data at larger masses, and this indeed is the reason why we were compelled to introduce the upper cut-off of
420\,MeV for this analysis~\cite{Allton:2008pn}.

\begin{figure}[hbt]
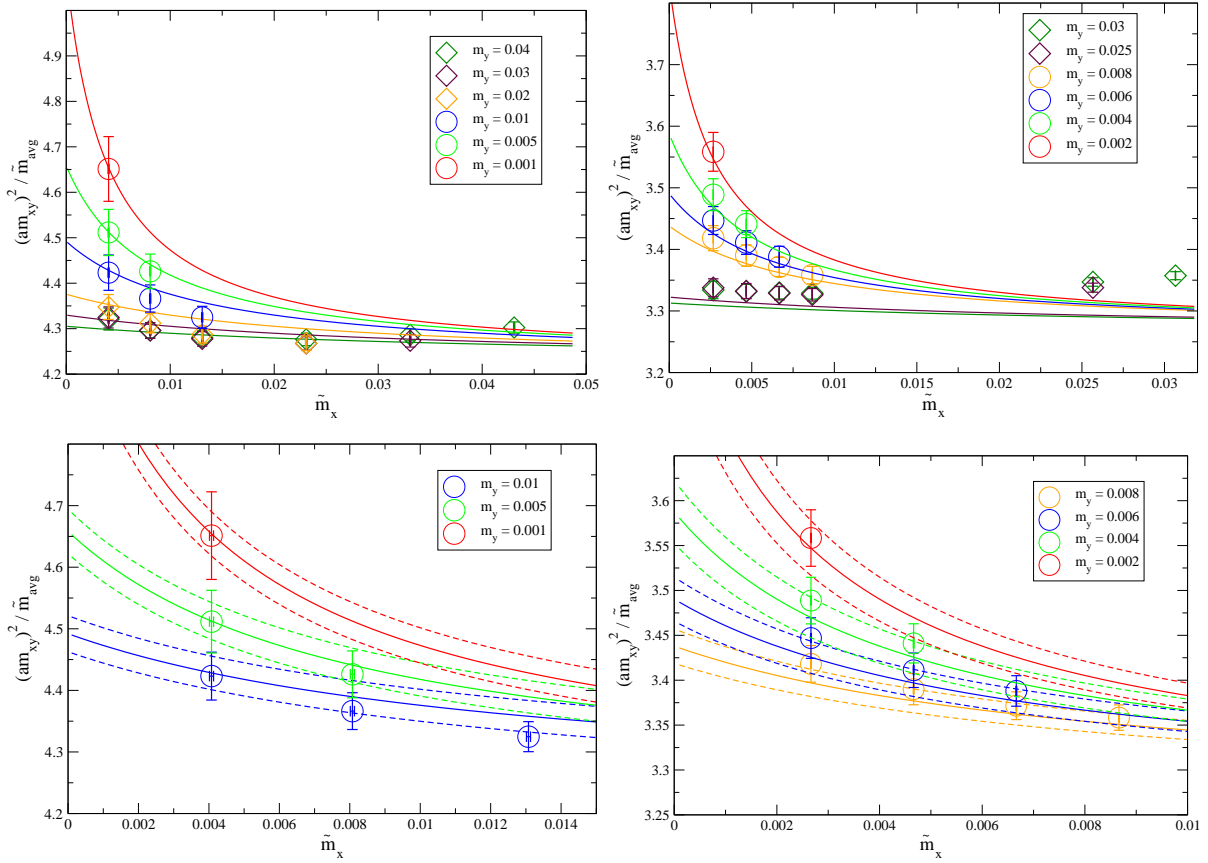

\centering
\includegraphics*[width=0.48\textwidth]{fig/Fig30a_mps2dml_PQ24_0.005_flavourexp.eps}
\includegraphics*[width=0.48\textwidth]{fig/Fig30b_mps2dml_PQ32_0.004_flavourexp.eps}\\[.1in]
\includegraphics*[width=0.48\textwidth]{fig/Fig30c_mps2dml_PQ24_0.005_flavourexp_closeup.eps}
\includegraphics*[width=0.48\textwidth]{fig/Fig30d_mps2dml_PQ32_0.004_flavourexp_closeup.eps}\\[.1in]
\caption{
\label{fig:mpi:on:m:analytic}
Global fit curves obtained using the analytic fit ansatz
(\ref{eq:CemANmll}) overlaying the simulated pion masses on the
$m_l=0.005$, $24^3$ ensemble (top-left) and the $m_l=0.004$, $32^3$ ensemble (top-right). Points marked by circles
were included in the fit, those marked by diamonds were not.
The simple linear expansion replicates the entire range of lattice
data reasonably well with the description being rather better than
NLO chiral perturbation theory at our larger masses. In the bottom two panels we zoom into the low-mass region, illustrating the fits to the points which were included ($24^3$ points on the left and $32^3$ points on the right).
(For fixed $\tilde{m}_x$, $m_y$ decreases as $(am_{xy})^2/\tilde{m}_{\textrm{avg}}$ increases.)}
\end{figure}

We now consider the chiral extrapolation of the pion mass
using the analytic form of Eq.\,(\ref{eq:CemANmll}) which is shown in Fig.\,\ref{fig:mpi:on:m:analytic}.
Comparing Figs.~\ref{fig:mpi:on:m:NLO} and \ref{fig:mpi:on:m:NLOfv}
with Fig.\,\ref{fig:mpi:on:m:analytic}
suggests that data at substantially larger masses
can be described by the analytic expansion,
without any curvature terms in the ansatz.
The division by the average valence quark mass in the plots, coupled to allowing
the tangent not to pass through the origin
(i.e. that the extrapolated $m_\pi^2$ at $m_x = m_y = 0$ may not be equal to zero)
allows the analytic fit to reproduce a structure that
might otherwise be attributed to chiral logarithms.

We emphasize that admitting the possibility that the constant term $C_0^{m_\pi}\neq 0$
allows for a pole in figure~\ref{fig:mpi:on:m:analytic} in the
unitary chiral limit. In fact we find that $C_0^{m_\pi}$ is numerically small and consistent with zero,
$C_0^{m_\pi} = -0.001(1) {\rm GeV}^2$. We stress again that while Goldstone's theorem implies the vanishing of the pion mass in the SU(2) chiral limit, this does not necessarily imply that $C_0^{m_\pi}=0$. Our model is that the linear ansatz is valid in the region between that where we have data and the physical point, and that if $C_0^{m_\pi}\neq 0$ then it is the  curvature due to chiral logarithms below the physical pion mass which will force the pion mass to zero in the chiral limit. Nevertheless, from the fits we found that $C_0^{m_\pi}$ is consistent with zero.
This is illustrated by the flat behaviour (within the statistical precision) for the chiral behaviour of the unitary points for $m_\pi^2/m_l$ in the continuum limit shown in the right panel in Fig.\,\ref{fig:mpi:on:m:contanalyticNLOcomparison}. Allowing for a non-zero value of $C_0^{m_\pi}$ does however lead to an amplified error for $m_\pi^2/m_l$ at the physical point. The left panel of Fig.\,\ref{fig:mpi:on:m:contanalyticNLOcomparison} shows the corresponding plots for the infinite and finite-volume ChPT fits.

\begin{figure}[hbt]
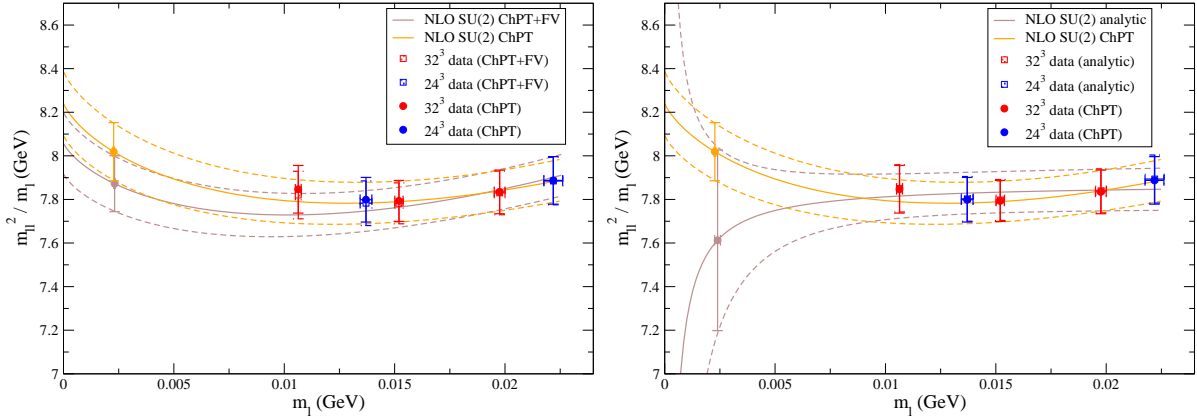

\centering
\includegraphics*[width=0.48\textwidth]{fig/Fig31a_mps2dml_unitary_su2_su2fv_comparison.eps}
\includegraphics*[width=0.48\textwidth]{fig/Fig31b_mps2dml_unitary_flavour_su2_comparison.eps}
\caption{
\label{fig:mpi:on:m:contanalyticNLOcomparison}
Left panel: Pion mass fit for the SU(2) NLO fit form in the continuum
limit, both with and without finite volume logarithms.
We adjust the data points to the continuum limit
using the $a^2$ dependence in our fit form
and overlay these. Right panel:
Chiral extrapolation of the pion mass using
the analytic (\ref{eq:CemANfll}) and infinite-volume NLO ChPT ans\"atze.
}
\end{figure}

Goldstone's theorem equally applies at vanishing valence-quark mass ($m_x=m_y=0$) but with a non-zero sea-quark mass
($m_l > 0$). In contrast with the unitary case discussed in the previous paragraph where $C_0^{m_\pi}$ was consistent with zero, in the partially quenched
direction we find that the corresponding constant $C_0^{m_\pi} + C_2^{m_\pi} m_l$ is non-zero,
specifically $C_2^{m_\pi} = 0.43(8) {\rm GeV}$. This value for
$C_2^{m_\pi}$ is much larger than might be created by propagating the
mass dependence in $m_{\rm res}^\prime(m)$ through
the term involving $C_1^{m_\pi}$; the greatest mass dependence in
$m_{\rm res}^\prime$ occurs on our $24^3$ ensembles in the
partially quenched direction, but can at most generate a 1\% correction
to $\tilde{m}$ and produces a term much smaller
than the measured $C_2^{m_\pi}$. Further, the
residual chiral symmetry breaking is four times
smaller for the $32^3$ ensemble which is also included in the global fit.
Our results from this global analytic fit therefore require a curvature,
most likely from partially-quenched chiral logarithms which are known
to be larger than in the unitary direction, in order for
Goldstone's theorem to be satisfied.

It is also worth emphasizing that the \textit{discovery} of chiral logarithms
in lattice data from plots such as those in Figs.\,\ref{fig:mpi:on:m:NLO} to \ref{fig:mpi:on:m:analytic}
is to a certain extent artificial. Inconsistency with LO chiral perturbation theory is certainly indicated.
Our linear fits suggest that the transformations made
in displaying the data render even conclusions of genuine curvature, let alone unambiguous
demonstration of logarithmic mass dependence, to be somewhat optimistic.
In order to prove logarithmic behaviour, one should
really change quark masses substantially on a logarithmic scale; our
present lattice data supports only the weaker claim of consistency with logarithmic behaviour in
the partially quenched direction.

\subsubsection{Chiral and continuum behaviour of the pion decay constant}\label{subsubsec:chiralfpi}

We now turn to the chiral behaviour of $f_\pi$ and the extrapolation to the physical point. The leading term in all the fits contains an $a^2$ correction and we display the fits performed at non-zero lattice spacing combined with the unmodified lattice data and also our continuum predictions combined with the
lattice data extrapolated to the continuum limit using the results of the fits.

We display our fits obtained using infinite volume NLO SU(2)
partially-quenched ChPT in Figure\,\ref{fig:fpiNLO}. The corresponding fits including finite-volume corrections
are shown in Figure\,\ref{fig:fpiNLOfv}. Finally
Figure\,\ref{fig:fpi:analytic} displays the fits obtained using our analytic ansatz. Having performed the fits, we adjust our unitary data to the continuum limit using the fitting functions with the determined parameters and display the adjusted data in Fig.\,\ref{fig:fpiNLOfvadjusted} together with the finite and infinite-volume NLO SU(2) ChPT fits (left panel) and the analytic fit (right panel). The effect of the adjustment to the continuum limit is illustrated in Figure\,\ref{fig:fpianalytic} where the fits are superimposed on the unadjusted unitary data. It can be seen from Figs.\,\ref{fig:fpiNLOfvadjusted} and \ref{fig:fpianalytic} that the adjustment to the continuum limit for the pion decay constant is very small.

\begin{figure}[hbt]
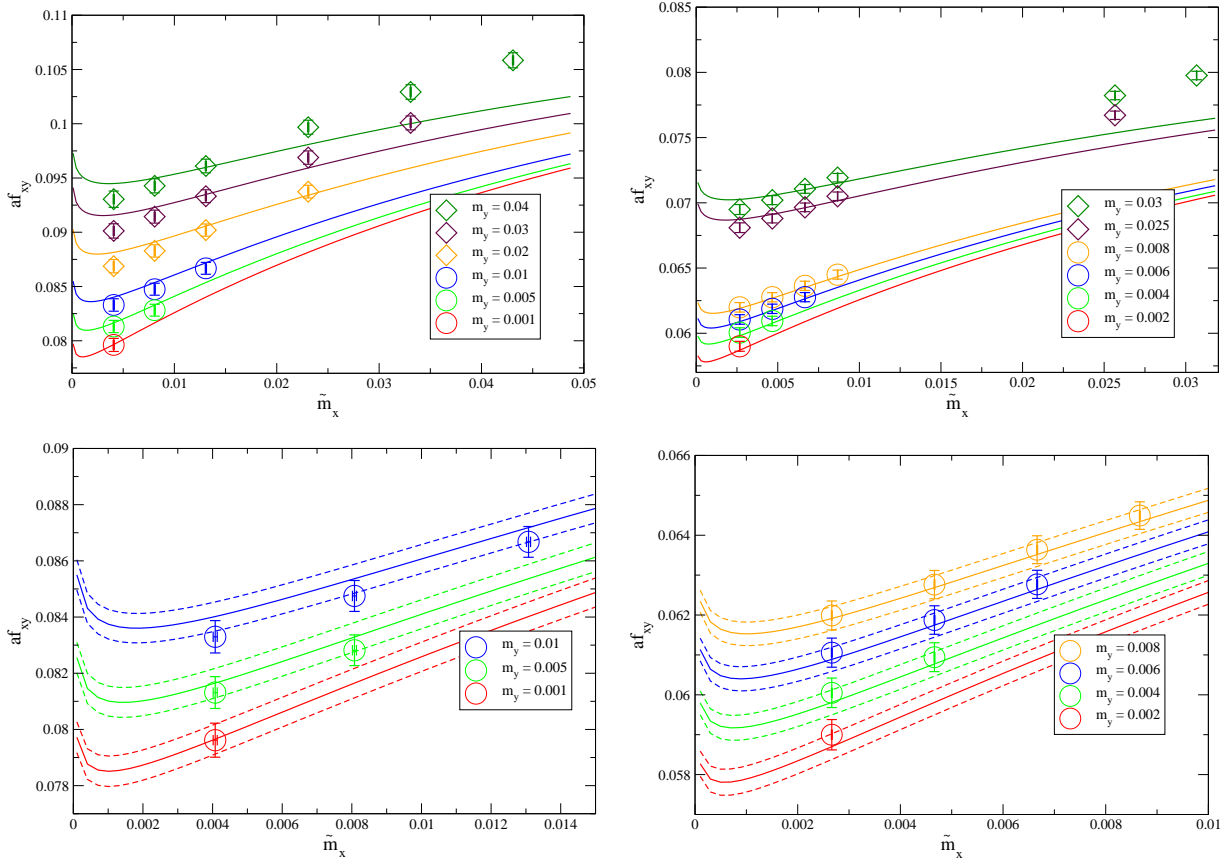

\includegraphics*[width=0.48\textwidth]{fig/Fig32a_fps_PQ24_0.005_chpt.eps}\quad
\includegraphics*[width=0.48\textwidth]{fig/Fig32b_fps_PQ32_0.004_chpt.eps}\\[.1in]
\includegraphics*[width=0.48\textwidth]{fig/Fig32c_fps_PQ24_0.005_chpt_closeup.eps}\quad
\includegraphics*[width=0.48\textwidth]{fig/Fig32d_fps_PQ32_0.004_chpt_closeup.eps}
\caption{
\label{fig:fpiNLO}
Global fits to the lattice data for the pion decay constant obtained using infinite-volume NLO SU(2) chiral perturbation theory. The top-left and top-right panels correspond to the $24^3$, $m_l=0.005$ and $32^3$, $m_l=0.004$ ensembles respectively. Points marked by circles are included in the fits, while those with heavier masses marked by diamonds are not. In the bottom two panels we zoom into the low-mass region, illustrating the fits to the points which were included ($24^3$ points on the left and $32^3$ points on the right).
(For fixed $\tilde{m}_x$, $m_y$ increases as $af_{xy}$ increases.)}
\end{figure}

\begin{figure}[hbt]
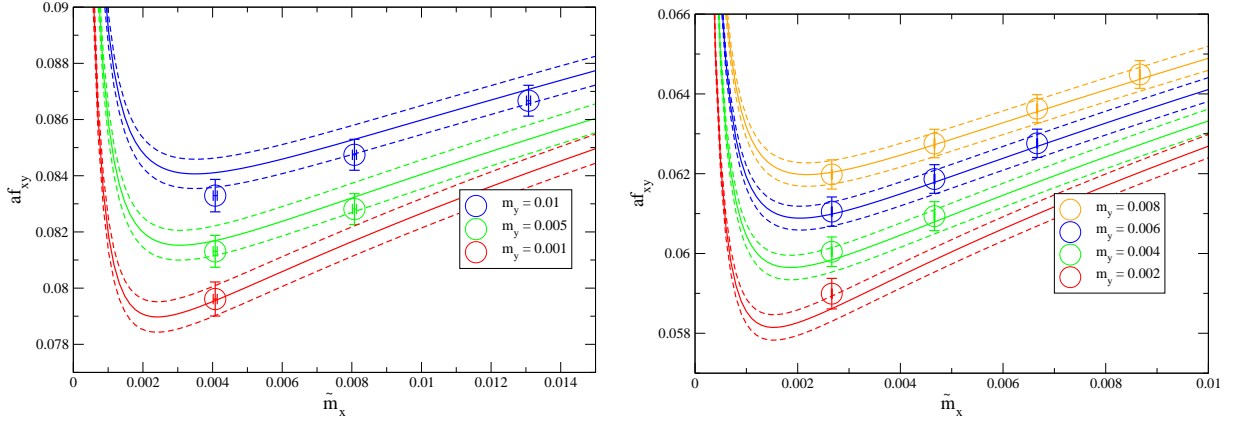

\includegraphics*[width=0.48\textwidth]{fig/Fig33a_fps_PQ24_0.005_chptfv_closeup.eps}\quad
\includegraphics*[width=0.48\textwidth]{fig/Fig33b_fps_PQ32_0.004_chptfv_closeup.eps}
\caption{
\label{fig:fpiNLOfv}
Global fits to the lattice data for the pion decay constant obtained using NLO SU(2) chiral perturbation theory with finite-volume corrections. In this case we only include the points which were included in the fit ($m_l=0.005$, $24^3$ points on the left and $m_l=0.004$, $32^3$ points on the right) since the finite-volume corrections at larger masses are small. (For fixed $\tilde{m}_x$, $m_y$ increases as $af_{xy}$ increases.)}
\end{figure}

\begin{figure}[hbt]
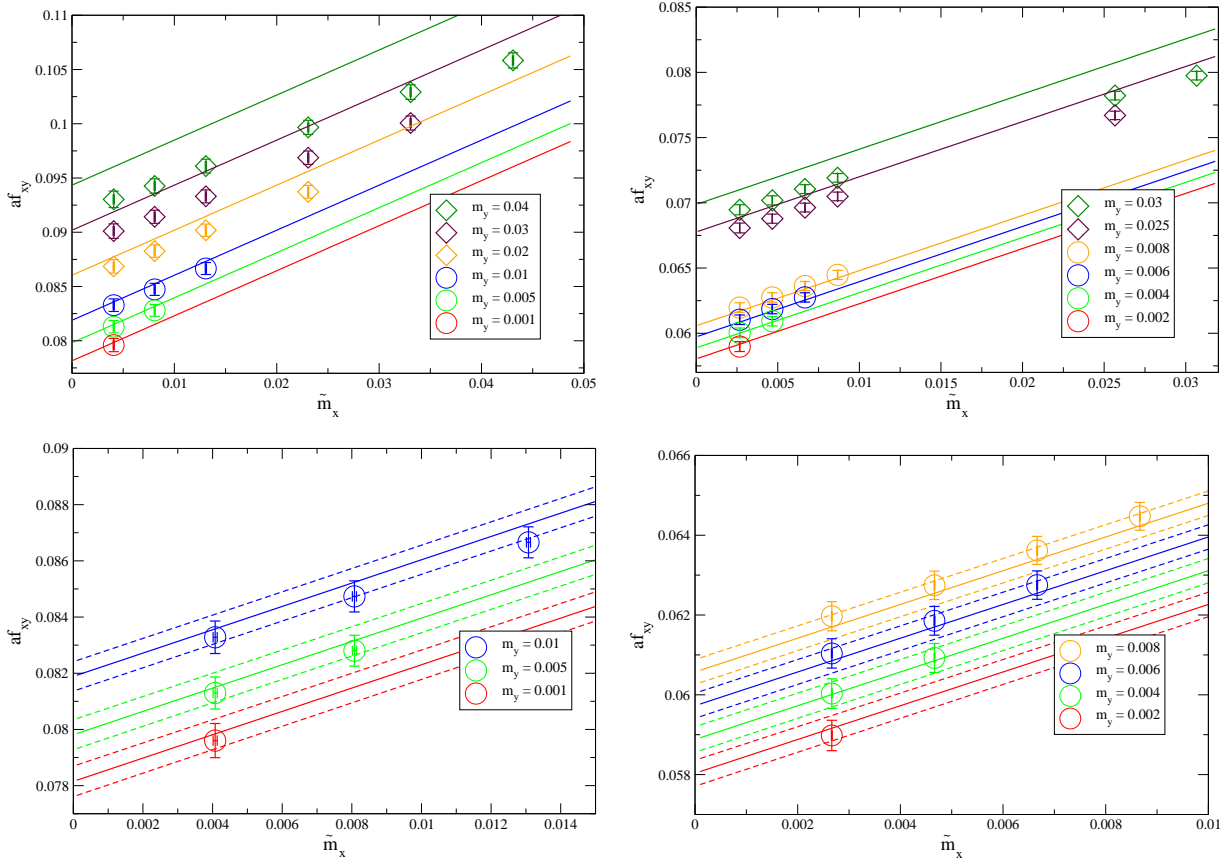

\centering
\includegraphics*[width=0.48\textwidth]{fig/Fig34a_fps_PQ24_0.005_flavourexp.eps}\quad
\includegraphics*[width=0.48\textwidth]{fig/Fig34b_fps_PQ32_0.004_flavourexp.eps}\\[0.1in]
\includegraphics*[width=0.48\textwidth]{fig/Fig34c_fps_PQ24_0.005_flavourexp_closeup.eps}\quad
\includegraphics*[width=0.48\textwidth]{fig/Fig34d_fps_PQ32_0.004_flavourexp_closeup.eps}
\caption{
\label{fig:fpi:analytic}
Global fits to the lattice data for the pion decay constant obtained using the analytic ansatz in Eq.\,(\ref{eq:CemANfll}). The top-left and top-right panels correspond to the $24^3$, $m_l=0.005$ and $32^3$, $m_l=0.004$ ensembles respectively. Points marked by circles are included in the fits, while those with heavier masses marked by diamonds are not. In the bottom two panels we zoom into the low-mass region, illustrating the fits to the points which were included ($24^3$ points on the left and $32^3$ points on the right). (For fixed $\tilde{m}_x$, $m_y$ increases as $af_{xy}$ increases.)}
\end{figure}

\begin{figure}[hbt]
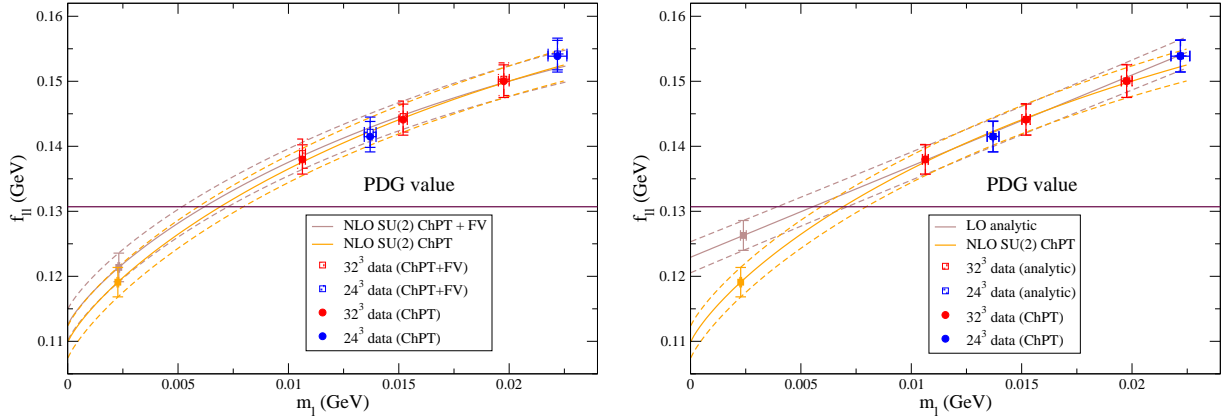

\includegraphics*[width=0.48\textwidth]{fig/Fig35a_fpi_unitary_su2_su2fv_comparison.eps}\quad
\includegraphics*[width=0.48\textwidth]{fig/Fig35b_fpi_unitary_flavour_su2_comparison.eps}
\caption{
\label{fig:fpiNLOfvadjusted}
Unitary data for $f_\pi$ adjusted
to the continuum limit using each of the fit ans\"atze.
The left panel compares the infinite volume and finite volume forms of the
NLO SU(2) fit, while the right panel compares the analytic fit to the infinite volume NLO
SU(2) fit. The horizontal solid line indicates the value $f_{\pi^-}$=130.4\,MeV (the authors of ref.~\cite{Amsler:2008zzb} quote $f_{\pi^-}=(130.4\pm 0.04\pm 0.2)\,$MeV).}
\end{figure}
\begin{figure}[hbt]
\centering
\includegraphics*[width=0.6\textwidth]{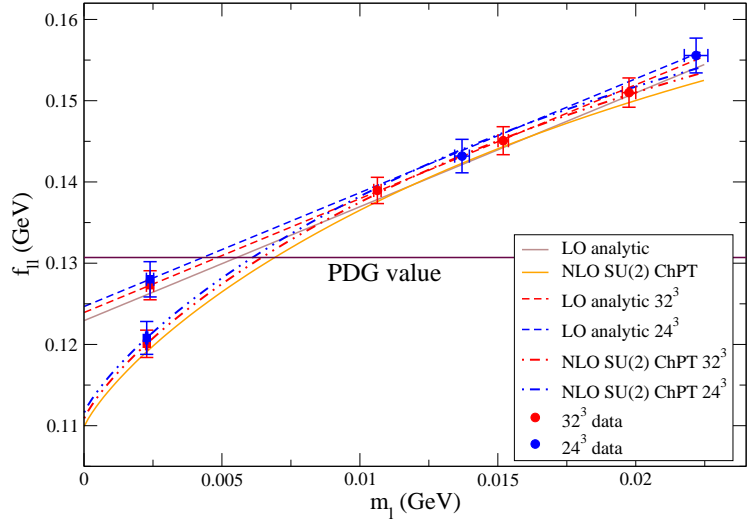}
\caption{
\label{fig:fpianalytic}
Chiral extrapolation of the pion decay constant using
the analytic~(\ref{eq:CemANfll}) and ChPT~(\ref{eq:chPTsu2:fPi:a2}) fit ans\"atze. Here, the lattice results from
the $24^3$ and $32^3$ ensembles are shown along with the mass dependence we
infer both at each lattice spacing and
in the continuum limit. The consistency of the two ensembles with each
other and with this continuum limit is indicative of the size of lattice
artefacts. The horizontal solid line indicates the value $f_{\pi^-}=(130.4\pm 0.04\pm 0.2)\,$MeV~\cite{Amsler:2008zzb}.}
\end{figure}

\begin{table}[hbt]
\centering
\begin{tabular}{c|c|c|c}
 & NLO & NLO fv & Analytic \\
\hline
$f_\pi^{\rm 24^3}$ & 0.121(2) & 0.123(2) & 0.128(2) \\
$f_\pi^{\rm 32^3}$ & 0.120(2) & 0.122(2) & 0.127(2) \\
$f_\pi^{\rm continuum}$ & 0.119(2) & 0.121(2) & 0.126(2) \\
\hline
\end{tabular}
\caption{
\label{tab:fpi:extrapolated}
Predictions for $f_\pi$ in GeV for each global fit ansatz 
at each simulated lattice spacing and in the continuum limit.}
\end{table}

\begin{table}[hbt]
\centering
\begin{tabular}{c|c|c|c}
 & NLO & NLO fv & Analytic \\
\hline
$f_K^{\rm 24^3}$ & 0.147(2) & 0.148(2) & 0.152(2) \\
$f_K^{\rm 32^3}$ & 0.147(2) & 0.148(2) & 0.151(2) \\
$f_K^{\rm continuum}$ & 0.146(2) & 0.147(2) & 0.151(2) \\
\hline
\end{tabular}
\caption{
\label{tab:fk:extrapolated}
Predictions for $f_K$ in GeV for each global fit ansatz 
at each simulated lattice spacing and in the continuum limit.}
\end{table}

\begin{table}[hbt]
\centering
\begin{tabular}{c|c|c|c}
 & NLO & NLO fv & Analytic \\
\hline
$(f_K/f_\pi)^{\rm 24^3}$ & 1.216(9) & 1.205(9) & 1.184(9) \\
$(f_K/f_\pi)^{\rm 32^3}$ & 1.221(6) & 1.209(6) & 1.188(6) \\
$(f_K/f_\pi)^{\rm continuum}$ & 1.229(8) & 1.215(7) & 1.194(7) \\
\hline
\end{tabular}
\caption{
\label{tab:fkoverfpi:extrapolated}
Predictions for $f_K/f_\pi$ for each global fit ansatz 
at each simulated lattice spacing and in the continuum limit.}
\end{table}

The predictions for $f_\pi$ extrapolated to the physical quark masses
for each of the fits is given in table~\ref{tab:fpi:extrapolated}.
We anticipate the discussion of the global fits for $f_K$ which are presented in Sec\,\ref{subsubsec:fkresults} and mention that
the predictions for $f_K$ extrapolated to the physical quark masses
are given in table~\ref{tab:fk:extrapolated}, and
the predictions for $f_K/f_\pi$ extrapolated to the physical quark masses
are given in table~\ref{tab:fkoverfpi:extrapolated}.

We find that the NLO SU(2) fits underestimate the physical value
at our simulated lattice spacings, and that this discrepancy is amplified a little by the
extrapolation to the continuum limit. At each of our two lattice spacings, the analytic ansatz extrapolates close to the physical value of $f_\pi$, but, with our ansatz for the
form of the $a^2$ effects, the result becomes statistically inconsistent in
the continuum limit.

From the above discussion we see that using NLO ChPT to perform the chiral extrapolation for
$f_\pi$ results in a value which is significantly smaller than the physical one. We recall that only data  limited to $m_\pi<420$\,MeV was used in the analysis and note that the fits were performed using the chiral expansion
with $f$, the decay constant in the SU(2) chiral limit, included in the expansion parameter $\chi_l/(4\pi f)^2$. The downward curvature at low masses seen in Figure\,\ref{fig:fpiNLOfvadjusted} can, of course, be reduced by replacing the mass-independent $f$ by an artificial larger parameter such as the physical $f_\pi$ or $f_{ll}(\wm_l)$ measured at each quark mass used in the simulation. The curvature can also be partially absorbed
by using a subset of terms that arise at NNLO. We have experimented with NNLO fits~\cite{Mawhinney:2009jy}
but find that the low-energy constants are insufficiently constrained by our data
to be of practical use. Thus the resulting predictions for the physical value of $f_\pi$ depend strongly on the model assumptions used at NNLO.

The observed $O(10\%)$ deviation found using NLO chiral perturbation theory is broadly consistent with the size of NNLO terms one might expect to be present at masses in the region of our data. Our data for $f_\pi$ vary from about 20\% to 40\% above the value of $f$ obtained from our extrapolations and the square of these terms can be taken as being indicative of the expected NNLO terms. We might therefore expect them to be around 5-15\% within our simulated mass range.

The discrepancy of the prediction for the physical value of $f_\pi$ from the analytic fits is smaller than that found with NLO ChPT, but is nevertheless visible. The results at each of the two lattice spacings are statistically consistent with $f_\pi$ but lead to an underestimate in the continuum limit. Given the sign of the chiral
logarithms at NLO, one might expect a linear ansatz to over-estimate
rather than underestimate the prediction for the physical value. It is nevertheless striking that
one cannot admit any significant non-linearity in this extrapolation
and retain consistency with the physical value for $f_\pi$. The simple analytic form used here appears to be a successful phenomenological model which is simpler and has fewer parameters than approaches based on ChPT with
arbitrarily chosen analytic subsets of NNLO and NNNLO terms.

It is of interest to pose the scientific question whether any of the fit ans\"atze could
in principal be consistent with the experimentally measured pion decay constant? To answer this question
we update the analysis of Ref.\,\cite{Scholz:2009yz} and include an artificially created data
point for each ensemble
that represents the experimental result in the continuum limit
but includes our fitted $a^2$ correction at each non-zero lattice spacing.
This is displayed in figure~\ref{fig:fpi_incphyspt} and
we find that the analytic ans\"atze could be consistent with an uncorrelated
$\chi^2/{\rm dof}=1.9(7)$,
while NLO ChPT would
fail to simultaneously fit our data and the physical point, with
$\chi^2/{\rm dof}=6(1)$ (infinite volume) and
$\chi^2/{\rm dof}=5(1)$ (finite volume).

\begin{figure}
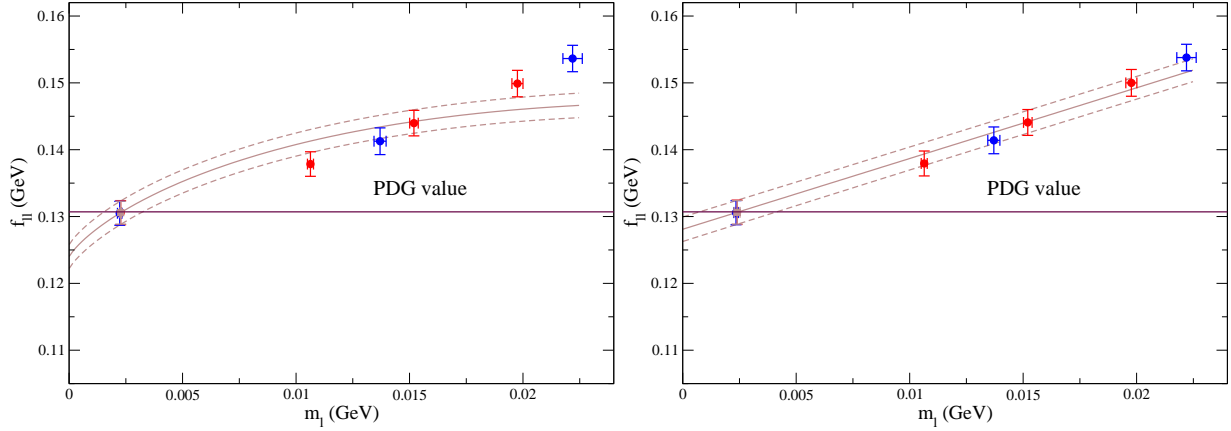

\includegraphics*[width=0.49\textwidth]{fig/Fig37a_fpi_incphyspt_chpt_continuum.eps}
\includegraphics*[width=0.49\textwidth]{fig/Fig37b_fpi_incphyspt_flavourexp_continuum.eps}
\caption{\label{fig:fpi_incphyspt}
An artificial data point (the left-most data point in each panel) corresponding to the physical value of $f_\pi$~\cite{Amsler:2008zzb}, but including our uncertainties in the lattice spacing, is added to the data for the pion decay constant from the five ensembles. The left-hand panel corresponds to the NLO SU(2) ChPT fits and the right-hand panel to the analytic ansatz.}
\end{figure}

Of course, improved statistical errors, simulations at a third lattice spacing
and larger physical volumes would give us better control of the continuum extrapolation and
finite-volume effects. However, our main conclusion is that it is imperative
to simulate with masses substantially nearer to the physical point; this will constrain both fit forms to give more consistent predictions. Ultimately simulations will be performed directly at physical quark masses and will eliminate this error completely. We are currently generating new ensembles with a coarser lattice
spacing, with a substantially larger volume and with very much lighter pion masses~(for a preliminary discussion of these configurations see~Ref.\,\cite{boblatt2010}) precisely to address this issue.

As an estimate of the systematic uncertainties in
physical quantities we take the difference between the results obtained using linear and finite-volume NLO ChPT analyses. This allows for the possible validity of the
full NLO non-analyticity in the region of masses between the data and the physical point but also recognises that part of this extrapolation may be outside the range
of validity of NLO ChPT as suggested by the observation that the present data is surprisingly consistent with linear behaviour. Guided by the results for $f_\pi$ discussed above, we take as our central values for phenomenological predictions the average of the results obtained from our finite-volume NLO ChPT fits and our analytic fits.

\subsubsection{Chiral and continuum behaviour of the mass of the kaon}

We display our fits using infinite volume NLO SU(2)
partially quenched ChPT in figure\,\ref{fig:mkNLO}.
Figure\,\ref{fig:mkNLOfv} displays the corresponding fits of the same data with the finite-volume corrections included, while the analytic fits are displayed in figure~\ref{fig:mkanalytic}. The corresponding unitary view of the data in the continuum limit is shown in figure~\ref{fig:mkcont}. All these plots are for results at the physical sea strange quark mass.

\begin{figure}[hbt]
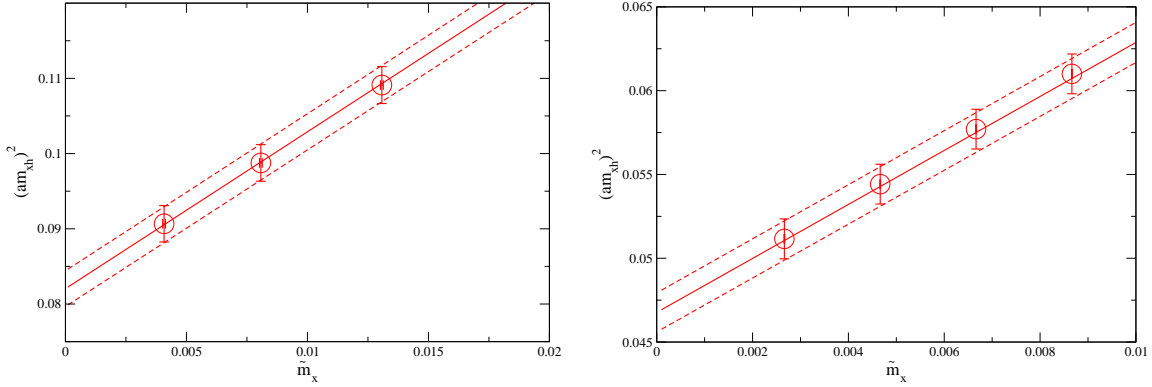

\centering
\includegraphics*[width=0.45\textwidth]{fig/Fig38a_mk2_PQ24_0.005_chpt.eps}\quad
\includegraphics*[width=0.45\textwidth]{fig/Fig38b_mk2_PQ32_0.004_chpt.eps}
\caption{
\label{fig:mkNLO}
Dependence of the kaon mass on the mass of the light valence quark with fits performed using infinite-volume NLO partially-quenched ChPT. The left panel shows the results from the $24^3$, $m_l=0.005$ ensemble and the right panel from the $32^3$, $m_l=0.004$ ensemble. In each case the results are for the physical strange-quark mass.}
\end{figure}

\begin{figure}[hbt]
\centering
\includegraphics*[width=0.45\textwidth]{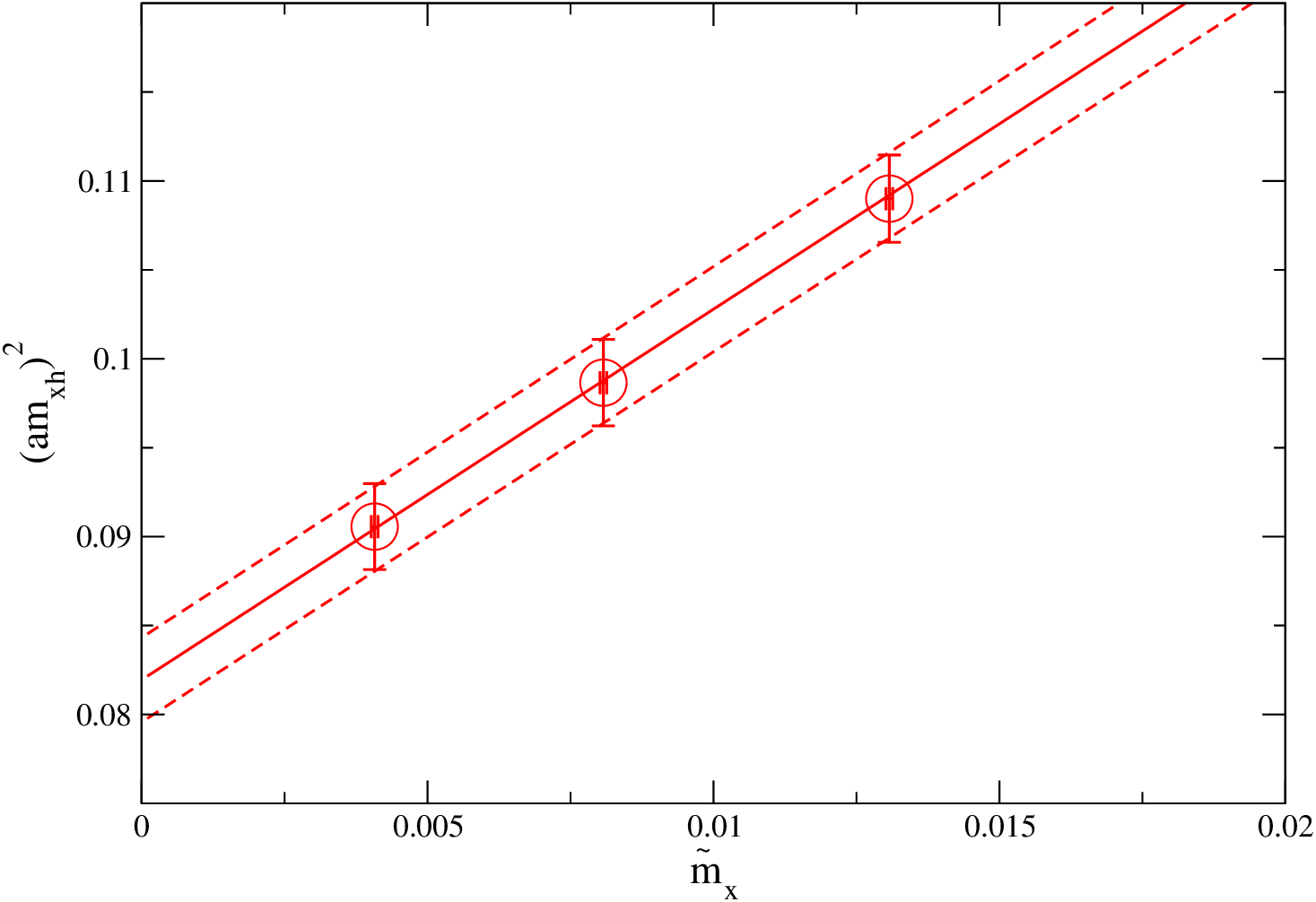}\quad
\includegraphics*[width=0.45\textwidth]{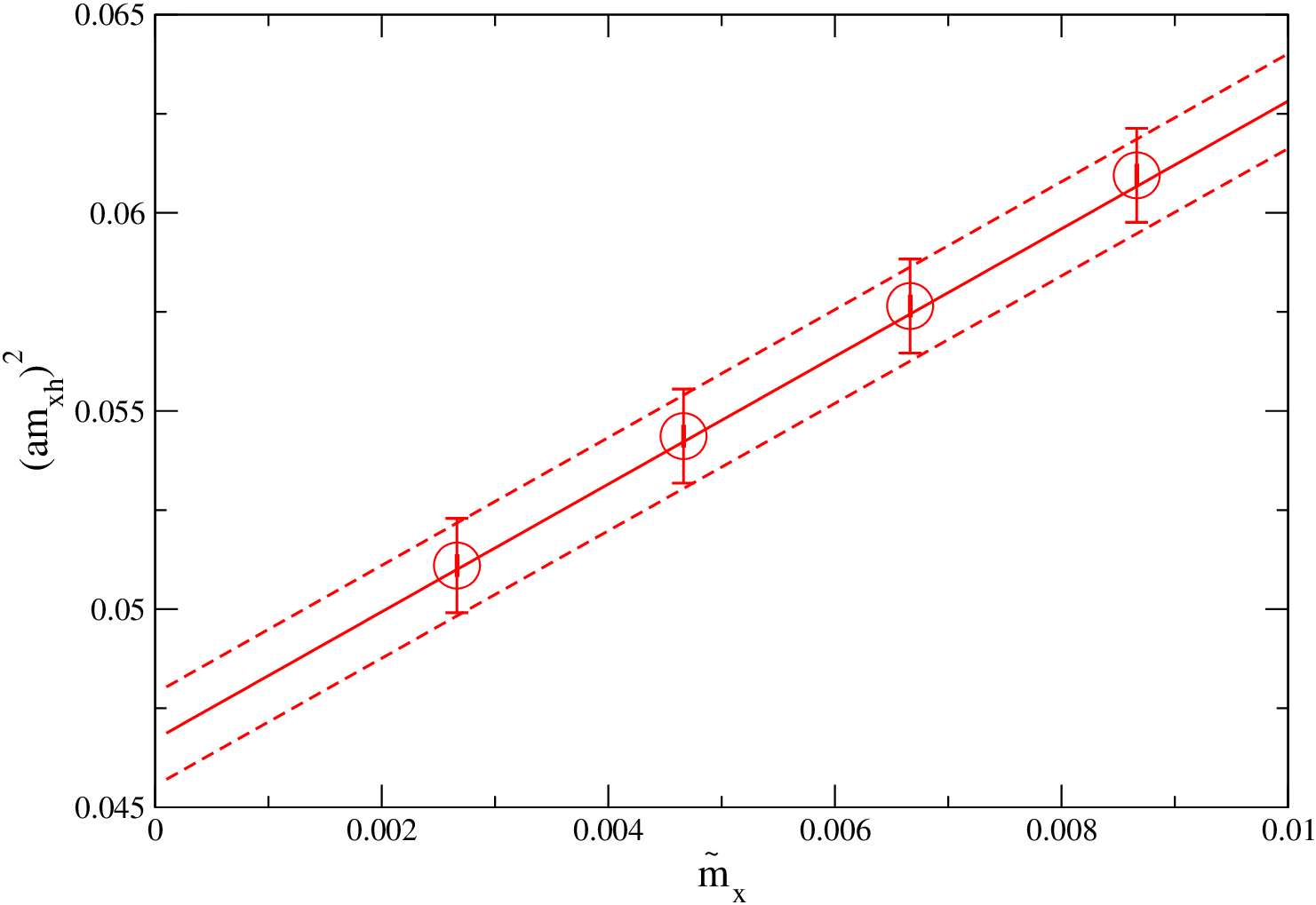}
\caption{
\label{fig:mkNLOfv}
Dependence of the kaon mass on the mass of the light valence quark with fits performed using finite-volume NLO partially-quenched ChPT. The left panel shows the results from the $24^3$, $m_l=0.005$ ensemble and the right panel from the $32^3$, $m_l=0.004$ ensemble. In each case the results are for the physical strange-quark mass.}
\end{figure}

\begin{figure}[hbt]
\centering
\includegraphics*[width=0.45\textwidth]{fig/Fig40a_mk2_PQ24_0.005_flavourexp.eps}
\includegraphics*[width=0.45\textwidth]{fig/Fig40b_mk2_PQ32_0.004_flavourexp.eps}
\caption{
\label{fig:mkanalytic}
Dependence of the kaon mass on the mass of the light valence quark with fits performed using the
analytic fit ansatz. The left panel shows the results from the $24^3$, $m_l=0.005$ ensemble and the right panel from the $32^3$, $m_l=0.004$ ensemble. In each case the results are for the physical strange quark mass.}
\end{figure}

\begin{figure}[hbt]
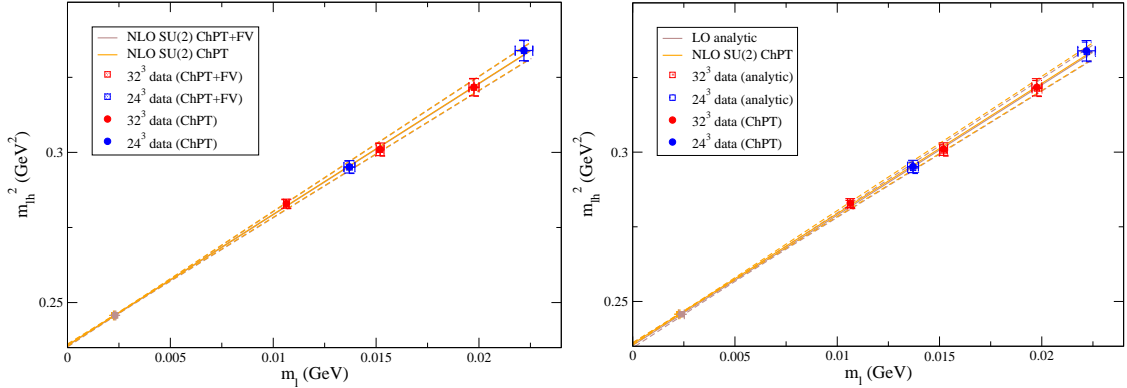

\centering
\includegraphics*[width=0.45\textwidth]{fig/Fig41a_mk2_unitary_su2_su2fv_comparison.eps}
\includegraphics*[width=0.45\textwidth]{fig/Fig41b_mk2_unitary_flavour_su2_comparison.eps}
\caption{
\label{fig:mkcont}
Chiral extrapolation of the kaon mass using unitary data points
adjusted to the continuum limit by the fitting ans\"atze.
Here we compare results obtained using the infinite-volume NLO ChPT ansatz to that using finite volume logarithms
(left panel) and to the analytic ansatz
(right panel).
}
\end{figure}

\subsubsection{Chiral and continuum behaviour of $f_K$}\label{subsubsec:fk}

We next discuss $f_K$, the decay constant of the kaon.
We display our fits using infinite-volume NLO SU(2)
partially quenched ChPT in Figure\,\ref{fig:fkNLO}. The following two figures display fits of the same partially quenched data to ChPT with finite-volume corrections (Figure\,\ref{fig:fkNLOfv}) and to the global analytic fit ansatz (Figure\,\ref{fig:fkanalytic}). The NLO ChPT fit ans\"atze, both with and without finite-volume
logarithms, are displayed for the unitary data adjusted to the continuum
limit in figure~\ref{fig:fkNLOadjusted}.

\begin{figure}[hbt]
\centering
\includegraphics*[width=0.45\textwidth]{fig/Fig42a_fk_PQ24_0.005_chpt.eps}\quad
\includegraphics*[width=0.45\textwidth]{fig/Fig42b_fk_PQ32_0.004_chpt.eps}
\caption{
\label{fig:fkNLO}
Dependence of the kaon decay constant on the mass of the light valence quark with fits performed using infinite-volume
partially quenched NLO ChPT. The left panel shows the results from the $24^3$, $m_l=0.005$ ensemble and the right panel from the $32^3$, $m_l=0.004$ ensemble. In each case the results are for the physical strange quark mass.}
\end{figure}

\begin{figure}[hbt]
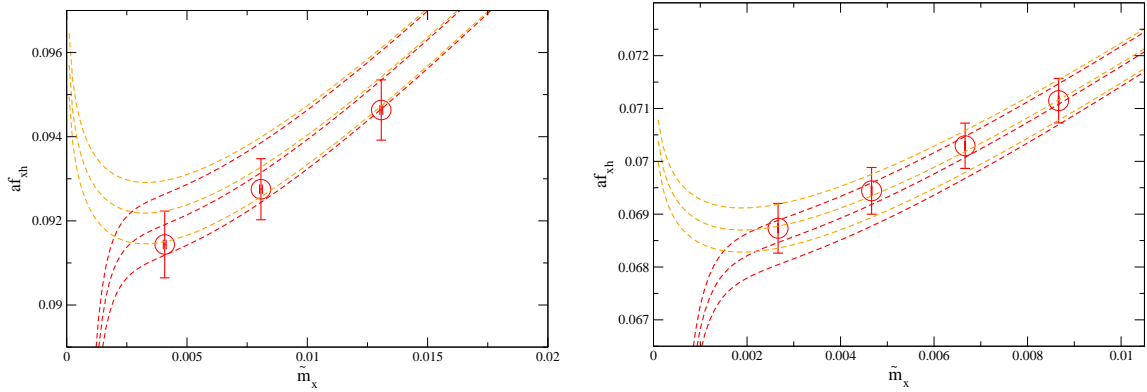

\centering
\includegraphics*[width=0.45\textwidth]{fig/Fig43a_fk_PQ24_0.005_chptfv.eps}\quad
\includegraphics*[width=0.45\textwidth]{fig/Fig43b_fk_PQ32_0.004_chptfv.eps}
\caption{
\label{fig:fkNLOfv}
Dependence of the kaon decay constant on the mass of the light valence quark.
The left panel shows the results from the $24^3$, $m_l=0.005$ ensemble
and the right panel from the $32^3$, $m_l=0.004$ ensemble. In each case the results are for the physical strange quark mass. There are two curves plotted. The orange curve is the result one infers for the
infinite volume, while the red curve is the result we obtain on the finite volume.
As we do not adjust our data for finite volume effects, the red curve should
go through our data. The orange curve also goes through our data which is an indication
that the finite volume effects in our data are substatistical, and the
difference between the orange and red curves at lighter masses indicates that
one should expect substantial finite volume effects \emph{if} one were to simulate
at these lighter masses without changing our present volume.
}
\end{figure}

\begin{figure}[hbt]
\centering
\includegraphics*[width=0.45\textwidth]{fig/Fig44a_fk_PQ24_0.005_flavourexp.eps}\quad
\includegraphics*[width=0.45\textwidth]{fig/Fig44b_fk_PQ32_0.004_flavourexp.eps}
\caption{
\label{fig:fkanalytic}
Dependence of the kaon decay constant on the mass of the light valence quark with fits performed using the analytic fit ansatz. The left panel shows the results from the $24^3$, $m_l=0.005$ ensemble and the right panel from the $32^3$, $m_l=0.004$ ensemble. In each case the results are for the physical strange quark mass.}
\end{figure}

\begin{figure}[hbt]
\centering
\includegraphics*[width=0.6\textwidth]{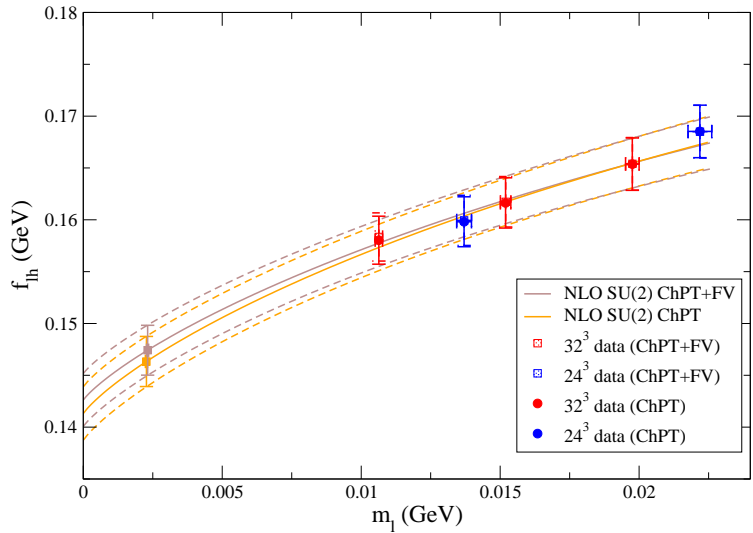}
\caption{
\label{fig:fkNLOadjusted}
Chiral extrapolation of the kaon decay constant for unitary data in
the continuum limit. We compare the NLO ChPT ansatz to the
corresponding ansatz with finite-volume logarithms.
}
\end{figure}

The two panels in Figure~\ref{fig:fkanalyticadjusted}
display the chiral behaviour of the actual unitary data from the two sets of ensembles (left panel)
as well as of the data adjusted to the continuum limit (right panel).

\begin{figure}[hbt]
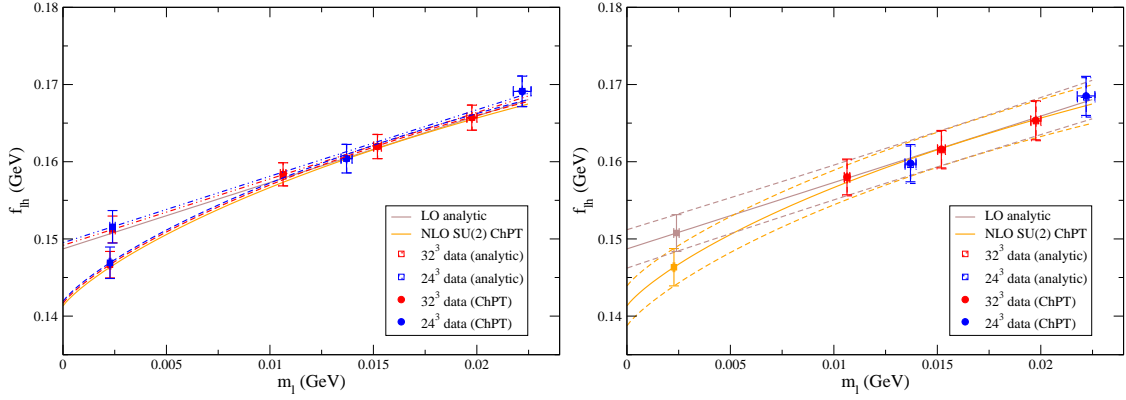

\centering
\includegraphics*[width=0.45\textwidth]{fig/Fig46a_fk_unitary_flavour_su2_comparison_uncorrected.eps}
\includegraphics*[width=0.45\textwidth]{fig/Fig46b_fk_unitary_flavour_su2_comparison.eps}
\caption{
\label{fig:fkanalyticadjusted}
Chiral extrapolation of the kaon decay constant for unitary data in
the continuum limit. We compare the NLO ChPT ansatz to the
analytic ansatz. The left panel displays the data and fits at non-zero lattice spacing,
while the right panel displays the predicted results and correspondingly
adjusted data points for the continuum limit.
}
\end{figure}

From these fits our final predictions for $f_K$ are given in
table~\ref{tab:fk:extrapolated}, and the corresponding results
for $\frac{f_K}{f_\pi}$ in table~\ref{tab:fkoverfpi:extrapolated}.

\subsubsection{Predictions}
\label{subsubsec:fkresults}

We now present our results for $f_\pi$, $f_K$ and their ratio as well as for the physical bare quark masses. As discussed above, our central value for any physical quantity is taken to be the average of the results obtained from analyses using the NLO SU(2) ChPT fit with finite volume corrections and those from the analytic fit. The difference between the analytic and finite-volume NLO SU(2) fits is taken as a systematic error. This procedure includes a NLO finite-volume correction, estimated from the difference between results obtained using NLO ChPT at infinite and finite volumes, and which is much smaller than the total systematic error here.

Our predictions for pseudoscalar decay constants therefore contain
systematic errors for finite volume effects, the chiral extrapolation,
and residual chiral symmetry breaking, while the discretisation
error is included indirectly by the fitting procedure:

\begin{eqnarray}
f_\pi^{\rm continuum}       &=& 124(2)(5)\,\textrm{MeV}\label{eq:fpifinal}\\
f_K^{\rm continuum}         &=& 149(2)(4)\,\textrm{MeV}\label{eq:fkfinal}\\
(f_K/f_\pi)^{\rm continuum} &=& 1.204(7)(25)\,,\label{eq:fkoverfpifinal}
\end{eqnarray}
where we display the statistical and systematic errors separately.
We note that the known, experimental value of $f_\pi$
influenced our choice to take the central value of physical quantities as the average of the results from the analytic and finite-volume NLO ChPT ans\"atze.
The prediction for $f_\pi$ cannot therefore
be considered unbiased, however as our aim is to select the most
likely central value for phenomenologically important
quantities such as $f_K/f_\pi$ and $B_K$ our procedure is both
appropriate and contains a prudent systematic error.

Applying the same procedure to obtain predictions for the physical bare
quark masses for the $\beta=2.25$ $32^3$ ensembles, we find:
\begin{equation}\label{eq:baremasses}
\tilde{m}_{ud}= 2.35(8)(9)\,\textrm{MeV}\qquad\textrm{and}\qquad
\tilde{m}_s= 63.7(9)(1)\textrm{MeV},
\end{equation}
and these will be renormalised in the following section. The corresponding bare masses for the $\beta=2.13$ $24^3$ ensembles can be obtained by dividing the results in (\ref{eq:baremasses}) by the values of $Z_l$ and $Z_h$ in Table\,\ref{tab:zlzhzacomparison}.

\subsubsection{Chiral and continuum behaviour of $r_0$ and $r_1$}\label{subsec:r0r1global}

\begin{table}[tp]
\centering
\subtable[\ $r_0$]{
\begin{tabular}{l|lll}
\hline\hline
Parameter & ChPT & ChPT-fv & Analytic\\
\hline
$c_{r_0}$ & 2.468(41) GeV$^{-1}$ & 2.468(41) GeV$^{-1}$ & 2.467(41) GeV$^{-1}$\\
$c_{r_0,a}$ & -0.25(14) GeV & -0.25(14) GeV& -0.25(14) GeV\\
$c_{r_0,m_l}$ & 0.42(1.23) GeV$^{-2}$& 0.44(1.23) GeV$^{-2}$& 0.47(1.23) GeV$^{-2}$
\end{tabular}
}
\subtable[\ $r_1$]{
\begin{tabular}{l|lll}
\hline\hline
Parameter & ChPT & ChPT-fv & Analytic\\
\hline
$c_{r_1}$ & 1.694(29) GeV$^{-1}$ & 1.694(29) GeV$^{-1}$ & 1.693(29) GeV$^{-1}$\\
$c_{r_1,a}$ & -0.15(11) GeV & -0.15(11) GeV & -0.15(12) GeV\\
$c_{r_1,m_l}$ & -1.76(64) GeV$^{-2}$& -1.76(64) GeV$^{-2}$& -1.76(64) GeV$^{-2}$
\end{tabular}
}
\caption{Parameters of the chiral/continuum fits to $r_0$ and $r_1$.}
\label{tab-r0r1fitparams}
\end{table}
\begin{table}[tp]
\centering
\begin{tabular}{l|ccc}
\hline\hline
Quantity & ChPT & ChPT-fv & Analytic\\
\hline
$r_0$ & 1.35(1.66) & 1.34(1.65) & 1.31(1.63)\\
$r_1$ & 2.69(2.39) & 2.68(2.38) & 2.66(2.37)
\end{tabular}
\caption{$\chi^2/\mathrm{d.o.f}$ of the chiral/continuum fits to $r_0$ and $r_1$.}
\label{tab-r0r1fitchisq}
\end{table}

Finally in this section we apply the combined chiral/continuum extrapolation procedure to the scales $r_0$ and $r_1$. Assuming a linear dependence for the light sea-quark mass dependence, and including a leading order $a^2$ term as before, the scales are independently fit to the form
\begin{equation}
r_i = c_{r_i} + c_{r_i,a} a^2 + c_{r_i,m_l}\tilde m_l\,,
\end{equation}
where $i=0,1$. Prior to the fit, the data are linearly interpolated to each of the physical strange quark masses obtained from the global fits and presented in Table\,\ref{tab:quarkmasseslatt}, and the fit and the subsequent extrapolation are performed using the corresponding physical light-quark mass and lattice spacings.

The parameters and $\chi^2/\mathrm{d.o.f}$ of the fits are given in Tables~\ref{tab-r0r1fitparams} and \ref{tab-r0r1fitchisq} respectively, and  plots showing the fits overlaying the data in the continuum limit are shown in figure~\ref{fig-r0r1chiralfits}. The fits to $r_0$ appear to describe the data well by eye, and have a reasonable (uncorrelated) $\chi^2/\mathrm{d.o.f}$ for the central value, but with a large deviation across the superjackknife distribution. The fits to $r_1$ also appear to describe the data reasonably well, although there does seem to be a tension with the heaviest point on the $24^3$ ensembles, which is likely responsible for the larger $\chi^2/\mathrm{d.o.f}$. As there are only five data points it is difficult to reach any stronger conclusions regarding the data: more ensembles and better statistics are needed. For the purpose of quoting a final result, we apply a PDG scale factor of $\sqrt{\chi^2/\mathrm{d.o.f}}$ to the statistical errors on each of the results. In order to retain the correlations between these quantities when the ratio is taken, the scale factor is applied to the difference of each jackknife sample from the mean.

The continuum results for $r_0$, $r_1$ and their ratio at physical quark masses are given in table~\ref{tab-r0r1fitcont}. Using the procedure for combining the results obtained using the different chiral ans\"atze outlined in Section~\ref{subsubsec:chiralfpi} and applying the PDG scale factor as above, gives:
\begin{equation}\label{eq:r0r1final}
\begin{array}{lclcl}
r_0 &=& 2.468(45)_{\mathrm{stat}}(1)_{\mathrm{FV}}(1)_{\chi}\,\mathrm{GeV}^{-1} &=& 0.4870(89)_{\mathrm{stat}}(2)_{\mathrm{FV}}(2)_{\chi}\,\mathrm{fm}\,,\\
r_1 &=& 1.689(47)_{\mathrm{stat}}(0)_{\mathrm{FV}}(1)_{\chi}\,\mathrm{GeV}^{-1} &=& 0.3333(93) _{\mathrm{stat}}(1)_{\mathrm{FV}}(2)_{\chi}\,\mathrm{fm}\,,\,\mathrm{and}\\
r_1/r_0 &=& 0.684(15)_{\mathrm{stat}}(0)_{\mathrm{FV}}(0)_{\chi}\,,
\end{array}
\end{equation}
where the finite volume error arising from the different determinations of the lattice spacings and quark masses is smaller than the quoted precision on the ratio. $\chi$ labels the error due to the chiral extrapolation. For comparison, the MILC collaboration recently obtained $r_1=0.3117(6)({}^{+12}_{-31})$~fm ($\simeq 1.580(3)(\mbox{}^{+6}_{-16})$\,GeV$^{-1}$)~\cite{Bazavov:2009tw}, and also $r_1=0.317(7)(3)$~fm ($\simeq 1.61(4)(2)$\,GeV$^{-1}$) and $r_0=0.462(11)(4)$ fm ($\simeq 2.34(6)(2)$\,GeV$^{-1}$) from an earlier study~\cite{Aubin:2004wf}. At this time we do not have an explanation of the discrepancy between our results in (\ref{eq:r0r1final}) and those of the MILC collaboration beyond noting the very different approaches to setting the scale and performing the chiral extrapolation.

\begin{table}[t]
\centering
\begin{tabular}{l|ccc}
\hline\hline
Quantity & ChPT & ChPT-fv & Analytic\\
\hline
$r_0$ & 2.469(39) GeV$^{-1}$ & 2.469(39) GeV$^{-1}$ & 2.468(39) GeV$^{-1}$\\
$r_1$ & 1.690(29) GeV$^{-1}$ & 1.690(29) GeV$^{-1}$ & 1.689(29) GeV$^{-1}$\\
$r_1/r_0$ & 0.6844(96) & 0.6844(97) & 0.6843(97)
\end{tabular}
\caption{Continuum values of $r_0$ and $r_1$ and the ratio $r_1/r_0$ at physical quark masses determined from a chiral/continuum fit using the lattice spacings and quark masses obtained from the global fits.}
\label{tab-r0r1fitcont}
\end{table}

\begin{figure}[tp]
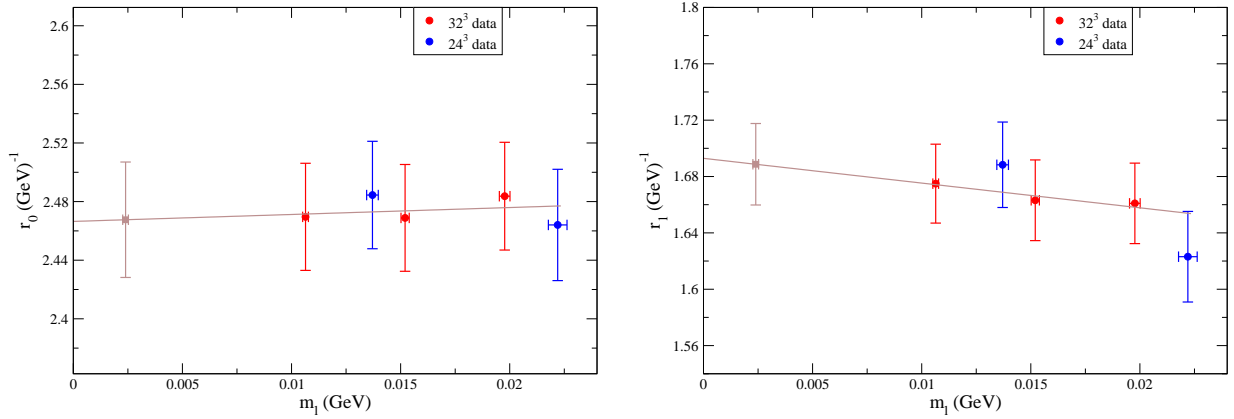

\centering
\includegraphics*[width=0.48\textwidth]{fig/Fig47a_r0_flavour.eps} \quad
\includegraphics*[width=0.48\textwidth]{fig/Fig47b_r1_flavour.eps}
\caption{The scales $r_0$ (left) and $r_1$ (right) corrected to the continuum limit, overlaid by the chiral/continuum fit. The extrapolated point at the physical light quark mass is shown as the grey cross. Here the lattice spacings and physical light quark mass were obtained from the global fits using the analytic ansatz. The fits using the quantities obtained with the ChPT and ChPT-fv global fit ans\"atze are almost indistinguishable from those shown in these figures.}
\label{fig-r0r1chiralfits}
\end{figure}

\fi

\section{Light-Quark Masses}
\label{sec:quark_masses}

\def\msbar{\overline{\mbox{\scriptsize MS}}}
\def\MOM{{\rm MOM}}
\def\SMOM{{\rm SMOM}}
\def\SMOMgm{{\rm SMOM}_{\gamma_\mu}}
\def\RIMOM{{\rm RI/MOM}}
\def\RI'MOM{{\rm RI'/MOM}}
\def\RISMOM{{\rm RI/SMOM}}
\def\RISMOMgm{{\rm RI/SMOM}_{\gamma_\mu}}
\def\mom{\mbox{\scriptsize MOM}}
\def\smom{\mbox{\scriptsize SMOM}}
\def\smomgm{\mbox{\scriptsize SMOM}_{\gamma_\mu}}
\def\rimom{\mbox{\scriptsize RI/MOM}}
\def\ri'mom{\mbox{\scriptsize RI'/MOM}}
\def\rismom{\mbox{\scriptsize RI/SMOM}}
\def\rismomgm{\mbox{\scriptsize RI/SMOM}_{\gamma_\mu}}
\def\Tr{{\rm Tr}}
\def\qhat{{\hat{q}}}
\def\qslash{{q\!\!\!\!/}}
\def\qhatslash{{\qhat\!\!\!\!/}}

The quark masses quoted in Eq.\,(\ref{eq:baremasses}) are the bare masses for the lattice action which we are using on the $32^3$ ensembles with $\beta=2.25$ corresponding to a lattice spacing $a^{-1}\simeq 2.28\,$GeV. In order to be useful in phenomenological applications these results must be translated into renormalized masses in some standard continuum scheme. Therefore in Subsection\,\ref{subsec:npr} we determine the renormalization constants relating the bare masses in (\ref{eq:baremasses}) to those renormalized in the $\MSb$ scheme at a renormalization scale of $2\,$GeV. In Subsection\,\ref{subsec:renmasses} we then combine these renormalization constants with the bare masses in (\ref{eq:baremasses}) to obtain the renormalized masses, the LO LEC $B^{\overline{\textrm{MS}}}(2\,\textrm{GeV})$ and the chiral condensate.

\subsection{Non-perturbative renormalization for quark masses}\label{subsec:npr}

The quark-mass renormalization factor which relates the lattice
bare quark mass to that in the $\MSb$ scheme is determined using non-perturbative
renormalization (NPR) with the RI/SMOM schemes proposed in
Ref.~\cite{Sturm:2009kb} as intermediate schemes. This is an extension of the Rome-Southampton NPR program in which the RI/MOM scheme was defined~\cite{Martinelli:1994ty}. Quark masses renormalized in the RI/SMOM or RI/MOM
schemes are obtained entirely non-perturbatively. Since it is not possible to simulate in a non-integer number of dimensions, continuum perturbation theory is needed to match the results in either the RI/SMOM or the RI/MOM scheme and the target $\MSb$ scheme. We stress however, that we completely avoid the use of lattice perturbation
theory which often converges more slowly than continuum perturbation theory (PT). Since RI/MOM and any of the schemes proposed in~\cite{Sturm:2009kb} are legitimate renormalization schemes, we exploit the freedom to choose an intermediate scheme to reduce its effect on the final result for the renormalized quark mass in the $\MSb$ scheme and to have a better understanding of this uncertainty.

Our earlier study~\cite{Aoki:2007xm}, used to normalize the quark mass on the $24^3$ ensembles,
applied the RI/MOM scheme to renormalize the quark masses and suffered from sizable
systematic errors with two dominant sources. One of these is the truncation error in the perturbative continuum matching between the RI/MOM and $\MSb$ schemes. This was estimated to be 6\% for $\mu=2$\,GeV from the relative size of the highest-order term used (3 loop). The other is a non-perturbative effect arising because the strange quark mass is fixed close to its physical value, and the chiral limit is not taken for this quark. We estimated the corresponding  systematic error on the quark-mass renormalization factor for
$a^{-1}=1.73$\,GeV and $\mu=2$\,GeV to be about 7\%. As the strange-quark mass and
the typical scale of spontaneous chiral symmetry breaking are
almost the same, this error can be viewed as a general error
due to contamination of non-perturbative effects (NPE). It was shown in Ref.~\cite{Aoki:2007xm}  that changing the
kinematics of momenta used to define the NPR scheme greatly reduces the contamination
from unwanted non-perturbative effects and this will be discussed below.
The actual implementation of the schemes with unconventional
kinematics has been done in Ref.~\cite{Sturm:2009kb} carefully ensuring that
the Ward-Takahashi chiral identities are satisfied. A pilot study~\cite{Aoki:2009ka} using the new schemes
demonstrated that it is a promising alternative to the conventional RI/MOM scheme with reduced systematic errors. In the present article we use two RI/SMOM schemes proposed in Ref.~\cite{Sturm:2009kb}. Preliminary results have been reviewed in Ref.~\cite{Aoki:2010yq}.

An important technical improvement introduced since the previous study \cite{Aoki:2007xm} is
the use of volume momentum sources for the quark propagators.
This helps to reduce the statistical error greatly and in addition reduces the
systematic error due to the dependence on the position of the local source used in
\cite{Aoki:2007xm}. More details about the use of momentum sources can be found in Ref.~\cite{bkpaper}.

The mass renormalization factor $Z_m$ is conveniently calculated using the
relation
\begin{equation}
 Z_m = 1/Z_S = 1/Z_P,\label{eq:ZmZsZp}
\end{equation}
where $Z_m$, $Z_S$, $Z_P$ are the quark mass, flavor non-singlet scalar
and pseudoscalar renormalization factors respectively. Here we are exploiting the important chiral symmetry properties of DWF.
Our convention is that the renormalization factors multiply the bare quantities to yield renormalized ones:
\begin{equation}\label{eq:mpsren}
 m_R = Z_m \wm,\quad
 P_R^a = Z_P P^a,\quad
 S_R^a = Z_S S^a,
\end{equation}
where the left-hand sides are the renormalized mass, pseudoscalar
and scalar densities and $a$ is a flavour label. $\wm$ in Equation\,(\ref{eq:mpsren}) is in physical units.
The relations in Eq.~(\ref{eq:ZmZsZp}) are necessary
for the Ward-Takahashi identities to hold for the renormalized operators.
The RI/MOM renormalization condition on the amputated scalar vertex $\Pi_S$
reads
\begin{equation}
 \frac{Z_S}{Z_q}\frac{1}{12} \Tr[\Pi_S\cdot I]  =  1. \\\label{eq:S_condition}
\end{equation}
$Z_q$ is the wave function renormalization factor, which can be determined
using the trace condition on the local vector operator,
\begin{equation}
 \frac{Z_V}{Z_q} \frac{1}{48}\Tr [\Pi_{V_\mu} \cdot\gamma_\mu]= 1.
  \label{eq:A_condition}
\end{equation}
The vertex functions $\Pi$ depend on the incoming and outgoing momenta
on the two fermion lines, $\Pi(p_{in},p_{out})$.
The conventional RI/MOM scheme is defined using the forward vertex with $p_{in}=p_{out}=p$.
The renormalization conditions Eqs.~(\ref{eq:S_condition}),
(\ref{eq:A_condition}) are applied by setting the
renormalization scale $\mu$ to be the off-shell external momentum,
$\mu^2=p^2$, in the chiral limit.

It is in principle possible to determine $Z_S\,(=Z_P)$ using the
pseudoscalar vertex function instead of the scalar one in
Eq.~(\ref{eq:S_condition}). However, with the original RI/MOM choice for the external momenta,
the pseudoscalar vertex couples to the zero-momentum pion, and the Green function diverges as $1/m_q$ as the quark mass $m_q\to 0$ at fixed $p$\,\cite{Cudell:1998ic}. Therefore the pseudoscalar vertex cannot
be used without some manipulation of the divergence (see e.g. \cite{Giusti:2000jr}) and has not been considered in our previous publication~\cite{Aoki:2007xm}. This is in contrast with the RI/SMOM schemes described below which do not have such a pole as $m_q\to 0$.
Similarly, the axial-vector vertex can be used to determine $Z_q$ because $Z_V=Z_A$. However, $Z_q$ obtained using the vector and axial-vector vertices at large but finite $p^2$ will differ because of the coupling of the axial current
to the Goldstone boson \cite{Martinelli:1994ty}. These differences are known to be of $O(1/p^2)$ at high momentum
from the operator product expansion~\cite{Martinelli:1994ty,Cudell:1998ic} or from Weinberg's theorem of power counting for a Feynman diagram ~\cite{Aoki:2007xm}. In Ref.~\cite{Aoki:2007xm}, the average of the vector and the axial-vector vertex was used to determine $Z_q$  and the difference was included in the systematic error, though the corresponding 1\% error is sub-dominant.

The caveats mentioned in the two preceding paragraphs are both connected to the RI/MOM scheme and its channel with an ``exceptional momentum''; specifically, the momentum transfer $q\equiv p_{in}-p_{out}=0$. This is the reason for the large NPE error. It was demonstrated that the use of non-exceptional momenta
$p_{in}-p_{out}\ne 0$ reduces the NPE effect significantly.
The RI/SMOM schemes are designed so that all channels have non-exceptional momenta. For quark bilinear operators we choose to have $p_{in}^2=p_{out}^2=q^2$ and hence introduce the name ``Symmetric Mom'' (SMOM) schemes. The two schemes
RI/SMOM and $\RISMOMgm$ are defined with this kinematical choice but differ in the
$\Gamma$-projection operators which are used to define the wave function
renormalization. For the vector (axial-vector) vertex function the projector
$\qslash q_\mu/q^2$ ($\gamma_5\qslash q_\mu/q^2$) is
used in the RI/SMOM scheme and $\gamma_\mu$ ($\gamma_5\gamma_\mu$) as in
Eq.~(\ref{eq:A_condition}) is used for $\RISMOMgm$.
The standard $I$ ($\gamma_5$) spinor projector is used
for the scalar (pseudoscalar) vertex in both new schemes.

The conversion factors from the RI/SMOM and $\RISMOMgm$ schemes
to $\MSb$ have been calculated at one-loop order in Ref.\,\cite{Sturm:2009kb}
and recently to two-loop order\,\cite{Gorbahn:2010bf,Almeida:2010ns}:
\begin{eqnarray}
 C_m(\RISMOM\to\MSb,\mu) & = &
  1 - \left(\frac{\alpha_s(\mu)}{4\pi}\right) 0.646
  - \left(\frac{\alpha_s(\mu)}{4\pi}\right)^2 (22.608+4.014 n_f) \cdots,
  \label{eq:CmSMOMatmu}\\
 C_m(\RISMOMgm\to\MSb,\mu) & = &
  1 - \left(\frac{\alpha_s(\mu)}{4\pi}\right) 1.979
  - \left(\frac{\alpha_s(\mu)}{4\pi}\right)^2 (55.032+6.162 n_f) \cdots,
  \label{eq:CmSMOMgmatmu}
\end{eqnarray}
where the coefficients have been rounded to the third decimal place.
Evaluating these factors at $\mu=2$\,GeV we have
\begin{eqnarray}
 C_m(\RISMOM\to\MSb,\mu=2{\rm GeV},n_f=3) & = &
  1 - 0.015 - 0.006 \cdots, \label{eq:CmSMOM}\\
 C_m(\RISMOMgm\to\MSb,\mu=2{\rm GeV},n_f=3) & = &
  1 - 0.046 - 0.020  \cdots\,. \label{eq:CmSMOMgm}
\end{eqnarray}
In the RI/MOM and RI$^\prime$/MOM schemes the conversion factors are known to three-loop order~\cite{Chetyrkin:1999pq,Gracey:2003yr}:
\begin{eqnarray}
 C_m(\RIMOM\to\MSb,\mu=2{\rm GeV},n_f=3) & = &
  1 - 0.123 - 0.070 - 0.048 + \cdots,\\
 C_m(\RI'MOM\to\MSb,\mu=2{\rm GeV},n_f=3) & = &
  1 - 0.123 - 0.065 - 0.044 + \cdots.
\end{eqnarray}
We note that, at least up to two-loop
order, the convergence of the series relating the new SMOM schemes to $\MSb$ is considerably
better than for the RI/MOM scheme. As already mentioned, the truncation error of the RI/MOM
scheme was estimated from the size of the highest order term available (3 loop).
Having in addition two intermediate SMOM schemes, we can expect to have a more reliable estimate of the
truncation error.

We now turn to the numerical evaluation of the renormalization factors. At each value of $\beta$, we use data obtained at the three light-quark masses: $m_l=0.004$, $0.006$ and
$0.008$ for the finer $32^3$ lattice and $m_l=0.005$, $0.01$ and $0.02$ for the coarser $24^3$ lattice.
20 configurations were analyzed for each point.
The ratio of quark wavefunction and local axial
current renormalization factors is calculated from the average
of vector and axial-vector vertex functions,
\begin{equation}
 \frac{Z_q}{Z_V} = \frac{1}{2} (\Lambda_V+\Lambda_A),\label{eq:ZqZA}
\end{equation}
with projected and traced vertex functions:
\begin{equation}
 \Lambda_V^{\rismom} =
  \frac{1}{12\qhat^2}\Tr[\Pi_{V_\mu}\cdot\qhatslash\qhat_\mu]\quad\textrm{and}\quad
 \Lambda_A^{\rismom}  =
  \frac{1}{12\qhat^2}\Tr[\Pi_{A_\mu}\cdot\gamma_5 \qhatslash\qhat_\mu],
\end{equation}
for the RI/SMOM scheme. Here $q_\mu$ in the continuum RI/SMOM scheme
\cite{Sturm:2009kb}
has been replaced with the $\qhat_\mu=\sin(q_\mu)$, as the
derivative for the divergence of the current in the continuum theory is
naturally replaced by the symmetric difference on the lattice.
A remarkable feature of the RI/SMOM scheme is that in the chiral limit
$\Lambda_V=\Lambda_A$ holds non-perturbatively,
in contrast to $\Lambda_V\ne \Lambda_A$ for RI/MOM scheme due to
spontaneous symmetry breaking (SSB).
In principle there could still be a small difference for the lattice RI/SMOM scheme
with non-zero $\mres$, which, however, is negligible in the momentum
range we use \cite{Aoki:2009ka}. Using the continuum Ward-Takahashi identities,
one can also show the equivalence of $Z_q$ in the RI/SMOM and RI$^\prime$/MOM schemes~\cite{Sturm:2009kb}.

The $\RISMOMgm$ scheme is defined using the conventional projectors,
\begin{equation}
 \Lambda_V^{\rismomgm}  =  \frac{1}{48}\Tr[\Pi_{V_\mu}\cdot\gamma_\mu]\quad\textrm{and}\quad
 \Lambda_A^{\rismomgm}  =
 \frac{1}{48}\Tr[\Pi_{A_\mu}\cdot\gamma_5\gamma_\mu]\,.
\end{equation}
Although these projectors are superficially the same as those used in the
RI/MOM scheme, it should be remembered that the kinematics is different in the two cases with no exceptional channels in the Green functions used to define the $\RISMOMgm$ scheme.

\begin{figure}
 \begin{center}
  \includegraphics[width=.5\textwidth]{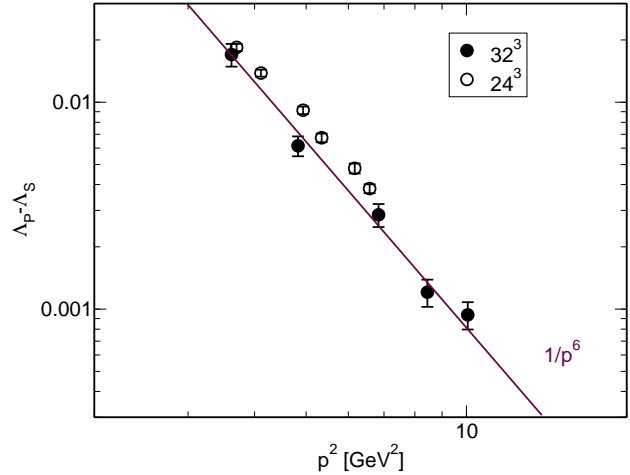}
 \end{center}
 \caption{$\Lambda_P-\Lambda_S$ as a function of $p^2$ [GeV$^2$]
 for fine ($32^3$) and coarse ($24^3$) lattices.
 A straight line with $1/p^6$ slope but arbitrary normalization is drawn to guide the eye.}
 \label{fig:P-S}
\end{figure}
The product of mass and wavefunction renormalization factors
is calculated from the average of scalar and pseudoscalar vertex
functions,
\begin{equation}
 Z_mZ_q = \frac{1}{2}(\Lambda_S+\Lambda_P),\label{eq:ZmZq}
\end{equation}
with
\begin{equation}
 \Lambda_S  =  \frac{1}{12}\Tr[\Pi_S\cdot 1]\quad\textrm{and}\quad
 \Lambda_P  =  \frac{1}{12}\Tr[\Pi_P\cdot\gamma_5],
\end{equation}
again defined with the SMOM kinematics for the vertex functions.
While $\Lambda_S=\Lambda_P$ holds to all orders in perturbation theory with
naive dimensional regularization, by using Weinberg's power-counting scheme we see that they can in general differ by terms of $O(1/p^6)$~\cite{Aoki:2007xm}. The difference
$\Lambda_P-\Lambda_S$ after the chiral extrapolation is plotted in
Fig.~\ref{fig:P-S} as a function of $p^2$ (in physical units) for both
the $24^3$ and $32^3$ lattices. The figure confirms the expected approximate $1/p^6$ scaling.
The unwanted non-perturbative effect from  SSB is small and the introduction of
non-exceptional momenta has had the expected effect.
This is in contrast  to the RI/MOM scheme with the exceptional channel, where
the same difference behaves as $1/(mp^2)$, and thus diverges in the chiral
limit at finite $p^2$.

The mass renormalization factor $Z_m^\sigma$, with $\sigma=\RISMOM$ or
$\RISMOMgm$, is given by combining
Eqs.~(\ref{eq:ZqZA}) and (\ref{eq:ZmZq}),
\begin{equation}
 Z_m^\sigma = \frac{1}{Z_V}
  \frac{\Lambda_S+\Lambda_P}{\Lambda_V^\sigma+\Lambda_A^\sigma}.\label{eq:Zm}
\end{equation}

In calculating the ratio of vertex functions in Eq.~(\ref{eq:Zm})
we take the average of $S$ and $P$ or $V$ and $A$ for each light-quark
mass and then fit with
a quadratic ($c+c'(m_l+\mres)^2$)
or linear $c+c''(m_l+\mres)$ formula to obtain the value $c$ in the
chiral limit for the numerator and denominator.
\begin{figure}
 \begin{center}
  \includegraphics[width=.5\textwidth]{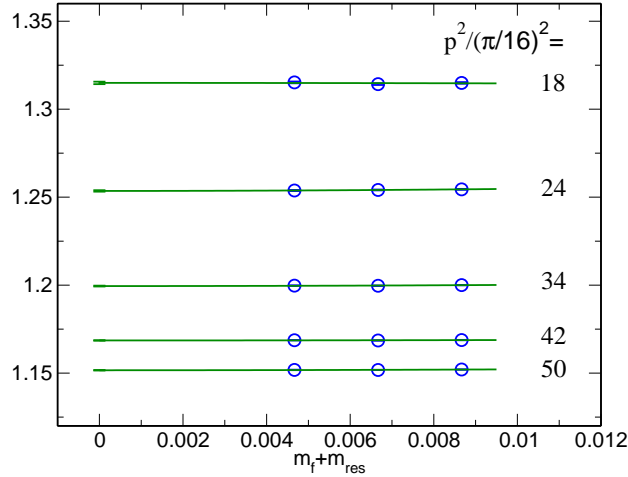}
 \end{center}
 \caption{
 Chiral extrapolation of $(\Lambda_P+\Lambda_S)/2$ for the fine ($32^3$)
 lattice for each $p^2$ point.
 \label{fig:P+S}}
 \end{figure}
For illustration, the extrapolation for the numerator using the quadratic formula is shown in Fig.~\ref{fig:P+S},
where the observed mass dependence is seen to be very small.
Because of the very mild mass dependence, to the precision with which we quote our results and errors,
the quadratic and linear extrapolation formulae lead to exactly the same quark-mass renormalization factor and error.
Finally taking the ratio and combining with $Z_V$ gives the mass renormalization factor
in the RI/SMOM schemes. The renormalization factor in the $\MSb$ scheme at a scale $\mu=2$ GeV
is obtained by first matching the scheme $\sigma$ to $\MSb$ at
$\mu^2=p_{in}^2=p_{out}^2=q^2$ using Eqs.~(\ref{eq:CmSMOMatmu}) and
(\ref{eq:CmSMOMgmatmu})
and then running to $2$ GeV using the
three-loop anomalous dimension in the $\MSb$ scheme.
We use the four-loop QCD beta functions \cite{vanRitbergen:1997va}
to calculate $\alpha_s^{(3)}(\mu)$ for running and matching
as shown in Appendix A of Ref.~\cite{Aoki:2007xm}. The relevant parameters
taken from the 2008 Particle Data Group~\cite{Amsler:2008zzb} are
\begin{equation}
 \alpha_s^{(5)}(m_Z) =  0.1176,\
 m_Z  =  91.1876\; {\rm GeV},\
 \overline{m}_b  =  4.20\; {\rm GeV}\ \textrm{and}\
 \overline{m}_c  =  1.27\; {\rm GeV},
\end{equation}
where the quark masses are in the $\MSb$ scheme at the scale of the mass itself, e.g. $\overline{m}_b=m_b^{\msbar}(\overline{m}_b)$\,.

\begin{figure}
 \begin{center}
  \includegraphics[width=.5\textwidth]{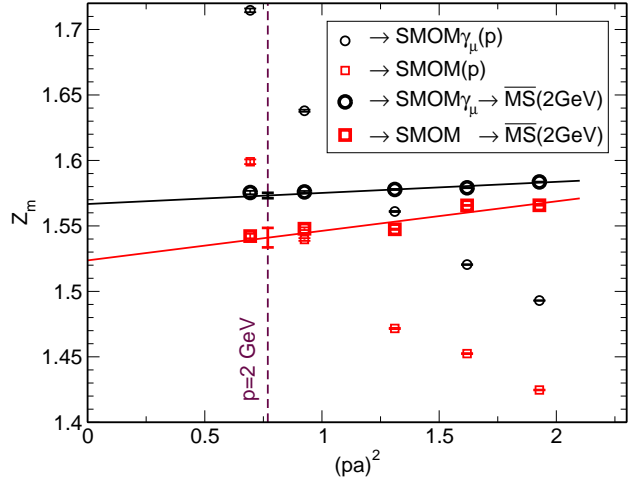}
 \end{center}
 \caption{
 $Z_m^{\smomgm}(\mu)$ and $Z_m^{\smom}(\mu)$ as functions of
 $\mu^2=p^2$,
 and $Z_m^{\msbar}(2 {\rm GeV})$ from the SMOM or $\SMOMgm$ schemes as
 function of matching scale squared $p^2$ for the fine
 lattice. The interpolation points are shown with the error bar at
 $p^2=(2\ {\rm GeV})^2$.
 \label{fig:Zm32}}
\end{figure}
In Fig.~\ref{fig:Zm32} we plot
$Z_m^{\smomgm}(\mu)$ and $Z_m^{\smom}(\mu)$ in the SU(2) chiral limit as functions of $\mu^2=p^2$ for the $32^3$ ensembles. In addition we also plot $Z_m^{\msbar}(2 {\rm GeV})$ as functions of the matching scale
$p^2$ obtained with SMOM and $\SMOMgm$ as the
intermediate schemes. In an ideal situation, i.e. one in which the errors due to NPE contamination, truncation of perturbation theory and lattice artifacts are all small, the results obtained using the
two intermediate schemes would give the same results for
$Z_m^{\msbar}(2 {\rm GeV})$, and the results would be independent of $(pa)^2$.
Since we have observed that the NPE error is small,
the difference between the two sets of results is mostly due to the truncation of perturbation theory and
lattice discretization errors.
The observed decrease in this difference as $p^2$ increases
is consistent with the expected behaviour of the truncation error. Conversely, since the truncation error increases as $p^2$ decreases, taking the limit $(pa)^2\to 0$, which is a typical treatment to eliminate the discretization error, is not an appropriate procedure. We therefore choose instead to evaluate $Z_m$ by taking an intermediate reference point
$p^2=(2\ {\rm GeV})^2$, for both the $24^3$ and $32^3$ lattices.
In this way, as we take the continuum limit of the renormalized quark mass,
the leading $(pa)^2$ discretization error associated with the non-perturbative
renormalization will be removed.

There is a subtlety due to lattice artefacts which are not $O(4)$ invariant and which are responsible for the non-smooth $(pa)^2$ dependence in the figure.  A term like $a^2\sum_\mu (p_\mu)^4/p^2$, whose presence has been demonstrated in the conventional RI/MOM scheme for Wilson quarks
\cite{Constantinou:2009tr}, could exist also
in the SMOM schemes. Such a term would manifest itself as scattered
data around a smooth curve in $p^2$, and the size of the scatter is expected to be comparable
to the leading $(pa)^2$ error as both are of the same order in $a^2$.
This appears to be compatible to what is shown in the figure. Of course, it would be very helpful to know these terms, but in the absence of this knowledge
we include this scatter in the systematic error by inflating the error
by a factor $\sqrt{\chi^2/{\rm dof}}$.
The results are
\begin{eqnarray}
 Z_m^{\msbar(32)}(\mu=2\ {\rm GeV},n_f=3; \smomgm) & = & 1.573(2),\label{eq:ZmSMOMgm}\\
 Z_m^{\msbar(32)}(\mu=2\ {\rm GeV},n_f=3; \smom) & = & 1.541(7).\label{eq:ZmSMOM}
\end{eqnarray}
The final arguments on the left-hand sides denote the choice of intermediate scheme.
The error on the right-hand sides is the combination of the statistical
fluctuations and the scatter of the points around the linear fit.
The central values and
errors are shown in the figure at the reference point,
$p^2=(2\,\textrm{GeV})^2$.
\begin{figure}
 \begin{center}
  \includegraphics[width=.5\textwidth]{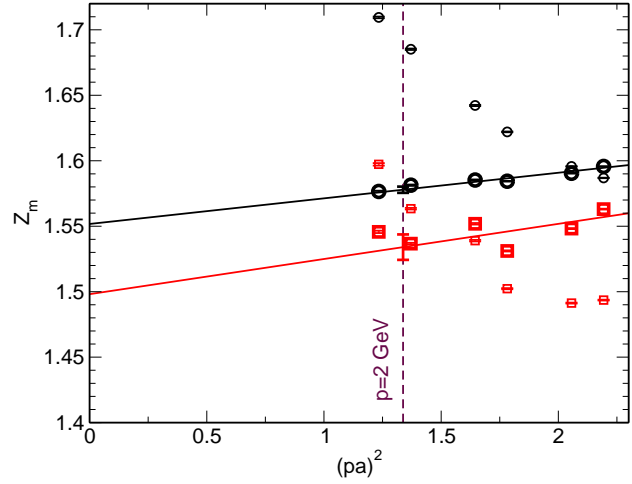}
 \end{center}
 \caption{
 Same figure as Fig.~\ref{fig:Zm32}, but for the coarse $24^3$ lattice.
}
 \label{fig:Zm24}
\end{figure}

The $24^3$ coarser lattice has been analyzed similarly for the $m_l=0.005$, $0.01$ and $0.02$ ensembles
and the results are shown in Fig.~\ref{fig:Zm24}. The mass renormalization factors on the $24^3$ lattice for the two intermediate SMOM schemes are:
\begin{eqnarray}
 Z_m^{\msbar(24)}(\mu=2\ {\rm GeV},n_f=3; \smomgm) & = & 1.578(2),\label{eq:Zm24SMOMgm}\\
 Z_m^{\msbar(24)}(\mu=2\ {\rm GeV},n_f=3; \smom) & = & 1.534(10).\label{eq:Zm24SMOM}
\end{eqnarray}

In Eq.\,(\ref{eq:baremasses}) we have presented the bare quark masses for the fine
$32^3$ lattice and in Table~\ref{tab:zlzhzacomparison} we give the ratios of equivalent bare masses on the $24^3$ and $32^3$ lattices. Because of the different
$O(a^2)$ artefacts for the light and heavy quark masses, there are two such
ratios $Z_l$ for the $ud$ quarks and $Z_h$ for the $s$ quark. These ratios $Z_l$ and $Z_h$ are also the scheme-independent ratios of the renormalization constants on the course and fine lattices. We now
use these ratios to estimate the difference of the $\MSb$ renormalized
masses with the $\SMOM$ and $\SMOMgm$ schemes in the continuum limit.
The continuum extrapolation of $Z_m^{(32)}$ and $Z_m^{(24)}/Z_l$ or
$Z_m^{(24)}/Z_h$ will remove the $(pa)^2$ error
in the non-perturbative renormalization. Thus, if a difference is
found, it can largely be attributed to the truncation error of the perturbative
matching. Performing such an extrapolation we find
\begin{eqnarray}
 Z_{ml}^{\msbar(32)c}(\mu=2\ {\rm GeV},n_f=3; \smomgm) & = & 1.527(6),\label{eq:ZmSMOMgm32cl}\\
 Z_{ml}^{\msbar(32)c}(\mu=2\ {\rm GeV},n_f=3; \smom) & = & 1.511(22),\label{eq:ZmSMOM32cl}
\end{eqnarray}
for the $ud$ quark, and
\begin{eqnarray}
 Z_{mh}^{\msbar(32)c}(\mu=2\ {\rm GeV},n_f=3; \smomgm) & = & 1.510(6),\label{eq:ZmSMOMgm32ch}\\
 Z_{mh}^{\msbar(32)c}(\mu=2\ {\rm GeV},n_f=3; \smom) & = & 1.495(22)\label{eq:ZmSMOM32ch}
\end{eqnarray}
for the $s$ quark.
Note that because these factors multiply
$\tilde{m}_{ud}(32^3)/a(32^3)$ or
$\tilde{m}_{s}(32^3)/a(32^3)$ presented in Eq.\,(\ref{eq:baremasses}) to give
the $\MSb$ mass in the continuum limit, they are made to
absorb the $O(a^2(32^3))$ discretization error in these bare quark
masses on the fine lattice.
Because of this, as well as the fact that the $Z_m$'s are free from $O(a^2)$ errors
originating from the $\SMOM$ non-perturbative renormalization,
we have put additional suffix ``$c$'' as ``continuum'' to distinguish
them from $Z_{m}^{\msbar(32)}$.
The existence of a mass dependent
contribution to the  $O(a^2)$ artefacts gives rise to the different $Z_m$ for the light
and heavy-quark masses.
From the two different estimates of the $\MSb$ renormalization
factors with the $\SMOM$ and $\SMOMgm$ intermediate non-perturbative
schemes, we choose to take $\SMOMgm$ for our central value. The reason is that
the scatter about the linear behaviour observed for the $\SMOM$ scheme in Figs.~\ref{fig:Zm32} and
\ref{fig:Zm24} is much larger. Although the effect of the scatter has
been taken into account in the error, we consider the continuum extrapolation from
the $\SMOM$ scheme to be less reliable. The difference in the central values of $ Z_{ml}^{\msbar(32)c}$ in Eqs.\,(\ref{eq:ZmSMOMgm32cl}) and (\ref{eq:ZmSMOM32cl}) is about 1\%, and this is also the case for the difference between the central values of  $Z_{mh}^{\msbar(32)c}$ in Eqs.\,(\ref{eq:ZmSMOMgm32ch}) and (\ref{eq:ZmSMOM32ch}). These differences of about 1\% give an indication of the possible size of the truncation error of the perturbative two-loop matching to $\MSb$ (it should be noted however, that the errors in the renormalization factors in the SMOM scheme are even a little larger). Another estimate of the truncation
error of the matching is obtained by evaluating the size of the two-loop
term in Eq.~(\ref{eq:CmSMOMgm}), resulting in $2.1\%$ for the $\SMOMgm$
scheme.  In order to be conservative, we shall take the latter as our estimate.
\begin{table}[t]
 \begin{tabular}{cccc}
  \hline
  \hline
  ensemble & fine ($32^2$) & coarse ($24^3$) & coarse ($16^3$)\cite{Aoki:2007xm}\\
  intermediate scheme & $\RISMOM$ & $\RISMOM$ & $\RIMOM$ \\
  \hline
  PT truncation error & 2.1\% & 2.1\% & 6\%\\
  $m_s\ne 0$ & 0.1\% & 0.2\% & 7\% \\
  $(\Lambda_P-\Lambda_S)/2$ & 0.5\% & 0.6\% & N.A. ($\infty$)\\
  $(\Lambda_A-\Lambda_V)/2$ & 0.0\% & 0.0\% & 1\% \\
  total & 2.2\% & 2.2\% & 9\%\\
  \hline
  \hline
 \end{tabular}
 \caption{Systematic error budget for $Z_m^{\MSb}(2 {\rm GeV})$ with
 intermediate  $\RISMOM$ schemes (this work) and $\RIMOM$ scheme
 \cite{Aoki:2007xm}.}
 \label{tab:Zm_sys_err}
\end{table}
Other systematic errors arise from the fact that the simulated strange mass is non-zero
and from the small difference in the scalar and pseudoscalar vertices due to the residual spontaneous symmetry breaking effects. The first error is estimated from the response of scalar and
pseudoscalar vertex functions to the variation of the light-quark mass\,\cite{Aoki:2007xm}. From the flat behaviour of $\Lambda_P+\Lambda_S$ on the light-quark mass in Fig.~\ref{fig:P+S} it can be seen that this uncertainty is small.
The error estimates are compiled in Table \ref{tab:Zm_sys_err}.
In the table, the corresponding errors from the  $\RIMOM$ analysis
\cite{Aoki:2007xm} are shown for comparison.
All errors have become significantly smaller for the new $\SMOM$ schemes.
Now our final values for the $\MSb$ renormalization factor read
\begin{eqnarray}
 Z_{ml}^{\msbar(32)c}(\mu=2\ {\rm GeV},n_f=3) & = & 1.527(6)(33),\label{eq:Zm32cl}\\
 Z_{mh}^{\msbar(32)c}(\mu=2\ {\rm GeV},n_f=3) & = & 1.510(6)(33),\label{eq:Zm32ch}
\end{eqnarray}
where the first error is the statistical uncertainty inflated to take into account the scatter about the linear behaviour due to $O(4)$ non-invariant effects (as explained above) and
the second is due to the remaining systematic effects and is
dominated by the $2.1\%$ truncation error of the perturbative matching.
Here we have not taken into account the statistical fluctuation of
$Z_V$, which will be properly included in the calculation of the
renormalized quark masses described in the next subsection.
The corresponding renormalization factor for the light-quark
mass on the coarse $24^3$ lattice is $Z_{ml}^{\msbar(24)c}(\mu=2\ {\rm GeV},n_f=3)=
Z_l\cdot Z_{ml}^{\msbar(32)c}(\mu=2\ {\rm GeV},n_f=3)=1.498(6)(33)$.
This value is consistent with our earlier estimate of the same
quantity using RI/MOM as the intermediate scheme, $1.656(157)$\,\cite{Aoki:2007xm},
but now with a considerably reduced error.

\subsection{Renormalized quark masses}\label{subsec:renmasses}

After the detailed discussion of the quark-mass renormalization, it is now straightforward to combine the renormalization constants in Eqs.\,(\ref{eq:Zm32cl}) and (\ref{eq:Zm32ch}) with the physical bare quark masses on the $32^3$ lattice in Eq.\,(\ref{eq:baremasses}) to obtain the light and strange quark masses
renormalized in $\MSb$ scheme:
\begin{eqnarray}\label{eq:mudfinal}
 m_{ud}^{\msbar}(2 {\rm GeV}) & = & Z_{ml}^{\msbar(32)c}(\mu=2\, {\rm
  GeV},n_f=3) \cdot \tilde{m}_{ud}(32^3) \cdot a^{-1}(32^3)\nonumber\\
  &=& 3.59 (13)_{\rm stat} (14)_{\rm sys} (8)_{\rm ren}\; {\rm MeV},\\
 m_s^{\msbar}(2\, {\rm GeV}) & = & Z_{mh}^{\msbar(32)c}(\mu=2 {\rm
  GeV},n_f=3) \cdot \tilde{m}_s(32^3) \cdot a^{-1}(32^3)\nonumber\\
  &=& 96.2 (1.6)_{\rm stat} (0.2)_{\rm sys} (2.1)_{\rm ren}\; {\rm MeV},
\label{eq:msfinal}\end{eqnarray}
where the three errors on the right-hand side correspond to the
statistical uncertainty, the systematic uncertainty due to the chiral
extrapolation and finite volume, and
the error in the renormalization factor. We recall that for the error due to the chiral extrapolation we conservatively take the full difference of the results obtained using the finite-volume NLO SU(2) and analytic fits and for the central value we take the average of these results. We estimate the finite-volume effects from the difference of the results obtained using finite volume and infinite-volume NLO ChPT fits and combine these errors in quadrature. The finite-volume errors prove to be small.
The error in the renormalization factor includes those in
Eqs.~(\ref{eq:Zm32cl}) and (\ref{eq:Zm32ch}).

The ratio of the $s$ and $ud$ quark masses is
\begin{equation}
  \frac{m_s}{m_{ud}} = 26.8
  (0.8)_{\rm stat}
  (1.1)_{\rm sys}.
\end{equation}

We end this section by presenting our results for the leading-order LEC $B$ and the chiral condensate. Using the finite-volume NLO ChPT fits we find
\begin{equation} \label{eq:QM_B_chpt}
  B^{\msbar}(2 {\rm GeV}) = Z_{ml}^{\msbar(32)-1}(\mu=2 {\rm
  GeV},n_f=3) \cdot  B(32^3) \cdot a^{-1}(32^3) =
  2.64(6)_{\rm stat}
  (6)_{\rm sys}
  (6)_{\rm ren}\; {\rm GeV}.\\
\end{equation}
Combining this result with the pion decay constant in the chiral limit,
also obtained using the finite-volume NLO ChPT fits the chiral condensate is found to be
\begin{equation}\label{eq:Sigmavalue}
 [\Sigma^{\msbar}(2 {\rm GeV})]^{1/3} =
  [f^2B(2 {\rm GeV})/2]^{1/3} = 256(5)_{\rm stat} (2)_{\rm sys} (2)_{\rm ren}\; {\rm MeV}.
\end{equation}
In Eqs.~(\ref{eq:QM_B_chpt}) and (\ref{eq:Sigmavalue}) the second error is
only due to finite volume corrections estimated from the difference of finite
and infinite volume NLO ChPT fits.

\section{Topological Susceptibility}
\label{sec:Topology}

\DeclareGraphicsRule{.tif}{png}{.png}{`convert #1 `dirname #1`/`basename #1 .tif`.png}

The topological charge $Q$, defined on a single Euclidean space-time
configuration, and its susceptibility, $\chi_Q$, are interesting
quantities to calculate.  While $Q$ depends only indirectly on the
quark masses, leading order SU(2) ChPT~\cite{DiVecchia:1980ve,Leutwyler:1992yt}
predicts a strong dependence of $\chi_Q$ on the light sea quark mass with $\chi_Q$
vanishing linearly as $m_l \rightarrow 0$, suggesting that $\chi_Q$
may show important dynamical quark mass effects.

In the continuum $Q$ and $\chi_Q$ are defined by
\begin{equation}
Q = \frac{g^2}{16\pi^2}\int d^4x\, G_{\mu\nu}(x)\tilde G_{\mu\nu}(x)\quad\textrm{and}\quad
\chi_Q = \langle Q^2\rangle / {V},
\label{eq:top charge}
\end{equation}
where $V$ is the four-volume of the lattice, $G_{\mu\nu}(x)$ is the
gluon field strength tensor and $\tilde G_{\mu\nu}(x)$, its dual.
In the continuum, $Q$ is integer valued and related to exact chiral
zero modes of the massless Dirac operator by the Atiyah-Singer
index theorem~\cite{Atiyah:1963}.   For sufficiently smooth gauge
fields it is possible to find a lattice expression which will
always evaluate to an integer~\cite{Luscher:1981zq}, as in the
continuum limit.  However, in the calculation reported here the
necessary smoothness condition is not obeyed and we instead
replace the right-hand side of Eq.~(\ref{eq:top charge}) by a sum
of Wilson loops chosen to approximate the $G_{\mu\nu}(x)\tilde
G_{\mu\nu}(x)$ product in Eq.~(\ref{eq:top charge}).  Specifically
we employ the ``five-loop improved" (5Li) definition of the
topological charge proposed in Ref.~\cite{deForcrand:1997sq}
which at tree level is accurate through order $a^4$.  However,
before evaluating this lattice expression for the topological
charge, we smooth the links in the lattice by performing a
series of APE smearing steps~\cite{Falcioni:1984ei,Albanese:1987ds}.
The smearing parameter was set to 0.45, and 60 smearing sweeps were
performed before measuring $Q$. The results are insensitive to the
choice of these parameters.

\begin{center}
\begin{figure}
\begin{minipage}{\textwidth}
\begin{center}
\includegraphics[width=.8\textwidth]{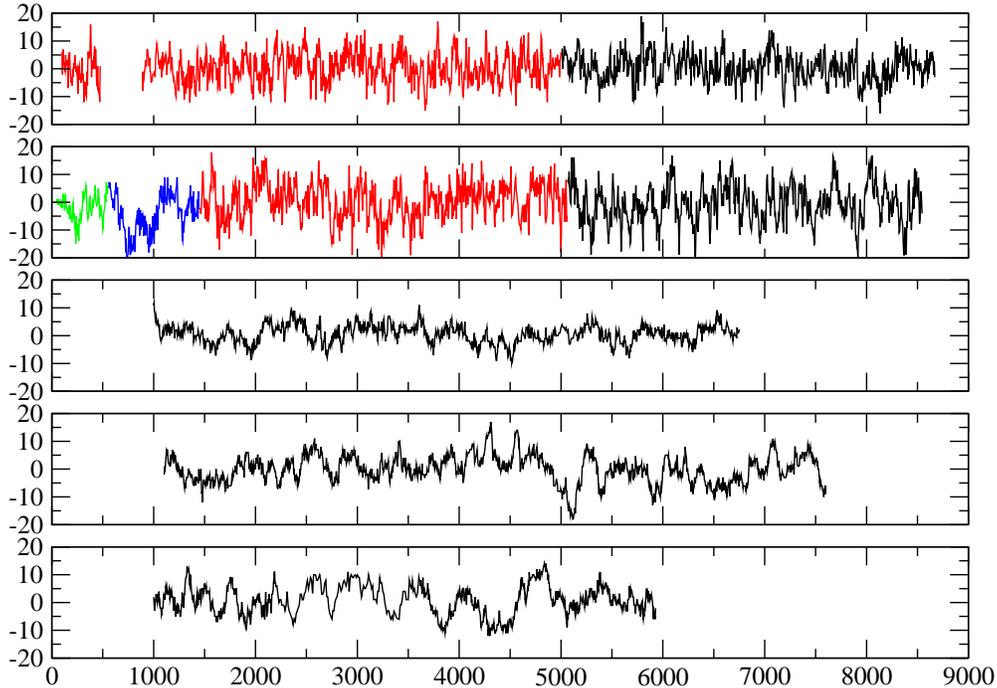}
\end{center}
\end{minipage}
\caption{\label{fig:top_history}
Monte Carlo time histories of the topological charge. The light sea quark mass increases from top to bottom, (0.005 and 0.01, $24^3$ (top two panels), and 0.004-0.008, $32^3$).
Data for the $24^3$ ensembles up to trajectory 5000 were reported originally in~\cite{Allton:2008pn} and the results from the new ensembles are plotted in black. Most of the data was generated using the RHMC II algorithm (red and black lines). The RHMC 0 (green line) and RHMC I (blue line) algorithms were used for trajectories up to 1455 for the $m_l=0.01$ ensemble. The small gap in the top panel represents missing measurements which are irrelevant since observables are always calculated starting from trajectory 1000.}
\end{figure}
\end{center}

In Fig.~\ref{fig:top_history} the Monte Carlo time history of
$Q$ is shown for each ensemble of gauge fields in our study.
For each case, the update algorithm RHMC II~\cite{Allton:2008pn}
was used, except for the first 1455 configurations for the
$m_{l}=0.01$ ensemble where the RHMC 0 and RHMC I algorithms
were used. In~\cite{Allton:2008pn} it was shown that RHMC II is
more effective in changing the gauge field topology, and therefore
produces shorter auto-correlation times. The data for the first
half (up to trajectory 5000) of both $24^{3}$ ensembles is
repeated from~\cite{Allton:2008pn}.  Figure~\ref{fig:top_history}
shows clearly the expected slowing of the rate of change of
topological charge when moving towards the continuum~\cite{Aoki:2005ga}
and, to a lesser degree, when decreasing the quark mass. The
integrated auto-correlation times for $Q$ for the smaller lattice
spacing ensembles are shown in Fig.~\ref{fig:ic_jk}.  {While this
figure is consistent with the autocorrelation times reaching a plateau
of about 80 time units when integrated over an interval of about 200
time units, the exploding errors make this conclusion highly uncertain.}
Scanning Fig.~\ref{fig:top_history} by eye, {one might argue
that} the auto-correlations could be 500 time units, or longer.
For example, note the large fluctuation to negative $Q$
beginning around time unit 4750 for $m_{l}=0.006$.

\begin{center}
\begin{figure}
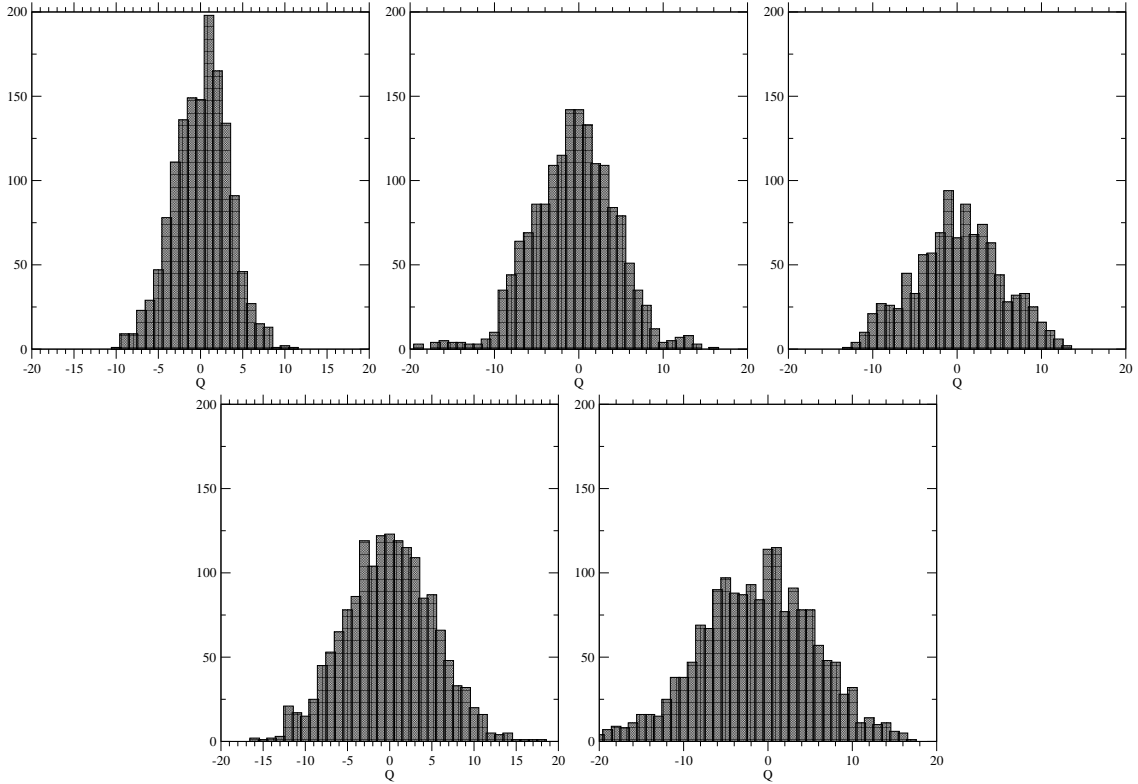

\begin{minipage}{\textwidth}
\begin{center}
\includegraphics[width=.3\textwidth]{fig/Fig53a_top-histogram-004.eps}
\includegraphics[width=.3\textwidth]{fig/Fig53b_top-histogram-006.eps}
\includegraphics[width=.3\textwidth]{fig/Fig53c_top-histogram-008.eps}
\includegraphics[width=.3\textwidth]{fig/Fig53d_top-histogram-005.eps}
\includegraphics[width=.3\textwidth]{fig/Fig53e_top-histogram-01.eps}
\end{center}
\end{minipage}
\caption{\label{fig:top histograms}
Topological charge distributions. Top: $32^3$, $m_l=0.004-0.008$,
left to right. Bottom: $24^3$, $m_l=0.005$ and 0.01.}
\end{figure}
\end{center}

The distributions of topological charge for each ensemble are shown
in Fig.~\ref{fig:top histograms}. The distributions become narrower
as the quark mass is decreased. For the smaller lattice spacing,
they also appear to exhibit non-Gaussian-like tails, or humps at
large $|Q|$.

Because of the parity symmetry of our calculation, the average of
the pseudo-scalar quantity $\langle Q\rangle$ vanishes.  However, $\chi_Q$
remains non-zero and at leading order in SU(2) chiral perturbation
theory~\cite{DiVecchia:1980ve,Leutwyler:1992yt} is given by
\begin{eqnarray}
\chi_Q &=& \Sigma \left( \frac{1}{m_u}+\frac{1}{m_d}\right)^{-1}
=~\Sigma \frac{m_u m_d}{m_u+ m_d},
	\label{eq:chiQ su2}
\end{eqnarray}
where $\Sigma = B f^2/2$ is the chiral condensate coming from a single
flavor in the limit of vanishing up and down quark mass.

At one-loop in chiral perturbation theory~\cite{Mao:2009sy},
\begin{eqnarray}
\chi_Q &=& \Sigma \left( \frac{1}{m_u}+\frac{1}{m_d}\right)^{-1}\times\nonumber\\
&&\left(
1-\frac{3}{(4\pi f)^2}m_{\pi}^2\log{\frac{m_{\pi}^2}{\Lambda^2}} +K_6(m_u+m_d)+2(2K_7+K_8)\frac{ m_u m_d}{m_u+m_d}\right),\label{eq:chiQ su2 NLO}\\
&=& \Sigma\frac{m_l}{2}\left(
1-\frac{3}{(4\pi f)^2}m_{ll}^2\log{\frac{m_{ll}^2}{\Lambda^2}} +(2 K_6+2K_7+K_8){m_l}
\right),
\label{eq:chiQ su2 NLO deg}
\end{eqnarray}
where $K_i=128\, \Sigma L_i/f^4$ are proportional to the Gasser-Leutwyler
NLO LEC's~\cite{Mao:2009sy}, and in the last line the formula is evaluated
for degenerate quarks. In contrast to other quantities considered in this
paper, we do not attempt to characterize or evaluate the corrections
to Eqs.~(\ref{eq:chiQ su2 NLO}) or~(\ref{eq:chiQ su2 NLO deg}) which come
from non-zero lattice spacing.  That interesting question is left for
future work.

In Tab.~\ref{tab:top summary} values of $\langle Q\rangle$ and $\chi_Q$
for each ensemble of configurations are summarized. To test for the
expected auto-correlations, the data were blocked into bins of various
sizes ranging from 10 to 600 time units. The quoted values of the
statistical errors resulted when the block sizes were taken large
enough that the errors no longer changed significantly. The block sizes
are given in Tab.~\ref{tab:top summary}.
For all cases the first 1000 time units were discarded for thermalization.

\begin{center}
\begin{figure}
\begin{minipage}{\textwidth}
\begin{center}
\includegraphics[width=.6\textwidth]{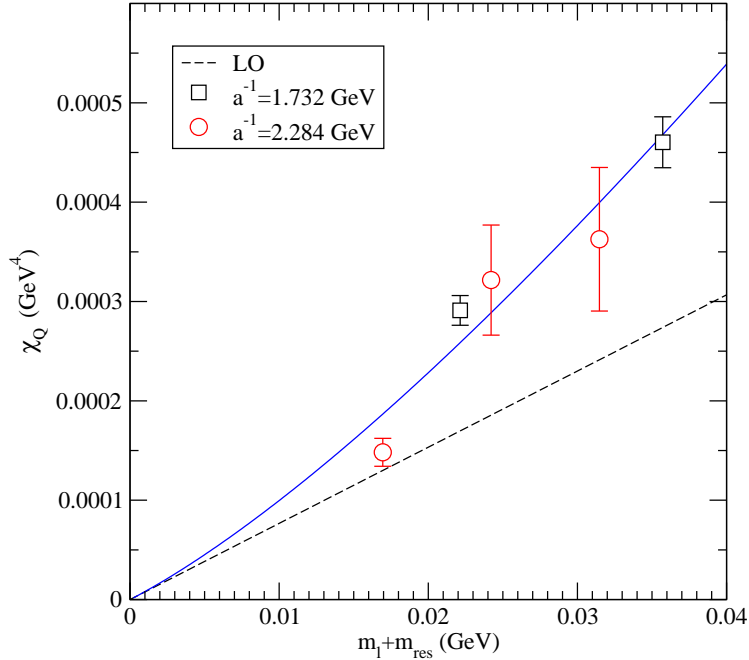}
\end{center}
\end{minipage}
\caption{\label{fig:chiQ}
Topological susceptibility ($24^3$ (squares), $32^3$ (circles)). The dashed
line is the prediction from LO SU(2) chiral perturbation theory
(Eq.~(\ref{eq:chiQ su2})) with the chiral condensate computed from the finite
volume LEC's given in Table~\ref{tab-NLOsu2fitparameters}.  The solid line
denotes the result of the single-parameter fit to the NLO formula given in
Eq.~(\ref{eq:chiQ su2 NLO deg}).}
\end{figure}
\end{center}

The dependence of $\chi_Q$ on the light quark mass is shown in Fig.~\ref{fig:chiQ}.
All of the data points lie above the LO curve (dashed line), all but the lightest
significantly so. The result of the fit ($\chi^2/$dof $\approx 13/4\approx 3$) to
the NLO formula Eq.~(\ref{eq:chiQ su2 NLO deg}) is also shown. Since we have not
determined $K_7$ in Eq.~(\ref{eq:chiQ su2 NLO deg}) from other means, we treat the
linear combination of LEC's as a single, new, free parameter in the fit and find
$(2 K_6+2K_7+K_8)=19.8(6.3)$.  Except for the lightest data point, there is scant
evidence for large $O(a^2)$ errors, though the statistical errors on the heavier
two points with $a^{-1}=2.284$ are somewhat large. Omitting the former point in
the fit leads to a more acceptable value of $\chi^2$/dof $\approx1.5$, suggesting
the lightest point may be systematically low due to long auto-correlations in $Q$
that are not well resolved in our finite Markov chain of configurations.  Despite
these limitations, the data appear to show a dependence on the light sea quark mass
that is consistent with the dictates of NLO SU(2) ChPT.

\begin{table}[htdp]
\caption{Topological charge and susceptibility. The measurement
frequency, ``meas. freq.'',   and ``block size'' are given in
units of Monte Carlo time.}
\begin{center}
\begin{tabular}{cccccc}
\hline
$m_l$ & meas. freq. & block size & $\langle Q\rangle$ & $\langle Q^2\rangle$ & $\chi$ (GeV$^{4}$) \\
\hline
0.005 & 5 & 50 & 0.49 (25) & 28.6 (1.4) & 0.000290 (14)\\
0.01 & 5 & 50 &  -0.22 (37) & 45.2 (2.5) & 0.000458 (25)\\
0.004 & 4 & 200 & 0.59 (42) & 11.4 (1.1) & 0.000148 (14)\\
0.006 & 4 & 200 & -0.07 (64) & 24.8 (4.3) & 0.000322 (55)\\
0.008 & 4 & 400 &0.64 (100) & 27.9 (5.6) & 0.000363 (72) \\
\hline
\end{tabular}
\end{center}
\label{tab:top summary}
\end{table}

\section{Conclusions}
\label{sec:Conclusions}

\ifnum\theConclusions=1
%
%

We have presented results from simulations using DWF and the Iwasaki gauge action for lattice QCD at two values of the lattice spacing ($a^{-1}$=\,1.73\,(3)\,GeV and $a^{-1}$=\,2.28\,(3)\,GeV) and for unitary pion masses in the range 290--420\,MeV (225--420\,MeV for the partially quenched pions). The raw data obtained at each of the two values of $\beta$ was presented in Sections\,\ref{sec:24cubed} and \ref{sec:32cubed} respectively and the chiral behaviour of physical quantities on the $24^3$ and $32^3$ lattices separately was studied in Appendix\,\ref{sec:appendix:separate_fits}. The main aim of this paper however, was to combine the data obtained at the two values of the lattice spacing into global chiral--continuum fits in order to obtain results in the continuum limit and at physical quark masses and we explain our procedure in Section\,\ref{sec:CombinedChiralFits}. In that section we define our scaling trajectory, explain how we match the parameters at the different lattice spacings so that they correspond to the same physics and discuss how we perform the extrapolations. We consider this discussion  to be a significant component of this paper and believe that this will prove to be a good approach in future efforts to obtain physical results from lattice data. Although we apply the procedures to our data at two values of the lattice spacing, we stress that the discussion is more general and can be used with data from simulations at an arbitrary number of different values of $\beta$. In the second half of Section\,\ref{sec:CombinedChiralFits} we then perform the combined continuum--chiral fits in order to obtain our physical results for the decay constants, physical bare quark masses (which are renormalized in Section\,\ref{sec:quark_masses}) and for the quantities $r_0$ and $r_1$ defined from the heavy-quark potential. For the discussion below, it is important to recall that we use the physical pion, kaon and $\Omega$ masses to determine the physical quark masses and the values of the lattice spacing and we then make predictions for other physical quantities.

In contrast to most other current lattice methods, the DWF formulation gives our simulations good control over chiral symmetry, non-perturbative renormalization factors and flavor symmetry.  This control allows us to measure and use, as either inputs or predictions:  pseudoscalar decay constants, as well as their ratios; pseudoscalar masses; baryon masses; weak matrix elements and static potential values, limited only by the statistics achievable for these observables.  The ability to predict many observables from the same simulations, provides evidence for the general reliability of the underlying methods.  The good properties of DWF also allow us to test scaling, over this wide range of observables, at unphysical quark masses, since there are no flavor or chiral symmetry breaking effects to distort a test of scaling.  We find scaling violations at the percent level, which supports including scaling corrections in only the leading order terms in our light-quark expansions.

As we reduce the quark masses used in the simulations, it is frustrating that there remains a doubt as to the best ansatz to use for the chiral extrapolation. We know of course that for sufficiently light $u$ and $d$ masses the behaviour is given by SU(2) ChPT; what we don't know is what "sufficiently light" means in practice. While in the range of quark masses accessible in our simulations, corresponding to 290\,-\,420\,MeV for unitary pions and 225\,-\,420\,MeV for partially quenched pions, our data are consistent with NLO SU(2) ChPT, we have seen that they are also consistent with a simple analytic ansatz leading to an inherent uncertainty in how best to perform the chiral extrapolation. This is particularly well illustrated in the study of $f_\pi$, see Fig.\,\ref{fig:fpiNLOfvadjusted} for example, where the data is well represented by all three ans\"atze (including NLO SU(2) ChPT with finite-volume corrections), but the extrapolated values differ as seen in Table\,\ref{tab:fpi:extrapolated} $f_\pi=121(2)$\,MeV from the NLO ChPT analysis with finite-volume corrections and $f_\pi=$126(2)\,MeV using the analytic ansatz. Since a complete NNLO ChPT analysis is not possible with the available data, we have resisted the temptation to introduce model dependence by including only some of the higher order corrections and for our current ``best" results we take the average of the two values and include the full difference in the systematic uncertainty obtaining $f_\pi=124(2)(5)$\,MeV. In Section\,\ref{subsubsec:chiralfpi} we investigated the increase in $\chi^2$/dof if the fits are required to pass through the physical value 130.7(4)\,MeV up to corrections from lattice artefacts and found $\chi^2$=1.9(7) for the analytic ansatz and an unacceptably large value of 5(1) for the NLO ChPT with finite volume corrections. In the future, it will be very interesting to see how the different ans\"atze for the chiral extrapolation become constrained or invalidated as we perform simulations with even lighter masses. We point out that the difference in the results from the analyses using the finite-volume ChPT and analytic ans\"atze is much smaller for the other quantities studied in this paper than for $f_\pi$.

The main physical results of this study are:
\begin{eqnarray*}
f_\pi=124(2)(5)\,\textrm{MeV}\ \{\textrm{Eq.}\,(\ref{eq:fpifinal})\};&\quad&
f_K=149(2)(4)\,\textrm{MeV}\ \{\textrm{Eq.}\,(\ref{eq:fkfinal})\};\\
&&\hspace{-1in}\frac{f_K}{f_\pi}=1.204(7)(25)\ \{\textrm{Eq.}\,(\ref{eq:fkoverfpifinal})\};
\end{eqnarray*}
\begin{equation*}
m_s^{\overline{\textrm{MS}}}(2\,\textrm{GeV})=
(96.2\pm 2.7)\,\textrm{MeV}\ \{\textrm{Eq.}\,(\ref{eq:msfinal})\};\ \
m_{ud}^{\overline{\textrm{MS}}}(2\,\textrm{GeV})=
(3.59\pm 0.21)\,\textrm{MeV}\ \{\textrm{Eq.}\,(\ref{eq:mudfinal})\};
\end{equation*}
\begin{equation*}
[\Sigma^{\msbar}(2 {\rm GeV})]^{1/3} = 256(6)\; {\rm MeV}
\ \{\textrm{Eq.}\,(\ref{eq:Sigmavalue})\};
\end{equation*}
\begin{equation}
r_0=0.487(9)\,\textrm{fm}\quad\textrm{and}\quad r_1=0.333(9)\,\textrm{fm}\ \ \{\textrm{Eq.}\,(\ref{eq:r0r1final})\}\,.\label{eq:finalresults}
\end{equation}
For convenience we also display the equation number where the results were presented earlier in this paper to help the reader find the corresponding discussion. All the results in Eq.\,(\ref{eq:finalresults}) were obtained after reweighting the strange-quark mass to its physical value at each $\beta$, and the renormalized quark masses were obtained using non-perturbative renormalization with non-exceptional momenta as described in Section\,\ref{sec:quark_masses}. The low-energy constants obtained by fitting our data to NLO chiral perturbation theory can be found in Sec.\,\ref{subsec:results}.

The configurations and results presented in this paper are being used in many of our current studies in particle physics phenomenology, including the determination of the $B_K$ parameter of neutral kaon mixing in the continuum limit~\cite{bkpaper}. In parallel to these studies we are exploiting configurations generated at almost physical pion masses on lattices with a large physical volume ($\sim$ 4.5~fm) but at the expense of an increased lattice spacing.  Preliminary results obtained for the meson spectrum and decay constants and for $\Delta I=3/2$ $K\to\pi\pi$ decay amplitudes were recently presented in Refs.\,\cite{boblatt2010,matthewlatt2010}. Having access to data with excellent chiral and flavor properties with a range of lattice spacings and quark masses makes this an exciting time indeed for studies in lattice phenomenology.

\fi

\ifnum\theAcknowledgments=1
\section*{Acknowledgments}

The calculations reported here were performed on the QCDOC computers \cite{Boyle:2005qc,Boyle:2003mj,Boyle:2005fb} at Columbia University, Edinburgh University, and at the Brookhaven National Laboratory (BNL).
At BNL, the QCDOC computers of the RIKEN-BNL Research Center and the USQCD Collaboration were used.  Most important were the computer resources of the Argonne Leadership Class Facility (ALCF) provided under the Incite Program of the US DOE.  The very large scale capability of the ALCF was critical for carrying out the challenging calculations reported here. We also thank the University of Southampton for access to the Iridis computer system used in the calculations of the non-perturbative renormalisation factors (with support from UK STFC grant ST/H008888/1). The software used includes: the CPS QCD codes http://qcdoc.phys.columbia.edu/chulwoo/index.html, supported in part by the USDOE SciDAC program; the BAGEL http://www.ph.ed.ac.uk/~paboyle/bagel/Bagel.html assembler kernel generator for many of the high-performance optimized kernels~\cite{Boyle:2009bagel}; and the UKHadron codes.

Y.A. is partially supported by JSPS Kakenhi grant No.~21540289. R.A.,
P.A.B., B.J.P. and J.M.Z. were partially supported by UK STFC grant ST/G000522/1.
T.B. and R.Z. were supported by US DOE grant DE-FG02-92ER40716. D.B., J.M.F. and C.T.S were partially supported by UK STFC Grant ST/G000557/1 and by EU contract MRTN-CT-2006-035482 (Flavianet). N.H.C., M.L. and R.D.M were supported by
US DOE grant number DE-FG02-92ER40699. C.J., T.I. and A.S. are partially supported by the US DOE under contract No. DE-AC02-98CH10886. E.E.S. is partly supported by DFG SFB/TR 55 and by the Research Executive Agency of the European Union under grant PITN-GA-2009-238353 (ITN STRONGnet).

\fi

\appendix
\ifnum\theAppendix=1
%
%
%
%
%
%
%

\section{Separate fits to \boldmath{$24^3$ and $32^3$} data}
\label{sec:appendix:separate_fits}

In this section we report on results obtained by fitting the data from the $24^3$ runs at $\beta=2.13$ and from the $32^3$ runs at $\beta=2.25$ separately to the predictions of $\SU(2)\times\SU(2)$ ChPT. This complements the material presented in Sections~\ref{sec:24cubed} and \ref{sec:32cubed} in which we presented the results for masses and decays constants at each set of quark masses but did not perform the chiral extrapolations and also that in Section~\ref{sec:CombinedChiralFits} in which we performed simultaneous chiral and continuum fits to the data at both lattice spacings. Our main motivation for studying separate fits here is to be able to compare directly our results obtained with the new data to those in our previous publication \cite{Allton:2008pn}. For that reason in this appendix we will be using the same renormalization constant $Z_A$ as in our previous publication, which differs from the one used in the global analysis presented in the main part of this paper, see the discussion in Sec.~\ref{sec:24cubed} and App.~\ref{sec:appendix:za} for details. We use the same method of iterated fits as outlined in our earlier publication \cite{Allton:2008pn}; at each lattice spacing we iterate the combined fits of the meson masses and decay constants with $m_x\leq 0.01$ to the $\SU(2)$-ChPT formulae, using kaon $\SU(2)$ ChPT to fit the kaon mass and decay constants and the extrapolation in the $\Omega$-baryon mass until convergence. The pion, kaon, and $\Omega$ masses are used to fix the physical bare quark masses $m_{ud}$, $m_s$ and the lattice scale $1/a$. Predictions for the remaining physical quantities are then obtained by extrapolation to these physical quark masses. For further details see \cite{Allton:2008pn}. In the case of the $24^3$ ensembles, the runs have been extended since the publication of \cite{Allton:2008pn} (see Sec.~\ref{sec:SimulationDetails} and especially Tab.~\ref{table:para} for details) so that a direct comparison of the results from the previous (smaller) data set with the new extended data set is possible.  We quote results from fits with and without corrections due to finite-volume effects. When including the finite volume corrections, the terms described in Appendix~C of \cite{Allton:2008pn} are included in the $\SU(2)$ ChPT in the pion sector (both for the meson masses and decay constants). We also include the correction terms containing the chiral logarithm of the light quark masses in the kaon decay constant \footnote{%
For completeness, since those have not been included in Appendix~C of \cite{Allton:2008pn}, the finite volume corrections for the kaon decay constant in $\SU(2)$-ChPT read:
\[ \Delta^{Lf_K}_{xy} \;=\; -\frac1{4\pi f^2} \left[ \frac{\chi_x+\chi_l}2\delta_1\left(\sqrt{\frac{\chi_x+\chi_l}2}L\right)+\frac{\chi_l-2\chi_x}4\delta_1(\sqrt{\chi_x}L) \right]\,. \]
See Appendix~C of \cite{Allton:2008pn} for an explanation of the notation.
} %
and note that up to NLO in the light-quark masses, no finite-volume corrections arise in the masses of the kaon and $\Omega$-baryon. Below we present the physical results in the infinite-volume limit, i.e. after removing the corrections. Finally, we will perform a na\"ive continuum extrapolation of the results obtained by the separate fits at the two lattice spacings, which can then be compared to results from the combined chiral-continuum extrapolations using the global fits described in Sec.~\ref{sec:CombinedChiralFits}. Note that in this appendix also for the combined chiral-continuum extrapolations we are going to quote results obtained using our previous definition of $Z_A$. For that reason the results reported here differ slightly from those in the main part of this paper.

\subsection{\label{subsec:appendix_separate_fits:24c}\boldmath SU(2)-ChPT fits to $24^3$ data\unboldmath}

In Tab.~\ref{tab:fit_separate24c} we summarize our results from the iterative fits to the masses and decay constants measured on the $24^3$ ensembles (see Sec.~\ref{sec:24cubed} for details) and compare them to our earlier results obtained with lower statistics~\cite{Allton:2008pn}. We have performed two kinds of fits: one including the $\Omega$-baryon masses determined at all the simulated light-quark masses, $m_l=0.005$, 0.01, 0.02, and 0.03, (as was done originally) and one where only the $\Omega$-baryon masses at the two lightest dynamical quark masses $m_l=0.005$ and 0.01 are included. The latter, limited range is also the one used in the combined chiral-continuum extrapolations in Section~\ref{sec:CombinedChiralFits} and in the separate fits to the $32^3$ data in the next subsection. In Fig.~\ref{fig:fit_separate24c} we plot the combined $\SU(2)$ ChPT fits (without finite-volume corrections) to the meson masses and decay constants in the pion sector. It is evident that over the fit range $(m_x+m_y)/2\leq0.01$, corresponding to a maximum meson mass of about 420 MeV, the data is well described by $\SU(2)$ ChPT. This is also true for the fits including the finite-volume corrections (not shown).

\begin{table}
\begin{center}
\begin{tabular}{c|c|ccc}
\hline\hline%
  & Allton et al.~\cite{Allton:2008pn}   &  \multicolumn{3}{c}{increased statistics}\\
  & no FV-corr. & \multicolumn{2}{c}{no FV-corr.} & incl.\ FV-corr.\\
  & $\Omega$: $m_l\leq0.03$ &  $\Omega$: $m_l\leq0.03$ & $\Omega$: $m_l\leq0.01$ & $\Omega$: $m_l\leq0.01$ \\\hline
$1/a$ $[{\rm GeV}]$ & 1.729(28) & 1.731(19) & 1.784(44) & 1.784(44) \\[10pt]
$B^{\overline{\rm MS}}(2\,{\rm GeV})$ $[{\rm GeV}]$ & $2.52(0.11)(0.23)_{\rm ren}$  & $2.63(0.06)(0.07)_{\rm ren}$ & $2.69(0.09)(0.08)_{\rm ren}$ & $2.63(0.09)(0.08)_{\rm ren}$ \\
$f$ $[{\rm MeV}]$ & 114.8(4.1) & 111.5(2.9) & 114.8(4.0) & 117.1(4.0) \\
$\bar{l}_3$       & 3.13(0.33) & 2.76(0.24) & 2.82(0.24) & 2.59(0.27) \\
$\bar{l}_4$       & 4.43(0.14) & 4.54(0.10) & 4.61(0.10) & 4.57(0.11) \\[10pt]
$f_\pi$ $[{\rm MeV}]$ & 124.1(3.6) & 121.2(2.5) & 124.4(3.6) & 126.4(3.6) \\
$f_K$ $[{\rm MeV}]$   & 149.6(3.6) & 147.9(2.6) & 151.0(3.7) & 152.1(3.7) \\
$f_K/f_\pi$            & 1.205(0.018) & 1.220(0.011) & 1.214(0.012) & 1.204(0.012) \\[10pt]
$m^{\overline{\rm MS}}_{ud}(2\,{\rm GeV})$ $[{\rm MeV}]$ & $3.72(0.16)(0.33)_{\rm ren}$ & $3.56(0.08)(0.10)_{\rm ren}$ & $3.48(0.12)(0.10)_{\rm ren}$ & $3.55(0.12)(0.11)_{\rm ren}$ \\
$m^{\overline{\rm MS}}_{s}(2\,{\rm GeV})$ $[{\rm MeV}]$ & $107.3(4.4)(9.7)_{\rm ren}$ & $101.0(1.9)(2.9)_{\rm ren}$ & $99.0(3.0)(3.0)_{\rm ren}$ & $98.8(3.0)(3.0)_{\rm ren}$\\
$\tilde{m}_{ud}:\tilde{m}_s$ & 1:28.8(0.4) & 1:28.37(0.27) & 1:28.44(0.26) & 1:27.89(0.28) \\[10pt]
$aB$ & 2.414(61) & 2.348(43) & 2.349(44) & 2.298(45)\\
$af$ & 0.0665(21) & 0.0644(14) & 0.0643(14) & 0.0656(14) \\
$L_4^{(2)} \times 10^{4}$ & 1.3(1.3) & 2.2(0.9) & 2.5(0.9) & 2.2(0.9)\\
$L_5^{(2)} \times 10^{4}$ & 5.16(0.73) & 5.00(0.47)  & 5.50(0.47) & 5.36(0.48) \\
$(2L_6^{(2)}-L_4^{(2)}) \times 10^{4}$ & -0.71(0.62) & -0.09(0.45) & 0.03(0.45) & 0.01(.49)\\
$(2L_8^{(2)}-L_5^{(2)}) \times 10^{4}$ & 4.64(0.43) & 4.86(0.30) & 4.36(0.38) & 5.34(0.33) \\
$a\tilde{m}_{ud}$ & 0.001300(58) & 0.001331(43) & 0.001251(71) & 0.001274(72)\\
$a\tilde{m}_s$ & 0.0375(16) & 0.0377(11) & 0.0356(19) & 0.0355(19) \\\hline\hline
\end{tabular}
\end{center}
\caption{Results from the $\SU(2)$ ChPT fits to the $24^3$ data (without and with finite-volume corrections) compared to those from~\cite{Allton:2008pn} obtained with lower statistics (without finite-volume corrections). We also quote in the lower part of the table the $\SU(2)$ ChPT fit parameters $aB$, $af$, $L_i^{(2)}$ (at the scale $\Lambda_\chi=1\,{\rm GeV}$) and bare quark masses $a\tilde{m}_{ud,s}$ in lattice units. Only statistical uncertainties are quoted except for quark masses and the LEC $B$ renormalized in the $\overline{\rm MS}$-scheme at 2 GeV where also the systematic uncertainty from the renormalization constant is quoted. (Mass renormalization constant at $1/a=1.731(19)\,{\rm GeV}$: $Z_m=1.546(0.002)_{\rm stat}(0.044)_{\rm ren}$ and at $1/a=1.784(44)\,{\rm GeV}$: $Z_m=1.559(0.003)_{\rm stat}(0.047)_{\rm ren}$.)}
\label{tab:fit_separate24c}
\end{table}

\begin{figure}
\begin{center}
\includegraphics[angle=-90,width=0.48\textwidth]{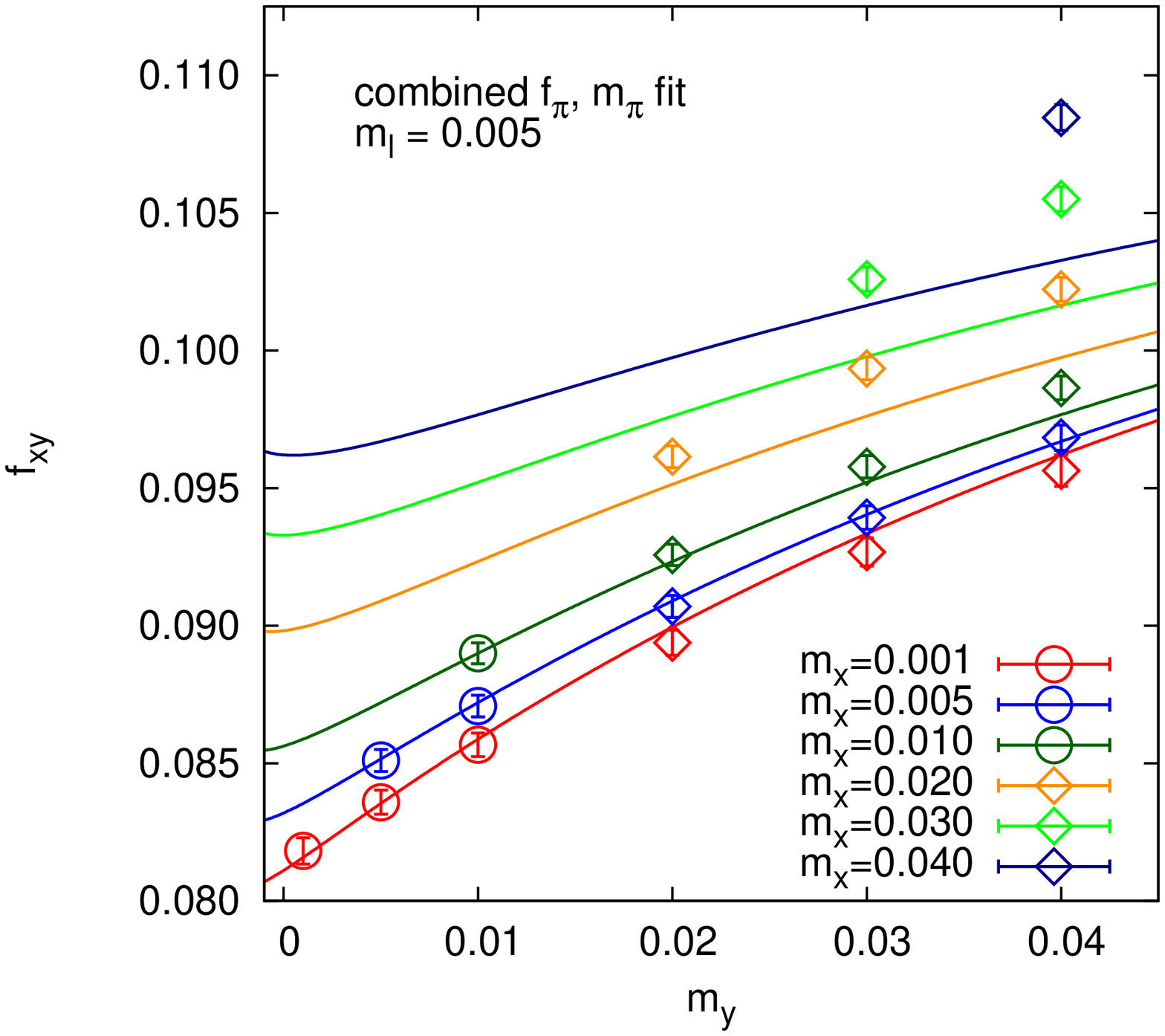}%
\vspace{.05\textwidth}%
\includegraphics[angle=-90,width=0.48\textwidth]{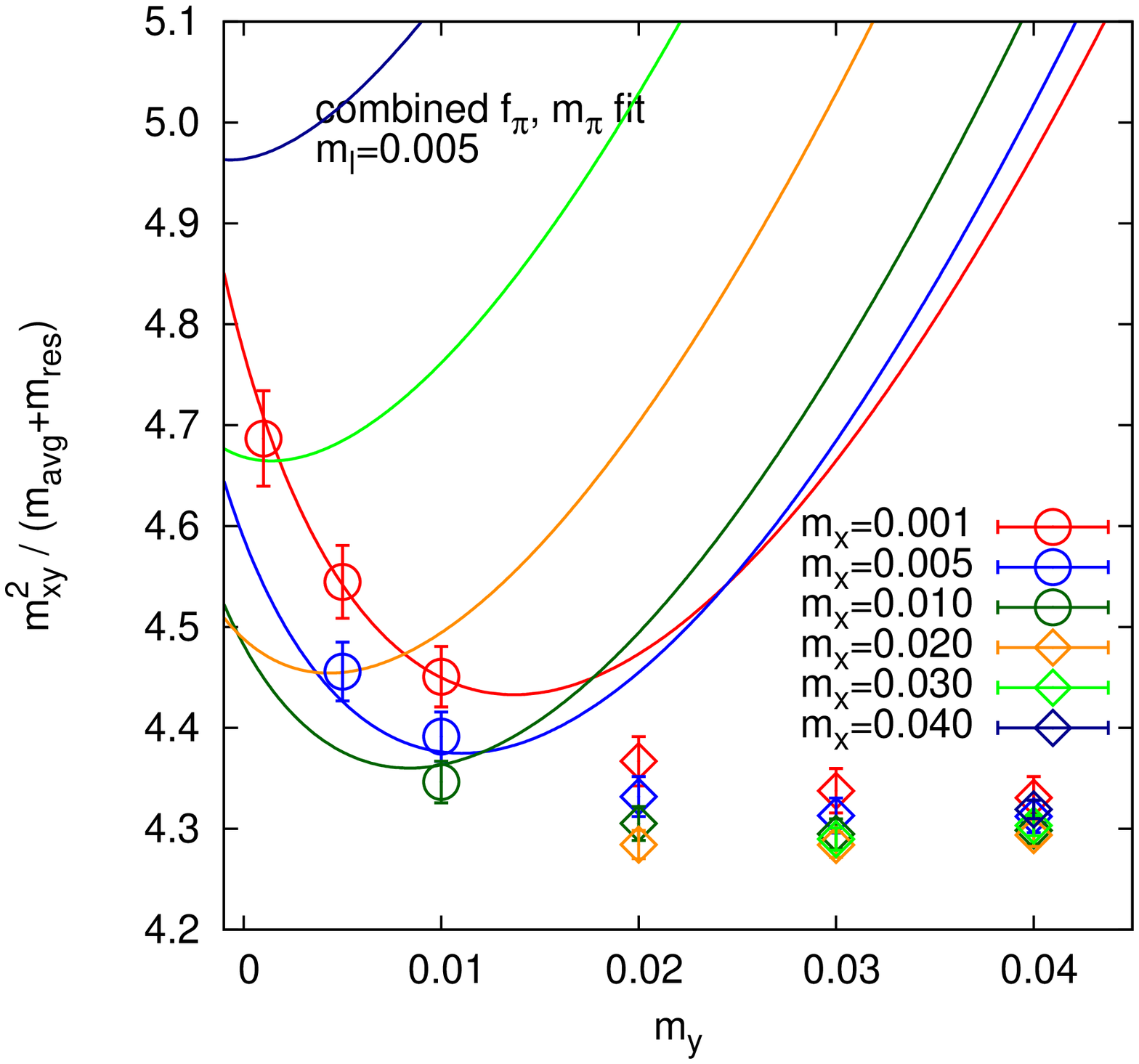}\\
\includegraphics[angle=-90,width=0.48\textwidth]{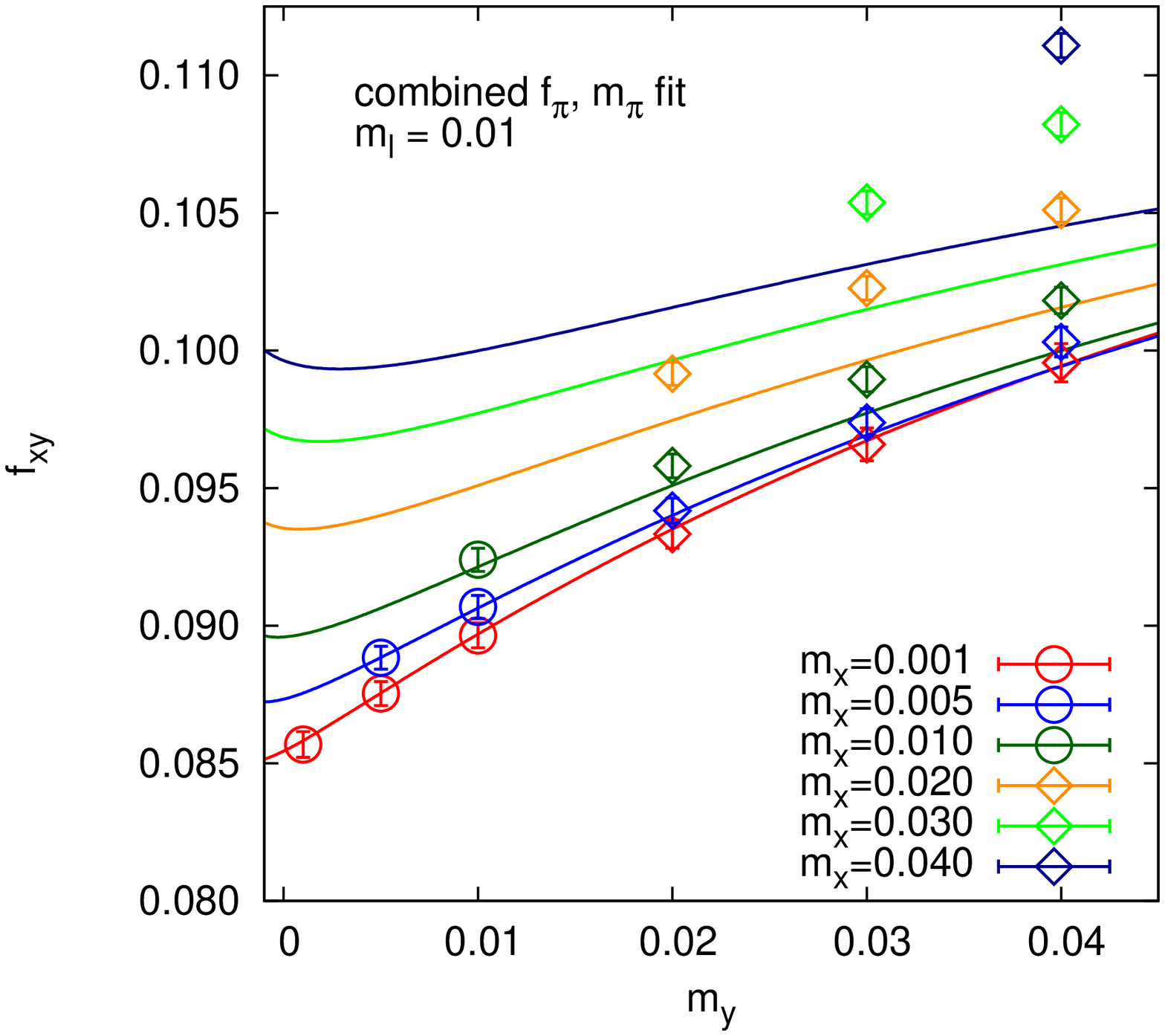}%
\vspace{.05\textwidth}%
\includegraphics[angle=-90,width=0.48\textwidth]{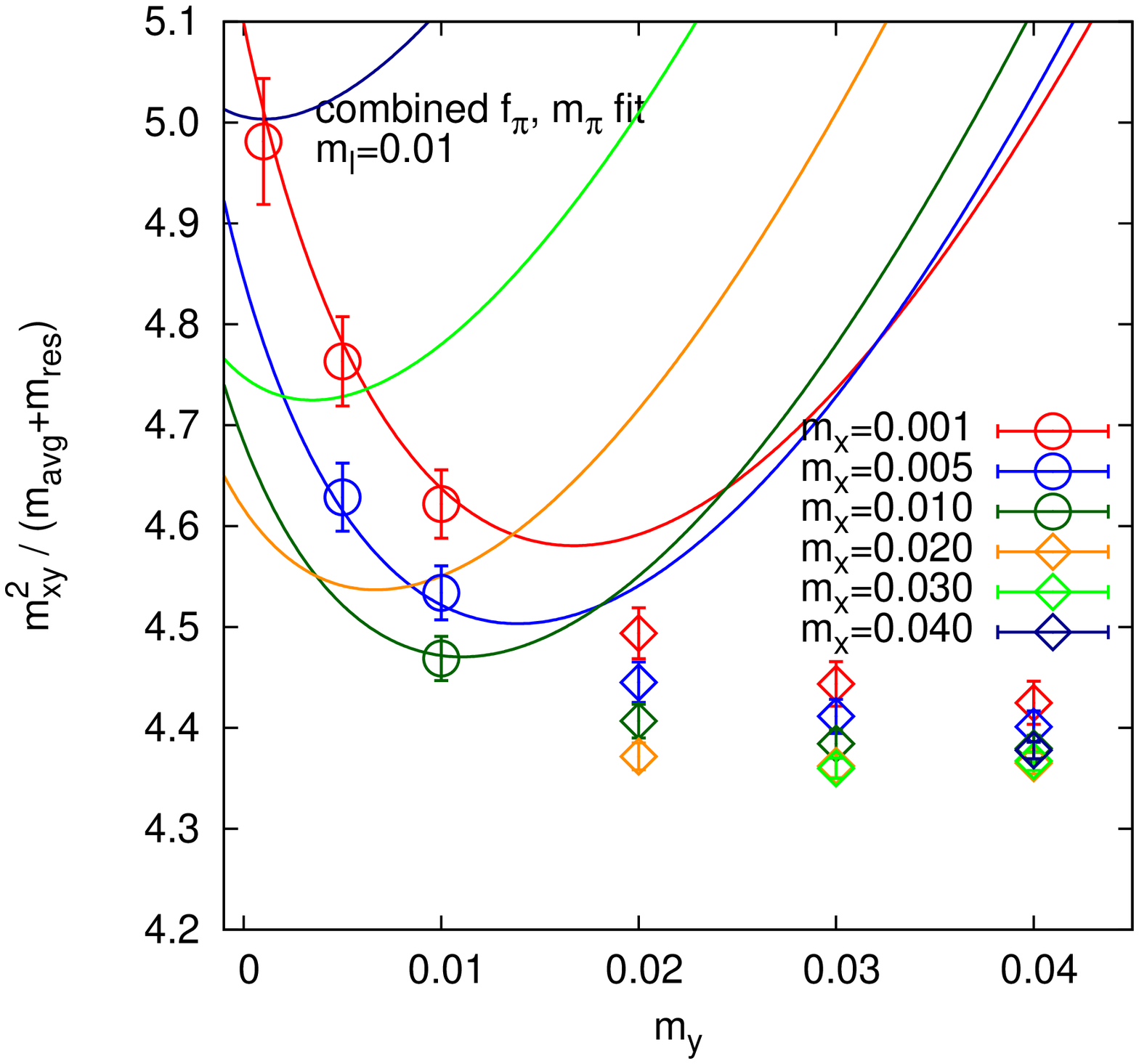}
\end{center}
\caption{Combined $\SU(2)$ ChPT fits (without finite-volume corrections) for the meson decay constants \textit{(left column)} and masses \textit{(right column)} on the $24^3$ data set at $m_l=0.005$ \textit{(top row)} and 0.01 \textit{(bottom row)}. Only points marked with \textit{circles}, corresponding to the range $(m_x+m_y)/2\leq0.01$ are included in the fits.}
\label{fig:fit_separate24c}
\end{figure}

We note that by comparing the results in the first two columns of Tab.~\ref{tab:fit_separate24c}, which have been obtained using the same (large) mass-range for the chiral extrapolation of the $\Omega$-baryon mass, the results obtained with the increased statistics (for each dynamical light-quark mass the statistics has nearly been doubled, see Section~\ref{sec:24cubed}) nicely agree with those from our previous publication~\cite{Allton:2008pn} within the statistical uncertainty. Furthermore, we observe the expected reduction in the statistical error. For the remainder of the discussion, we focus on the fits in which only the two lightest dynamical masses have been included in the extrapolation of the $\Omega$-baryon mass, i.e.\ the last two columns of Tab.~\ref{tab:fit_separate24c}. The major difference resulting from this change in the fit range is in the value of the lattice scale $1/a$, but within 1.4 standard deviations (statistical error only, taking into account correlations) the results still show agreement.  Including the finite-volume effects results in higher values for the decay constants (both in the chiral limit and at the physical point), which is a statistically significant effects (taking the correlations into account). In Tab.~\ref{tab:decay_separateVSglobal} we compare the decay constants and their ratio obtained from the separate fits with the corresponding results from the global analysis at the simulated, finite value of the lattice spacing (i.e.\ not extrapolated to the continuum, see Sec.~\ref{sec:CombinedChiralFits} and especially Tabs.~\ref{tab:fpi:extrapolated}, \ref{tab:fk:extrapolated}, \ref{tab:fkoverfpi:extrapolated} but note the difference due to the use of our previous definition of $Z_A$ here). We are reassured by the observed agreement between the results obtained using the global fits with those obtained using our previous strategy in Ref.\,\cite{Allton:2008pn} which was developed at that time to describe data at only a single lattice spacing.

\begin{table}
\begin{center}
\begin{tabular}{lll|ccc}
\hline\hline%
&       &          & $f_\pi$ $[{\rm MeV}]$ & $f_K$ $[{\rm MeV}]$ & $f_K/f_\pi$ \\\hline
no FV-corr.& $24^3$, $\beta=2.13$ & separate & 124.4(3.6)           & 151.0(3.7)          & 1.214(0.012) \\
           &                      & global   & 123(2)               & 150(2)              & 1.215(0.009) \\ \hhline{~*{5}{-}}
           & $32^3$, $\beta=2.25$ & separate & 120.4(1.9)           & 147.1(2.0)          & 1.222(0.007) \\
           &                      & global   & 121(2)               & 147(2)              & 1.222(0.006) \\\hline
incl.\ FV-corr. & $24^3$, $\beta=2.13$ & separate & 126.4(3.6)           & 152.1(3.7)          & 1.204(0.012) \\
                &                      & global   & 126(2)               & 151(2)              & 1.204(0.009) \\ \hhline{~*{5}{-}}
                & $32^3$, $\beta=2.25$ & separate & 122.3(1.9)           & 148.1(2.0)          & 1.212(0.007) \\
                &                      & global   & 123(2)               & 149(2)              & 1.210(0.006) \\ \hline\hline
\end{tabular}
\end{center}
\caption{Comparison of the pion and kaon decay constants and their ratios at finite lattice spacing from separate (see Tabs.~\ref{tab:fit_separate24c}, \ref{tab:fit_separate32c}) and global fits using our previous definition of $Z_A$.} 
\label{tab:decay_separateVSglobal}
\end{table}

\subsection{\label{subsec:appendix_separate_fits:32c}\boldmath SU(2)-ChPT fits to $32^3$ data\unboldmath}

The results of a separate fit on the $32^3$ data set are summarized in Tab.~\ref{tab:fit_separate32c}. Here we only included the $\Omega$-baryon masses from the $m_l=0.004$, 0.006, and 0.008 ensembles. In Fig.~\ref{fig:fit_separate32c} we show the fits for the meson masses and decay constants in the pion sector (without finite-volume corrections). Again, over the fit range ($(m_x+m_y)/2\leq0.008$), corresponding to a maximum pion mass of about 400~MeV, the data is well described by $\SU(2)$ ChPT.

\begin{table}
\begin{center}
\begin{tabular}{c|cc}
\hline\hline%
 & no FV-corr. & FV-corr. incl.\\\hline
$1/a$ $[{\rm GeV}]$ & 2.221(29) & 2.221(29) \\[10pt]
$B^{\overline{\rm MS}}(2\,{\rm GeV})$ $[{\rm GeV}]$ & $2.62(0.05)(0.06)_{\rm ren}$ & $2.57(0.05)(0.06)_{\rm ren}$ \\
$f$ $[{\rm MeV}]$ &  111.4(2.2) & 113.7(2.2) \\
$\bar{l}_3$       &  2.84(0.21) & 2.61(0.24)\\
$\bar{l}_4$       &  4.18(0.09) & 4.10(0.09)\\[10pt]
$f_\pi$ $[{\rm MeV}]$ & 120.4(1.9) & 122.3(1.9) \\
$f_K$ $[{\rm MeV}]$   & 147.1(2.0) & 148.1(2.0) \\
$f_K/f_\pi$            & 1.222(0.007) & 1.212(0.007) \\[10pt]
$m^{\overline{\rm MS}}_{ud}(2\,{\rm GeV})$ $[{\rm MeV}]$ & $3.58(0.07)(0.08)_{\rm ren}$ & $3.64(0.07)(0.08)_{\rm ren}$ \\
$m^{\overline{\rm MS}}_{s}(2\,{\rm GeV})$ $[{\rm MeV}]$ & $100.6(1.7)(2.2)_{\rm ren}$ & $100.4(1.7)(2.2)_{\rm ren}$\\
$\tilde{m}_{ud}:\tilde{m}_s$ & 1:28.08(0.19) & 1:27.60(0.20) \\[10pt]
$aB$ & 1.826(0.024) & 1.790(0.025)\\
$af$ & 0.0502(0.0007) & 0.0512(0.0007)\\
$L_4^{(2)} \times 10^{4}$ & -0.75(0.79) & -1.21(.82) \\
$L_5^{(2)} \times 10^{4}$ & 5.14(0.40) & 4.87(0.41) \\
$(2L_6^{(2)}-L_4^{(2)}) \times 10^{4}$ & -0.93(0.42) & -1.03(0.45) \\
$(2L_8^{(2)}-L_5^{(2)}) \times 10^{4}$ & 6.22(0.23) & 7.37(0.24) \\
$a\tilde{m}_{ud}$ & 0.001040(31)& 0.001057(32) \\
$a\tilde{m}_s$ & 0.0292(08) & 0.0292(08) \\\hline\hline
\end{tabular}
\end{center}
\caption{Results from the $\SU(2)$ ChPT fits to the $32^3$ data (without and with finite-volume corrections). We also quote in the lower part of the table the $\SU(2)$ ChPT fit parameters $aB$, $af$, $L_i^{(2)}$ (at the scale $\Lambda_\chi=1\,{\rm GeV}$) and quark masses $a\tilde{m}_{ud,s}$ in lattice units. Only statistical uncertainties are quoted except for quark masses and the LEC $B$ renormalized in the $\overline{\rm MS}$-scheme at 2\,GeV where also the systematic uncertainty from the renormalization constant is quoted. (Mass renormalization constant at $1/a=2.221(29)\,{\rm GeV}$: $Z_m=1.550(0.002)_{\rm stat}(0.034)_{\rm ren}$.)}
\label{tab:fit_separate32c}
\end{table}

\begin{figure}
\begin{center}
\includegraphics[angle=-90,width=0.48\textwidth]{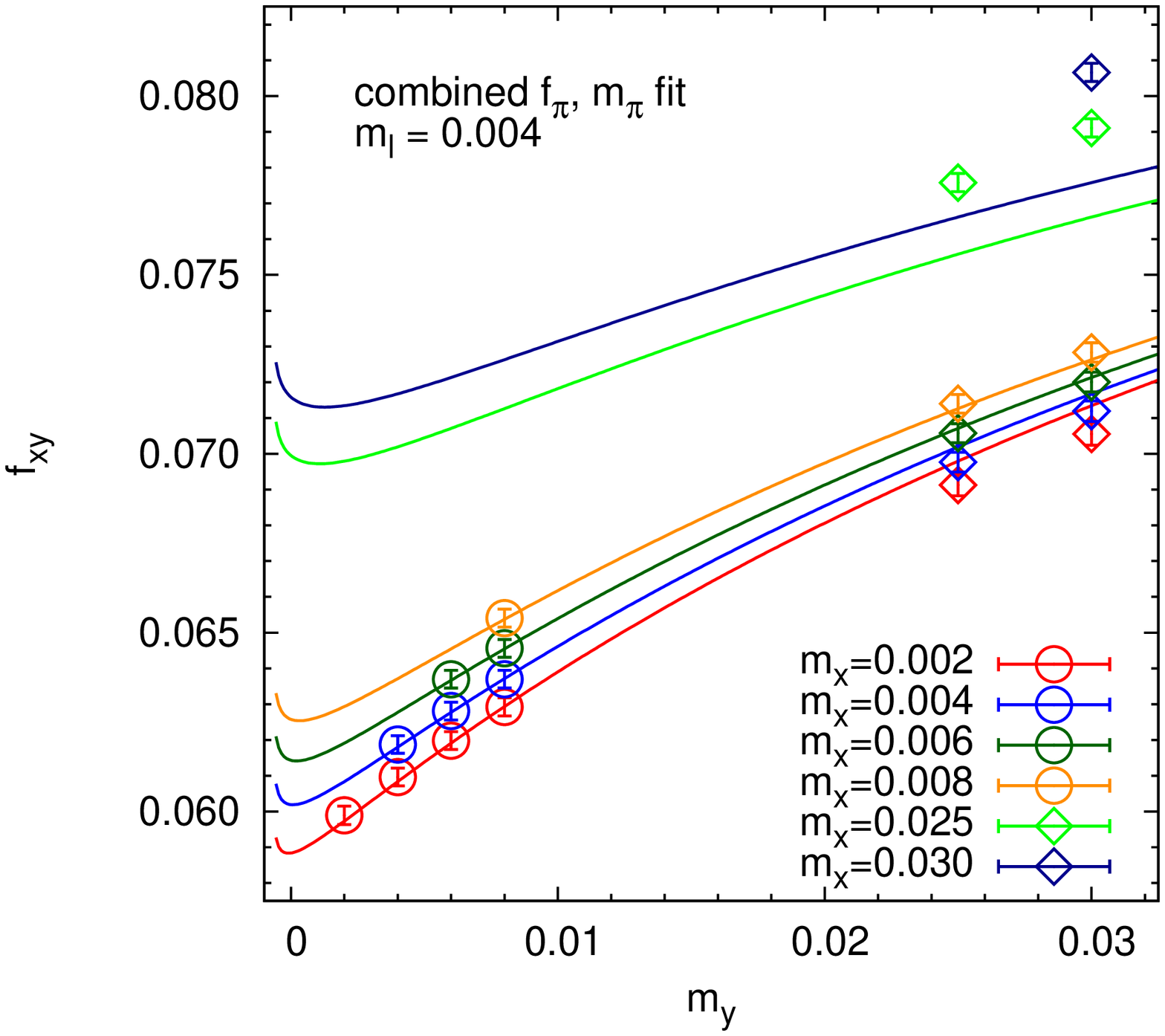}%
\vspace{.05\textwidth}%
\includegraphics[angle=-90,width=0.48\textwidth]{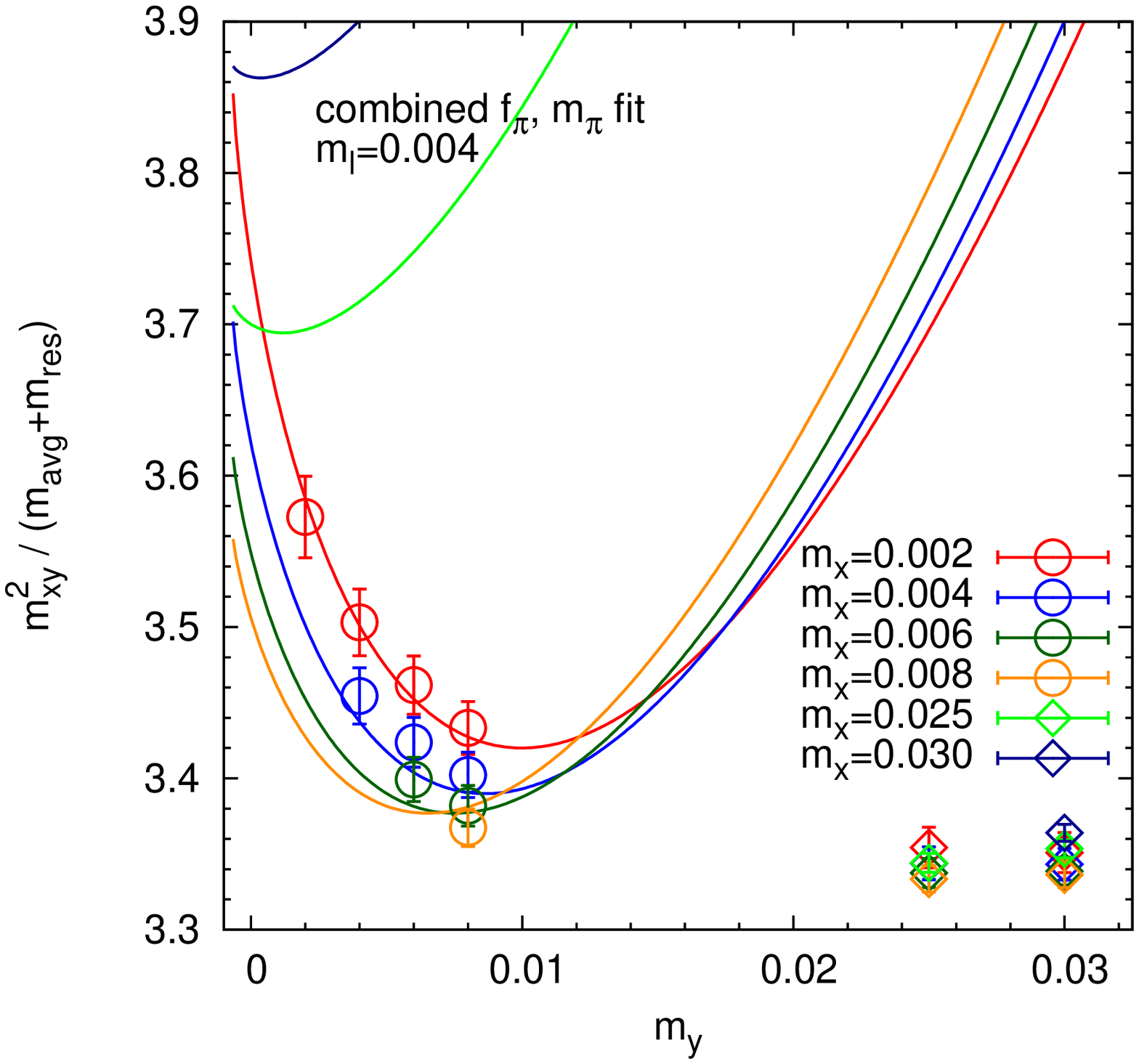}\\
\includegraphics[angle=-90,width=0.48\textwidth]{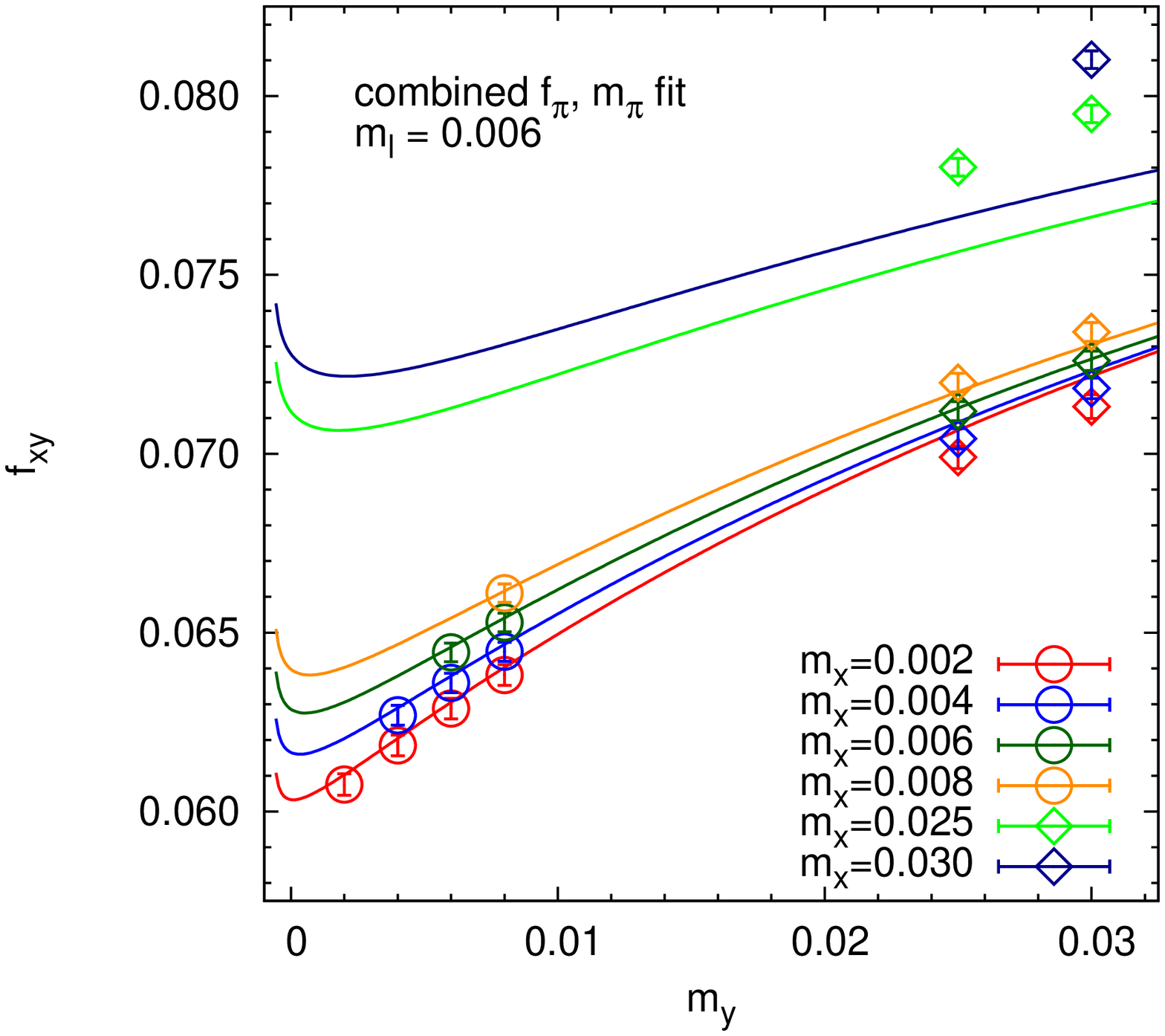}%
\vspace{.05\textwidth}%
\includegraphics[angle=-90,width=0.48\textwidth]{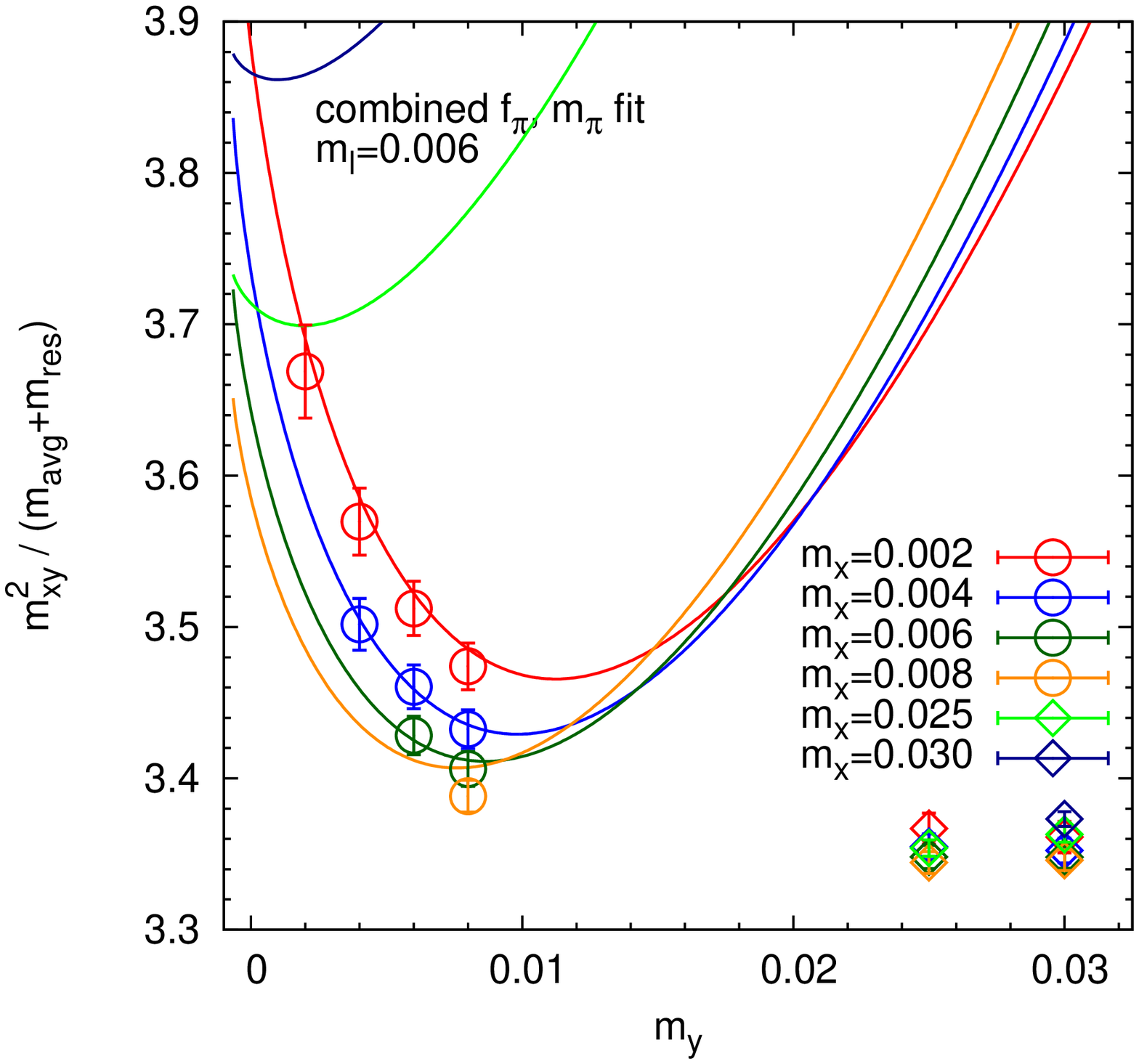}\\
\includegraphics[angle=-90,width=0.48\textwidth]{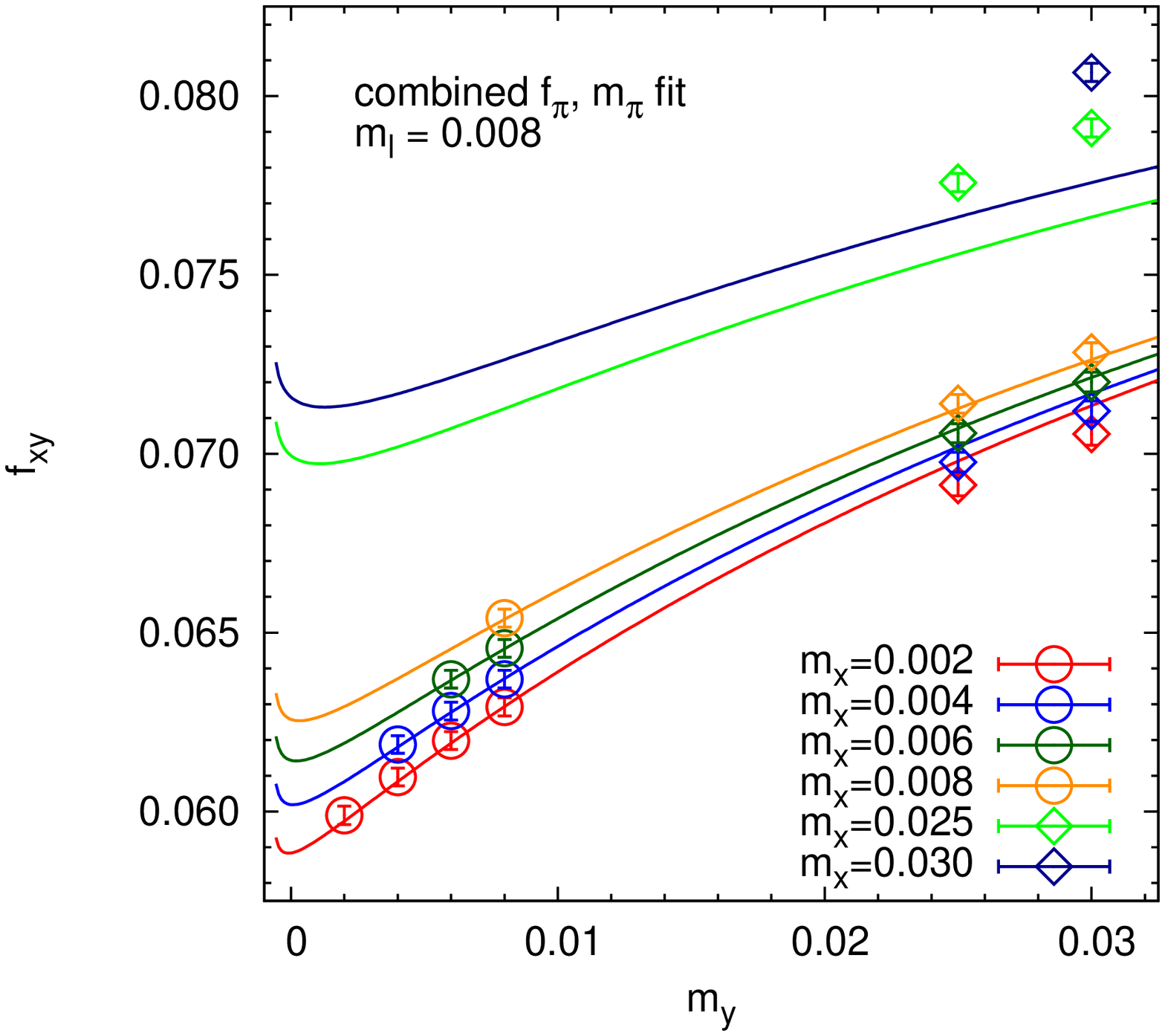}%
\vspace{.05\textwidth}%
\includegraphics[angle=-90,width=0.48\textwidth]{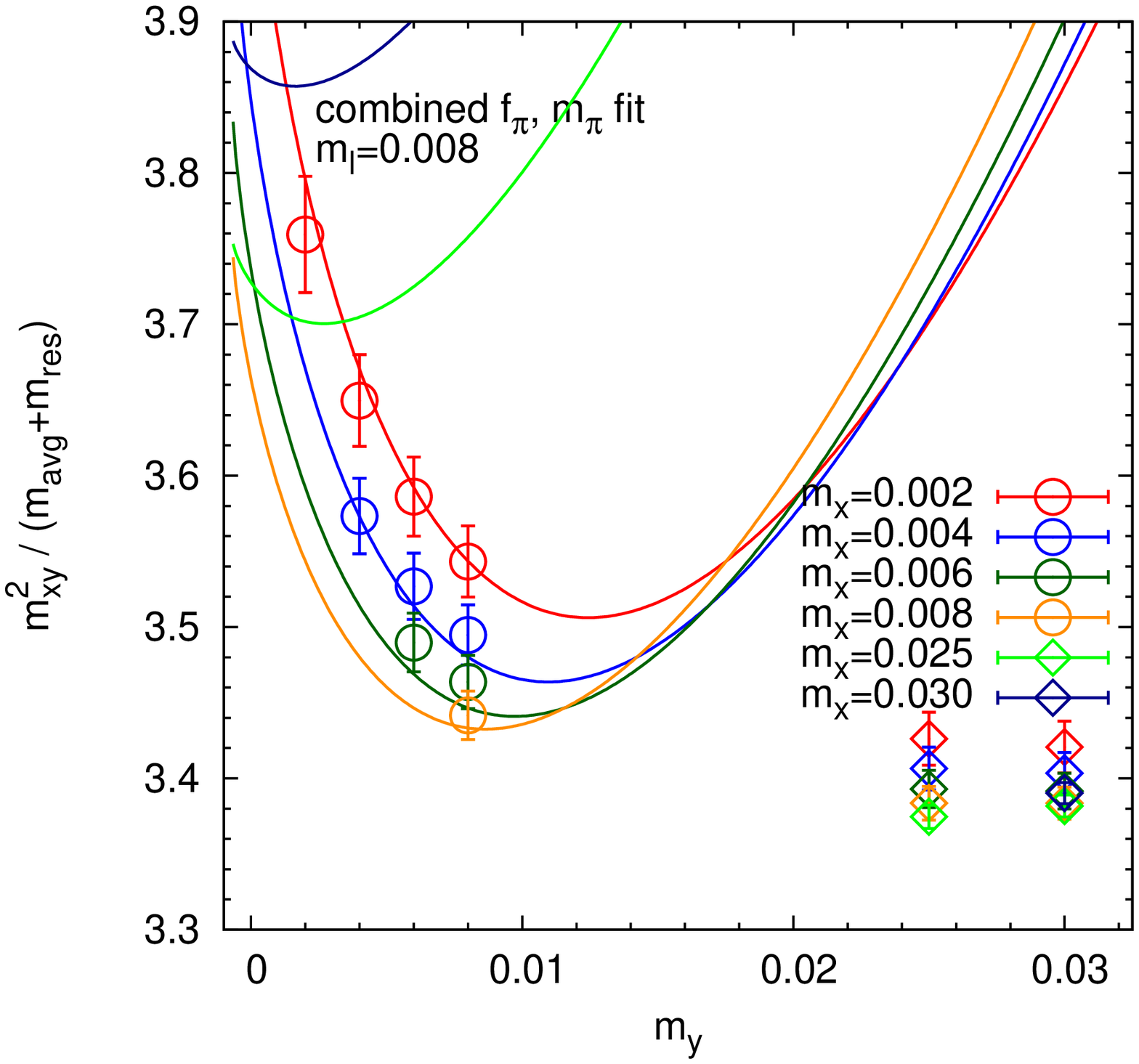}
\end{center}
\caption{Combined $\SU(2)$ ChPT fits (without finite-volume corrections) for the meson decay constants \textit{(left column)} and masses \textit{(right column)} on the $32^3$ data set at $m_l=0.004$ \textit{(top row)}, 0.006 \textit{(middle row)}, and 0.008 \textit{(bottom row)}. Only points marked with \textit{circles},
corresponding to the range $(m_x+m_y)/2\leq0.008$ are included in the fits.}
\label{fig:fit_separate32c}
\end{figure}

As was already the case for the $24^3$ ensembles, taking finite-volume corrections into account also leads to a good description of the data and results in higher values for the decay constants at the physical point and in the chiral limit. Again, taking the correlations into account, we note that this is a statistically significant effect. As was also the case on the $24^3$ ensembles, we observe a good agreement for the decay constants and their ratio between the results of the separate fits to the $32^3$ data and the results from the global fits at finite lattice spacing, see Tab.~\ref{tab:decay_separateVSglobal}.

\subsection{\label{subsec:appendix_separate_fits:naiveCL}\boldmath Extrapolation to the Continuum Limit \unboldmath}

With the results obtained from separate chiral extrapolations on the $24^3$ (extended statistics) and the $32^3$ data sets (see the two previous subsections, respectively) we can perform a na\"ive continuum limit extrapolation assuming $a^2$-scaling. Of course, with only two lattice spacings available, we are not able to confirm this scaling behaviour. Further caveats include the fact that here, for simplicity, we did not use reweighting and so the dynamical strange-quark mass is not tuned to exactly the same value on the two data sets and indeed is not exactly the physical one on either set. Also, the dynamical light-quark mass ranges are a little different at the two lattice spacings, corresponding to unitary pion masses in the range 330--420\,MeV on the coarser $24^3$ lattices and 290--400 MeV on the finer $32^3$ lattices (a similar statement is true for the partially-quenched masses).
One might therefore expect a larger uncertainty in the chiral extrapolation of the $24^3$ results. In the na\"ive continuum ansatz followed here, we are not taking into account this effect. Because of this, and maybe more importantly, since two separate chiral extrapolations have been performed (one at each of the two values of the lattice spacing), the continuum extrapolation is not completely disentangled from the chiral extrapolation. Recall that in our procedure for the global fits described in the main part of this paper, these two extrapolations are indeed disentangled. There this is achieved by adding $O(a^2)$ terms into the two functions, such that the chiral and continuum extrapolations are performed simultaneously and independently from each other.

\begin{table}
\begin{center}
\begin{tabular}{c|cc|c||c} 
\hline\hline
 & \multicolumn{4}{c}{no FV-corr.} \\
 & \multicolumn{2}{c|}{separate fits} & na\"ive CL &  comb.\ chiral/CL \\
 & $24^3$, $\beta=2.13$ & $32^3$, $\beta=2.25$ & & \\\hline
$a$ $[{\rm fm}]$      & 0.1106(27) & 0.0888(12) & $\to0$      & $\to0$  \\[10pt] 
$f$ $[{\rm MeV}]$     & 114.8(4.0) & 111.4(2.2) & 105.2(10.4) & 107(2) \\
$\bar{l}_3$           & 2.82(0.24) & 2.84(0.21) & 2.87(0.74)  & 2.81(0.16) \\
$\bar{l}_4$           & 4.61(0.10) & 4.18(0.09) & 3.39(0.36)  & 3.76(0.08) \\[10pt] 
$f_\pi$ $[{\rm MeV}]$ & 124.4(3.6) & 120.4(1.9) & 113.0(9.5)  & 117(2) \\
$f_K$   $[{\rm MeV}]$ & 151.0(3.7) & 147.1(2.0) & 139.9(9.6)  & 144(2) \\
$f_K/f_\pi$           & 1.214(0.012) & 1.222(0.007) & 1.236(0.030) & 1.233(0.008) \\\hline\hline
& \multicolumn{4}{c}{including FV-corr.}  \\
 & \multicolumn{2}{c|}{separate fits} & na\"ive CL &  comb.\ chiral/CL \\
& $24^3$, $\beta=2.13$ & $32^3$, $\beta=2.25$ & & \\\hline
$a$ $[{\rm fm}]$      &0.1106(27) & 0.0888(12) & $\to 0$ & $\to0$ \\[10pt]
$f$ $[{\rm MeV}]$     & 117.1(4.0) & 113.7(2.2) & 107.4(10.3) & 110(2)  \\
$\bar{l}_3$           & 2.59(0.27) & 2.61(0.24) & 2.64(0.83) & 2.55(0.18) \\
$\bar{l}_4$           & 4.57(0.11) & 4.10(0.09) & 3.26(0.38) & 3.83(0.09) \\[10pt]
$f_\pi$ $[{\rm MeV}]$ & 126.4(3.6) & 122.3(1.9) & 114.8(9.4) & 119(2) \\
$f_K$   $[{\rm MeV}]$ & 152.1(3.7) & 148.1(2.0) & 140.9(9.6) & 145(2) \\
$f_K/f_\pi$           & 1.204(0.012) & 1.212(0.007) & 1.226(0.029) & 1.219(0.007) \\\hline\hline
\end{tabular}
\end{center}
\caption{Selected results from separate fits to the $24^3$ and $32^3$ data sets ($\Omega$ masses from $m_l\leq0.1$ for $24^3$ data set, cf.\ Tabs.~\ref{tab:fit_separate24c} and \ref{tab:fit_separate32c}) and their na\"ive continuum limit assuming $a^2$-scaling (see Fig.~\ref{fig:naiveCL}) compared to results from the combined chiral-continuum extrapolation using the previous definition of $Z_A$. The top table contains results without finite-volume corrections whereas the results in the bottom table were obtained by including finite-volume effects.}
\label{tab:naiveCL}
\end{table}

\begin{figure}
\begin{center}
\includegraphics[angle=-90,width=.4\textwidth]{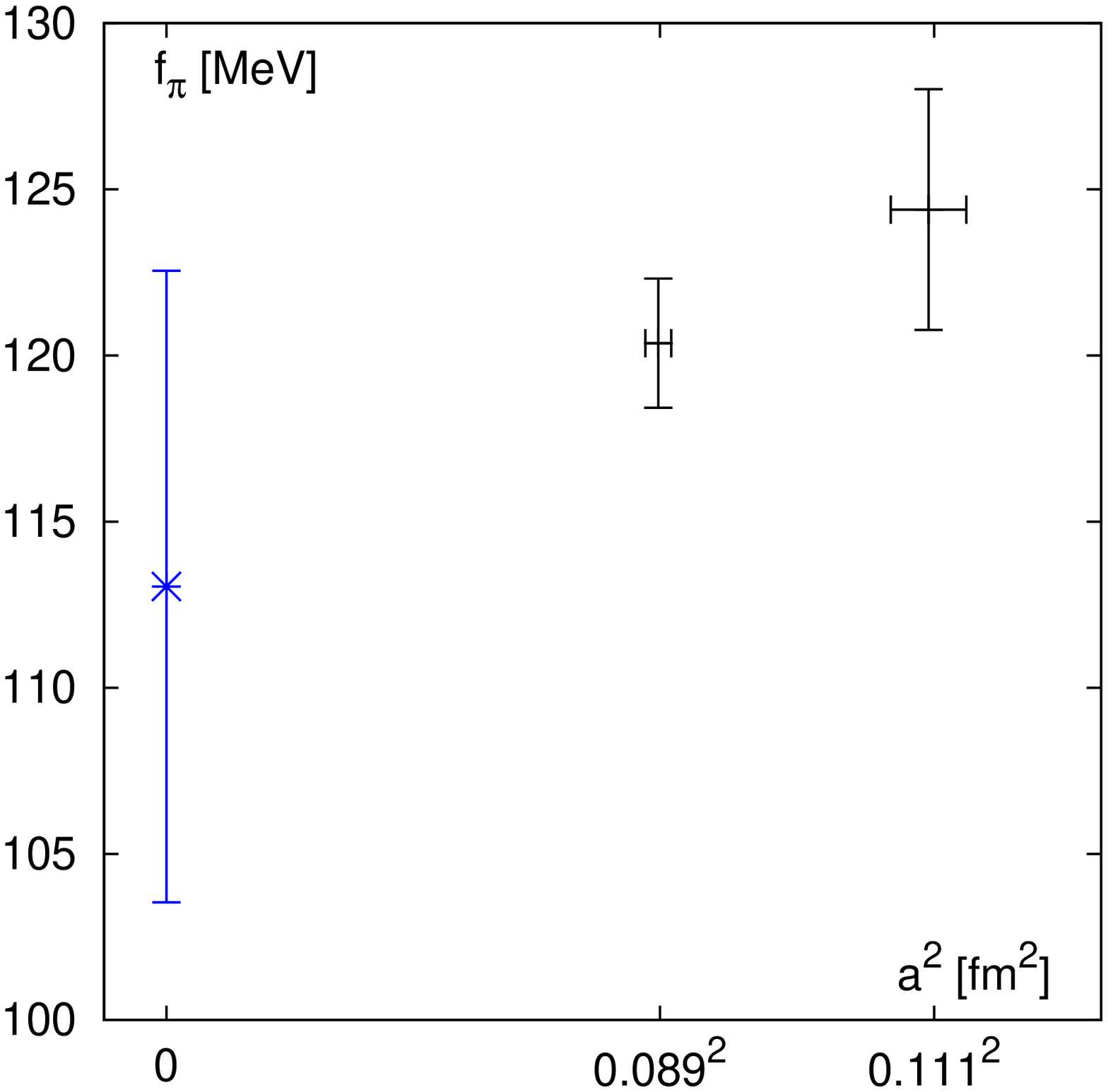}%
\hspace*{.05\textwidth}%
\includegraphics[angle=-90,width=.4\textwidth]{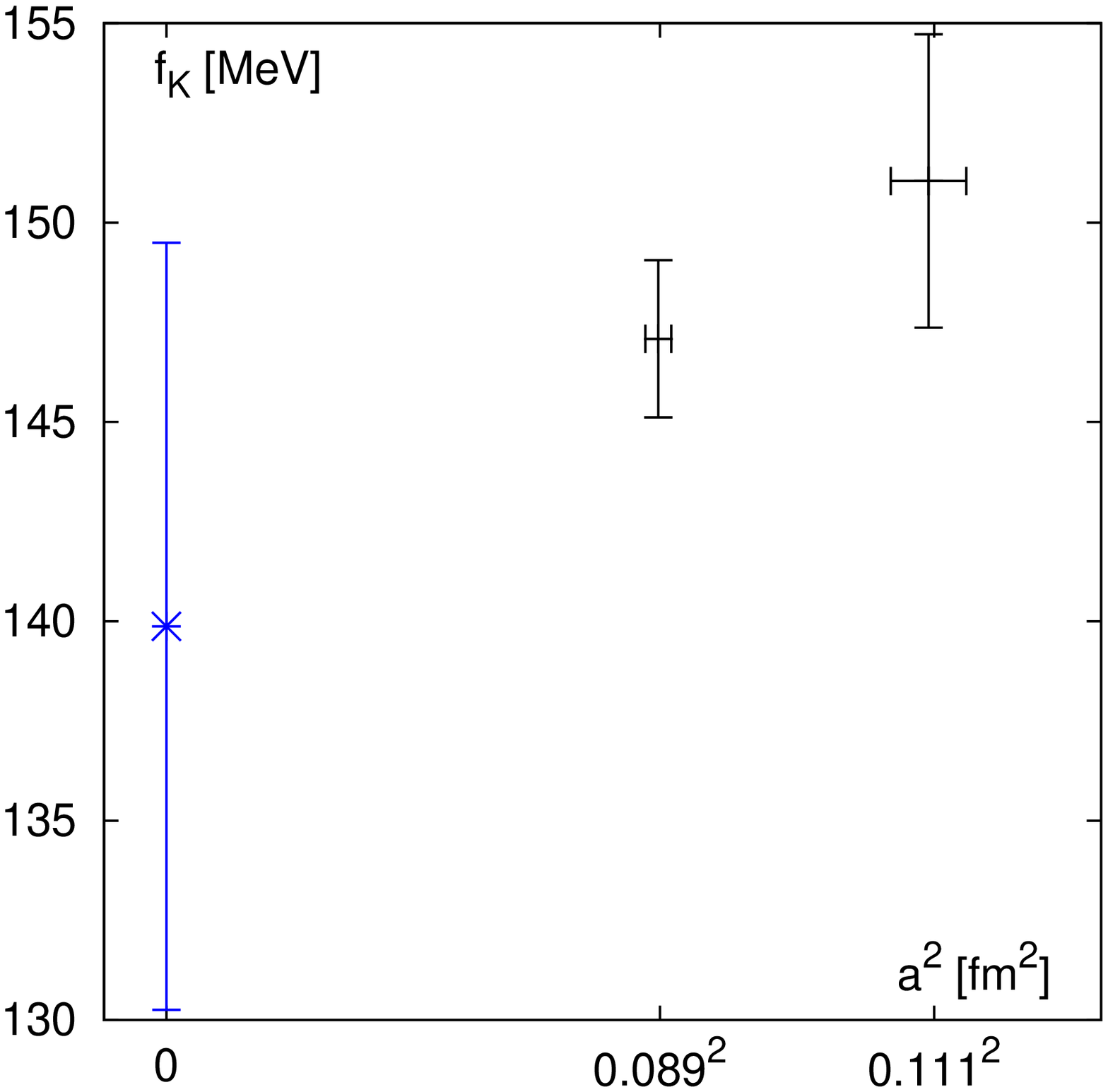}\\
\includegraphics[angle=-90,width=.4\textwidth]{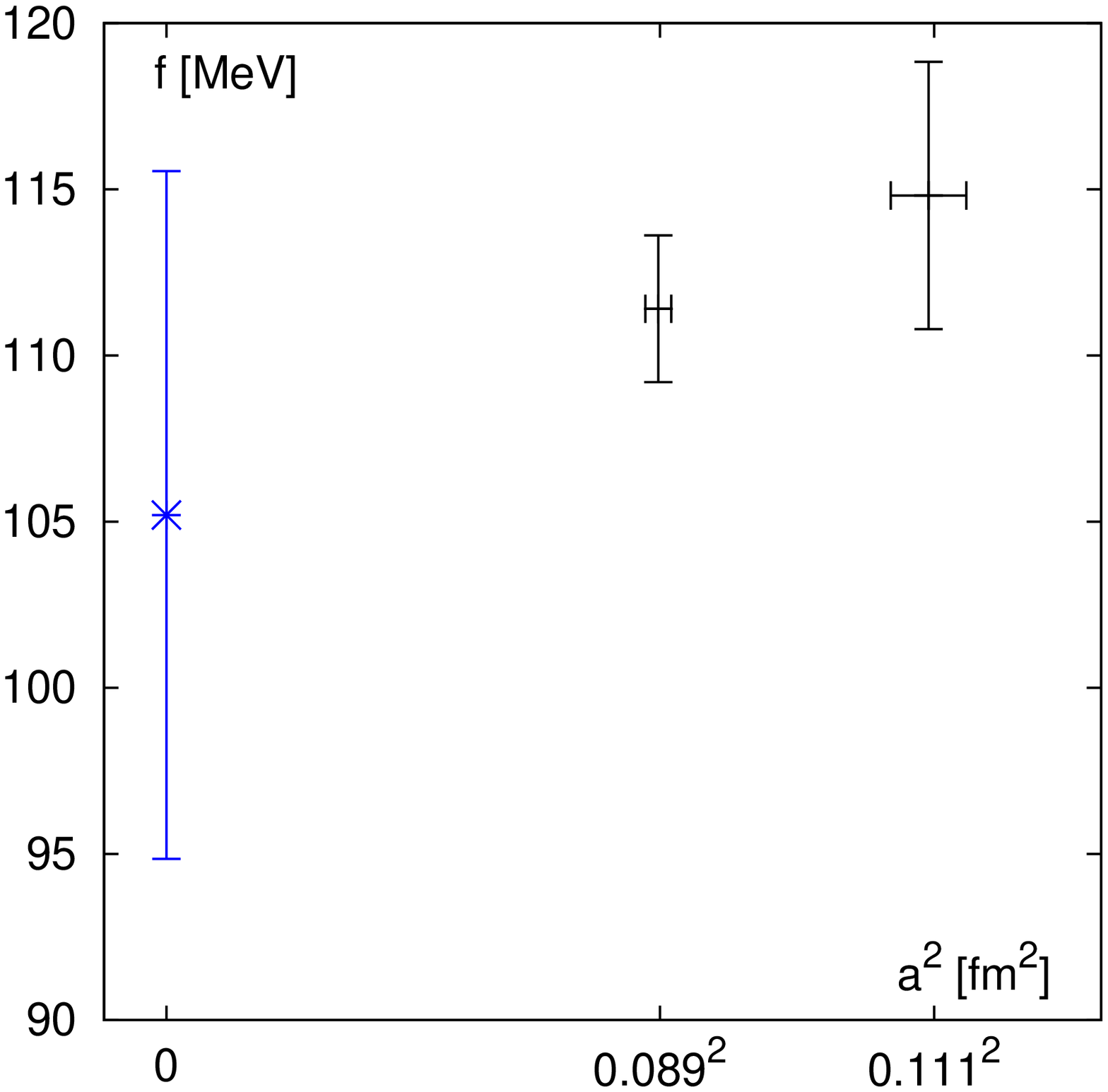}%
\hspace*{.05\textwidth}%
\includegraphics[angle=-90,width=.4\textwidth]{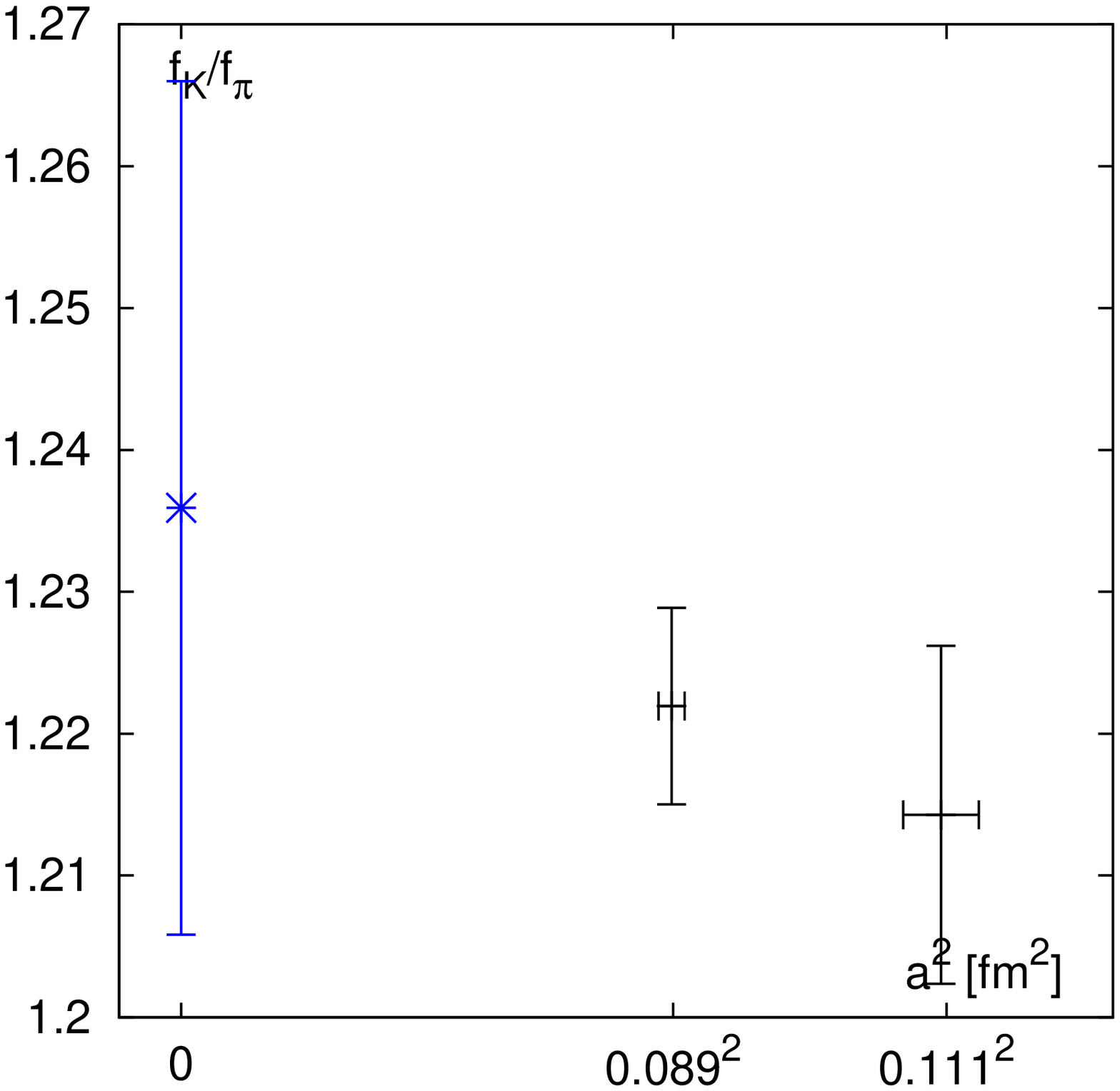}\\
\includegraphics[angle=-90,width=.4\textwidth]{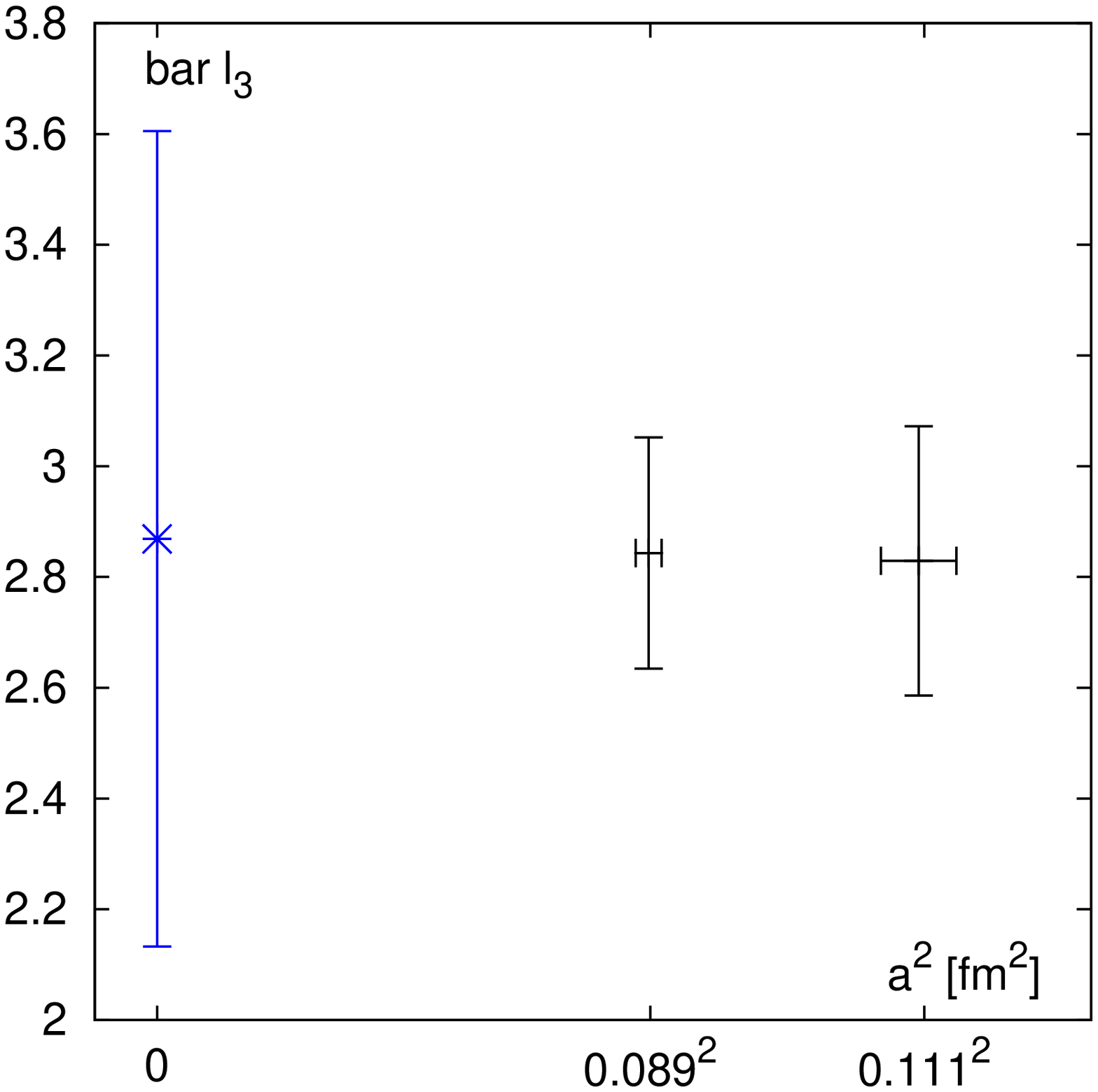}
\hspace*{.05\textwidth}
\includegraphics[angle=-90,width=.4\textwidth]{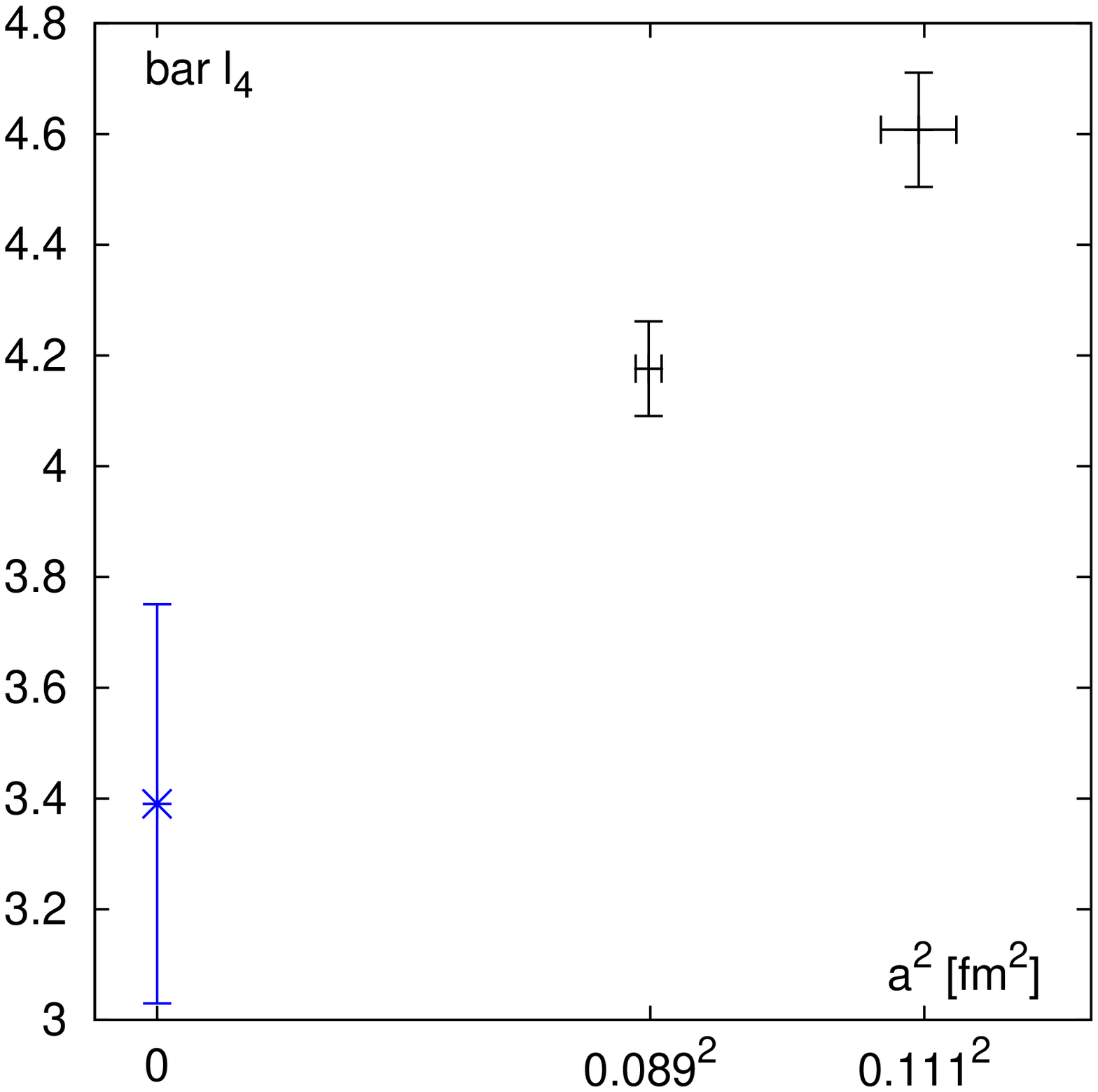}%
\end{center}
\caption{Results from separate fits (without finite-volume corrections) to the $24^3$ and $32^3$ data sets \textit{(black points)} and the na\"ive continuum-limit extrapolation \textit{(blue asterisks)} for selected quantities assuming $a^2$-scaling. For details see Subsec.~\ref{subsec:appendix_separate_fits:naiveCL} and Tab.~\ref{tab:naiveCL}.}
\label{fig:naiveCL}
\end{figure}

In Tab.~\ref{tab:naiveCL} we repeat the results obtained at the two different lattice spacings (with and without finite-volume corrections) and give the values extrapolated to the continuum limit assuming $a^2$ scaling. Fig.~\ref{fig:naiveCL} illustrates the continuum extrapolation of the various quantities (only results obtained without taking into account finite-volume corrections are shown there). Note, that the two points at the different lattice spacings are completely uncorrelated, the only correlation in the data for the continuum extrapolation is between the uncertainty in the lattice spacing (the ``x''-datum) and the quantity itself at that lattice spacing (the ``y''-datum). These correlations were treated by the super-jackknife method which we have been using in our earlier work and which is clearly explained in ~\cite{DelDebbio:2007pz,Bratt:2010jn}. For comparison, Tab.~\ref{tab:naiveCL} also contains our results from the combined continuum-chiral extrapolation as described in the main part of this paper but here using our previous definition of $Z_A$. As one can see, the combined continuum-chiral extrapolation gives a substantially smaller (up to a factor of 5) statistical uncertainty compared to the na\"ive continuum extrapolation. The main reason, of course, is the correlation in the combined fits between the two data sets at different lattice spacings. This correlation occurs because we require the fitted parameters to be the same on both data sets and only include $O(a^2)$ corrections for the leading-order terms, as is consistent with our power counting scheme. In this way, the continuum extrapolation in the combined fits is also more constrained, leading to a smaller statistical uncertainty. Comparing the results of the na\"ive continuum extrapolation and the combined continuum-chiral extrapolation for the quantities in Tab.~\ref{tab:naiveCL} we observe agreement better than 0.5-$\sigma$ (taking into account correlations) for all quantities except for $\bar{l}_4$, where the agreement still holds at the 1- or 1.5-$\sigma$ level (without and with taking FV-corrections into account, respectively). It is reassuring, that the results from the two methods agree well, although the value of this statement is limited, given the large (statistical) uncertainty of almost 10\% for the decay constants or even more in case of the LECs from the na\"ive method. However, it should be noted that the same agreement holds, not only for the continuum values, but also for the results obtained in the separate fits as compared to the predictions of the global fit made for the finite lattice spacings. This has already been discussed in the previous subsections and is shown in Tab.~\ref{tab:decay_separateVSglobal}.

\fi


\section{Determining $Z_{\cal A}$}\label{sec:appendix:za}

As pointed out by Sharpe~\cite{Sharpe:2007yd} and refined in
Ref.~\cite{Allton:2008pn}, the normalization of the partially
conserved axial current defined for domain wall fermions~\cite{Furman:1995ky}
is expected to deviate from that of the conventionally normalized
continuum current by an amount of order $m_{\rm res}a$.   Here and
below when making such estimates we will introduce the explicit lattice
spacing $a$ and express the residual mass in physical units in order
to make the comparison of various terms in a Symanzik expansion in
powers of $a$ easier to recognize.  Since such a deviation can be
viewed as $O(m a)$ which is formally larger than the $O(m a^2)$ which
we neglect in our power counting scheme and because the normalization
of this axial current plays a central role in our determination of
the important quantities $f_\pi$ and $f_K$, we have calculated this
normalization factor $Z_{\cal A}$ numerically.  We explain our method
and result in this appendix.  The first subsection contains a discussion
of the theoretical issues and explains the basis for our method of
determining $Z_{\cal A}$.  The second subsection describes the
actual calculation and results.

\subsection{Determining the normalization of ${\cal A}_\mu$}

To determine the normalization of ${\cal A}_\mu$ we compare the matrix element of four distinct
domain wall fermion currents.  The first two are the conserved/partially
conserved vector and axial currents ${\cal V}^a_\mu(x)$ and
${\cal A}^a_\mu(x)$ respectively, where $a$ and $\mu$ are flavor and space-time
indices.  These currents were introduced by Furman and Shamir
\cite{Furman:1995ky} and involve fermion fields evaluated on each of
the $L_s$ 4-dimensional hyperplanes and at both the space-time
points $x$ and $x+\hat e_\mu$ where $\hat e_\mu$ is a unit vector
pointing the $\mu^{th}$ direction.  Thus, these currents are
local but distributed in the fifth dimension and one-link non-local
in space-time.  While this vector current is exactly conserved,
the divergence of the axial current contains the usual mass
term and a mid-point term $J_{5q}^a$.  In the long-distance limit
this midpoint term can be decomposed into the residual mass term,
a piece that is conveniently written as $(1-Z_{\cal A})$ times the
divergence of the same axial current and a final term of dimension
five which we write out explicitly as the sum of the
dimension-five, chiral rotation of the usual clover term and the
four-dimensional Laplacian applied to the pseudoscalar density:
\begin{equation}
J_{5q}^a = m_{\rm res} \overline{q}\gamma^5 \lambda^a q
           + \frac{1-Z_{\cal A}}{2}\Delta_\mu {\cal A}_\mu^a
           + c_1 \overline{q}\sigma^{\mu\nu} F^{\mu\nu}\lambda^a q
           + c_2 \partial_\mu \partial_\mu \overline{q}\gamma^5 \lambda^a q.
\label{eq:axial_divergence}
\end{equation}
In Equation\,(\ref{eq:axial_divergence}) $\lambda^a$ is the generator which acts on the fermion fields
corresponding to the flavor index $a$ while $q(x)$ and
$\overline{q}(x)$ are the ``physical'', four-dimensional quark fields
obtained by evaluating the five-dimensional domain wall fields on the
$s=0$ and $s=L_s-1$ boundaries.  (See Eqs. (11) and (12) in
Ref.~\cite{Allton:2008pn}.)

The second pair of currents which we will need in this appendix is
the local vector and axial currents, $V^a_\mu(x)$ and $A^a_\mu(x)$,
constructed in the standard way from the four-dimensional
quark fields, $q(x)$ and $\overline{q}(x)$.  These currents are localized
in all five dimensions and neither is conserved.

Finally it will also be convenient to introduce the scalar densities
$\overline q (x) q(x)$, $\overline q (x) \lambda^a q(x)$ from which
the domain fermion mass is constructed and their chiral transforms
$\overline q (x)\gamma^5 q(x)$, $\overline q (x) \lambda^a \gamma^5 q(x)$.
These four classes of operators
will be labeled $S(x)$, $S^a(x)$, $P(x)$ and $P^a(x)$.

Following Symanzik, we can add improvement terms to each of these
six operators to insure that their Green's functions, when evaluated
with an appropriately improved action, will agree with the corresponding
continuum Green's functions up to errors of order $a^n$.  For our
present purposes, accuracy up to $O(a m)$ where $m$ is a quark mass
in physical units, will be sufficient.  Since $m_{\rm res}$ and $m$
have a similar size, we are explicitly attempting to control the
$m_{\rm res} a$ corrections described above.  We do not attempt to
explicitly remove $O(a^2)$ terms since these will be eliminated by the
final linear extrapolation $a^2 \rightarrow 0$.

In the discussion to follow we will recognize constraints on the
required Symanzik improvement terms and relations between the various
renormalization constants by applying the approximate chiral symmetry
of domain wall fermions to Green's functions containing these various
operators.  For such arguments to be valid we will assume that these
Green's functions are evaluated at sufficiently small distances that
the effects of the vacuum chiral symmetry breaking of QCD can be ignored
but at sufficiently large distances that the Symanzik improvement program
can be applied.  Since this discussion is a theoretical one, constraining
the form of the Symanzik improvement terms, we need not be concerned
about practical questions regarding the degree to which such conditions
can be realized in our present calculation.

Using the notation $V^{{\rm S}a}_\mu$, $A^{{\rm S}a}_\mu$, $S^{{\rm S}a}$
and $P^{{\rm S}a}$ for the Symanzik-improved vector current, axial
current, scalar density and pseudoscalar density respectively,
keeping improvement terms which are nominally of order $a$ and imposing
charge conjugation symmetry, we find:
\begin{eqnarray}
V^{{\rm S}a}_\mu &=& Z_{\cal V} {\cal V}^a_\mu
       + C_{\cal V} \partial_\nu \overline{q}\sigma^{\mu\nu}\lambda^a q
\label{eq:V_c} \\
A^{{\rm S}a}_\mu &=& Z_{\cal A} {\cal A}^a_\mu + C_{\cal A} \partial_\mu P^a
\label{eq:A_pc} \\
V^{{\rm S}a}_\mu &=& Z_V V^a_\mu
       + C_V \partial_\nu \overline{q}\sigma^{\mu\nu}\lambda^a q
\label{eq:V_l}  \\
A^{{\rm S}a}_\mu &=& Z_A A^a_\mu + C_A \partial_\mu P^a
\label{eq:A_l}  \\
S^{{\rm S}a} &=& Z_S S^a
\label{eq:S}\\
P^{{\rm S}a} &=& Z_P P^a
\label{eq:P}.
\end{eqnarray}
In contrast to the Symanzik-improved current operators, we have
not specified a normalization convention for the operators
$S^{{\rm S}a}$ and $P^{{\rm S}a}$.  Adopting definitive conventions
for $S^{{\rm S}a}$ and $P^{{\rm S}a}$ is not needed here beyond
the requirement that those conventions are consistent with
$S^{{\rm S}a} \pm P^{{\rm S}a}$ belonging to the
$(\overline{3},3)/(3,\overline{3})$ representations of the
SU(3)$_L\times$SU(3)$_R$ flavor symmetry.

Because the operators $S$ and $P$ contain no vector indices, any
correction terms must increase the dimension by two and we have
chosen to neglect such $O(a^2)$ contributions.  Thus, Eqs.~(\ref{eq:S})
and (\ref{eq:P}) are particularly simple.  However, we can also drop
the dimension four, $O(a)$ correction terms to
Eqs.~(\ref{eq:V_c})-(\ref{eq:A_l}).  This can be established by considering
the chiral structure of the Symanzik and conserved/partially conserved current
operators.  Ignoring effects of order $m$, the Symanzik currents
will couple to pairs of quarks which are either left- or right-handed.
Likewise the domain wall conserved/partially conserved current operators
couple to a pair of quarks with the same value of the coordinate $s$
in the fifth dimension.  For $s=0$ these are left-handed fermions
while for $s=L_s-1$ they are right-handed.  As the coordinate $s$
moves into the fifth-dimensional bulk, the amplitude for coupling to
such physical modes decreases until when $s \approx L_s/2$ the amplitude
will be suppressed by two traversals half-way through the fifth dimension
which implies a suppression of order $m_{\rm res} a$.  Of course, the
$s \approx 0$ and $s \approx L_s-1$ terms will dominate.  The character
of the local vector and axial currents is simpler since they contain
quark field strictly limited to $s=0$ and $L_s-1$.   Since the
four, dimension-four improvement terms included in
Eqs.~(\ref{eq:V_c})-(\ref{eq:A_l}) involve pairs of quarks with opposite
handedness, such terms require a complete propagation across the
fifth dimension if they are to couple to the conserved/partially
conserved or local currents.  This is true even for the terms with
general $s$ which appear in the former currents.  Thus, these correction
terms involve an additional power of $m_{\rm res} a$ and are of order
$m_{\rm res} a^2$ and can be neglected in our power counting scheme.

With this simplification, we can demonstrate that to this order the
following relations hold:
\begin{eqnarray}
Z_{\cal V} &=& 1 \label{eq:Z_V_c}\\
Z_V &=& Z_A \label{eq:Z_VA_l}\\
Z_S &=& Z_P \label{eq:Z_SP}.
\end{eqnarray}
Equation (\ref{eq:Z_V_c}) follows easily from the fact that ${\cal V}^a_\mu$
is conserved at finite lattice spacing and has been given the conventional
normalization.  Equations (\ref{eq:Z_VA_l}) and (\ref{eq:Z_SP}) can each be
shown using essentially the same argument which we will now review.

In the massless continuum theory the operators
$\overline q^c \lambda ^a \gamma^\mu(1\pm \gamma^5)q^c$ are independent
involving only right-handed/left-handed degrees of freedom. Here the label $c$ indicates \textit{continuum}. This implies
the vanishing of the Symanzik-improved Green's function:
\begin{equation}
\left\langle (V_\mu^{{\bf S}a}+A_\mu^{{\bf S}a})(x) (V_\nu^{{\bf S}a}-A_\nu^{{\bf S}a})(y)
                \right\rangle = 0.
\label{eq:sym_no_mix}
\end{equation}

This same property is obeyed by the local domain wall currents up to order
$(m_{\rm res} a)^2$ since non-vanishing terms which can contribute to the DWF
version of Eq.~(\ref{eq:sym_no_mix}) must connect both fermion degrees of
freedom between the left and right walls requiring two-traversals of the
fifth dimension and hence are of order $(m_{\rm res}a)^2$~\cite{Christ:2005xh,
Sharpe:2007yd}.  It is then easy to see that these two behaviors
can be consistent through order $m_{\rm res}a$ only if $Z_V=Z_A$ through
order $m_{\rm res}a$.  We need only examine the mixing between
$V_\mu^{{\bf S}a} \pm A_\mu^{{\bf S}a}$ that is generated by $Z_V-Z_A$:
\begin{eqnarray}
\left\langle (V_\mu^{{\rm S}a} +  A_\mu^{{\rm S}a})(x) \cdot
                (V_\nu^{{\rm S}a} -  A_\nu^{{\rm S}a})(y)\right\rangle &&
\label{eq:S_P_no-mix} \\
  && \hskip -2.5in
= \left\langle (Z_V V_\mu^a + Z_A A_\mu^a)(x) \cdot
               (Z_V V_\nu^a - Z_A A_\nu^a)(y)\right\rangle
\nonumber \\
  && \hskip -2.5in
 = \frac{1}{4}\,\Big\langle\,\Big[\, (Z_V+Z_A)(V_\mu^a+A_\mu^a)(x)
      +(Z_V-Z_A)(V_\mu^a -A_\mu^a)(x)\,\Big]
\nonumber \\
  &&\hskip -1.9in
 \cdot \Big[\, (Z_V+Z_A)(V_\nu^a-A_\nu^a)(y)
      +(Z_V-Z_A)(V_\nu^a+A_\nu^a)(y)\,\Big]\,\Big\rangle.
\nonumber
\end{eqnarray}

The product of the left-most operators in the square brackets on the
right-hand side of Eq.~(\ref{eq:S_P_no-mix}) cannot mix at order $m_{\rm res}$
because of their construction from domain wall quark fields as explained above.
Likewise the product of the right-most terms also vanishes.  However, the
two cross terms have non-zero correlators implying that for the entire
expression to be of order $m_{\rm res}^2$, the difference $Z_V-Z_A$ must be
of order $(m_{\rm res} a)^2$, demonstrating the intended result.  A very
similar argument can be constructed which shows that $Z_S = Z_P$ through
order $m_{\rm res} a$.  One must invoke the flavor structure and, for example,
consider correlators between $(S^1-iS^2)(x) + (P^1-iP^2)(x))$ and
$(S^1+iS^2)(y) + (P^1+iP^2)(y))$ which also must vanish in the chiral
limit.  Here $a=1,2$ is a specific choice of the eight octet indices $a=1-8$.

The relations in Eqs.~(\ref{eq:Z_V_c}), (\ref{eq:Z_VA_l}) and (\ref{eq:Z_SP})
were established by considering the domain wall and continuum theories
in a limit in which the physical quark masses could be neglected,
at sufficiently short distances that vacuum chiral symmetry breaking
could be ignored but at sufficiently long distances that the Symanzik
effective theory could be applied.  While this is an excellent regime
in which to establish these theoretical constraints, it is not a
practical one for calculations.  Thus, we will now employ these
relations at low energies where vacuum chiral symmetry breaking is
important in order to provide a practical method to compute $Z_{\cal A}$.

Since at low energies the left- and right-hand sides of Eqs.~(\ref{eq:V_l})
and (\ref{eq:A_l}) must have identical matrix elements, the ratio of
long-distance correlators computed with the Symanzik and local currents must
give identical constants: $Z_V = Z_A$.  Thus, we have established:
\begin{equation}
  \frac{\left\langle V_i^{{\rm S}a}(x) V_i^a(y)\right\rangle}
     {\left\langle V^a_i(x) V_i^a(y)\right\rangle}
= \frac{\left\langle A_0^{{\rm S}a}(x) P^a(y)\right\rangle}
       {\left\langle A^a_0(x) P^a(y)\right\rangle}
\label{eq:Z_v=Z_A}
\end{equation}
where we have introduced the fixed spatial index $i$, the temporal index $0$
and sources $V_i^a(y)$ and $P^a(y)$ that will correspond to those used in
our actual calculation.  Next we can use the long-distance equality
represented by Eqs.~(\ref{eq:V_c}) and (\ref{eq:A_pc}) to write
\begin{eqnarray}
 1          &=&  \frac{\left\langle V_i^{{\rm S}a}(x) V_i^a(y)\right\rangle}
                      {\left\langle {\cal V}^a_i(x) V_i^a(y)\right\rangle}
\label{eq:1} \\
 Z_{\cal A} &=&  \frac{\left\langle A_0^{{\rm S}a}(x) P^a(y)\right\rangle}
                      {\left\langle {\cal A}_0^a  (x) P^a(y)\right\rangle}.
\label{eq:Z_A_pc}
\end{eqnarray}
Then we can combine Eqs.~(\ref{eq:Z_v=Z_A}), (\ref{eq:1}) and (\ref{eq:Z_A_pc})
to yield an equation for $Z_{\cal A}$ which does not involve the Symanzik
currents:
\begin{equation}
Z_{\cal A} =  \frac{\left\langle A_0^a(x) P^a(y)\right\rangle}
                   {\left\langle {\cal A}_0^a(x) P^a(y)\right\rangle}
         \cdot\frac{\left\langle {\cal V}^a_i(x) V_i^a(y)\right\rangle}
                   {\left\langle V^a_i(x) V_i^a(y)\right\rangle}\,,
\label{eq:result}
\end{equation}
which determines $Z_{\cal A}$ in terms of four correlators which we
have evaluated directly in our lattice calculation.

In order to relate the discussion of the Symanzik improved operators
given in Eqs.~(\ref{eq:V_c})-(\ref{eq:P}) with the operators appearing in
Eq.~(\ref{eq:axial_divergence}),
we should recognize that the quantity $Z_{\cal A}$ has been
introduced in two places.  The most important is in the relation between
the Symanzik current and the partially conserved domain wall operator
in Eq.~(\ref{eq:A_pc}).  It is this quantity that is determined in
Eq.~(\ref{eq:result}) and which is needed to give a physical normalization
to the axial current matrix elements determined in our calculation.
However, the quantity $Z_{\cal A}$ also appears in the expression for
$J_{5q}$ given in Eq.~(\ref{eq:axial_divergence}).  For completeness,
we will now demonstrate that these two quantities are in fact the same
up to order $(m_{\rm res} a)^2$.

This is easily done by introducing a flavor-breaking mass term
$\overline{q}M q$ into the DWF action, examining the divergence equations
obeyed by ${\cal V}^a_\mu$ and ${\cal A}^a_\mu$ and using the relation
$Z_S = Z_P$ established above.  With the additional mass term the
conserved/partially conserved vector and axial currents obey the
lattice divergence equations, through $O(m_{\rm res} a)$:
\begin{eqnarray}
\Delta^\mu {\cal V}^a_\mu &=& \overline{q}[\lambda^a,M]q
\label{eq:V_c_div} \\
\Delta^\mu {\cal A}^a_\mu &=& \overline{q}\{\lambda^a,M\}\gamma^5 q
                  + 2 m_{\rm res} \overline{q}\gamma^5 q
                  - (Z_{\cal A}-1)\Delta^\mu {\cal A}^a_\mu.
\end{eqnarray}
Taking the $Z_{\cal A}-1$ term to the left hand side and recognizing
that the scalar and pseudoscalar operators $S^a$ and $P^a$ are
symmetrically normalized $(Z_S = Z_P)$, we can conclude that the
operators ${\cal V}^a_\mu$ and $Z_{\cal A} {\cal A}^a_\mu$ must
be related to the corresponding Symanzik currents by the same factor.
This establishes that our two definitions of $Z_{\cal A}$ are consistent.

We will conclude this analysis with a brief discussion
of the effects of the explicit quark mass, $m_f$, on the operator
product expansion represented by Eq.~(\ref{eq:axial_divergence})
and on the Symanzik-improved operators given in
Eqs.~(\ref{eq:V_c})-(\ref{eq:P}).  Although $m_f$ explicitly
connects the $s=0$ and $s=L_s-1$ walls, it can combine with the
midpoint operator $J_{5q}$ appearing on the left hand side of
Eq.~(\ref{eq:axial_divergence}) to create effects with arbitrary
chiral properties.  Thus, we expect multiplicative corrections
of the form $(1+b_i m_f a)_{1 \le i \le 4}$ to each of the four
terms on the right hand side of Eq.~(\ref{eq:axial_divergence}).
In the case of the left-most term the correction is of order $m_fm_{\textrm{res}} a$ while for the remaining three terms the corrections are of order $m_f m_{\rm res} a^2$ or $m_fm_{\rm res}a^3$, all
beyond the level of accuracy of the current paper.  The
conclusion that $Z_{\cal V}=1$ through order $m_{\rm res} a^2$
(and order $m_f a^2$) prevents the appearance of a factor
$1+b(m_f a)$ multiplying the $Z_{\cal V}$ in Eq.~(\ref{eq:V_c}).
The argument that $Z_A=Z_V$ and $Z_S=S_P$ with corrections of
order $(m_{\rm res} a)^2$ applies equally well to the left-right
mixings created by $m_f$ but again the allowed $m_f m_{\rm res} a^2$
and $(m_f a)^2$ terms are negligible within our present power
counting scheme so Eqs.~(\ref{eq:V_l})-(\ref{eq:P}) need no $O(m_f a)$
corrections.  Lastly, consider adding a factor of the form $(1+b(mf a))$
multiplying the $Z_{\cal A}$ on the right-hand side of
Eq.~\ref{eq:A_pc}. As explained above, a similar correction to $Z_{\cal A}$ appearing in
Eq.~(\ref{eq:axial_divergence}) carries the additional suppression
of one power of $m_{\rm res} a$.  Since the equality derived above
between the $Z_{\cal A}$ factors appearing in the divergence
equation, Eq.~(\ref{eq:axial_divergence}), and the Symanzik-improved
current ${\cal A}_\mu^a$, in Eq.~(\ref{eq:A_pc}), holds at order
$m_f a$ such a $1+b (m_fa)$ factor is not allowed in
Eq.~(\ref{eq:A_pc}).  Thus, no $m_f a$ terms need to be introduced
into the equations presented in this appendix.

\subsection{Computational method and results}

We have evaluated the two factors in Eq.~(\ref{eq:result}) to determine
$Z_{\cal A}$ on both the $32^3 \times 64$, $\beta=2.25$ ($m_l=0.004$,
0.006 and 0.008) and the $24^3\times 64$, $\beta=2.13$ ($m_l=0.005$,
0.01 and 0.02) ensembles.  We used a small subset of these six ensembles
and obtained the results given in Tab.~\ref{tab:numerical_results}.
The results presented for $Z_A/Z_{\cal A}$ duplicate those from the
calculation of $Z_A$ described in Sections III and IV.  In this appendix
we add the factor $Z_{\cal A}$ in the denominator because we are now
determining the deviation of this factor from unity.  We do not simply
use the results presented earlier in the paper because our calculation
of $Z_V/Z_{\cal V}$ has been performed on a subset of the configurations
analyzed earlier and results for $Z_A/Z_{\cal A}$ are needed on this
same subset of configurations if ratios with meaningful jackknife errors
are to be determined.

\begin{figure}
\includegraphics[width=0.47\hsize]{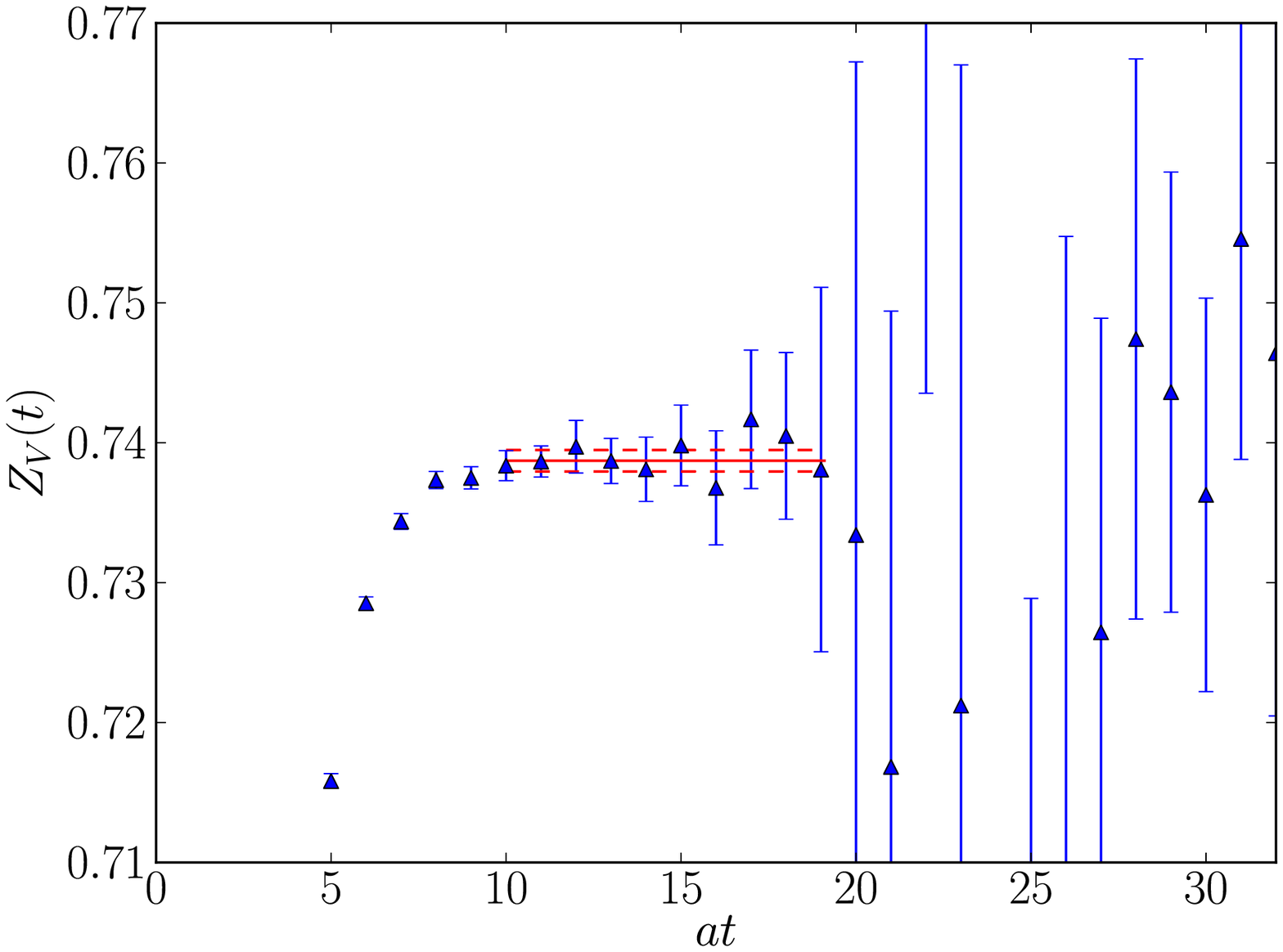}\qquad
\includegraphics[width=0.47\hsize]{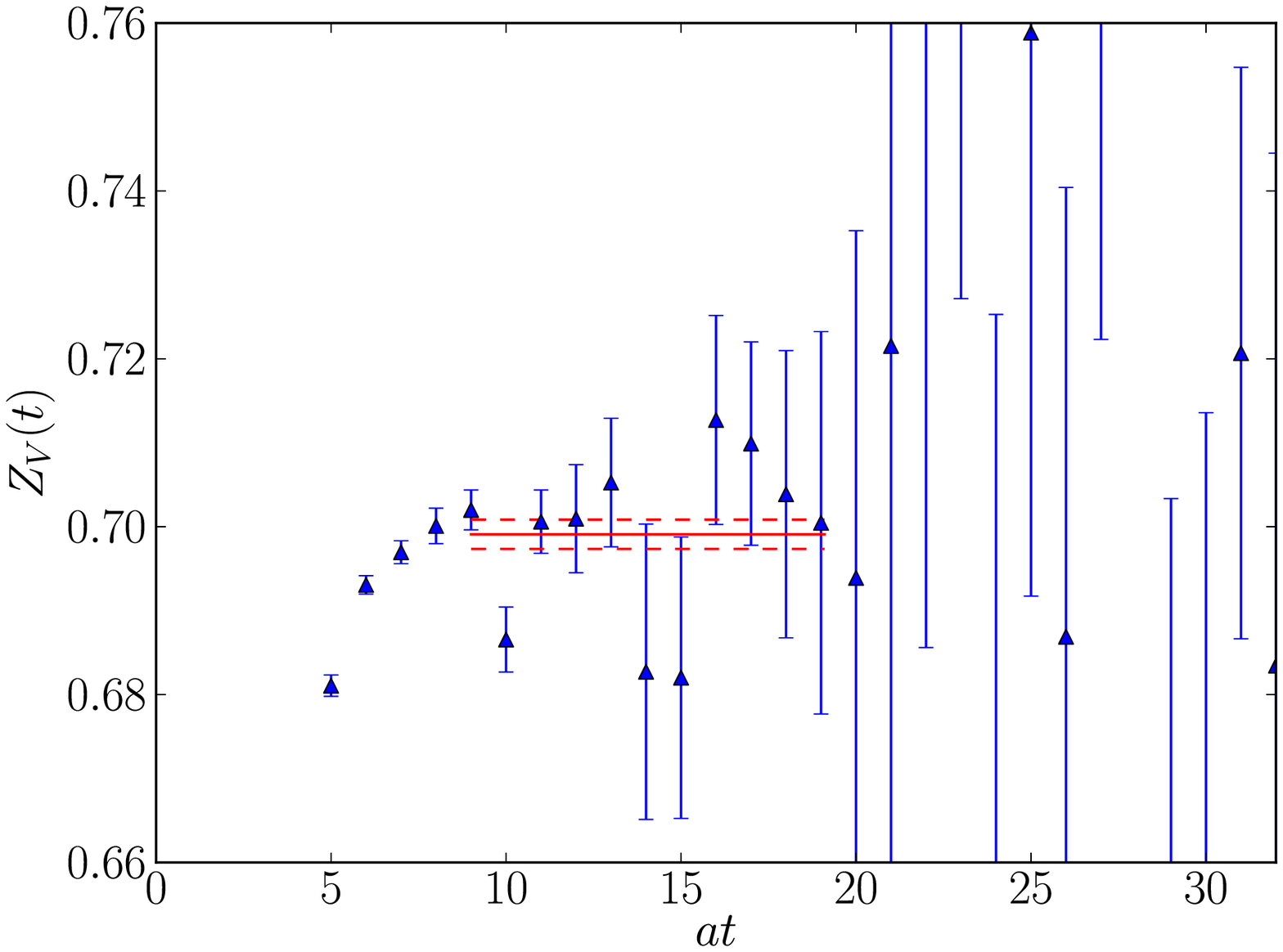} \\
\caption{Plots of the correlator ratio which determines the renormalization
factor $Z_V/Z_{\cal V}$ as a function of time.  The left panel shows results
from the $32^3$, $m_l=0.004$ ensemble while the right panel the result from
the $24^3$, $m_l=0.005$ ensemble.  The horizontal line with error bands in
each panel shows the fitting range and the result obtained in each case.}
\label{fig:ZV_data}
\end{figure}

\begin{figure}
\includegraphics[width=0.47\hsize]{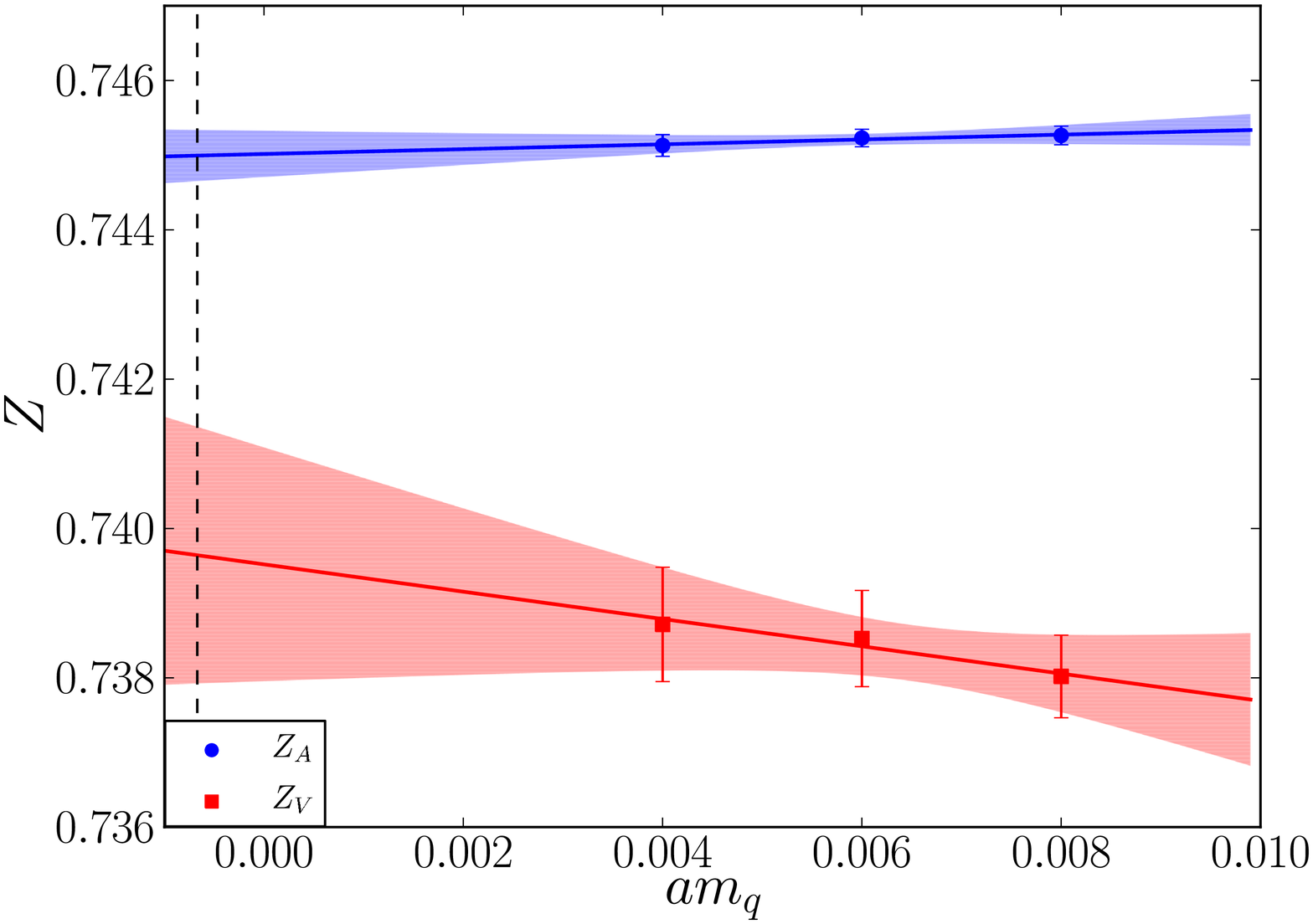}\qquad
\includegraphics[width=0.47\hsize]{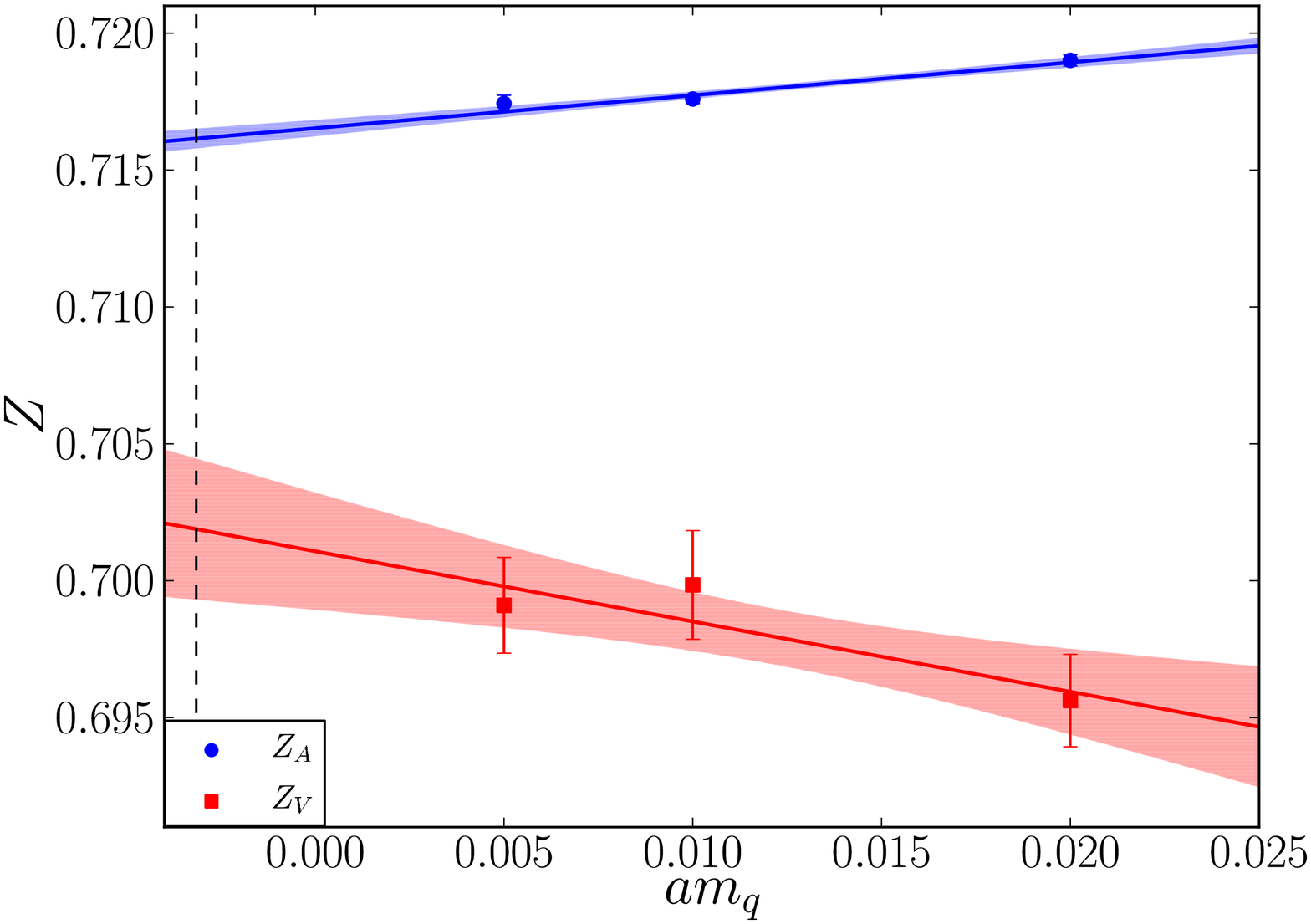} \\
\caption{The quantities $Z_A/Z_{\cal A}$ and $Z_V/Z_{\cal V}$
extrapolated to the chiral limit for the $32^3$ (left panel) and
$24^3$ (right panel) ensembles.}
\label{fig:ZA-ZV_chiral_extrap}
\end{figure}

The ratio $Z_A/Z_{\cal A}$ was computed from the same ratio of
current-pseudoscalar correlators studied in Sections III and IV,
using the method specified in Ref.~\cite{Blum:2000kn}.  Similar
methods are used to compute $Z_V/Z_{\cal V}$ using the ratio of
vector correlators
\begin{equation}
\frac{Z_V}{Z_{\cal V}} = \frac{\sum_{i=1}^3 \sum_{\vec x}
               \left\langle{\cal V}_i^a(\vec x,t) V_i^a(\vec 0,0)\right\rangle}
                              {\sum_{i=1}^3 \sum_{\vec x}
               \left\langle{V}_i^a(\vec x,t) V_i^a(\vec 0,0)\right\rangle},
\label{eq:Z_V_calc}
\end{equation}
an equation expected to be valid for time separations $t$ much larger
than one lattice spacing: $t \gg a$.  Figure\,\ref{fig:ZV_data}
shows the right-hand side of Eq.~(\ref{eq:Z_V_calc}) as a function of time
for the case of the lightest mass for each of the $32^3$ and $24^3$
ensembles.  A constant fit to plateau regions identified by the horizontal
lines was then used to determine the $Z_V/Z_{\cal V}$ on the left-hand side
of this equation. Fig.\,\ref{fig:ZA-ZV_chiral_extrap} displays the chiral extrapolation of the two quantities
$Z_A/Z_{{\cal A}}$ and $Z_V/Z_{{\cal V}}$ on both sets of ensembles.

\begin{table}
\begin{tabular}{ccccccc}
$\beta$& $m_l$ & $Z_A/Z_{\cal A}$
                             & $Z_V/Z_{\cal V}$
                                           & $Z_{\cal V}/Z_{\cal A}$
                                                        & Fit range  & $N_{\rm meas}$ \\
\hline\hline
2.13   & 0.02  & 0.71900(20) & 0.6956(17)  & 1.0336(25) & 9-54/9-17  & 50  \\
2.13   & 0.01  & 0.71759(16) & 0.6998(20)  & 1.0254(29) & 9-54/9-17  & 50  \\
2.13   & 0.005 & 0.71743(30) & 0.6991(17)  & 1.0262(25) & 9-54/10-19 & 105 \\
2.13   & $-m_{\rm res}$
               & 0.71615(36) & 0.7019(26)  & 1.0208(40) &            &     \\
\hline
2.25   & 0.008 & 0.74526(12) & 0.73802(55) & 1.0098(7)  & 9-54/9-20  & 85  \\
2.25   & 0.006 & 0.74523(12) & 0.73853(64) & 1.0090(9)  & 9-54/9-18  & 76  \\
2.25   & 0.004 & 0.74513(15) & 0.73871(77) & 1.0087(10) & 9-54/10-19 & 166 \\
2.25   & $-m_{\rm res}$
               & 0.74499(34) & 0.7396(17)  & 1.0073(23) &            &     \\
\end{tabular}
\caption{Results for the ratios $Z_A/Z_{\cal A}$, $Z_V/Z_{\cal V}$  and
$Z_{\cal V}/Z_{\cal A}$ computed on six ensembles.  The rows with quark
mass $-m_{\rm res}$ contain the chiral extrapolation to the light quark
mass $m_l=-m_{\rm res}$.  The left-hand portion of the fit range gives
that used for the axial current ratio while the right hand portion that
for the vector current.  For the $Z_V/Z_{\cal V}$ calculation the data
at $t$ and $63-t$ were combined for $ 0 \le t < 32$.}
\label{tab:numerical_results}
\end{table}

Two useful results follow from this Appendix.  First the ratio
$Z_{\cal V}/Z_{\cal A}$ differs from unity on our two ensembles
and that difference decreases more rapidly than $a^2$ with
increasing $\beta$.  Thus, we will obtain more accurate results
in our continuum extrapolation from both matrix elements of the
local axial current and our NPR calculations which are normalized using
off-shell Green's functions containing the local vector and
axial currents if we convert the normalization of these local
currents to the usual continuum normalization by using the ratio
$Z_V/Z_{\cal V}$ instead of the ratio $Z_A/Z_{\cal A}$, the quantity
which we have used in previous work for such conversions.  The
values of $Z_V/Z_{\cal V}$ presented in Table~\ref{tab:numerical_results}
are therefore used to normalize the results presented in the
current paper and are the second result obtained in this appendix. Because these ratios were calculated on a smaller subset of configurations than were used for our main results, we have included their statistical fluctuations as independent within our superjackknife, statistical error analysis.  Since these fluctuations are at or below the 0.5\% level, this omission of possible statistical correlations is unimportant.


\newcommand{\LD}{\Bigl\langle\!\!\Bigl\langle}
\newcommand{\RD}{\Bigr\rangle\!\!\Bigr\rangle}


\section{Statistical errors of reweighted quantities}
\label{sec:reweighting_appendix}

In this appendix we discuss the statistical errors that should be expected
when Monte Carlo data is reweighted to obtain results for a gauge or
fermion action that is different from that used to generate the data.
Throughout this discussion we will make the assumption that the reweighting
factors are not correlated with the data.  Of course, if this assumption
were exactly true then the reweighting would not be needed.  However,
the correlation between the data and reweighting factors is often
small in practice and neglecting this correlation may well provide
a reasonably accurate view of the resulting errors.  As we will show,
with this assumption the usual analysis of the statistical errors applies
easily to reweighted data and yields simple, useful formula which we
present here.

Consider a quantity $x$ and the corresponding ordered ensemble of $N$
Monte Carlo configurations with corresponding measured values $\{x_n\}$,
$1 \le n \le N$.  For each of these $N$ configurations we will determine
a reweighting factor $w_n$ so that the final, reweighted quantity of
interest is given by
\begin{equation}
\left< x \right>_N = \frac{\sum_{n=1}^N x_n w_n}{\sum_{n=1}^N w_n}.
\label{eq:reweighting}
\end{equation}
Here the single brackets $\left< \ldots \right>_N$ indicate an average
over a single Monte Carlo ensemble of $N$ samples.  In this appendix we
are interested in how the statistical fluctuations in the quantity
$\left< x \right>_N$ are affected by the operation of reweighting.
We can then express the true value for $x_N$ as
\begin{equation}
\overline{x_N} = \LD \left< x \right>_N \RD
\label{eq:meta_reweighting}
\end{equation}
where the double brackets $\left<\!\left< \ldots \right>\!\right>$ indicate
a ``meta'' average over many equivalent Monte Carlo ensembles.  The
statistical fluctuation present in a particular result $\left< x \right>_N$
can then be characterized by the average fluctuation of $\left< x \right>_N$
about $\overline{x_N}$:
\begin{equation}
\mathrm{Error}(x) =
   \sqrt{\LD(\left< x \right>_N -\overline{x_N})^2 \RD}.
\label{eq:error_def}
\end{equation}

A quantity such as $\left< x_N \right>$, defined in Eq.~(\ref{eq:reweighting})
as a ratio of averages, will be a biased estimator of the physical result
which must be determined in the limit $N\rightarrow\infty$.  Thus, the meta
average $\overline{x_N}=\left<\!\left< \left< x \right>_N \right>\!\right>$
will differ from the true result by terms of order $1/N$.  While these
$1/N$ corrections are not difficult to enumerate and estimate from our data,
these corrections are not the subject of the present appendix and will not
be considered further here.  Instead we will study how the size of the
statistical fluctuations of $\left< x_N \right>$ about $\overline{x_N}$
is affected by the reweighting.  Thus, the quantity Error$(x)$ defined
in Eq.~(\ref{eq:error_def}) describes the average deviation of
$\left< x_N \right>$ from $\overline{x_N}$ not from the
$N \rightarrow \infty$ limit of $\overline{x_N}$.

We will now work out an expression for $\mathrm{Error}(x)$ in the
case that nearby measurements $x_n$ and $x_{n+l}$ in a single Markov
chain (or reweighting factors $w_n$ and $w_{n+l}$) are correlated
but with the assumption that $x_n$ and $w_{n+l}$ are not:
\begin{eqnarray}
\LD\Bigl(\left<x\right>_N - \overline{x_N}\Bigr)^2\RD &=&
  \left<\!\!\left<\left(\frac{\sum_{n=1}^N x_n w_n}{\sum_n w_n}
                                          - \overline{x_N}\right)
              \left(\frac{\sum_{n'=1}^N x_{n'} w_{n'}}{\sum_{n'} w_{n'}}
                                          - \overline{x_N}\right)\right>\!\!\right> \\
  && \hskip -1.4in =
 \left<\!\!\left<\frac{\left(\sum_{n=1}^N x_n w_n
                  - \overline{x_N}\sum_{n=1}^N w_n\right)
                        \left(\sum_{n'=1}^N x_{n'} w_{n'}
                  - \overline{x_N}\sum_{n'=1}^N w_{n'}\right)}
                  {\left(\sum_{n=1}^N w_n \right)\left(\sum_{n'=1}^N w_{n'}\right)}\right>\!\!\right> \\
  && \hskip -1.4in =
 \left<\!\!\left<\frac{\left(\sum_{n=1}^N (x_n - \overline{x_N}) w_n)\right)
                   \left(\sum_{n'=1}^N (x_{n'}-\overline{x_N})w_{n'}\right)}
                  {\left(\sum_{n=1}^N w_n \right)\left(\sum_{n'=1}^N w_{n'}\right)}\right>\!\!\right> \\
  && \hskip -1.4in =
 \frac{\sum_{n=1}^N \sum_{l=1-n}^{N-n} \left\{ \LD( x_n - \overline{x_N})
                   (x_{n+l}-\overline{x_N})\RD \LD w_n w_{n+l}\RD\right\}}
                  {\LD\sum_{n=1}^N w_n\RD^2},
\label{eq:error_4}
\end{eqnarray}
where in the last line we have used our assumption of the lack of
correlation between the $x_n$ and $w_n$ to write the average of
their product as the product of their separate averages.  We have
also assumed that our sample size $N$ is sufficiently large that
correlated fluctuations of the averages in the numerator and denominator
will be sufficiently small that the average of the original ratios and
products can be replaced by the corresponding ratios and products of
the individual averages.

This result can be cast in a simple form if we define the three
averages:
\begin{eqnarray}
\delta x^2 &=& \LD (x_n -\overline{x_N})^2 \RD \\
\overline{w}  &=& \LD w_n \RD \\
\overline{w^2} &=& \LD w_n^2 \RD
\end{eqnarray}
(where $\delta x^2$ is the usual width of the distribution of
the measured quantity $x_n$) and the two autocorrelation
functions:
\begin{eqnarray}
C(l) &=& \frac{\LD ( x_n - \overline{x_N})(x_{n+l}-\overline{x_N}) \RD}
              {\delta x^2} \\
W(l) &=& \frac{\LD w_n  w_{n+l} \RD}{\overline{w^2}},
\end{eqnarray}
defined so that $C(0)=W(0)=1$.  Making the conventional assumption that
the range of $l$ over which the correlation function $C(l)$ is non-zero
is small compared to the sample size $N$ and using the quantities defined
above, we can rewrite Eq.~(\ref{eq:error_4}) as
\begin{eqnarray}
\LD\Bigl(\left<x\right>_N - \overline{x_N}\Bigr)^2\RD &=&
 \frac{\delta x^2 \sum_{l=-L_{\rm max}}^{L_{\rm max}} C(l)W(l) \overline{w^2}}
                  {N \left(\overline{w}\right)^2} \\
  &=& \delta x^2 \frac{\tau_{\rm corr}}{N}
      \frac{\overline{w^2}}{\left(\overline{w}\right)^2}
\label{eq:next-to-final}
\end{eqnarray}
where the autocorrelation time $\tau_{\rm corr}$ is defined as
\begin{equation}
\tau_{\rm corr} = \sum_{l=-L_{\rm max}}^{L_{\rm max}} C(l)W(l).
\label{eq:tau_corr}
\end{equation}
The limit $L_{\rm max}$ is chosen to be larger than the region within
which $C(l)$ is non-zero and has been introduced as a reminder that
when working with a single finite sample, one must take care to evaluate
the limit of large $N$ before the limit of large $L_{\rm max}$.
Finally, Eq.~(\ref{eq:next-to-final}) can be written in the conventional form
\begin{equation}
\mathrm{Error}(x) = \sqrt{\frac{\delta x^2}{N_{\rm eff}}}
\label{eq:result_1}
\end{equation}
where the effective number of configurations $N_{\rm eff}$ is given by:
\begin{equation}
N_{\rm eff} = \frac{N}{\tau_{\rm corr}} \frac{\;\overline{w}^2\;}{\overline{w^2}}.
\label{eq:result_2}
\end{equation}

This result makes precise a number of aspects of reweighting that may
be useful to understand.  In the case that there are no autocorrelations
so $\tau_{\rm corr} =1$, the ratio $\overline{w}^2/\overline{w^2}$ expresses
the degree to which the reweighting process selectively samples the original
data and degrades the initial statistics.  The general inequality
$\overline{w}^2/\overline{w^2} \le 1$ (a consequence of the Schwartz inequality)
is saturated only in the case that the reweighting factors $w_n$ do not vary
with $n$.  In the extreme case that a single sample $w_n$ dominates the
averages then $\overline{w}^2/\overline{w^2} = 1/N$ and $N_{\rm eff}=1$.
Thus, in the case of uncorrelated data (which is the case for most of the
results presented here) we should expect the statistical fluctuations to
grow as the degree of reweighting increases by the factor
$\overline{w^2}/\overline{w}^2$.

Including autocorrelations makes the effects of reweighting on the size
of the statistical fluctuations less certain because the behavior of
the factors $1/\tau_{\rm corr}$ and $\overline{w}^2/\overline{w^2}$ in
Eq.~(\ref{eq:result_2}) become entangled.  In the limit in which the
autocorrelation time associated with the measured quantity $x_n$ alone,
\begin{equation}
\tau_x = \sum_{l=-L_{\rm max}}^{L_{\rm max}} C(l),
\end{equation}
becomes much larger than that of the reweighting factor $w_n$, then
the majority of the sum in Eq.~(\ref{eq:tau_corr}) contributing to
$\tau_{\rm corr}$ will come from values of $l$ where
$\LD w_n w_{n+l} \RD \approx \LD w \RD^2$ so that
\begin{equation}
\tau_{\rm corr} \approx \tau_x \frac{\;\overline{w}^2\;}{\overline{w^2}}.
\end{equation}
In this case the error given by Eq.~(\ref{eq:result_1}) reduces to the
standard expression $\sqrt{\delta x^2 \tau_x/N}$ that holds if no
reweighting is performed!  Of course, this is easy to understand.  When
such long autocorrelation times are involved, the average over the
autocorrelation time is providing an average over the reweighting factors
$w_n$ which is sufficiently precise that the error-enhancing fluctuations
in the reweighting factors are averaged away.  Given the large size of
the fluctuations between the reweighting factors and the relatively short
autocorrelation times seen in our data, it is unlikely that this averaging
would be seen in the results presented here.

A second type of behavior for $\tau_{\rm corr}$ occurs if the $w_n$ are
relatively uncorrelated and $\overline{w^2} \gg \overline{w}^2$ so that
only the $l=0$ term contributes to the sum in Eq.~(\ref{eq:tau_corr}) giving
$\tau_{\rm corr}=1$.  In this case reweighting has removed the effects of autocorrelation but increased the statistical fluctuations by the factor
$\overline{w^2}/\overline{w}^2$ which was assumed to be large.  Here the
fluctuation-enhancing effects of autocorrelations and reweighting are not
compounded.

\bibliography{paper}

\end{document}